\documentclass[preprint, 12pt]{elsarticle}
\usepackage[T1]{fontenc}    %
\usepackage[margin=1in]{geometry} %
\date{}

\usepackage{comment}
\usepackage{ulem}

\usepackage{fancyhdr}

\usepackage{enumerate}

\usepackage{amsmath,amsthm,amsfonts,amssymb,amscd}
\usepackage{upgreek} %
\usepackage{mathrsfs} %
\usepackage{mathtools} %
\usepackage{derivative}

\usepackage{siunitx}

\usepackage{thmtools,thm-restate}

\usepackage{xcolor}

\usepackage{float}
\usepackage{placeins} %

\usepackage{graphicx}
\usepackage{tikz,tikz-cd}
\usetikzlibrary{arrows, shapes}
\usetikzlibrary{mindmap}
\tikzcdset{scale cd/.style={every label/.append style={scale=#1}, cells={nodes={scale=#1}}}}
\tikzcdset{trapezium stretches=true}
\tikzcdset{arrow style=tikz, diagrams={>=stealth'}}

\usepackage{booktabs}
\usepackage{multirow}
\usepackage{makecell}

\usepackage{caption}
\usepackage{subcaption}

\usepackage[linesnumbered,ruled]{algorithm2e}
\usepackage{listings}

\usepackage[colorlinks=true, citecolor=blue, filecolor=black, linkcolor=blue, urlcolor=gray]{hyperref}

\usepackage{natbib}

\usepackage{lastpage}

\usepackage[capitalise,nameinlink,noabbrev]{cleveref}
\crefname{section}{Sec.}{Sec.}
\Crefname{section}{Section}{Sections}
\crefname{figure}{Fig.}{Fig.}
\Crefname{figure}{Figure}{Figures}
\crefname{table}{Table}{Tables}
\Crefname{table}{Table}{Tables}
\crefname{equation}{Eq.}{Eq.}
\Crefname{equation}{Equation}{Equations}
\crefname{algocf}{Alg.}{Alg.}
\Crefname{algocf}{Algorithm}{Algorithms}
\crefname{appendix}{Appendix}{Appendices}
\Crefname{appendix}{Appendix}{Appendices}

\DeclareMathOperator*{\argmin}{arg\,min}

\newcommand{\nospaceleft}{\mathopen{}\mathclose\bgroup\left}
\newcommand{\nospaceright}{\aftergroup\egroup\right}

\newcommand{\pp}[2]{\frac{\partial #1}{\partial #2}}
\newcommand{\ppp}[2]{\frac{\partial^2 #1}{\partial #2^2}}

\newcommand{\grad}[1]{\nabla #1}
\newcommand{\norm}[2]{\left\|\, #1 \,\right\|_{#2}}

\newcommand{\Ndof}{n_{\mathrm{dof}}}

\newcommand{\bszero}{\boldsymbol{0}}
\newcommand{\bsb}{\boldsymbol{b}}
\newcommand{\bsc}{\boldsymbol{c}}
\newcommand{\bse}{\boldsymbol{e}}
\newcommand{\bsn}{\boldsymbol{n}}
\newcommand{\bsp}{\boldsymbol{p}}
\newcommand{\bsq}{\boldsymbol{q}}
\newcommand{\bsr}{\boldsymbol{r}}
\newcommand{\bsu}{\boldsymbol{u}}
\newcommand{\bsx}{\boldsymbol{x}}
\newcommand{\bsy}{\boldsymbol{y}}
\newcommand{\bsA}{\boldsymbol{A}}
\newcommand{\bsB}{\boldsymbol{B}}
\newcommand{\bsC}{\boldsymbol{C}}
\newcommand{\bsI}{\boldsymbol{I}}
\newcommand{\bsJ}{\boldsymbol{J}}
\newcommand{\bsK}{\boldsymbol{K}}
\newcommand{\bsM}{\boldsymbol{M}}
\newcommand{\bsW}{\boldsymbol{W}}
\newcommand{\bsX}{\boldsymbol{X}}
\newcommand{\bsepsilon}{\boldsymbol{\epsilon}}
\newcommand{\bstheta}{\boldsymbol{\theta}}
\newcommand{\bsmu}{\boldsymbol{\mu}}
\newcommand{\bsphi}{\boldsymbol{\phi}}
\newcommand{\bschi}{\boldsymbol{\chi}}
\newcommand{\bspsi}{\boldsymbol{\psi}}
\newcommand{\bsXi}{\boldsymbol{\Xi}}
\newcommand{\bsSigma}{\boldsymbol{\Sigma}}
\newcommand{\bsPhi}{\boldsymbol{\Phi}}

\newcommand{\RR}{\mathbb{R}}

\newcommand{\CalC}{\mathcal{C}}
\newcommand{\CalD}{\mathcal{D}}
\newcommand{\CalL}{\mathcal{L}}
\newcommand{\CalM}{\mathcal{M}}
\newcommand{\CalN}{\mathcal{N}}
\newcommand{\CalO}{\mathcal{O}}
\newcommand{\CalP}{\mathcal{P}}
\newcommand{\CalR}{\mathcal{R}}
\newcommand{\CalS}{\mathcal{S}}
\newcommand{\CalT}{\mathcal{T}}
\newcommand{\CalU}{\mathcal{U}}
\newcommand{\CalV}{\mathcal{V}}

\newcommand{\hu}{\hat{u}}
\newcommand{\hpsi}{\widehat{\psi}}

\newcommand{\hbsc}{\boldsymbol{\hat{c}}}
\newcommand{\hbsp}{\boldsymbol{\hat{p}}}
\newcommand{\hbsu}{\boldsymbol{\hat{u}}}
\newcommand{\hbstheta}{\boldsymbol{\widehat{\theta}}}

\newcommand{\barbsu}{\boldsymbol{\bar{u}}}

\newcommand{\dotu}{\dot{u}}

\begin{document}

\normalem

\begin{frontmatter}
\title{Bayesian Variational System Identification with Weak-Form Residual Likelihoods}

\author[UM_ME,USC]{Chengyang Huang}
\author[Auburn]{Siddhartha Srivastava}
\author[USC]{Krishna Garikipati}
\author[UM_ME]{Xun Huan}

\affiliation[UM_ME]{
    organization={Department of Mechanical Engineering, University of Michigan},
    city={Ann Arbor},
    postcode={MI 48109},
    country={United States}
}

\affiliation[Auburn]{
    organization={Department of Aerospace Engineering, Auburn University},
    city={Auburn},
    postcode={AL 36849},
    country={United States}
}

\affiliation[USC]{
    organization={Department of Aerospace and Mechanical Engineering, University of Southern California},
    city={Los Angeles},
    postcode={CA 90089},
    country={United States}
}

\begin{abstract}
We consider system identification for discovering parameterized operators in governing partial differential equations (PDEs) from noisy spatiotemporal data. Building on variational system identification (VSI), which identifies PDEs through Galerkin weak-form residuals, we develop a Bayesian VSI (B-VSI) framework for operator selection, parameter estimation, and uncertainty quantification. The central idea is to define the likelihood directly in weak-form residual space by propagating observation uncertainty through the weak-form residual map. The resulting likelihood captures heteroscedastic and correlated residual errors while avoiding repeated forward PDE solves during inference. For efficient computation, we use lagged-covariance updates that yield generalized least-squares estimates and conjugate posterior approximations when applicable, together with gradient-based and particle-based methods for more general priors and posterior structures. Model-form uncertainty is handled through sequential operator elimination guided by a residual-space Bayesian information criterion. We demonstrate the framework on state-linear and nonlinear PDEs, including the Fokker--Planck equation and a two-field Cahn--Hilliard equation. The results show that B-VSI accurately recovers active operators and coefficients from noisy data, improves robustness relative to classical VSI, and provides posterior uncertainty estimates for coefficients and derived physical quantities.
\end{abstract}

\begin{keyword}
    uncertainty quantification \sep
    equation discovery \sep
    model selection \sep
    finite element method \sep
    Fokker--Planck equation \sep
    Cahn--Hilliard equation
    \vspace{1em}
    \MSC[2020] 65M32 \sep 62F15 \sep 93B30 \sep 65M60 \sep 35R30
\end{keyword}

\begin{highlights}
    \item Emphasizes the advantages of the weak form for operator learning with Bayesian UQ
    \item Novel residual-space likelihood avoids repeated forward solves of PDEs
    \item Information-theoretic model selection
    \item Compute-efficient inference with lagged-covariance and particle-based methods
    \item Works for linear and  nonlinear PDEs; e.g. Fokker–Planck and Cahn–Hilliard equations

\end{highlights}

\end{frontmatter}

\section{Introduction}
\label{sec:introduction}

Understanding and predicting the behavior of complex physical systems fundamentally relies on identifying the governing equations---often in the form of partial differential equations (PDEs)---that encode the underlying physical principles~\cite{neto2012introduction}.
The task of identifying both the structure and parameters of these governing equations directly from observational data is commonly referred to as system identification (SI)~\cite{neto2012introduction,Brunton2022PIML}.

A broad class of SI approaches formulates the problem as PDE-constrained optimization~\cite{DelosReyes2015Numerical}.
In this setting, parameters are estimated by minimizing the discrepancy between observational data and simulated system states subject to the governing PDE.
The dominant computational cost typically arises from repeated forward PDE solves, for example by finite element or finite volume methods, because model-data comparisons are performed directly in the solution space.
When gradient-based inference is required, additional sensitivity or adjoint calculations may further increase the computational burden.

An alternative class of methods identifies governing equations through residuals evaluated directly on observational data.
Sparse Identification of Nonlinear Dynamical Systems (SINDy) and its variants~\cite{Brunton2016SINDy,Rudy2017SINDyPDE} bypass full forward solves by applying candidate differential operators to the observed data and solving the resulting sparse regression problem.
From this perspective, PDE-constrained optimization assesses candidate models through the quality of their simulated solutions, whereas SINDy-type methods assess how well the observed data satisfy the proposed governing equations themselves.

Building on this residual-based viewpoint, Variational System Identification (VSI)\footnote{Throughout this paper, \emph{Variational System Identification} denotes the weak-form SI framework and is distinct from variational inference used for posterior approximation.}~\cite{Wang2019Variational,Wang2020Perspective,Wang2021Variational} and weak-form SINDy~\cite{Messenger2021WeakSINDy} employ weak formulations of the governing PDEs rather than operating in strong form.
The weak form shifts differentiation onto test functions, improving robustness to noisy data while retaining the sparsity-promoting structure central to SINDy-like methods.
Residual-based SI methods are especially attractive for PDE discovery because they identify governing terms directly from data without repeatedly simulating candidate models.
VSI and weak-form SINDy retain this computational advantage while improving robustness to noisy data through the weak formulation.

Although weak-form residual methods provide an efficient route to PDE discovery, their standard formulations are typically deterministic.
Measurement noise, limited data, and model-form ambiguity can significantly affect the selected governing equation, the estimated coefficients, and derived physical quantities.
Bayesian uncertainty quantification (UQ) provides a systematic probabilistic framework for representing and updating these uncertainties~\cite{Smith2013UQ,gelman2013bayesian}.
In this framework, prior distributions encode initial beliefs about model structure and parameters, which are updated to posterior distributions through Bayes' rule as data become available.
The likelihood function plays a central role in this update by quantifying the probability of the observed data under candidate models and parameters.

Recent work has incorporated UQ into several SI frameworks.
Ensemble-SINDy adopts a frequentist approach, using bagging to generate model ensembles and quantify uncertainty through empirical coefficient distributions~\cite{Fasel2022EnsembleSINDy}.
UQ-SINDy uses sparsity-promoting priors and Markov chain Monte Carlo (MCMC) sampling to characterize parametric and structural uncertainty in ordinary differential equation models~\cite{Hirsh2022UQSINDy}.
Bayesian Dynamical System Identification formulates sparse regression for dynamical systems as Bayesian maximum a posteriori estimation and uses joint maximum a posteriori and variational Bayesian approximations to quantify model uncertainty and rank candidate models~\cite{Niven2024BDSI}.
Bayesian-SINDy incorporates likelihood modeling through noise propagation, yielding a coefficient-dependent residual covariance while approximating the propagated covariance by its diagonal for computational tractability~\cite{Fung2025BayesianSINDy}. B-VSI similarly propagates observation uncertainty to construct a coefficient-dependent residual likelihood, but formulates it for Galerkin weak-form PDE residuals while retaining the induced spatial and temporal residual covariance.
Bayesian Identification of Nonlinear Dynamics jointly infers governing-equation terms and their coefficients through reversible-jump MCMC over a variable-dimensional posterior~\cite{Champneys2025BINDy}. In contrast, B-VSI separates operator selection from coefficient inference, with posterior inference performed conditional on the selected model.
Beyond sparse equation discovery, Bayesian neural PDE methods have been developed for forward, inverse, and operator-learning problems.
These approaches provide complementary tools for uncertainty-aware PDE inference, but they rely on learned state or operator representations rather than the weak-form residual regression structure used by VSI.

These developments still leave a methodological gap between weak-form residual-based SI and Bayesian UQ. Classical VSI and weak-form SINDy preserve computational efficiency by evaluating candidate equations directly on observed data, but their residual-minimization formulations do not specify how measurement noise propagates through the weak form. Consequently, they do not directly provide posterior distributions over PDE coefficients or derived physical quantities, and their operator-selection procedures are not derived from an explicit residual-space likelihood. Conversely, many Bayesian SI and inverse-problem formulations define likelihoods through discrepancies between simulated and observed states, which typically requires repeated forward PDE solves and, for gradient-based methods, sensitivity or adjoint calculations. A Bayesian extension of VSI therefore requires more than placing priors on regression coefficients. It requires a likelihood defined directly in weak-form residual space, induced by the observational noise model, and tractable enough to preserve the computational advantages of VSI.

Constructing such a residual-space likelihood is nontrivial because weak-form residuals are not generally independent and identically distributed (i.i.d.), even when the observational noise is i.i.d.
In a Galerkin weak formulation, each residual entry depends on weighted combinations of neighboring nodal values and on the candidate differential operators.
As a result, measurement noise propagates through the weak-form residual map to produce residual uncertainties that are generally heteroscedastic and correlated.
Accounting for this covariance structure yields a principled residual-space likelihood, replacing the implicit isotropic-error assumption underlying unweighted least-squares residual minimization.

In this work, we address this gap by developing Bayesian Variational System Identification (B-VSI), a Bayesian extension of VSI based on weak-form PDE residual likelihoods.
For a candidate model form, the proposed framework propagates observational uncertainty through the VSI residual map and constructs an explicit likelihood in residual space.
This enables operator selection, parameter uncertainty quantification, and posterior predictive uncertainty without requiring repeated forward PDE solves during inference.

The primary contributions of this work are as follows.
\begin{itemize}
    \item We derive a weak-form residual likelihood by propagating observational uncertainty through the VSI residual map. This produces an explicit residual-space probabilistic model and avoids the repeated forward PDE solves required by observation-space likelihoods.

    \item We characterize the induced residual covariance and show that it captures heteroscedastic and correlated uncertainty arising from the weak-form discretization.

    \item We use the residual-space likelihood for model comparison and sequential operator selection, enabling uncertainty-aware identification of active PDE operators and their coefficients from noisy spatiotemporal data.

    \item We develop computational strategies for posterior inference by adapting and integrating established statistical inference tools, including an expectation-maximization (EM)-style lagged-covariance update for approximately conjugate settings, gradient-based optimization for more general priors and likelihoods, and a hybrid strategy for non-Gaussian posterior approximation.

    \item We demonstrate the framework on a linear Fokker--Planck (FP) problem and a nonlinear two-field Cahn--Hilliard (CH) problem, including posterior predictive uncertainty in physically meaningful quantities such as the driving potential function and free-energy derivatives that define the total chemical potential.
\end{itemize}

The remainder of the paper is organized as follows.
\Cref{sec:preliminaries} reviews the Galerkin finite element method and the classical VSI framework.
\Cref{sec:B-VSI} presents the proposed B-VSI formulation, including residual-likelihood construction, posterior inference, model selection, and posterior predictive uncertainty.
\Cref{sec:examples} demonstrates the method on both linear and nonlinear PDEs, specifically the FP and CH equations.
\Cref{sec:discussion} compares different likelihood formulations, analyzes computational complexity, and offers a Bayesian interpretation of classical VSI.
\Cref{sec:conclusion} concludes with a summary of the main findings and contributions.

\section{Preliminaries}
\label{sec:preliminaries}

We consider a class of time-dependent PDEs that are linear in an unknown coefficient vector but may be nonlinear in the state:
\begin{align}
    \label{eq:strong_form_pde}
    \pp{u(\bsx,t)}{t} - \CalM
    \big(
        u(\bsx,t)
    \big)^{\top}\bstheta = 0,
\end{align}
where $u(\bsx,t):\Omega \times \CalT \to \RR$ is a real-valued state variable defined on the spatiotemporal domain $\Omega \times \CalT$.
The vector $\CalM(u) = [F_1(u),\ldots,F_{d_{\bstheta}}(u)]^{\top}$ denotes the active model form, whose entries are algebraic or differential operators acting on $u$.
The vector $\bstheta = [\theta_1,\ldots,\theta_{d_{\bstheta}}]^{\top}$ contains the corresponding coefficients.
For example, the diffusion equation with a constant source term,
\begin{align}
    \label{eq:diffusion_example}
    \pp{u}{t} - \beta^{-1}\grad^2 u - \gamma = 0,
\end{align}
corresponds to $\CalM(u) = [\grad^2 u, 1]^{\top}$ and $\bstheta = [\beta^{-1},\gamma]^{\top}$.
Throughout this work, $u$ denotes the physical quantity of interest described by the governing PDE.
In the examples in \cref{sec:examples}, $u$ may represent the probability density function ($u \equiv p$) in the FP equation or a concentration field ($u \equiv c_i$) in the CH equation.
Although the notation in this section is written for a scalar state, the same construction extends to multi-field systems by stacking the residual equations for each state variable.

\subsection{Galerkin finite element method}
\label{sec:FEM}

For completeness, we present our treatment of finite element methods. The Galerkin weak form of \cref{eq:strong_form_pde} is obtained by multiplying the strong form by a test function $v(\bsx):\Omega \to \RR$ and integrating over the spatial domain $\Omega$~\cite{Brenner2008FEM}:
\begin{align}
    \label{eq:weak_form_pde}
    \int_{\Omega} v(\bsx)
    \left(
        \pp{u(\bsx,t)}{t} - \CalM
            \big(
                u(\bsx,t)
            \big)^{\top}\bstheta
    \right)
    \mathrm{d}\bsx = 0.
\end{align}

Let $\Omega^h = \bigcup_{e=1}^{n_e}\Omega_e$ denote a finite element mesh of $\Omega$, where the interiors of distinct elements are disjoint.
The mesh defines a finite-dimensional function space
\begin{align}
    \CalV^h = \mathrm{span}\{N_i\}_{i=1}^{\Ndof},
\end{align}
where $\{N_i\}_{i=1}^{\Ndof}$ are finite element shape functions and $\Ndof$ is the total number of degrees of freedom.
For standard nodal Lagrange elements, the degrees of freedom are associated with nodal locations $\{\bsx_i\}_{i=1}^{\Ndof}$.

The finite element approximations of the state and the test function are written as
\begin{subequations}
    \label{eq:approximate_functions}
    \begin{align}
        \label{eq:approximate_solution}
        u^h(\bsx,t) &= \sum_{i=1}^{\Ndof} u_i(t) N_i(\bsx), \\
        v^h(\bsx) &= \sum_{i=1}^{\Ndof} v_i N_i(\bsx),
    \end{align}
\end{subequations}
where $u_i(t)$ are the time-dependent degrees of freedom of the approximate solution and $v_i$ are arbitrary test-function coefficients.
The shape functions $N_i$ are typically piecewise polynomial functions with local support on elements adjacent to the corresponding degree of freedom.

The trial and test spaces are chosen according to the differential order of the PDE and the prescribed boundary conditions.
For second-order operators after integration by parts, an $H^1(\Omega)$-conforming space is typically sufficient.
Higher-order operators may require additional regularity or a discontinuous Galerkin treatment, as in the CH example in \cref{sec:examples}.

Substituting \cref{eq:approximate_functions} into \cref{eq:weak_form_pde} gives
\begin{align}
    \label{eq:discrete_weak_form}
    \int_{\Omega}
    \sum_{i=1}^{\Ndof} v_i N_i(\bsx)
    \left(
        \pp{u^h(\bsx,t)}{t}
        - \CalM \big(u^h(\bsx,t)\big)^{\top}\bstheta
    \right)
    \mathrm{d}\bsx = 0.
\end{align}
After integration by parts where needed, application of boundary conditions, and use of the arbitrariness of the coefficients $v_i$, the weak form yields one residual equation per degree of freedom:
\begin{align}
    \label{eq:residual}
    r_i(t;u^h,\bstheta)
    =
    \int_{\Omega} N_i \pp{u^h}{t}\,\mathrm{d}\bsx
    -
    \sum_{j=1}^{d_{\bstheta}}
    \theta_j \mathscr{F}_j(N_i,u^h),
    \quad i=1,\ldots,\Ndof.
\end{align}
Here, $\mathscr{F}_j(N_i,u^h)$ denotes the finite-dimensional weak-form contribution of the operator $F_j(u)$ when tested against $N_i$, including any boundary terms that arise from integration by parts.
For a forward simulation problem, an appropriate time discretization is added, and the Galerkin approximation is obtained by solving the residual equations.

\subsection{Variational system identification}
\label{sec:VSI}

SI, formulated as an inverse problem, seeks to learn the PDE model form and its parameters from observational data.
VSI identifies the governing equation in weak form by selecting a subset of active operators $\CalM$ from a candidate dictionary $\bschi$, evaluating them against a spatiotemporal dataset $\CalD$, and estimating their coefficients $\bstheta$~\cite{Wang2019Variational,Wang2021Variational}.

Let
\begin{align}
    \CalD = \left\{\hbsu^{(k)}\right\}_{k=0}^{n_t},
    \quad
    \hbsu^{(k)} =
    \left[\hu_1^{(k)},\ldots,\hu_{\Ndof}^{(k)}\right]^{\top}
    \in \RR^{\Ndof},
\end{align}
denote measurements collected at times $\left\{t^{(k)}\right\}_{k=0}^{n_t}$.
The finite element mesh used to evaluate the weak form is chosen so that its degrees of freedom correspond to the measurement locations.
If the measurement locations do not coincide with the finite element degrees of freedom, the data are first interpolated or projected onto the finite element space.
At time $t^{(k)}$, the measured field is represented as
\begin{align}
    \hat{u}^{h,(k)}(\bsx)
    =
    \sum_{j=1}^{\Ndof} \hu_j^{(k)} N_j(\bsx).
\end{align}

For each time interval $\left[t^{(k-1)},t^{(k)}\right]$, with uniform step size $\Delta t = t^{(k)}-t^{(k-1)}$, the residual equations in \cref{eq:residual} are assembled into
\begin{align}
    \label{eq:residual_vector}
    \bsr^{(k)}
    =
    {\bsXi^{\dotu}}^{(k)}
    -
    \bsXi^{(k)}\bstheta,
    \quad k=1,\ldots,n_t.
\end{align}
The residual vector satisfies $\bsr^{(k)}\in\mathbb{R}^{n_{\mathrm{eq}}}$, where $n_{\mathrm{eq}}$ denotes the number of residual equations at each time step. For the scalar state considered above, $n_{\mathrm{eq}}=\Ndof$; for a multi-field system, $n_{\mathrm{eq}}$ is the combined dimension of the residual blocks across all fields.
The vector ${\bsXi^{\dotu}}^{(k)} \in \RR^{\Ndof}$ is the weak-form representation of the time-derivative term.
Using a backward-difference approximation,
\begin{align}
    \label{eq:time_derivative_vector}
    \left({\bsXi^{\dotu}}^{(k)}\right)_i
    =
    \int_{\Omega}
    N_i
    \sum_{j=1}^{\Ndof}
    \frac{\hu_j^{(k)}-\hu_j^{(k-1)}}{\Delta t}
    N_j
    \,\mathrm{d}\bsx,
    \quad i=1,\ldots,\Ndof.
\end{align}
The matrix $\bsXi^{(k)} \in \RR^{\Ndof \times d_{\bstheta}}$ is assembled from the active model $\CalM \subseteq \bschi$, with $d_{\bstheta}=|\CalM|$.
Its entries are
\begin{align}
    \label{eq:basis_matrix_entries}
    \left(\bsXi^{(k)}\right)_{ij}
    =
    \mathscr{F}_j\nospaceleft(N_i,\hat{u}^{h,(k)}\nospaceright),
    \quad
    i=1,\ldots,\Ndof,\quad
    j=1,\ldots,d_{\bstheta}.
\end{align}
For example, for $\CalM(u)=[\grad^2 u,1]^{\top}$ from \cref{eq:diffusion_example}, the basis matrix at time step $k$ is
\begin{align}
    \label{eq:basis_matrix}
    \bsXi^{(k)}
    =
    \begin{bmatrix}
        {\bsXi^{\grad^2 u}}^{(k)}
        &
        {\bsXi^{\mathrm{const}}}^{(k)}
    \end{bmatrix},
\end{align}
where
\begin{align}
    \left({\bsXi^{\grad^2 u}}^{(k)}\right)_i
    &=
    \mathscr{F}_{\grad^2 u}\nospaceleft(N_i,\hat{u}^{h,(k)}\nospaceright),
    \\
    \left({\bsXi^{\mathrm{const}}}^{(k)}\right)_i
    &=
    \int_{\Omega} N_i\,\mathrm{d}\bsx.
\end{align}
Further details on basis construction and the treatment of boundary conditions are provided in \cref{sec:examples} and in~\citep{Wang2019Variational,Wang2021Variational}.

Stacking all time intervals gives
\begin{align}
    \bsy
    =
    \begin{bmatrix}
        {\bsXi^{\dotu}}^{(1)} \\
        {\bsXi^{\dotu}}^{(2)} \\
        \vdots \\
        {\bsXi^{\dotu}}^{(n_t)}
    \end{bmatrix},
    \quad
    \bsX
    =
    \begin{bmatrix}
        \bsXi^{(1)} \\
        \bsXi^{(2)} \\
        \vdots \\
        \bsXi^{(n_t)}
    \end{bmatrix},
    \quad
    \bsr = \bsy - \bsX\bstheta.
    \label{eq:yXdef}
\end{align}
The classical VSI estimate is obtained by minimizing the unweighted residual norm:
\begin{align}
    \label{eq:VSI_objective}
    \bstheta_{\mathrm{VSI}}
    =
    \argmin_{\bstheta}
    \norm{\bsy-\bsX\bstheta}{2}^{2}.
\end{align}
When $\bsX^{\top}\bsX$ is nonsingular, the least-squares solution is
\begin{align}
    \label{eq:VSI_solution}
    \bstheta_{\mathrm{VSI}}
    =
    \left(\bsX^{\top}\bsX\right)^{-1}\bsX^{\top}\bsy.
\end{align}
If $\bsX^{\top}\bsX$ is singular or ill-conditioned, the Moore--Penrose pseudoinverse or a regularized least-squares formulation can be used instead.

For computational efficiency, the normal-equation terms can be assembled without explicitly forming the full stacked matrix $\bsX$:
\begin{align}
    \label{eq:normal_equation_sums}
    \bsX^{\top}\bsX
    &=
    \sum_{k=1}^{n_t}
    {\bsXi^{(k)}}^{\top}\bsXi^{(k)},
    \\
    \bsX^{\top}\bsy
    &=
    \sum_{k=1}^{n_t}
    {\bsXi^{(k)}}^{\top}{\bsXi^{\dotu}}^{(k)}.
\end{align}
To mitigate overfitting, $\ell_1$- or $\ell_2$-norm regularization may be added to \cref{eq:VSI_objective}.
Operator selection is commonly performed using a statistical criterion, such as the F-statistic, to assess the significance of candidate operators and construct a parsimonious governing model, following the procedure introduced by~\citet{Wang2019Variational,Wang2021Variational}.

The key computational feature of VSI is that all quantities in \cref{eq:residual_vector} are assembled directly from observational data.
The method uses finite element shape functions to evaluate weak-form residuals, but it does not require solving the forward PDE for each candidate parameter vector $\bstheta$.

\section{Bayesian Variational System Identification}
\label{sec:B-VSI}

\subsection{Problem setup and Bayesian target}
\label{sec:BVSI_setup}

We now formulate B-VSI for noisy spatiotemporal measurements of a PDE-governed system.
Using the notation introduced in \cref{sec:VSI}, let
\begin{align}
    \CalD
    =
    \left\{
        \hbsu^{(k)}
    \right\}_{k=0}^{n_t},
    \qquad
    \hbsu^{(k)} \in \RR^{\Ndof},
\end{align}
denote measurements collected at discrete times $\left\{t^{(k)}\right\}_{k=0}^{n_t}$.
The data are assumed to have been interpolated or projected onto the finite element space used to evaluate the weak-form residuals.

Conditioned on a candidate model $\CalM$ and coefficient vector $\bstheta$, let
$\barbsu_{\bstheta,\CalM}^{(k)} \in \RR^{\Ndof}$
denote the model-induced noiseless state at time $t^{(k)}$, represented on the same spatial degrees of freedom as the data.
This quantity plays the role of the forward-model prediction in a standard Bayesian inverse problem.
The conditional observation model is
\begin{align}
    \label{eq:observation_model}
    \hbsu^{(k)}
    =
    \barbsu_{\bstheta,\CalM}^{(k)}
    +
    \bsepsilon^{(k)},
    \qquad
    k=0,\ldots,n_t,
\end{align}
where $\bsepsilon^{(k)}$ denotes measurement noise.
We assume that the measurement errors are independent across time levels and focus on the Gaussian model
\begin{align}
    \label{eq:observation_noise_model}
    \bsepsilon^{(k)}
    \sim
    \CalN\nospaceleft(\bszero,\bsSigma_{\bsu}^{(k)}\nospaceright).
\end{align}
Equivalently,
\begin{align}
    \label{eq:conditional_observation_model}
    \hbsu^{(k)} \mid \bstheta,\CalM
    \sim
    \CalN\nospaceleft(
        \barbsu_{\bstheta,\CalM}^{(k)},
        \bsSigma_{\bsu}^{(k)}
    \nospaceright).
\end{align}
The covariance $\bsSigma_{\bsu}^{(k)}$ denotes the observational noise covariance and may vary with time.
The homoscedastic i.i.d. Gaussian noise model is recovered when
$\bsSigma_{\bsu}^{(k)}=\sigma_u^2\bsI$ for all $k$.
In what follows, $\bsSigma_r^{(k)}$ denotes the covariance induced in the weak-form residuals.

Model-form ambiguity is represented through a dictionary of candidate algebraic and differential operators,
\begin{align}
    \bschi
    =
    \left[
        F_1(u),
        F_2(u),
        \ldots,
        F_d(u)
    \right]^{\top}.
\end{align}
A candidate model $\CalM \subseteq \bschi$ is a subset of these operators.
Its coefficient vector is denoted by $\bstheta \in \RR^{d_{\bstheta}}$, where $d_{\bstheta}=|\CalM|$.
For a fixed candidate model $\CalM$, the Bayesian target is the posterior distribution
\begin{align}
    \label{eq:bayesian_target}
    p(\bstheta \mid \CalD,\CalM)
    \propto
    p(\CalD \mid \bstheta,\CalM) \,
    p(\bstheta \mid \CalM),
\end{align}
where $p(\bstheta \mid \CalM)$ is the prior and $p(\CalD \mid \bstheta,\CalM)$ is the likelihood.
The omitted proportionality constant is the model evidence $p(\CalD \mid \CalM)$, which is discussed in \cref{sec:model_selection} in the context of model comparison.

The central modeling choice is the likelihood.
In many Bayesian PDE inverse problems, the likelihood is defined by comparing measured states with forward-model predictions.
Under \cref{eq:conditional_observation_model}, this yields an observation-space likelihood that quantifies the resulting data discrepancy:
\begin{align}
    \label{eq:likelihood_data_discrepancy}
    p_{\mathrm{obs}}(\CalD \mid \bstheta,\CalM)
    =
    \prod_{k=0}^{n_t}
    \CalN\nospaceleft(
        \hbsu^{(k)}
        ;
        \barbsu_{\bstheta,\CalM}^{(k)},
        \bsSigma_{\bsu}^{(k)}
    \nospaceright).
\end{align}
Although natural, this likelihood requires computing
$\barbsu_{\bstheta,\CalM}^{(k)}$ by solving the candidate PDE for each proposed $\bstheta$ and $\CalM$.
Gradient-based posterior inference also requires differentiating through the forward solver by sensitivity, adjoint, or automatic-differentiation calculations.
A detailed comparison between this observation-space likelihood and the residual-space likelihood proposed below is provided in \cref{sec:discussion_comparison}.

B-VSI instead uses the weak-form residual map introduced in \cref{sec:VSI}.
For a fixed model $\CalM$, define
\begin{align}
    \label{eq:residual_map_intro}
    \bsr_{\bstheta,\CalM}
    =
    \CalR_{\bstheta,\CalM}(\CalD),
\end{align}
where $\CalR_{\bstheta,\CalM}$ maps the observed fields to the weak-form PDE residuals associated with $\CalM$ and $\bstheta$.
When the model is clear from context, we write $\bsr_{\bstheta}$ for compactness.
For the coefficient-linear PDE form considered in this work, the residual can be written as
\begin{align}
    \label{eq:residual_regression_intro}
    \bsr_{\bstheta}
    =
    \bsy(\CalD)
    -
    \bsX(\CalD;\CalM)\bstheta,
\end{align}
where $\bsy$ and $\bsX$ are constructed from the weak-form time-derivative and candidate-operator terms following \cref{eq:yXdef}.
The matrix $\bsX$ may depend nonlinearly on the observed fields when nonlinear operators are included in $\CalM$.

The model-induced trajectory
$\big\{\barbsu_{\bstheta,\CalM}^{(k)}\big\}_{k=0}^{n_t}$
satisfies the candidate weak-form equations for the same $\bstheta$ and $\CalM$, up to the chosen temporal discretization and solver tolerance.
Therefore,
\begin{align}
    \label{eq:model_prediction_zero_residual}
    \CalR_{\bstheta,\CalM}
    \nospaceleft(
        \left\{
        \barbsu_{\bstheta,\CalM}^{(k)}
        \right\}_{k=0}^{n_t}
    \nospaceright)
    =
    \bszero.
\end{align}
Since the model-induced noiseless trajectory has zero weak-form residual, the residual evaluated on the observed data arises from the measurement perturbations in \cref{eq:observation_model}; its distribution is therefore obtained by propagating those perturbations through $\CalR_{\bstheta,\CalM}$.
B-VSI uses this induced residual distribution as its likelihood,
\begin{align}
    \label{eq:residual_likelihood_intro}
    p_{\mathrm{res}}(\CalD \mid \bstheta,\CalM)
    :=
    p\nospaceleft(
        \CalR_{\bstheta,\CalM}(\CalD)
        \mid \bstheta,\CalM
    \nospaceright)
    =
    p(\bsr_{\bstheta,\CalM} \mid \bstheta,\CalM).
\end{align}
The corresponding residual-space posterior is
\begin{align}
    \label{eq:residual_space_posterior_intro}
    p_{\mathrm{res}}(\bstheta \mid \CalD,\CalM)
    \propto
    p_{\mathrm{res}}(\CalD \mid \bstheta,\CalM) \,
    p(\bstheta \mid \CalM).
\end{align}
For notational simplicity, we write this residual-space posterior as
$p(\bstheta \mid \CalD,\CalM)$ in the remainder of the paper unless the observation-space likelihood is explicitly discussed.

The remainder of this section follows the B-VSI workflow.
\Cref{sec:residual_likelihood} derives the weak-form residual likelihood for linear and nonlinear operators.
\Cref{sec:posterior_inference_fixed_model} describes posterior inference for a fixed candidate model.
\Cref{sec:model_selection} presents residual-space model comparison and sequential operator selection.
\Cref{sec:posterior_predictive} describes posterior predictive uncertainty for derived quantities of interest.
\Cref{sec:algorithm_summary} summarizes the complete B-VSI procedure.

\subsection{Weak-form residual likelihood}
\label{sec:residual_likelihood}

We now derive the residual-space likelihood
$p(\bsr_{\bstheta,\CalM} \mid \bstheta,\CalM)$ introduced in \cref{eq:residual_likelihood_intro}.
The goal is not to assign an \textit{ad hoc} probability model to the weak-form residuals.
Instead, we obtain the residual distribution by propagating the conditional observation model in \cref{eq:observation_model} through the weak-form residual map $\CalR_{\bstheta,\CalM}$.

For a fixed candidate model $\CalM$ and coefficient vector $\bstheta$, let
\begin{align}
    \label{eq:per_step_residual_map}
    \bsr_{\bstheta,\CalM}^{(k)}
    =
    \CalR_{\bstheta,\CalM}^{(k)}
    \nospaceleft(
        \hbsu^{(k-1)},\hbsu^{(k)}
    \nospaceright),
    \qquad
    k=1,\ldots,n_t,
\end{align}
denote the weak-form residual over the time interval
$\big[t^{(k-1)},t^{(k)}\big]$.
When the model is clear from context, we write
$\bsr_{\bstheta}^{(k)}$ for compactness.
The model-induced trajectory
$\big\{\barbsu_{\bstheta,\CalM}^{(k)}\big\}_{k=0}^{n_t}$
satisfies the candidate weak-form equations for the same $\bstheta$ and $\CalM$.
Thus,
\begin{align}
    \label{eq:model_induced_per_step_zero_residual}
    \CalR_{\bstheta,\CalM}^{(k)}
    \nospaceleft(
        \barbsu_{\bstheta,\CalM}^{(k-1)},
        \barbsu_{\bstheta,\CalM}^{(k)}
    \nospaceright)
    =
    \bszero,
    \qquad
    k=1,\ldots,n_t,
\end{align}
up to the chosen temporal discretization and solver tolerance.
Under the observation model, the observed residual
$\bsr_{\bstheta}^{(k)}$ is therefore driven by the measurement perturbations
$\bsepsilon^{(k-1)}$ and $\bsepsilon^{(k)}$.
The following subsections make this perturbation explicit and derive its induced covariance.

A Monte Carlo construction could be used by sampling observational data from
\cref{eq:conditional_observation_model} and evaluating the weak-form residual for each realization.
This approach accommodates nonlinear residual maps and non-Gaussian observation models, but it yields an implicit residual distribution rather than an analytically evaluable density.
For efficient posterior inference and model comparison, we instead construct explicit Gaussian residual likelihoods.
For residual maps that are linear in the observed state, this propagation is exact under Gaussian observational noise.
For nonlinear residual maps, we use a first-order local Gaussian approximation.
Although the present framework focuses on Gaussian likelihoods, the underlying propagation methodology extends naturally to other distribution families that are closed under linear transformations, including the Student-$t$ distribution and Gaussian mixture models.

\subsubsection{Operators linear in the state}
\label{sec:linear_residual_likelihood}

Consider an operator $L \in \bschi$ that is linear in the state variable $u$.
The corresponding weak-form basis vector can be written as a linear map of the nodal data,
\begin{align}
    \label{eq:linear_operator_matrix_form}
    \bsXi^{L}
    =
    \bsA^{L}\hbsu,
\end{align}
where $\bsA^{L}\in\RR^{\Ndof\times\Ndof}$ is the Galerkin matrix associated with $L$.
Consistent with the weak-form notation in \cref{eq:residual}, its entries are
\begin{align}
    \label{eq:stiffness_matrix}
    A_{ij}^{L}
    =
    \mathscr{F}_{L}(N_i,N_j),
    \qquad
    i,j=1,\ldots,\Ndof.
\end{align}
Equivalently, before integration by parts this expression takes the form
$\int_{\Omega} N_i L(N_j)\,\mathrm{d}\bsx$.
The notation in \cref{eq:stiffness_matrix} also includes boundary terms and sign changes introduced by integration by parts.
Indeed, for a measured field
$\hat{u}^{h}(\bsx)=\sum_{j=1}^{\Ndof}\hu_jN_j(\bsx)$, the $i$-th entry of the basis vector is
\begin{align}
    \label{eq:linear_operator_basis}
    \Xi_i^{L}
    &=
    \mathscr{F}_{L}
    \nospaceleft(
        N_i,
        \sum_{j=1}^{\Ndof}\hu_jN_j
    \nospaceright)
    =
    \sum_{j=1}^{\Ndof}
    \hu_j
    \mathscr{F}_{L}(N_i,N_j)
    =
    A_{i,:}^{L}\hbsu .
\end{align}

Let $\bsM\in\RR^{\Ndof\times\Ndof}$ denote the mass matrix,
\begin{align}
    \label{eq:mass_matrix}
    M_{ij}
    =
    \int_{\Omega} N_i(\bsx)N_j(\bsx)\,\mathrm{d}\bsx,
\end{align}
and define
\begin{align}
    \label{eq:B_matrix_time_derivative}
    \bsB
    =
    \frac{1}{\Delta t}\bsM .
\end{align}
Using the backward-difference approximation from \cref{eq:time_derivative_vector}, the weak-form time-derivative vector is
\begin{align}
    \label{eq:time_derivative_matrix_form}
    {\bsXi^{\dotu}}^{(k)}
    =
    \bsB
    \left(
        \hbsu^{(k)}-\hbsu^{(k-1)}
    \right).
\end{align}

Suppose first that all data-dependent candidate operators in $\CalM$ are linear in the state.
Data-independent terms, such as constant source terms, may also be included.
They enter the residual through deterministic vectors and do not contribute to the residual covariance.
The residual
can then be written as
\begin{align}
\bsr_{\bstheta}^{(k)} &= {\bsXi^{\dotu}}^{(k)}-\bsXi^{(k)}\bstheta \nonumber\\
&=
\bsB\left(\hbsu^{(k)}-\hbsu^{(k-1)}\right)
-
\sum_{L_i\in\CalM_{\mathrm{lin}}}
    \theta_i\bsA^{L_i}\hbsu^{(k)}
-
\bsq_{\bstheta} \nonumber\\
    \label{eq:residual_vector_linear}
    &=
    \bsK_{\bstheta}\hbsu^{(k)}
    -
    \bsB\hbsu^{(k-1)}
    -
    \bsq_{\bstheta},
\end{align}
where
\begin{align}
    \label{eq:linear_current_sensitivity}
    \bsK_{\bstheta}
    =
    \bsB
    -
    \sum_{L_i\in\CalM_{\mathrm{lin}}}
    \theta_i\bsA^{L_i}.
\end{align}
Here, $\CalM_{\mathrm{lin}}$ denotes the set of candidate operators that are linear in $u$, and $\bsq_{\bstheta}$ collects any data-independent weak-form contributions.

Conditioned on $\bstheta$ and $\CalM$, the observations satisfy
\begin{align}
    \hbsu^{(j)}
    =
    \barbsu_{\bstheta,\CalM}^{(j)}
    +
    \bsepsilon^{(j)},
    \qquad
    j=k-1,k.
\end{align}
Substituting the conditional observation model into
\cref{eq:residual_vector_linear} gives
\begin{align}
    \bsr_{\bstheta}^{(k)}
    &=
    \bsK_{\bstheta}
    \left(
        \barbsu_{\bstheta,\CalM}^{(k)}
        +
        \bsepsilon^{(k)}
    \right)
    -
    \bsB
    \left(
        \barbsu_{\bstheta,\CalM}^{(k-1)}
        +
        \bsepsilon^{(k-1)}
    \right)
    -
    \bsq_{\bstheta}
    \nonumber\\
    &=
    \left[
        \bsK_{\bstheta}\barbsu_{\bstheta,\CalM}^{(k)}
        -
        \bsB\barbsu_{\bstheta,\CalM}^{(k-1)}
        -
        \bsq_{\bstheta}
    \right]
    +
    \bsK_{\bstheta}\bsepsilon^{(k)}
    -
    \bsB\bsepsilon^{(k-1)}.
\end{align}
Because the model-induced trajectory satisfies the candidate weak-form residual equations for the same $\bstheta$ and $\CalM$, the bracketed term vanishes.
Hence,
\begin{align}
    \label{eq:linear_residual_noise_map}
    \bsr_{\bstheta}^{(k)}
    =
    \bsK_{\bstheta}\bsepsilon^{(k)}
    -
    \bsB\bsepsilon^{(k-1)}.
\end{align}

If the observational noise is Gaussian and independent across time levels, as in \cref{eq:observation_noise_model},
\begin{align}
    \bsepsilon^{(k)}
    \sim
    \CalN\nospaceleft(\bszero,\bsSigma_{\bsu}^{(k)}\nospaceright),
    \nonumber
\end{align}
then the residual at time step $k$ is Gaussian,
\begin{align}
    \label{eq:likelihood_linear}
    p\nospaceleft( \bsr_{\bstheta}^{(k)} \mid \bstheta,\CalM\nospaceright)
    =
    \CalN
    \nospaceleft(
        \bsr_{\bstheta}^{(k)}
        ;
        \bszero,
        \bsSigma_{r,\bstheta}^{(k)}
    \nospaceright),
\end{align}
with covariance
\begin{align}
    \label{eq:covariance_linear}
    \bsSigma_{r,\bstheta}^{(k)}
    =
    \bsK_{\bstheta}
    \bsSigma_{\bsu}^{(k)}
    \bsK_{\bstheta}^{\top}
    +
    \bsB
    \bsSigma_{\bsu}^{(k-1)}
    \bsB^{\top}.
\end{align}
The residual covariance depends on $\bstheta$ through $\bsK_{\bstheta}$.

\subsubsection{Operators nonlinear in the state}
\label{sec:nonlinear_residual_likelihood}

When $\CalM$ contains operators that are nonlinear in the state, the weak-form residual remains linear in the previous time-step data through the backward-difference term, but it is generally nonlinear in the current time-step data.
Let $\CalM_{\mathrm{state}}\subseteq\CalM$ denote the operators whose weak-form basis vectors depend on the observed state.
The residual can be written as
\begin{align}
    \label{eq:nonlinear_residual_map}
    \bsr_{\bstheta}^{(k)}
    &= {\bsXi^{\dotu}}^{(k)}-\bsXi^{(k)}\bstheta \nonumber\\
    &=
    \bsB
    \left(
        \hbsu^{(k)}-\hbsu^{(k-1)}
    \right)
    -
    \sum_{F_i\in\CalM_{\mathrm{state}}}
    \theta_i
    \bsXi^{F_i}\nospaceleft(\hbsu^{(k)}\nospaceright)
    -
    \bsq_{\bstheta},
\end{align}
where $\bsXi^{F_i}\big(\hbsu^{(k)}\big)$ denotes the weak-form basis vector associated with $F_i$ at time $t^{(k)}$.
As before, $\bsq_{\bstheta}$ collects data-independent weak-form contributions.

The exact distribution of
$\bsr_{\bstheta}^{(k)}$
is generally non-Gaussian because
\cref{eq:nonlinear_residual_map}
is a nonlinear transformation of the observed data.
We approximate this distribution using a first-order Taylor expansion of the residual map with respect to the current state.
The current-time sensitivity is
\begin{align}
    \label{eq:overall_jacobian}
    \bsJ_{\bstheta}^{(k)}
    =
    \left.
    \frac{
        \partial
        \CalR_{\bstheta,\CalM}^{(k)}
    }{
        \partial \bsu^{(k)}
    }
    \right|_{\bsu^{(k)}=\hbsu^{(k)}}
    =
    \bsB
    -
    \sum_{F_i\in\CalM_{\mathrm{state}}}
    \theta_i
    \bsJ_{F_i}^{(k)},
\end{align}
where
\begin{align}
    \label{eq:operator_jacobian}
    \bsJ_{F_i}^{(k)}
    =
    \left.
    \frac{
        \partial
        \bsXi^{F_i}(\bsu)
    }{
        \partial \bsu
    }
    \right|_{\bsu=\hbsu^{(k)}}.
\end{align}
For a candidate model containing both state-linear and nonlinear operators, $\bsJ_{F_i}^{(k)}$ is assembled operator-wise. In particular,
if $F_i$ is linear in $u$, then $\bsJ_{F_i}^{(k)}=\bsA^{F_i}$.

Formally, the delta-method linearization would use the Jacobian evaluated at the model-induced state
$\barbsu_{\bstheta,\CalM}^{(k)}$.
Computing that state would require a forward PDE solve and would therefore forfeit the computational efficiency that motivates the residual-based formulation.
B-VSI therefore uses the plug-in approximation in \cref{eq:overall_jacobian}, evaluated at the observed data.
After this plug-in evaluation, $\bsJ_{\bstheta}^{(k)}$ is held fixed within each parameter-update step when constructing the local Gaussian residual likelihood.
With this approximation, the residual perturbation is
\begin{align}
    \label{eq:nonlinear_residual_noise_map}
    \bsr_{\bstheta}^{(k)}
    \approx
    \bsJ_{\bstheta}^{(k)}
    \bsepsilon^{(k)}
    -
    \bsB
    \bsepsilon^{(k-1)}.
\end{align}
The resulting local Gaussian residual likelihood is
\begin{align}
    \label{eq:likelihood_nonlinear}
    p\nospaceleft( \bsr_{\bstheta}^{(k)} \mid \bstheta,\CalM\nospaceright)
    \approx
    \CalN
    \nospaceleft(
        \bsr_{\bstheta}^{(k)}
        ;
        \bszero,
        \bsSigma_{r,\bstheta}^{(k)}
    \nospaceright),
\end{align}
where
\begin{align}
    \label{eq:covariance_nonlinear}
    \bsSigma_{r,\bstheta}^{(k)}
    =
    \bsJ_{\bstheta}^{(k)}
    \bsSigma_{\bsu}^{(k)}
    {\bsJ_{\bstheta}^{(k)}}^{\top}
    +
    \bsB
    \bsSigma_{\bsu}^{(k-1)}
    \bsB^{\top}.
\end{align}
The approximation in \cref{eq:covariance_nonlinear} is a first-order delta-method approximation.
The second-order Taylor terms generally produce a nonzero residual-mean correction of order $\CalO(\|\bsSigma_{\bsu}^{(k)}\|)$ and additional covariance corrections of higher order.
We neglect these terms and use the zero-mean Gaussian likelihood in \cref{eq:likelihood_nonlinear}, which is accurate when the observation noise is sufficiently small and the residual map is locally well approximated by its Jacobian.
The effect of this approximation is further examined in \cref{sec:successive_approximations}.

For operators linear in the state, \cref{eq:covariance_nonlinear} reduces to \cref{eq:covariance_linear} with
$\bsJ_{\bstheta}^{(k)}=\bsK_{\bstheta}$.

\subsubsection{Stacked residual likelihood}
\label{sec:stacked_residual_likelihood}
Let
\begin{align}
    \bsr_{\bstheta}
    =
    \begin{bmatrix}
        {\bsr_{\bstheta}^{(1)}}^{\top}
        &
        {\bsr_{\bstheta}^{(2)}}^{\top}
        &
        \cdots
        &
        {\bsr_{\bstheta}^{(n_t)}}^{\top}
    \end{bmatrix}^{\top}
\end{align}
denote the stacked residual vector.
Because $\bsr_{\bstheta}^{(k)}$ depends on both
$\hbsu^{(k-1)}$ and $\hbsu^{(k)}$, adjacent residuals generally share observational noise at a common time level.
Thus, even when observational errors are independent across time, the stacked residual covariance is not generally block diagonal.

For both the state-linear case and the nonlinear local approximation, the residual perturbation can be written as
\begin{align}
    \label{eq:unified_residual_noise_map}
    \bsr_{\bstheta}^{(k)}
    \approx
    \bsC_{\bstheta}^{(k)}\bsepsilon^{(k)}
    -
    \bsB\bsepsilon^{(k-1)},
\end{align}
where
\begin{align}
    \label{eq:C_definition}
    \bsC_{\bstheta}^{(k)}
    =
    \begin{cases}
        \bsK_{\bstheta}, & \text{for operators linear in the state}, \\
        \bsJ_{\bstheta}^{(k)}, & \text{for nonlinear operators under the local approximation}.
    \end{cases}
\end{align}
Under this linearized representation, the covariance has diagonal blocks
\begin{align}
    \label{eq:stacked_covariance_diagonal_block}
    \boldsymbol{D}_k
    =
    \bsSigma_{r,\bstheta}^{(k)}
    =
    \operatorname{Cov}
    \nospaceleft(
        \bsr_{\bstheta}^{(k)},
        \bsr_{\bstheta}^{(k)}
    \nospaceright)
    =
    \bsC_{\bstheta}^{(k)}
    \bsSigma_{\bsu}^{(k)}
    {\bsC_{\bstheta}^{(k)}}^{\top}
    +
    \bsB
    \bsSigma_{\bsu}^{(k-1)}
    \bsB^{\top},
\end{align}
and adjacent off-diagonal blocks
\begin{align}
    \label{eq:stacked_covariance_off_diagonal_block}
    \boldsymbol{E}_k
    =
    \operatorname{Cov}
    \nospaceleft(
        \bsr_{\bstheta}^{(k)},
        \bsr_{\bstheta}^{(k+1)}
    \nospaceright)
    =
    -
    \bsC_{\bstheta}^{(k)}
    \bsSigma_{\bsu}^{(k)}
    \bsB^{\top}.
\end{align}
All non-adjacent blocks vanish under the assumed independence of observational errors across time.

The exact first-order stacked covariance induced by the assumed observation model is therefore block tridiagonal (BT):
\begin{align}
    \label{eq:block_tridiagonal_residual_covariance}
    \bsSigma_{r,\bstheta}^{\mathrm{BT}}
    =
    \begin{bmatrix}
        \boldsymbol{D}_1 & \boldsymbol{E}_1 & \bszero & \cdots & \bszero \\
        \boldsymbol{E}_1^{\top} & \boldsymbol{D}_2 & \boldsymbol{E}_2 & \ddots & \vdots \\
        \bszero & \boldsymbol{E}_2^{\top} & \boldsymbol{D}_3 & \ddots & \bszero \\
        \vdots & \ddots & \ddots & \ddots & \boldsymbol{E}_{n_t-1} \\
        \bszero & \cdots & \bszero & \boldsymbol{E}_{n_t-1}^{\top} & \boldsymbol{D}_{n_t}
    \end{bmatrix}.
\end{align}%
Retaining the off-diagonal blocks couples adjacent time steps but requires computationally intensive sequential factorization of the stacked covariance.
For computational efficiency, the block-diagonal (BD) approximation retains the marginal residual covariance at each time step while neglecting the covariance between adjacent time steps:
\begin{align}
    \label{eq:block_diagonal_residual_covariance}
    \bsSigma_{r,\bstheta}^{\mathrm{BD}}
    =
    \operatorname{blockdiag}
    \nospaceleft(
        \boldsymbol{D}_1,\ldots,\boldsymbol{D}_{n_t}
    \nospaceright).
\end{align}
This approximation is consistent with the local residual treatment of classical VSI and enables computationally tractable likelihood evaluation through independent covariance factorizations and solves that can be parallelized over time steps.
\Cref{eq:block_tridiagonal_residual_covariance,eq:block_diagonal_residual_covariance} thus provide two alternative treatments of residual covariance within the same likelihood framework, with their respective likelihood evaluations detailed in \cref{sec:residual_space_objective}.

\subsection{Posterior inference for a fixed model}
\label{sec:posterior_inference_fixed_model}

We now describe inference for the coefficient vector $\bstheta$ under a fixed candidate model $\CalM$.
Model comparison and operator selection are addressed later in \cref{sec:model_selection}.
Let $\CalC\in\{\mathrm{BT},\mathrm{BD}\}$ select either the block-tridiagonal covariance in \cref{eq:block_tridiagonal_residual_covariance} or the computationally efficient block-diagonal approximation in \cref{eq:block_diagonal_residual_covariance}.
They are two treatments of the residual covariance in this same Gaussian residual-space likelihood, not different likelihood models.
For either choice of $\CalC$, the residual-space likelihood is
\begin{align}
    \label{eq:residual_likelihood_theta_dependent}
    p_{\mathrm{res}}(\CalD \mid \bstheta,\CalM)
    =
    p^{\CalC}(\bsr_{\bstheta} \mid \bstheta,\CalM)
    =
    (2\pi)^{-\frac{n}{2}}
    \det\left(
        \bsSigma_{r,\bstheta}^{\CalC}
    \right)^{-\frac{1}{2}}
    \exp\left[
        -\frac{1}{2}
        \bsr_{\bstheta}^{\top}
        \left(
            \bsSigma_{r,\bstheta}^{\CalC}
        \right)^{-1}
        \bsr_{\bstheta}
    \right],
\end{align}
where $n=\dim(\bsr_{\bstheta})=n_t n_{\mathrm{eq}}$. Although \cref{eq:residual_likelihood_theta_dependent} gives a global representation of the stacked-residual likelihood, its covariance need not be stored or factored as a dense $n\times n$ matrix. The block-diagonal structure permits independent factorizations over time intervals, while the block-tridiagonal structure admits a sequential block recursion. The likelihood can therefore be evaluated using only block-sized matrices, as detailed in \cref{sec:residual_space_objective}.

The selected covariance $\bsSigma_{r,\bstheta}^{\CalC}$ is induced by the weak-form residual map and generally depends on $\bstheta$ through the residual sensitivity matrices in \cref{eq:stacked_covariance_diagonal_block} (and \cref{eq:stacked_covariance_off_diagonal_block} if the block-tridiagonal structure is used).
This dependence prevents direct closed-form posterior updates, even when the residual likelihood and prior are Gaussian.
This motivates the optional lagged-covariance approximation in \cref{sec:lagged_covariance_inference}, which holds $\bsSigma_{r,\bstheta}^{\CalC}$ fixed during each parameter update.

We distinguish two levels of approximation for posterior inference.
The first is the block-diagonal residual covariance approximation if selected.
The second is an optional lagged-covariance approximation used to obtain closed-form parameter updates in \cref{sec:lagged_covariance_inference}.
For either covariance structure, one may either retain the full $\bstheta$-dependence of $\bsSigma_{r,\bstheta}^{\CalC}$ or hold the covariance fixed during each parameter update.

\subsubsection{Residual-space log density}
\label{sec:residual_space_objective}

The residual-space likelihood in \cref{eq:residual_likelihood_theta_dependent} provides the common probabilistic object underlying both point estimation and posterior inference.
Up to an additive constant, its negative log-likelihood is
\begin{align}
    \label{eq:theta_dependent_negative_log_likelihood}
    \CalL_{\mathrm{res}}(\bstheta)
    =
    \frac{1}{2}
    \bsr_{\bstheta}^{\top}
    \left(
        \bsSigma_{r,\bstheta}^{\CalC}
    \right)^{-1}
    \bsr_{\bstheta}
    +
    \frac{1}{2}
    \log\det
    \left(
        \bsSigma_{r,\bstheta}^{\CalC}
    \right).
\end{align}

\paragraph{Log-likelihood with the block-diagonal structure}
Under the block-diagonal covariance, the residual blocks are independent and the likelihood can be factorized over time intervals:
\begin{align}
    \label{eq:residual_likelihood_factorized}
    p^{\mathrm{BD}}\nospaceleft(
        \bsr_{\bstheta}\mid\bstheta,\CalM
    \nospaceright)
    =
    \prod_{k=1}^{n_t}
    \CalN\nospaceleft(
        \bsr_{\bstheta}^{(k)};
        \bszero,\boldsymbol{D}_k
    \nospaceright).
\end{align}%
Consequently, the quadratic-form and log-determinant terms in \cref{eq:theta_dependent_negative_log_likelihood} can be evaluated independently for each time interval and then summed:
\begin{subequations}
    \label{eq:bd_residual_likelihood_terms}
    \begin{align}
        \bsr_{\bstheta}^{\top}
        \left(\bsSigma_{r,\bstheta}^{\mathrm{BD}}\right)^{-1}
        \bsr_{\bstheta}
        &=
        \sum_{k=1}^{n_t}
        {\bsr_{\bstheta}^{(k)}}^{\top}
        \boldsymbol{D}_k^{-1}
        \bsr_{\bstheta}^{(k)},\\
        \log\det\bsSigma_{r,\bstheta}^{\mathrm{BD}}
        &=
        \sum_{k=1}^{n_t}\log\det\boldsymbol{D}_k.
    \end{align}
\end{subequations}%

\paragraph{Log-likelihood with the block-tridiagonal structure}
For the stacked residual distribution, the probability chain rule gives
\begin{align}
    \label{eq:bt_chain_rule_factorization}
    p^{\mathrm{BT}}\nospaceleft(\bsr_{\bstheta} \mid \bstheta,\CalM\nospaceright)
    =
    p\nospaceleft(\bsr_{\bstheta}^{(1)} \mid \bstheta,\CalM\nospaceright)
    \prod_{k=2}^{n_t}
    p\nospaceleft(
        \bsr_{\bstheta}^{(k)}
        \mid
        \bsr_{\bstheta}^{(1:k-1)},\bstheta,\CalM
    \nospaceright),
\end{align}%
where $\bsr_{\bstheta}^{(1:k-1)}$ denotes the residual blocks from intervals $1$ through $k-1$. For the Gaussian block-tridiagonal covariance, block Gaussian elimination yields the conditional factors sequentially, with the Schur complement $\boldsymbol{S}_k$ giving the conditional covariance at each interval. The recursion uses only block-sized matrices and begins with the first diagonal block:
\begin{align}
    \label{eq:bt_schur_initialization}
    \boldsymbol{S}_1=\boldsymbol{D}_1.
\end{align}%
For $k=2,\ldots,n_t$, sequential block elimination gives
\begin{align}
    \label{eq:bt_schur_recursion}
    \boldsymbol{S}_k
    =
    \boldsymbol{D}_k
    -
    \boldsymbol{E}_{k-1}^{\top}
    \boldsymbol{S}_{k-1}^{-1}
    \boldsymbol{E}_{k-1}.
\end{align}%
For likelihood evaluation, initialize $\widetilde{\bsr}_{\bstheta}^{(1)}=\bsr_{\bstheta}^{(1)}$ and apply the same elimination to the residual right-hand side:
\begin{align}
    \label{eq:bt_residual_recursion}
    \widetilde{\bsr}_{\bstheta}^{(k)}
    =
    \bsr_{\bstheta}^{(k)}
    -
    \boldsymbol{E}_{k-1}^{\top}
    \boldsymbol{S}_{k-1}^{-1}
    \widetilde{\bsr}_{\bstheta}^{(k-1)},
    \qquad k=2,\ldots,n_t.
\end{align}%
The transformed block $\widetilde{\bsr}_{\bstheta}^{(k)}$ is the block innovation, namely the residual remaining after conditioning on the preceding blocks. The conditional factorization in \cref{eq:bt_chain_rule_factorization} therefore becomes
\begin{align}
    \label{eq:bt_residual_likelihood_factorized}
    p^{\mathrm{BT}}\nospaceleft(
        \bsr_{\bstheta}\mid\bstheta,\CalM
    \nospaceright)
    =
    \prod_{k=1}^{n_t}
    \CalN\nospaceleft(
        \widetilde{\bsr}_{\bstheta}^{(k)};
        \bszero,\boldsymbol{S}_k
    \nospaceright).
\end{align}%
The factors in \cref{eq:bt_residual_likelihood_factorized} are conditional Gaussian densities, not independent marginal likelihoods.
The exact block-tridiagonal quadratic form and log determinant are then
\begin{subequations}
    \label{eq:bt_likelihood_terms}
    \begin{align}
        \bsr_{\bstheta}^{\top}
        \left(\bsSigma_{r,\bstheta}^{\mathrm{BT}}\right)^{-1}
        \bsr_{\bstheta}
        &=
        \sum_{k=1}^{n_t}
        \left(\widetilde{\bsr}_{\bstheta}^{(k)}\right)^{\top}
        \boldsymbol{S}_k^{-1}
        \widetilde{\bsr}_{\bstheta}^{(k)},\\
        \log\det\bsSigma_{r,\bstheta}^{\mathrm{BT}}
        &=
        \sum_{k=1}^{n_t}\log\det\boldsymbol{S}_k.
    \end{align}
\end{subequations}%
Here and below, inverse notation denotes linear solves, implemented in practice with factorizations such as Cholesky rather than explicit matrix inversion.

\paragraph{Log-posterior}
When a prior $p(\bstheta \mid \CalM)$ is specified, the same likelihood defines the posterior density through Bayes' rule.
Equivalently, the negative log-posterior is
\begin{align}
    \label{eq:theta_dependent_negative_log_posterior}
    \CalL_{\mathrm{post}}(\bstheta)
    =
    \CalL_{\mathrm{res}}(\bstheta)
    -
    \log p(\bstheta \mid \CalM).
\end{align}
These quantities can be used in two complementary ways.
Minimizing $\CalL_{\mathrm{res}}$ or $\CalL_{\mathrm{post}}$ gives residual-space maximum likelihood (MLE) or maximum a posteriori (MAP) point estimates, respectively.
Alternatively, the same likelihood and prior define the posterior density
\begin{align}
    p_{\mathrm{res}}(\bstheta \mid \CalD,\CalM)
    \propto
    \exp\left[-\CalL_{\mathrm{res}}(\bstheta)\right]
    p(\bstheta \mid \CalM),
\end{align}
which can be approximated using Gaussian, conjugate, gradient-based, or particle-based methods.
Both $\CalL_{\mathrm{res}}$ and $\CalL_{\mathrm{post}}$ include the weighted residual term and the log-determinant term, and both generally depend on $\bstheta$ through the selected residual covariance structure.

The inference methods below are organized according to how this $\bstheta$-dependent covariance is handled.
The lagged-covariance approach freezes the covariance at the current iterate and yields generalized least-squares estimates and conjugate posterior updates when applicable.
Gradient-based and particle-based methods can instead target the full residual-space likelihood or posterior density.
A hybrid strategy uses the lagged-covariance approximation to initialize the more general methods.

\subsubsection{Lagged-covariance inference}
\label{sec:lagged_covariance_inference}

The lagged-covariance approximation uses a fixed-point iteration.
Given a current parameter estimate $\bstheta^{(l)}$ and covariance structure choice $\CalC\in\{\mathrm{BD},\mathrm{BT}\}$, define
\begin{align}
    \label{eq:lagged_covariance}
    \bsSigma_l^{\CalC}
    =
    \bsSigma_{r,\bstheta^{(l)}}^{\CalC},
    \qquad
    \bsW_l^{\CalC}
    =
    \left(\bsSigma_l^{\CalC}\right)^{-1}.
\end{align}
The covariance is then held fixed while updating $\bstheta$.
After the parameter update, the covariance is recomputed at the new estimate.
The iteration is typically initialized with the classical VSI estimate in \cref{eq:VSI_solution}.

\paragraph{Point estimation with lagged covariance}

With $\bsSigma_l$ fixed, the log-determinant term is constant with respect to $\bstheta$.
The residual-space MLE update is therefore obtained from the generalized least-squares problem
\begin{align}
    \label{eq:lagged_covariance_mle_objective}
    \bstheta_{\mathrm{MLE}}^{(l+1)}
    =
    \argmin_{\bstheta}
    \frac{1}{2}
    \left(
        \bsy-\bsX\bstheta
    \right)^{\top}
    \bsW_l^{\CalC}
    \left(
        \bsy-\bsX\bstheta
    \right).
\end{align}
When $\bsX^{\top}\bsW_l^{\CalC}\bsX$ is nonsingular, the closed-form update is
\begin{align}
    \label{eq:lagged_covariance_mle_update}
    \bstheta_{\mathrm{MLE}}^{(l+1)}
    =
    \left(
        \bsX^{\top}\bsW_l^{\CalC}\bsX
    \right)^{-1}
    \bsX^{\top}\bsW_l^{\CalC}\bsy.
\end{align}
If this matrix is singular or ill-conditioned, a pseudoinverse or regularized least-squares formulation may be used.

The same lagged-covariance idea also gives a MAP update when a prior is specified.
For example, with a Gaussian prior $p(\bstheta \mid \CalM)=\CalN(\bstheta;\bsmu_0,\bsSigma_0)$, the fixed-covariance MAP problem is
\begin{align}
    \label{eq:lagged_covariance_map_objective}
    \bstheta_{\mathrm{MAP}}^{(l+1)}
    =
    \argmin_{\bstheta}
    \frac{1}{2}
    \left(
        \bsy-\bsX\bstheta
    \right)^{\top}
    \bsW_l^{\CalC}
    \left(
        \bsy-\bsX\bstheta
    \right)
    +
    \frac{1}{2}
    \left(
        \bstheta-\bsmu_0
    \right)^{\top}
    \bsSigma_0^{-1}
    \left(
        \bstheta-\bsmu_0
    \right).
\end{align}
When $\bsX^{\top}\bsW_l^{\CalC}\bsX+\bsSigma_0^{-1}$ is nonsingular, this yields the closed-form ridge-type update
\begin{align}
    \label{eq:lagged_covariance_map_update}
    \bstheta_{\mathrm{MAP}}^{(l+1)}
    =
    \left(
        \bsX^{\top}\bsW_l^{\CalC}\bsX+\bsSigma_0^{-1}
    \right)^{-1}
    \left(
        \bsX^{\top}\bsW_l^{\CalC}\bsy+\bsSigma_0^{-1}\bsmu_0
    \right).
\end{align}
More general priors lead to regularized optimization problems that may not have closed-form updates.

\subparagraph{Block-diagonal implementation}
Using the block-diagonal covariance in \cref{eq:block_diagonal_residual_covariance}, the likelihood normal-equation terms are independent time-step contributions:
\begin{subequations}
    \label{eq:lagged_covariance_normal_equations}
    \begin{align}
        \bsX^{\top}\bsW_l^{\mathrm{BD}}\bsX
        &=
        \sum_{k=1}^{n_t}
        {\bsXi^{(k)}}^{\top}
        \boldsymbol{D}_k^{-1}
        \bsXi^{(k)},
        \\
        \bsX^{\top}\bsW_l^{\mathrm{BD}}\bsy
        &=
        \sum_{k=1}^{n_t}
        {\bsXi^{(k)}}^{\top}
        \boldsymbol{D}_k^{-1}
        \bsXi_{\dot u}^{(k)}.
    \end{align}
\end{subequations}

\subparagraph{Block-tridiagonal implementation}
For the block-tridiagonal structure, initialize the transformed regression blocks as
\begin{align}
    \label{eq:bt_regression_initialization}
    \widetilde{\bsXi}^{(1)}&=\bsXi^{(1)},
    &
    \widetilde{\bsXi}_{\dot u}^{(1)}&=\bsXi_{\dot u}^{(1)}.
\end{align}%
For $k=2,\ldots,n_t$, apply the same elimination to the design and time-derivative blocks using the Schur-complement blocks from \cref{eq:bt_schur_initialization,eq:bt_schur_recursion}:
\begin{subequations}
    \label{eq:bt_regression_recursion}
    \begin{align}
        \widetilde{\bsXi}^{(k)}
        &=
        \bsXi^{(k)}
        -
        \boldsymbol{E}_{k-1}^{\top}
        \boldsymbol{S}_{k-1}^{-1}
        \widetilde{\bsXi}^{(k-1)},
        \\
        \widetilde{\bsXi}_{\dot u}^{(k)}
        &=
        \bsXi_{\dot u}^{(k)}
        -
        \boldsymbol{E}_{k-1}^{\top}
        \boldsymbol{S}_{k-1}^{-1}
        \widetilde{\bsXi}_{\dot u}^{(k-1)}.
    \end{align}
\end{subequations}%
The likelihood normal-equation terms are then
\begin{subequations}
    \label{eq:bt_lagged_covariance_normal_equations}
    \begin{align}
        \bsX^{\top}\bsW_l^{\mathrm{BT}}\bsX
        &=
        \sum_{k=1}^{n_t}
        \left(\widetilde{\bsXi}^{(k)}\right)^{\top}
        \boldsymbol{S}_k^{-1}
        \widetilde{\bsXi}^{(k)},
        \\
        \bsX^{\top}\bsW_l^{\mathrm{BT}}\bsy
        &=
        \sum_{k=1}^{n_t}
        \left(\widetilde{\bsXi}^{(k)}\right)^{\top}
        \boldsymbol{S}_k^{-1}
        \widetilde{\bsXi}_{\dot u}^{(k)}.
    \end{align}
\end{subequations}%
In general, $\boldsymbol{D}_k,\boldsymbol{E}_k,\boldsymbol{S}_k\in\mathbb{R}^{n_{\mathrm{eq}}\times n_{\mathrm{eq}}}$, $\widetilde{\bsXi}^{(k)}\in\mathbb{R}^{n_{\mathrm{eq}}\times d_{\bstheta}}$, and $\widetilde{\bsXi}_{\dot u}^{(k)}\in\mathbb{R}^{n_{\mathrm{eq}}}$. Thus, no global $(n_t n_{\mathrm{eq}})\times(n_t n_{\mathrm{eq}})$ covariance is formed: only $n_{\mathrm{eq}}\times n_{\mathrm{eq}}$ block operations and solves are required.

These terms are also used in the Gaussian-prior MAP update in \cref{eq:lagged_covariance_map_update}, together with the prior precision $\bsSigma_0^{-1}$ and prior-weighted mean $\bsSigma_0^{-1}\bsmu_0$.

This update has an EM-like structure because the covariance is updated using the current parameter estimate and then held fixed during the subsequent parameter update. We thus also refer to it as an EM-style lagged-covariance update.
However, it should be interpreted as a fixed-point approximation to minimizing the full negative log-likelihood or negative log-posterior objective in \cref{eq:theta_dependent_negative_log_likelihood,eq:theta_dependent_negative_log_posterior}, not as a standard latent-variable EM algorithm.

We use the classical VSI estimate as a deterministic, data-informed initialization to the same weak-form regression, avoiding arbitrary coefficient scales or physically inadmissible starting values. In applications, convergence of both the coefficient iterates and the residual objective should be monitored.

\paragraph{Gaussian posterior with known observation-noise covariance}

Suppose that the observation-noise covariance $\bsSigma_{\bsu}^{(k)}$ is known for all time levels and that the coefficient prior is Gaussian,
\begin{align}
    \label{eq:gaussian_prior_known_noise}
    p(\bstheta \mid \CalM)
    =
    \CalN
    \nospaceleft(
        \bstheta;
        \bsmu_0,
        \bsSigma_0
    \nospaceright).
\end{align}
The posterior update procedure itself is identical for either covariance structure; only the likelihood normal-equation terms differ.
With the lagged covariance $\bsSigma_l^{\CalC}$ fixed, the residual likelihood is Gaussian in $\bstheta$.
The approximate posterior is also Gaussian~\cite{Diaconis1979Conjugate}:
\begin{align}
    \label{eq:lagged_gaussian_posterior}
    p^{(l+1)}(\bstheta \mid \CalD,\CalM)
    =
    \CalN
    \nospaceleft(
        \bstheta;
        \bsmu_{\bstheta}^{(l+1)},
        \bsSigma_{\bstheta}^{(l+1)}
    \nospaceright),
\end{align}
where
\begin{subequations}
    \label{eq:lagged_gaussian_posterior_parameters}
    \begin{align}
        \bsSigma_{\bstheta}^{(l+1)}
        &=
        \left(
            \bsSigma_0^{-1}
            +
            \bsX^{\top}\bsW_l^{\CalC}\bsX
        \right)^{-1},
        \\
        \bsmu_{\bstheta}^{(l+1)}
        &=
        \bsSigma_{\bstheta}^{(l+1)}
        \left(
            \bsSigma_0^{-1}\bsmu_0
            +
            \bsX^{\top}\bsW_l^{\CalC}\bsy
        \right).
    \end{align}
\end{subequations}
For the next lagged-covariance iteration, we use the posterior mean as the representative parameter value and set
\begin{align}
    \label{eq:lagged_posterior_mean_update}
    \bstheta^{(l+1)}
    =
    \bsmu_{\bstheta}^{(l+1)}.
\end{align}
After convergence, the final Gaussian distribution provides an approximate local posterior for the fixed model $\CalM$.

\paragraph{Normal-inverse-gamma posterior with unknown scalar observation-noise variance}

We next consider the case in which the observation-noise covariance has an unknown scalar variance:
\begin{align}
    \label{eq:unknown_noise_observation_covariance}
    \bsSigma_{\bsu}^{(k)}
    =
    \sigma_u^2\bsI,
    \qquad
    k=0,\ldots,n_t.
\end{align}
For either covariance structure, the residual covariance can be written as
\begin{align}
    \label{eq:scaled_residual_covariance}
    \bsSigma_{r,\bstheta}^{\CalC}
    =
    \sigma_u^2
    \widetilde{\bsSigma}_{r,\bstheta}^{\CalC}.
\end{align}

For the block-diagonal structure, the covariance shape is
\begin{align}
    \label{eq:scaled_residual_covariance_blockdiag}
    \widetilde{\bsSigma}_{r,\bstheta}^{\mathrm{BD}}
    =
    \operatorname{blockdiag}
    \nospaceleft(
        \widetilde{\bsSigma}_{r,\bstheta}^{(1)},
        \ldots,
        \widetilde{\bsSigma}_{r,\bstheta}^{(n_t)}
    \nospaceright).
\end{align}
For each time interval,
\begin{align}
    \label{eq:scaled_residual_covariance_per_step}
    \widetilde{\bsSigma}_{r,\bstheta}^{(k)}
    =
    \bsC_{\bstheta}^{(k)}
    {\bsC_{\bstheta}^{(k)}}^{\top}
    +
    \bsB\bsB^{\top},
\end{align}
where $\bsC_{\bstheta}^{(k)}$ is defined in \cref{eq:C_definition}.

For the block-tridiagonal structure, the same diagonal shape blocks are supplemented by adjacent blocks $\widetilde{\boldsymbol{E}}_k=-\bsC_{\bstheta}^{(k)}\bsB^{\top}$, as follows from \cref{eq:stacked_covariance_off_diagonal_block}.

At iteration $l$, define the lagged covariance shape and its inverse by
\begin{align}
    \label{eq:lagged_scaled_covariance}
    \widetilde{\bsSigma}_l^{\CalC}
    =
    \widetilde{\bsSigma}_{r,\bstheta^{(l)}}^{\CalC},
    \qquad
    \widetilde{\bsW}_l^{\CalC}
    =
    \left(\widetilde{\bsSigma}_l^{\CalC}\right)^{-1}.
\end{align}
With this covariance shape held fixed, the residual model is approximated by
\begin{align}
    \label{eq:unknown_noise_linear_model}
    \bsy \mid \bstheta,\sigma_u^2,\CalM
    \approx
    \CalN
    \nospaceleft(
        \bsX\bstheta,
        \sigma_u^2
        \widetilde{\bsSigma}_l^{\CalC}
    \nospaceright).
\end{align}
We use the conjugate normal-inverse-gamma prior
\begin{subequations}
    \label{eq:NIG_prior}
    \begin{align}
        \sigma_u^2
        &\sim
        \operatorname{Inv\text{-}Gamma}
        \nospaceleft(
            \alpha_0,
            \beta_0
        \nospaceright),
        \\
        \bstheta \mid \sigma_u^2,\CalM
        &\sim
        \CalN
        \nospaceleft(
            \bsmu_0,
            \sigma_u^2\bsSigma_0
        \nospaceright).
    \end{align}
\end{subequations}
The approximate posterior is normal-inverse-gamma~\cite{Diaconis1979Conjugate}:
\begin{subequations}
    \label{eq:NIG_posterior}
    \begin{align}
        \sigma_u^2 \mid \CalD,\CalM
        &\sim
        \operatorname{Inv\text{-}Gamma}
        \nospaceleft(
            \alpha^{(l+1)},
            \beta^{(l+1)}
        \nospaceright),
        \\
        \bstheta \mid \sigma_u^2,\CalD,\CalM
        &\sim
        \CalN
        \nospaceleft(
            \bsmu_{\bstheta}^{(l+1)},
            \sigma_u^2
            \bsSigma_{\bstheta}^{(l+1)}
        \nospaceright),
    \end{align}
\end{subequations}
where
\begin{subequations}
    \label{eq:NIG_posterior_parameters}
    \begin{align}
        \bsSigma_{\bstheta}^{(l+1)}
        &=
        \left(
            \bsSigma_0^{-1}
            +
            \bsX^{\top}\widetilde{\bsW}_l^{\CalC}\bsX
        \right)^{-1},
        \\
        \bsmu_{\bstheta}^{(l+1)}
        &=
        \bsSigma_{\bstheta}^{(l+1)}
        \left(
            \bsSigma_0^{-1}\bsmu_0
            +
            \bsX^{\top}\widetilde{\bsW}_l^{\CalC}\bsy
        \right),
        \\
        \alpha^{(l+1)}
        &=
        \alpha_0
        +
        \frac{n}{2},
        \\
        \beta^{(l+1)}
        &=
        \beta_0
        +
        \frac{1}{2}
        \left[
            \bsy^{\top}\widetilde{\bsW}_l^{\CalC}\bsy
            +
            \bsmu_0^{\top}\bsSigma_0^{-1}\bsmu_0
            -
            {\bsmu_{\bstheta}^{(l+1)}}^{\top}
            \left(
                \bsSigma_{\bstheta}^{(l+1)}
            \right)^{-1}
            \bsmu_{\bstheta}^{(l+1)}
        \right].
    \end{align}
\end{subequations}
These normal-inverse-gamma formulas are unchanged between the two covariance structures. For the block-tridiagonal structure, the weighted normal-equation terms and the weighted quadratic form $\bsy^{\top}\widetilde{\bsW}_l^{\mathrm{BT}}\bsy$ are obtained by the recursions in \cref{eq:bt_schur_recursion,eq:bt_regression_recursion}, using the relevant right-hand sides.
The marginal posterior of $\bstheta$ is a multivariate Student-$t$ distribution.
When only posterior moments are needed, the mean of $\bstheta$ is
$\bsmu_{\bstheta}^{(l+1)}$ and, for $\alpha^{(l+1)}>1$, the marginal covariance is
\begin{align}
    \label{eq:theta_marginal_covariance_unknown_noise}
    \operatorname{Cov}
    \left[
        \bstheta \mid \CalD,\CalM
    \right]
    \approx
    \frac{
        \beta^{(l+1)}
    }{
        \alpha^{(l+1)}-1
    }
    \bsSigma_{\bstheta}^{(l+1)}.
\end{align}
The posterior mean of the scalar observation-noise variance is
\begin{align}
    \label{eq:sigma2_posterior_mean_unknown_noise}
    \mathbb{E}
    \left[
        \sigma_u^2 \mid \CalD,\CalM
    \right]
    =
    \frac{
        \beta^{(l+1)}
    }{
        \alpha^{(l+1)}-1
    },
    \qquad
    \alpha^{(l+1)}>1.
\end{align}
Its posterior variance is
\begin{align}
    \label{eq:sigma2_posterior_variance_unknown_noise}
    \operatorname{Var}
    \left[
        \sigma_u^2 \mid \CalD,\CalM
    \right]
    =
    \frac{
        \left(\beta^{(l+1)}\right)^2
    }{
        \left(\alpha^{(l+1)}-1\right)^2
        \left(\alpha^{(l+1)}-2\right)
    },
    \qquad
    \alpha^{(l+1)}>2.
\end{align}
For the next lagged-covariance iteration, we use the posterior mean of $\bstheta$ as the representative parameter value and set, as in \cref{eq:lagged_posterior_mean_update},
\begin{align}
    \bstheta^{(l+1)}
    =
    \bsmu_{\bstheta}^{(l+1)}.
    \nonumber
\end{align}

\subsubsection{Full \texorpdfstring{$\bstheta$}{theta}-dependent inference}
\label{sec:full_theta_dependent_inference}

The lagged-covariance updates are efficient, but they approximate the full residual-space objective by holding the covariance fixed during each parameter update.
For non-Gaussian priors, non-Gaussian posteriors, or direct optimization of the full objective in \cref{eq:theta_dependent_negative_log_likelihood,eq:theta_dependent_negative_log_posterior}, we use gradient-based and particle-based methods.

\paragraph{Gradient-based MLE and MAP}

For point estimation, MLE and MAP estimates are obtained by minimizing
\cref{eq:theta_dependent_negative_log_likelihood} and
\cref{eq:theta_dependent_negative_log_posterior}, respectively.
A basic gradient descent update is
\begin{align}
    \label{eq:gradient_descent}
    \bstheta^{(l+1)}
    =
    \bstheta^{(l)}
    -
    \lambda
    \nabla_{\bstheta}
    \CalL
    \nospaceleft(
        \bstheta^{(l)}
    \nospaceright),
\end{align}
where $\lambda$ is the learning rate and $\CalL$ denotes either the negative log-likelihood or the negative log-posterior.
More advanced first-order or quasi-Newton optimizers may be used in practice.
When the prior is non-smooth, such as a Laplace prior, subgradient, proximal, or smoothed-prior variants may be used.

The residual-space formulation avoids repeated forward PDE solves during these updates.
Both the residual vector and the residual covariance are assembled from the observed fields and the weak-form operators.
If analytic derivatives are available, the $i$-th component of the gradient of the negative log-likelihood in \cref{eq:theta_dependent_negative_log_likelihood} can be written as
\begin{align}
    \label{eq:full_likelihood_gradient_component}
    \frac{\partial \CalL_{\mathrm{res}}}{\partial \theta_i}
    =
    -
    \bsX_i^{\top}
    \bsW_{\bstheta}
    \bsr_{\bstheta}
    +
    \frac{1}{2}
    \operatorname{tr}
    \nospaceleft(
        \bsW_{\bstheta}
        \frac{\partial \bsSigma_{r,\bstheta}^{\CalC}}{\partial \theta_i}
    \nospaceright)
    -
    \frac{1}{2}
    \bsr_{\bstheta}^{\top}
    \bsW_{\bstheta}
    \frac{\partial \bsSigma_{r,\bstheta}^{\CalC}}{\partial \theta_i}
    \bsW_{\bstheta}
    \bsr_{\bstheta},
\end{align}
where
\begin{align}
    \bsW_{\bstheta}
    =
    \left(
        \bsSigma_{r,\bstheta}^{\CalC}
    \right)^{-1},
\end{align}
and $\bsX_i$ denotes the $i$-th column of $\bsX$.
Automatic differentiation may also be used.

\paragraph{SVGD posterior approximation}

For posterior approximation, we use Stein variational gradient descent (SVGD)~\cite{Liu2016SVGD} when a particle representation is preferred.
Let
$\big\{\bstheta_i^{(l)}\big\}_{i=1}^{n_p}$
denote particles at iteration $l$.
SVGD updates the particles according to
\begin{subequations}
    \label{eq:SVGD}
    \begin{align}
        \bstheta_i^{(l+1)}
        &=
        \bstheta_i^{(l)}
        +
        \lambda
        \widehat{\bsphi}
        \nospaceleft(
            \bstheta_i^{(l)}
        \nospaceright),
        \\
        \widehat{\bsphi}(\bstheta)
        &=
        \frac{1}{n_p}
        \sum_{j=1}^{n_p}
        \left[
            \kappa
            \nospaceleft(
                \bstheta_j^{(l)},
                \bstheta
            \nospaceright)
            \nabla_{\bstheta_j}
            \log
            p_{\mathrm{res}}
            \nospaceleft(
                \bstheta_j^{(l)} \mid \CalD,\CalM
            \nospaceright)
            +
            \nabla_{\bstheta_j}
            \kappa
            \nospaceleft(
                \bstheta_j^{(l)},
                \bstheta
            \nospaceright)
        \right],
    \end{align}
\end{subequations}
where $\kappa$ is a positive-definite kernel, such as a radial basis function.
The first term drives particles toward regions of high posterior probability, while the second term promotes particle diversity.
The resulting target density is the residual-space posterior $p_{\mathrm{res}}(\bstheta \mid \CalD,\CalM)$ from \cref{eq:residual_space_posterior_intro}.

\subsubsection{Hybrid initialization and constrained inference}
\label{sec:hybrid_constraints}

The lagged-covariance approximation provides an efficient initialization for gradient-based inference.
For point estimation, the converged MLE or MAP estimate from \cref{sec:lagged_covariance_inference} can be used to initialize optimization of the full negative log-likelihood or negative log-posterior where the $\bstheta$-dependence of the residual covariance is retained.
For particle-based posterior inference, particles can be initialized from the approximate Gaussian posterior,
\begin{align}
    \label{eq:hybrid_particle_initialization}
    \bstheta_i^{(0)}
    \sim
    \CalN
    \nospaceleft(
        \bsmu_{\bstheta},
        \bsSigma_{\bstheta}
    \nospaceright),
    \qquad
    i=1,\ldots,n_p,
\end{align}
and then refined using SVGD under the desired prior and residual-space likelihood.
This strategy reduces sensitivity to random initialization and improves convergence when the target posterior is non-Gaussian.

When the desired prior is non-Gaussian but has finite moments, one may use a Gaussian moment-matched approximation in the lagged-covariance phase and then refine the posterior using the true prior in the gradient-based or SVGD phase.
This is the hybrid strategy used in the examples below.

Physical constraints on the parameters can be incorporated through priors with constrained support, reparameterization, projected particle updates, or constrained optimization.
For example, diffusion coefficients are often required to be positive.
When using the lagged-covariance Gaussian approximation, linear inequality constraints can be enforced by solving a quadratic program at each iteration:
\begin{subequations}
    \label{eq:constrained_lagged_update}
    \begin{align}
        \bstheta_{\mathrm{MAP}}^{(l+1)}
        =
        \argmin_{\bstheta}
        \quad
        &
        \frac{1}{2}
        \left(
            \bsy-\bsX\bstheta
        \right)^{\top}
        \bsW_l
        \left(
            \bsy-\bsX\bstheta
        \right)
        +
        \frac{1}{2}
        \left(
            \bstheta-\bsmu_0
        \right)^{\top}
        \bsSigma_0^{-1}
        \left(
            \bstheta-\bsmu_0
        \right),
        \\
        \text{subject to}
        \quad
        &
        \bsA_c\bstheta
        \leq
        \bsb_c .
    \end{align}
\end{subequations}
Here, $\bsA_c$ and $\bsb_c$ define the constraint set.
The prior term can be omitted for constrained MLE.

\subsubsection{Fixed-model inference algorithm}
\label{sec:parameter_inference_algorithm}

\Cref{alg:parameter_inference_fixed_model} summarizes posterior inference for a fixed model $\CalM$.

\begin{algorithm}[htbp]
    \caption{Posterior inference for a fixed model}
    \label{alg:parameter_inference_fixed_model}
    \SetKwInOut{Input}{Input}
    \SetKwInOut{Output}{Output}

    \Input{dataset $\CalD=\{\hbsu^{(k)}\}_{k=0}^{n_t}$; candidate model $\CalM$; prior $p(\bstheta \mid \CalM)$; observation-noise model; covariance structure $\CalC\in\{\mathrm{BD},\mathrm{BT}\}$;}
    \Output{point estimate $\bstheta_{\mathrm{MLE}}$ or $\bstheta_{\mathrm{MAP}}$, approximate posterior, or particle approximation;}

    Construct $\bsy$, $\bsX$, and the residual sensitivity matrices needed for $\bsSigma_{r,\bstheta}^{(k)}$\;
    Initialize $\bstheta^{(0)}$ using the VSI estimate in \cref{eq:VSI_solution}\;

    \tcp{Lagged-covariance phase}
    \While{$\bstheta^{(l)}$ not converged}{
        \eIf{$\CalC=\mathrm{BD}$}{
            Assemble the diagonal blocks $\{\boldsymbol{D}_k\}_{k=1}^{n_t}$ and compute the covariance-weighted sufficient statistics using \cref{eq:lagged_covariance_normal_equations}\;
        }{
            Assemble $\{\boldsymbol{D}_k\}_{k=1}^{n_t}$ and $\{\boldsymbol{E}_k\}_{k=1}^{n_t-1}$, perform the sequential Schur-complement recursion in \cref{eq:bt_schur_recursion}, and compute the sufficient statistics using \cref{eq:bt_lagged_covariance_normal_equations}\;
        }

        \eIf{MLE or MAP is desired}{
            Update $\bstheta_{\mathrm{MLE}}^{(l+1)}$ using \cref{eq:lagged_covariance_mle_update}, or $\bstheta_{\mathrm{MAP}}^{(l+1)}$ using \cref{eq:lagged_covariance_map_update}\;
            Set $\bstheta^{(l+1)}=\bstheta_{\mathrm{MLE}}^{(l+1)}$, or $\bstheta^{(l+1)}=\bstheta_{\mathrm{MAP}}^{(l+1)}$\;
        }{
            \eIf{the observation-noise covariance is known and a Gaussian prior is used}{
                Update the Gaussian posterior parameters using \cref{eq:lagged_gaussian_posterior_parameters}\;
                Set $\bstheta^{(l+1)}=\bsmu_{\bstheta}^{(l+1)}$\;
            }{
                Update the normal-inverse-gamma posterior parameters using \cref{eq:NIG_posterior_parameters}\;
                Set $\bstheta^{(l+1)}=\bsmu_{\bstheta}^{(l+1)}$\;
            }
        }
    }

    \tcp{Optional full-objective refinement}
    \If{the full $\bstheta$-dependent likelihood, a non-Gaussian prior, or a particle posterior is desired}{
        Initialize the optimizer or particles from the lagged-covariance estimate\;
        \While{the optimizer or particles have not converged}{
            Update point estimates using \cref{eq:gradient_descent}, or update posterior particles using \cref{eq:SVGD}\;
        }
    }
\end{algorithm}

\subsection{Posterior predictive uncertainty}
\label{sec:posterior_predictive}

Given a selected model (by the procedure in \cref{sec:model_selection}) and a coefficient posterior obtained by the methods in \cref{sec:posterior_inference_fixed_model}, we often wish to propagate parameter uncertainty to derived physical quantities.
In the examples in \cref{sec:examples}, these quantities include the potential function in the FP equation and the free-energy derivatives in the CH equation.
For a fixed model $\CalM$, let
\begin{align}
    \bspsi
    =
    g(\bstheta;\CalM)
\end{align}
denote a quantity of interest determined by the model coefficients.
The posterior predictive distribution is the pushforward of the coefficient posterior through $g$:
\begin{align}
    \label{eq:posterior_predictive_general}
    p(\bspsi \mid \CalD,\CalM)
    =
    \int
    p(\bspsi \mid \bstheta,\CalM) \,
    p(\bstheta \mid \CalD,\CalM)
    \,\mathrm{d}\bstheta .
\end{align}
When $\bspsi$ is a deterministic function of $\bstheta$, the conditional distribution
$p(\bspsi \mid \bstheta,\CalM)$ is a Dirac measure centered at $g(\bstheta;\CalM)$.
After model selection, the posterior predictive distribution is evaluated with $\CalM=\CalM^*$.

\paragraph{Gaussian predictive}

Suppose the coefficient posterior is Gaussian,
\begin{align}
    p(\bstheta \mid \CalD,\CalM)
    =
    \CalN
    \nospaceleft(
        \bstheta;
        \bsmu_{\bstheta},
        \bsSigma_{\bstheta}
    \nospaceright),
\end{align}
and the quantity of interest depends linearly on the coefficients:
\begin{align}
    \label{eq:linear_predictive_map}
    \bspsi
    =
    \bsPhi\bstheta .
\end{align}
Here, $\bsPhi$ is the evaluation or basis matrix that maps coefficients to the desired physical quantity.
For example, $\bsPhi$ may evaluate a polynomial potential on a spatial grid or evaluate free-energy derivatives on a grid in concentration space.
Then the posterior predictive distribution is Gaussian:
\begin{align}
    \label{eq:posterior_predictive_gaussian}
    p(\bspsi \mid \CalD,\CalM)
    =
    \CalN
    \nospaceleft(
        \bspsi;
        \bsPhi\bsmu_{\bstheta},
        \bsPhi\bsSigma_{\bstheta}\bsPhi^{\top}
    \nospaceright).
\end{align}

\paragraph{Monte Carlo approximation}

For nonlinear quantities of interest or non-Gaussian coefficient posteriors, the posterior predictive distribution can be approximated using posterior samples or particles.
Let $\{\bstheta^{(i)}\}_{i=1}^{n_p}$ denote samples or particles from $p(\bstheta \mid \CalD,\CalM)$.
For SVGD, these correspond to the final particle ensemble.
Each coefficient sample is pushed forward through the quantity-of-interest map:
\begin{align}
    \label{eq:posterior_predictive_samples}
    \bspsi^{(i)}
    =
    g
    \nospaceleft(
        \bstheta^{(i)};\CalM
    \nospaceright),
    \qquad
    i=1,\ldots,n_p .
\end{align}
The empirical distribution of $\{\bspsi^{(i)}\}_{i=1}^{n_p}$ then approximates $p(\bspsi \mid \CalD,\CalM)$.
Posterior predictive summaries, such as means, variances, credible intervals, and marginal densities, can be computed directly from this empirical distribution.
For field-valued quantities, these summaries are evaluated pointwise or componentwise over the field.
Histograms, kernel density estimates, or empirical quantiles can also be constructed when a more detailed representation of the predictive distribution is desired.

In this work, posterior prediction focuses on coefficient-derived quantities, including the FP potential and CH free-energy derivatives. Predictive uncertainty in the PDE state can also be quantified through Monte Carlo propagation of posterior coefficient samples through the forward PDE model, although this requires additional computationally expensive forward solves, together with initial and boundary conditions and an observation- or process-error model.

\subsection{Residual-space model comparison and operator selection}
\label{sec:model_selection}

A candidate model $\CalM \subseteq \bschi$ is a subset of operators selected from the dictionary.
The model dimension is the number of active coefficients,
$d_{\bstheta}=|\CalM|$.
An exhaustive search over all subsets of a dictionary with $d=|\bschi|$ candidate operators requires evaluating $\CalO(2^d)$ models, which is generally impractical for large libraries.
We therefore use a greedy sequential operator-elimination procedure guided by a residual-space Bayesian information criterion (BIC)~\cite{Schwarz1978BIC}.

The goal of this step is to select a parsimonious model form.
For computational efficiency, model comparison is performed using point estimates under each candidate model.
Full posterior inference for $\bstheta$ is then carried out only after the final model $\CalM^*$ has been selected.

\subsubsection{Residual-space Bayesian information criterion}
\label{sec:residual_bic}

For model comparison, the Bayesian posterior probability of a candidate model is
\begin{align}
    p(\CalM \mid \CalD)
    =
    \frac{
        p(\CalD \mid \CalM)\,p(\CalM)
    }{
        \sum_{\CalM_i}
        p(\CalD \mid \CalM_i)\,p(\CalM_i)
    }.
\end{align}
Under a uniform prior over candidate models, selecting the most probable model is equivalent to maximizing the model evidence,
\begin{align}
    p(\CalD \mid \CalM)
    =
    \int
    p(\CalD \mid \bstheta,\CalM)
    \,p(\bstheta \mid \CalM)
    \,\mathrm{d}\bstheta .
\end{align}

In B-VSI, we consider instead a residual-space evidence, defined through the residual-space likelihood.
Specifically, we define the residual-space model evidence as
\begin{align}
    \label{eq:residual_model_evidence}
    p_{\mathrm{res}}(\CalD \mid \CalM)
    =
    \int
    p_{\mathrm{res}}(\CalD \mid \bstheta,\CalM) \,
    p(\bstheta \mid \CalM)
    \,\mathrm{d}\bstheta ,
\end{align}
where
$p_{\mathrm{res}}(\CalD \mid \bstheta,\CalM)
=
p^{\CalC}(\bsr_{\bstheta,\CalM} \mid \bstheta,\CalM)$
is the residual-space likelihood introduced in \cref{eq:residual_likelihood_theta_dependent}, evaluated with a specified covariance structure $\CalC\in\{\mathrm{BD},\mathrm{BT}\}$.

Under the usual regularity assumptions for the large-sample Laplace/BIC approximation, applying this approximation to \cref{eq:residual_model_evidence} gives
\begin{align}
    \label{eq:log_residual_model_evidence_bic}
    \log p_{\mathrm{res}}(\CalD \mid \CalM)
    \approx
    \log
    p_{\mathrm{res}}
    \nospaceleft(
        \CalD \mid \widehat{\bstheta}_{\mathrm{MLE},\CalM},\CalM
    \nospaceright)
    -
    \frac{d_{\bstheta}}{2}
    \log n_r,
\end{align}
where
$\widehat{\bstheta}_{\mathrm{MLE},\CalM}$
is the residual-space MLE under model $\CalM$ and
$n_r=\dim(\bsr_{\bstheta,\CalM})$ is the number of stacked residual equations.
For a scalar state with $n_t$ time intervals and $\Ndof$ spatial degrees of freedom, $n_r=\Ndof n_t$.

Equivalently, maximizing the Laplace approximation to the residual-space evidence in \cref{eq:log_residual_model_evidence_bic} corresponds to minimizing the residual-space BIC:
\begin{align}
    \mathrm{BIC}_{\mathrm{res}}(\CalM)
&    =
    -2
    \log
    p_{\mathrm{res}}
    \nospaceleft(
        \CalD \mid \widehat{\bstheta}_{\mathrm{MLE},\CalM},\CalM
    \nospaceright)
    +
    d_{\bstheta}
    \log n_r
    \nonumber\\
    \label{eq:residual_bic}
&    =
    -2
    \log
    p^{\CalC}
    \nospaceleft(
        \bsr_{\widehat{\bstheta}_{\mathrm{MLE},\CalM},\CalM}
        \mid
        \widehat{\bstheta}_{\mathrm{MLE},\CalM},
        \CalM
    \nospaceright)
    +
    d_{\bstheta}
    \log n_r .
\end{align}
The first term rewards agreement between the observed data and the candidate weak-form equation in residual space.
The second term penalizes model complexity.
Thus, $\mathrm{BIC}_{\mathrm{res}}$ favors models that explain the residuals well while avoiding unnecessary operators.

When evaluating \cref{eq:residual_bic}, the likelihood includes the covariance and log-determinant terms from \cref{eq:residual_likelihood_theta_dependent}.
In practice, $\widehat{\bstheta}_{\mathrm{MLE},\CalM}$ can be obtained using the lagged-covariance MLE update in \cref{eq:lagged_covariance_mle_update}.
If the full $\bstheta$-dependent likelihood is targeted, the MLE may instead be refined using the gradient-based methods in \cref{sec:full_theta_dependent_inference}.

The two covariance structures provide alternative evaluations of the residual-space MLE and likelihood entering this criterion; the block-diagonal structure introduces a covariance approximation but does not alter the model-comparison construction.
Model comparison is performed using this criterion under a choice of covariance structure, and its interpretation is specific to the residual-space formulation rather than observation-space model evidence.

\subsubsection{Sequential operator elimination}
\label{sec:sequential_operator_elimination}

Sequential operator elimination provides a computationally tractable approximation to exhaustive model search.
The procedure begins with the full candidate model
\begin{align}
    \CalM^{(0)}=\bschi .
\end{align}
If some operators are known a priori to be part of the governing equation, they may be placed in a protected set and excluded from elimination.
For example, a diffusion operator may be retained throughout model selection when the physical form of the PDE requires it.

At iteration $s$, let $\CalM^{(s)}$ denote the current model.
For each removable operator $F_j\in\CalM^{(s)}$, define the reduced model
\begin{align}
    \CalM_{-j}^{(s)}
    =
    \CalM^{(s)}
    \setminus
    \{F_j\}.
\end{align}
Each reduced model is fitted to the data by computing
$\widehat{\bstheta}_{\mathrm{MLE},\CalM_{-j}^{(s)}}$,
and its residual-space BIC is evaluated using \cref{eq:residual_bic}.
The operator removal that yields the smallest BIC is accepted if it improves upon the current model:
\begin{align}
    \label{eq:sequential_elimination_step}
    j^*
    =
    \argmin_j
    \mathrm{BIC}_{\mathrm{res}}
    \nospaceleft(
        \CalM_{-j}^{(s)}
    \nospaceright).
\end{align}
If
\begin{align}
    \mathrm{BIC}_{\mathrm{res}}
    \nospaceleft(
        \CalM_{-j^*}^{(s)}
    \nospaceright)
    <
    \mathrm{BIC}_{\mathrm{res}}
    \nospaceleft(
        \CalM^{(s)}
    \nospaceright),
\end{align}
we set
\begin{align}
    \CalM^{(s+1)}
    =
    \CalM_{-j^*}^{(s)}.
\end{align}
Otherwise, the elimination process terminates.

The selected model is the model with the lowest residual-space BIC encountered during this procedure:
\begin{align}
    \label{eq:selected_model_bic}
    \CalM^*
    =
    \argmin_{\CalM \in \CalS_{\mathrm{seq}}}
    \mathrm{BIC}_{\mathrm{res}}(\CalM),
\end{align}
where $\CalS_{\mathrm{seq}}$ denotes the set of models visited by the sequential elimination path.
After $\CalM^*$ is selected, posterior inference for the coefficients is performed under the fixed model $\CalM^*$ using the methods in \cref{sec:posterior_inference_fixed_model}.

Stepwise operator elimination follows earlier VSI work, where statistical tests were used to guide elimination under varying data fidelity and noise~\cite{Wang2019Variational}, with subsequent work introducing a confirmation test for inferred operators~\cite{Wang2021Variational}. In B-VSI, we retain this sequential elimination strategy but use the residual-space BIC as the model-selection criterion.

The resulting backward-elimination procedure reduces the search from exhaustive enumeration of $2^d$ candidate models to $\CalO(d^2)$ model evaluations, at the cost of exploring only a restricted path through the model space. Consequently, the selected model is not guaranteed to recover the globally optimal or true operator support, particularly when candidate operators are strongly correlated, weakly excited, or poorly resolved by the available data.

Nevertheless, the operator search is modular: the residual-space likelihood and BIC criterion in \cref{eq:residual_bic} can be combined with alternative search strategies without changing their formulation. More flexible approaches, such as floating search that allows previously eliminated operators to re-enter the model~\cite{Pudil1994Floating}, exact subset selection~\cite{FurnivalWilson1974,BertsimasKingMazumder2016}, or Bayesian optimization~\cite{Baptista2018BOCS,Oh2019COMBO}, can explore the model space more thoroughly at additional computational cost. \Cref{app:large-dictionary-nested} demonstrates that floating search improves model recovery over greedy elimination in enlarged FP dictionaries with correlated operators, at the cost of additional model evaluations.

\subsection{Complete B-VSI workflow}
\label{sec:algorithm_summary}

\Cref{alg:B-VSI} summarizes the complete B-VSI workflow.
The procedure has three main stages.
First, a parsimonious model form is selected by sequential operator elimination using the residual-space BIC described in \cref{sec:model_selection}.
Second, posterior inference is performed for the coefficients of the selected model $\CalM^*$ using the fixed-model methods in \cref{sec:posterior_inference_fixed_model}.
Third, when quantities of interest are specified, the coefficient posterior is propagated to posterior predictive distributions using the procedures in \cref{sec:posterior_predictive}.
The covariance structure is user-specified for both model selection and fixed-model inference, allowing a trade-off between computational cost and approximation error arising from the truncation of temporal correlations.

During model selection, full posterior inference is not required for every candidate model.
Only a point estimate $\widehat{\bstheta}_{\mathrm{MLE},\CalM}$ is needed to evaluate $\mathrm{BIC}_{\mathrm{res}}(\CalM)$.
In the examples below, this estimate is obtained using the lagged-covariance MLE update in \cref{eq:lagged_covariance_mle_update}.
If the full $\bstheta$-dependent residual likelihood is desired for model comparison, this estimate may be refined using the gradient-based MLE procedure in \cref{sec:full_theta_dependent_inference}.
Posterior inference is then carried out only once, after $\CalM^*$ has been selected.
This final inference step may use the Gaussian approximation, the normal-inverse-gamma update for unknown scalar noise variance, SVGD, or the hybrid strategy, depending on the prior and likelihood assumptions.
The resulting coefficient posterior can then be pushed forward through a quantity-of-interest map to obtain posterior predictive uncertainty in derived physical quantities, such as potentials or free-energy derivatives.

\begin{algorithm}[htbp]
    \caption{Bayesian Variational System Identification}
    \label{alg:B-VSI}
    \SetKwInOut{Input}{Input}
    \SetKwInOut{Output}{Output}

    \Input{dataset $\CalD=\{\hbsu^{(k)}\}_{k=0}^{n_t}$; candidate operator dictionary $\bschi$; observation-noise model; prior $p(\bstheta \mid \CalM)$; covariance structure $\CalC\in\{\mathrm{BD},\mathrm{BT}\}$; optional protected operator set $\CalM_{\mathrm{fix}}\subseteq\bschi$;}
    \Output{selected model $\CalM^*$; coefficient posterior $p(\bstheta \mid \CalD,\CalM^*)$; optional posterior predictive distribution;}

    \tcp{Residual-space model selection}
    Set $s=0$ and initialize $\CalM^{(0)}=\bschi$\;
    Fit $\CalM^{(0)}$ with covariance structure $\CalC$ by computing $\widehat{\bstheta}_{\mathrm{MLE},\CalM^{(0)}}$\;
    Compute $c^{(0)}=\mathrm{BIC}_{\mathrm{res}}(\CalM^{(0)})$ using \cref{eq:residual_bic}\;
    Set $\CalS_{\mathrm{seq}}=\{(\CalM^{(0)},c^{(0)})\}$\;
    Set $\mathrm{improved}=\mathrm{true}$\;

    \While{$\mathrm{improved}$}{
        Set $\CalM_{\mathrm{best}}=\CalM^{(s)}$ and $c_{\mathrm{best}}=c^{(s)}$\;

        \ForEach{$F_j\in \CalM^{(s)}\setminus\CalM_{\mathrm{fix}}$}{
            Define the reduced model
            $\CalM_{-j}^{(s)}=\CalM^{(s)}\setminus\{F_j\}$\;
            Fit $\CalM_{-j}^{(s)}$ with covariance structure $\CalC$ by computing
            $\widehat{\bstheta}_{\mathrm{MLE},\CalM_{-j}^{(s)}}$\;
            Compute
            $c_{-j}^{(s)}=\mathrm{BIC}_{\mathrm{res}}(\CalM_{-j}^{(s)})$
            using \cref{eq:residual_bic}\;
            Add $(\CalM_{-j}^{(s)},c_{-j}^{(s)})$ to $\CalS_{\mathrm{seq}}$\;

            \If{$c_{-j}^{(s)} < c_{\mathrm{best}}$}{
                Set $\CalM_{\mathrm{best}}=\CalM_{-j}^{(s)}$\;
                Set $c_{\mathrm{best}}=c_{-j}^{(s)}$\;
            }
        }

        \eIf{$c_{\mathrm{best}} < c^{(s)}$}{
            Set $s=s+1$\;
            Set $\CalM^{(s)}=\CalM_{\mathrm{best}}$\;
            Set $c^{(s)}=c_{\mathrm{best}}$\;
        }{
            Set $\mathrm{improved}=\mathrm{false}$\;
        }
    }

    Select
    $\CalM^*=\argmin_{\CalM\in\CalS_{\mathrm{seq}}}
    \mathrm{BIC}_{\mathrm{res}}(\CalM)$\;

    \tcp{Posterior inference under the selected model}
    Construct $\bsy$, $\bsX$, and the residual sensitivity matrices for $\CalM^*$\;
    Approximate $p(\bstheta \mid \CalD,\CalM^*)$ using covariance structure $\CalC$ and the fixed-model inference methods in \cref{alg:parameter_inference_fixed_model}\;

    \tcp{Optional posterior prediction}
    \If{a quantity of interest $\bspsi=g(\bstheta;\CalM^*)$ is specified}{
        Propagate $p(\bstheta \mid \CalD,\CalM^*)$ to
        $p(\bspsi \mid \CalD,\CalM^*)$ using \cref{sec:posterior_predictive}\;
    }
\end{algorithm}

\section{Numerical Examples}
\label{sec:examples}

We demonstrate the B-VSI framework on two representative PDE discovery problems.
The first is a state-linear Fokker--Planck equation, and the second is a nonlinear two-field Cahn--Hilliard equation.
These examples test the proposed residual-space likelihood, posterior inference, and model-selection procedures in both linear and nonlinear settings.

For each problem, we construct a candidate operator dictionary, carry out residual-space model selection and posterior inference for the selected coefficients, and propagate parameter uncertainty to physically meaningful quantities of interest.
For the FP equation, this quantity is the potential function.
For the CH equation, the quantities are the free-energy derivatives.

To further assess the robustness and flexibility of the framework, \cref{app:repeated-noise} examines the variability in model selection, coefficient inference, and posterior uncertainty across independent observation-noise realizations for both examples, with noise-free trajectories and numerical settings held fixed.
\Cref{app:computational_performance} reports the wall-clock times of the computations in this section.
\Cref{app:large-dictionary} extends the FP example to candidate dictionaries of up to 57 operators, including a non-polynomial dictionary that cannot exactly represent the true potential.

The examples are implemented in Python using the DOLFINx finite element library~\cite{baratta_2023_10447666} for PDE solves.

\subsection{State-linear Fokker--Planck equation}
\label{sec:example_FP}

We first consider a state-linear example governed by the FP equation.
The FP equation describes the time evolution of the probability density function of a stochastic process~\cite{Risken1996FP}.
It is commonly used to model particles subject to deterministic drift and random fluctuations.
We consider the form
\begin{align}
    \label{eq:FP_equation}
    \pp{p(\bsx,t)}{t}
    =
    \grad \cdot
    \big(
        p(\bsx,t)\grad\psi(\bsx)
    \big)
    +
    \beta^{-1}\grad^2 p(\bsx,t),
\end{align}
where $p(\bsx,t):\Omega\times\CalT\to\RR^{+}$ is the probability density,
$\psi(\bsx):\Omega\to\RR$ is the potential function, and $\beta>0$ is the inverse temperature.
In the notation of \cref{eq:strong_form_pde}, the state variable is $u\equiv p$.
For a fixed potential $\psi$, the FP equation is linear in the state $p$.
After expressing $\psi$ in a basis that is linear in unknown coefficients, the governing equation is also linear in the coefficient vector to be estimated.

\subsubsection{Data generation}
\label{sec:FP_data_generation}

\begin{figure}[htbp]
    \centering
    \includegraphics[width=0.5\textwidth]{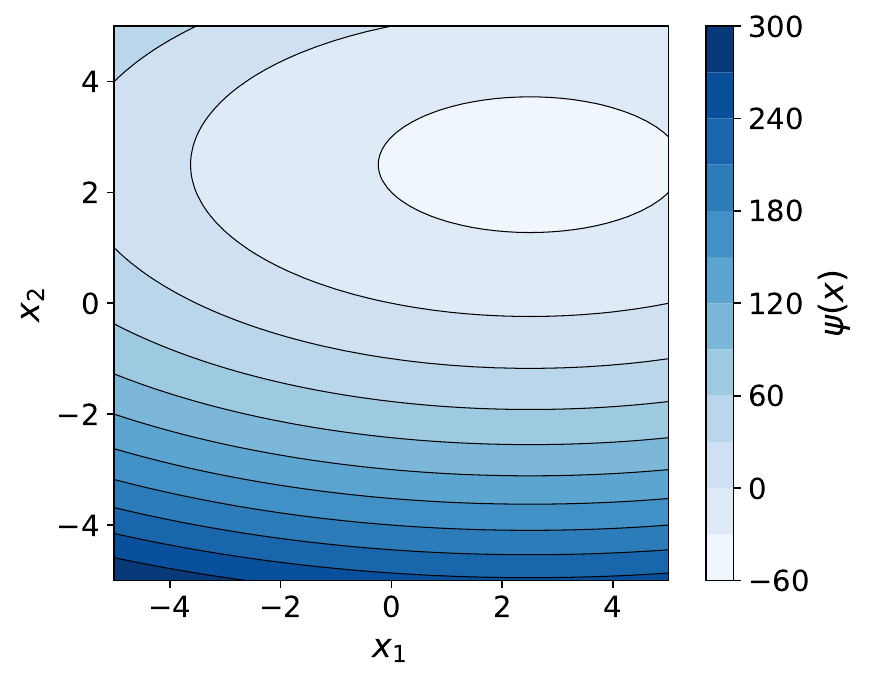}
    \caption{Ground-truth potential function $\psi_{\mathrm{GT}}$ in the FP equation.}
    \label{fig:FP_potential_function}
\end{figure}

The spatial domain is $\Omega=(-5,5)^2$.
We prescribe the convex quadratic potential
\begin{align}
    \label{eq:FP_true_potential}
    \psi_{\mathrm{GT}}(\bsx)
    =
    (\bsx-\bsc)^{\top}\bsW(\bsx-\bsc),
    \qquad
    \bsc=
    \begin{bmatrix}
        2.5\\
        2.5
    \end{bmatrix},
    \qquad
    \bsW=
    \begin{bmatrix}
        1 & 0\\
        0 & 5
    \end{bmatrix}.
\end{align}
The potential achieves its minimum at $\bsx=\bsc$ and is shown in \cref{fig:FP_potential_function}.
Although $\psi_{\mathrm{GT}}$ is quadratic in $\bsx$, the FP equation remains linear in $p$ because the differential operators act linearly on the state.
Since only $\grad\psi$ enters \cref{eq:FP_equation}, the potential is identifiable only up to an additive constant.
All comparisons of recovered potentials below use a common additive-constant convention.

The initial density is a Gaussian distribution,
\begin{align}
    \label{eq:FP_initial_condition}
    p(\bsx,0)
    =
    \CalN
    \nospaceleft(
        \bsx;
        \bsx_0,
        \bsSigma_{\bsx_0}
    \nospaceright),
    \qquad
    \bsx_0=
    \begin{bmatrix}
        -2\\
        -2
    \end{bmatrix},
    \qquad
    \bsSigma_{\bsx_0}
    =
    \begin{bmatrix}
        0.2 & -0.3\\
        -0.3 & 0.5
    \end{bmatrix}.
\end{align}
We impose homogeneous no-flux boundary conditions on the density,
\begin{align}
    \label{eq:FP_boundary_condition}
    (p\nabla \psi+\beta^{-1}\nabla p)\cdot\bsn = 0
    \qquad
    \text{on } \partial\Omega,
\end{align}
where $\bsn$ is the outward unit normal on $\partial\Omega$.

\begin{figure}[htbp]
    \centering
    \begin{subfigure}[t]{0.18\textwidth}
        \centering
        \includegraphics[height=3cm, trim={0 0.45cm 1.3cm 0.95cm}, clip]{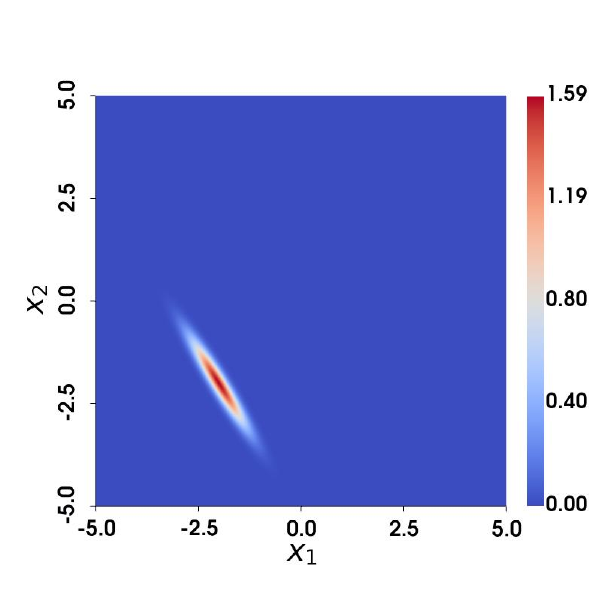}
        \caption{$p^{(0)}$ at $t=0$}
    \end{subfigure}
    \hfill
    \begin{subfigure}[t]{0.18\textwidth}
        \centering
        \includegraphics[height=3cm, trim={0 0.45cm 1.3cm 0.95cm}, clip]{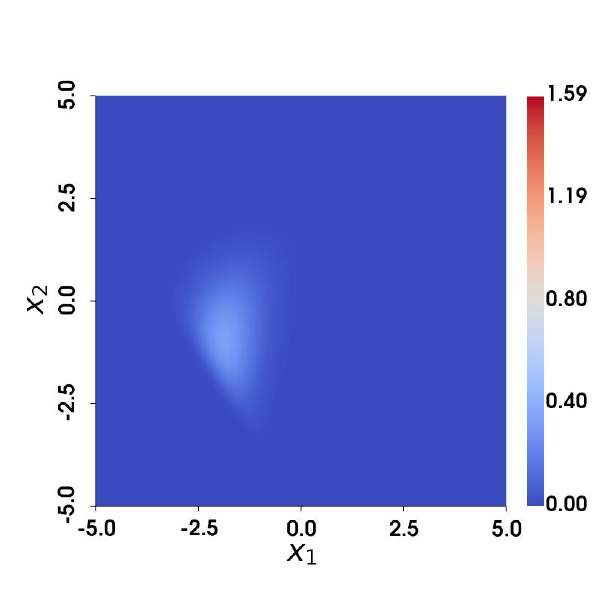}
        \caption{$p^{(2)}$ at $t=0.04$}
    \end{subfigure}
    \hfill
    \begin{subfigure}[t]{0.18\textwidth}
        \centering
        \includegraphics[height=3cm, trim={0 0.45cm 1.3cm 0.95cm}, clip]{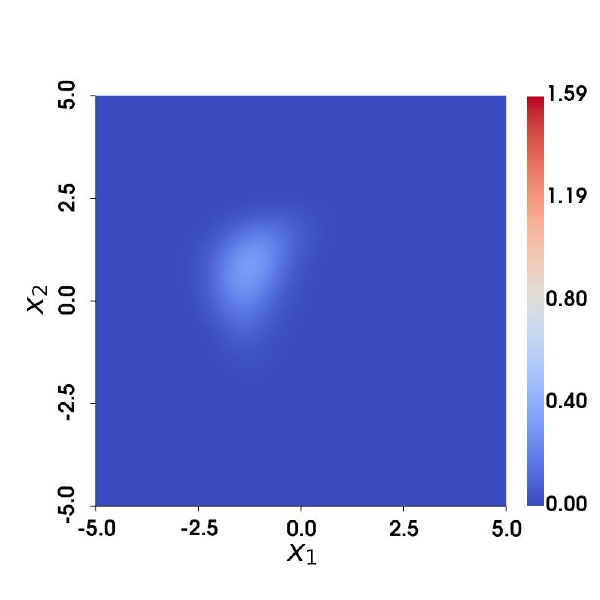}
        \caption{$p^{(5)}$ at $t=0.1$}
    \end{subfigure}
    \hfill
    \begin{subfigure}[t]{0.18\textwidth}
        \centering
        \includegraphics[height=3cm, trim={0 0.45cm 1.3cm 0.95cm}, clip]{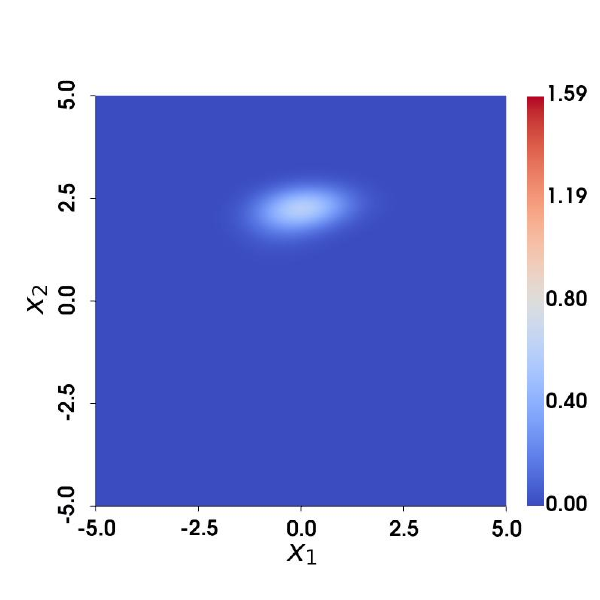}
        \caption{$p^{(15)}$ at $t=0.3$}
    \end{subfigure}
    \hfill
    \begin{subfigure}[t]{0.225\textwidth}
        \centering
        \includegraphics[height=3cm, trim={0 0.45cm 0 0.95cm}, clip]{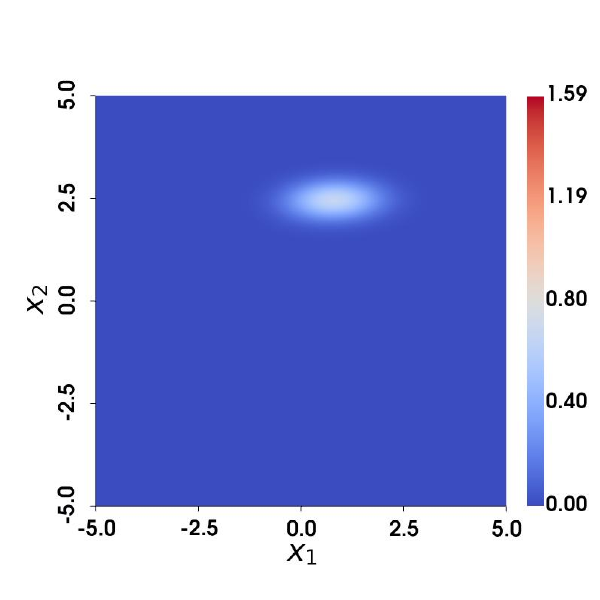}
        \caption{$p^{(25)}$ at $t=0.5$}
    \end{subfigure}
    \caption{Time evolution of the probability density $p(\bsx,t)$ in the FP equation at selected time steps.}
    \label{fig:FP_data}
\end{figure}

The FP equation is solved using the finite element method described in \cref{sec:FEM} with $\beta=1$.
The noise-free solution is computed on a $400\times 400$ mesh over the time interval $[0,0.5]$ with $n_t=25$ time steps, using triangular elements with first-order Lagrange polynomials.
The resulting solution is shown in \cref{fig:FP_data}.
To generate the observed dataset, the solution is downsampled to a $50\times 50$ B-VSI mesh, and additive independent Gaussian noise is added to all nodal values at each time step:
\begin{align}
    \label{eq:FP_observation_noise}
    \hbsp^{(k)}
    =
    \bsp^{(k)}
    +
    \bsepsilon_p^{(k)},
    \qquad
    \bsepsilon_p^{(k)}
    \sim
    \CalN
    \nospaceleft(
        \bszero,
        10^{-6}\bsI
    \nospaceright),
    \qquad
    k=0,\ldots,n_t .
\end{align}
The final dataset is
$\CalD
    =
    \left\{
        \hbsp^{(k)}
    \right\}_{k=0}^{n_t}$.

\subsubsection{Candidate dictionary and model selection}
\label{sec:FP_model_selection}

We assume that the data are generated by an FP equation of the form in \cref{eq:FP_equation}, with a known observation-noise model and boundary conditions.
The model-selection task is to identify a parsimonious representation of the potential function $\psi$.
We approximate $\psi$ using polynomial basis functions up to degree three:
\begin{align}
    \label{eq:FP_psi}
    \psi(\bsx)
    =
    \sum_{i=1}^{9}
    \theta_{\phi_i}
    \phi_i(\bsx)
    +
    \text{constant},
\end{align}
where
\begin{align}
    \label{eq:FP_polynomial_basis}
    \left\{
        \phi_i
    \right\}_{i=1}^{9}
    =
    \left\{
        x_1,\,
        x_2,\,
        x_1^2,\,
        x_2^2,\,
        x_1x_2,\,
        x_1^3,\,
        x_2^3,\,
        x_1^2x_2,\,
        x_1x_2^2
    \right\}.
\end{align}
The additive constant in \cref{eq:FP_psi} is not identifiable because its gradient is zero.
For the ground-truth potential in \cref{eq:FP_true_potential}, the identifiable nonzero coefficients are
\begin{align}
    \label{eq:FP_true_coefficients}
    \theta_{x_1}=-5,
    \qquad
    \theta_{x_2}=-25,
    \qquad
    \theta_{x_1^2}=1,
    \qquad
    \theta_{x_2^2}=5.
\end{align}

The candidate operator dictionary is
\begin{align}
    \label{eq:FP_candidate_dictionary}
    \bschi
    =
    \left[
        \grad^2 p,\,
        \grad\cdot(p\grad\phi_1),\,
        \ldots,\,
        \grad\cdot(p\grad\phi_9)
    \right]^{\top}.
\end{align}
The Laplace term $\grad^2 p$ is retained throughout model selection in order to remain consistent with the FP form.
The coefficient vector is
\begin{align}
    \label{eq:FP_coefficient_vector}
    \bstheta
    =
    \left[
        \beta^{-1},
        \theta_{\phi_1},
        \ldots,
        \theta_{\phi_9}
    \right]^{\top}.
\end{align}
Other differentiable basis families, such as Hermite polynomials, may be used in place of \cref{eq:FP_polynomial_basis}.

By integration by parts, the finite-dimensional weak forms of the diffusion and drift operators are
\begin{subequations}
    \label{eq:FP_weak_forms}
    \begin{align}
        \int_{\Omega}
        v^h \grad^2 p^h
        \,\mathrm{d}\bsx
        &=
        -
        \int_{\Omega}
        \grad v^h\cdot\grad p^h
        \,\mathrm{d}\bsx
        +
        \int_{\partial\Omega}
        v^h \grad p^h\cdot\bsn
        \,\mathrm{d}S,
        \\
        \int_{\Omega}
        v^h
        \grad\cdot
        \left(
            p^h\grad\phi_i
        \right)
        \,\mathrm{d}\bsx
        &=
        -
        \int_{\Omega}
        p^h\grad v^h\cdot\grad\phi_i
        \,\mathrm{d}\bsx
        +
        \int_{\partial\Omega}
        v^h p^h\grad\phi_i\cdot\bsn
        \,\mathrm{d}S .
    \end{align}
\end{subequations}
For this example, each candidate operator is linear in the state $p$.
The corresponding basis vectors and residual covariance are therefore constructed using the state-linear formulas in \cref{sec:linear_residual_likelihood}.
The boundary terms in the diffusion and drift weak forms are omitted, since their sum for the true coefficients vanishes under \cref{eq:FP_boundary_condition}.

\begin{figure}[htbp]
    \centering
    \captionsetup[subfigure]{justification=centering}
    \includegraphics[width=0.65\textwidth]{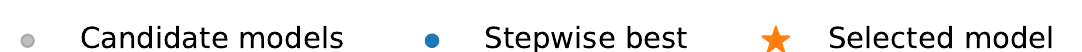}\par\smallskip
    \begin{subfigure}[t]{0.495\textwidth}
        \centering
        \includegraphics[height=5.5cm]{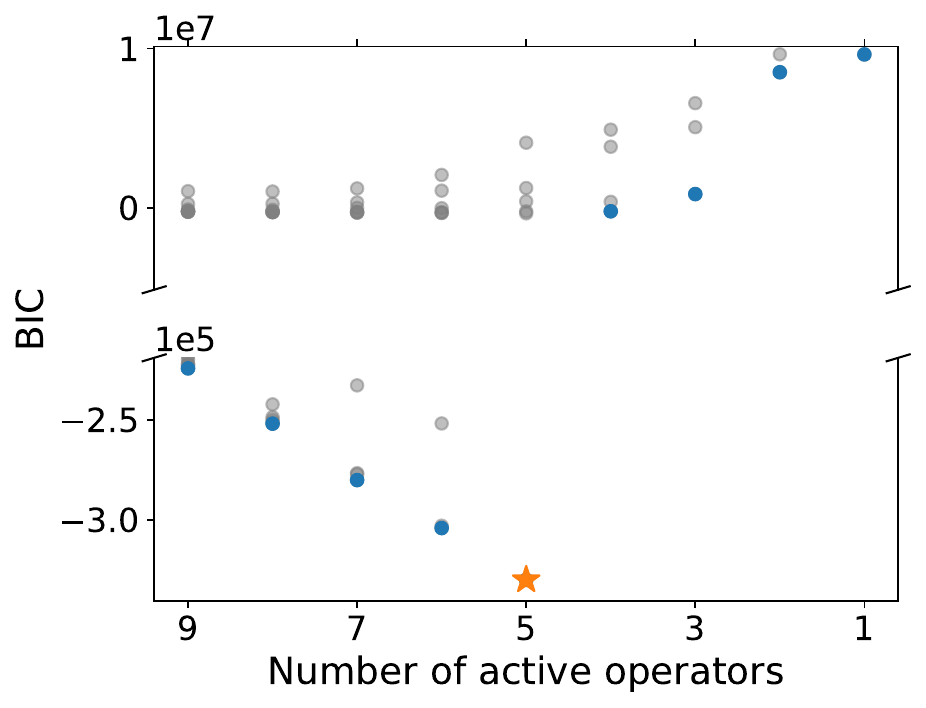}
        \caption{Block-diagonal likelihood covariance}
        \label{fig:FP_model_selection_bd}
    \end{subfigure}
    \hfill
    \begin{subfigure}[t]{0.48\textwidth}
        \centering
        \includegraphics[height=5.5cm]{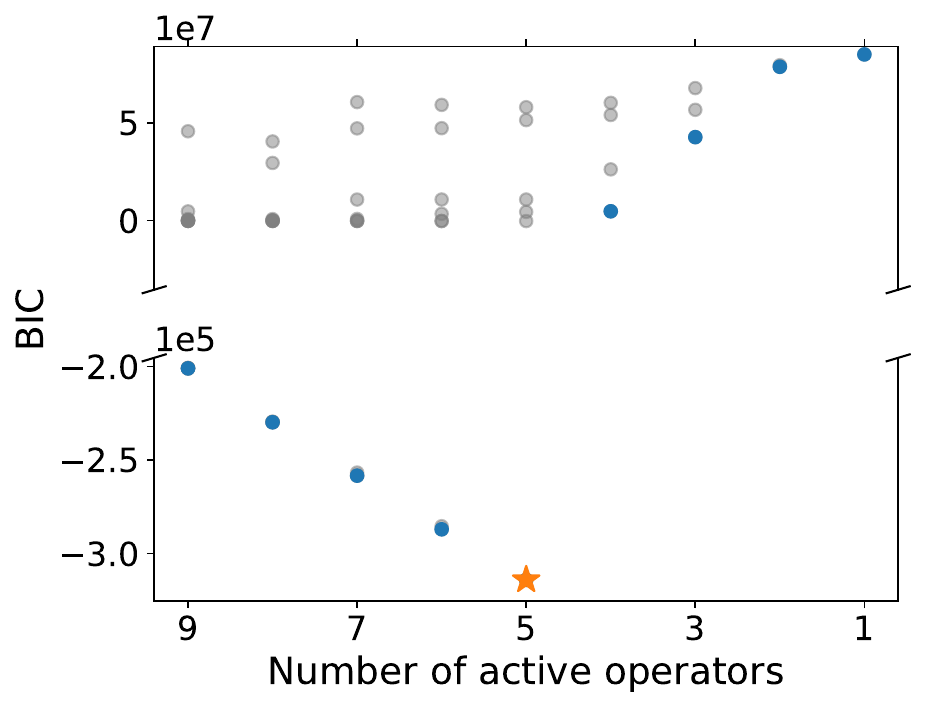}
        \caption{Block-tridiagonal likelihood covariance}
        \label{fig:FP_model_selection_bt}
    \end{subfigure}
    \caption{
    Residual-space BIC along the sequential operator-elimination paths for the FP equation using (a) block-diagonal and (b) block-tridiagonal likelihood covariance structures.
    Each vertical group corresponds to the candidate models obtained by removing one operator from the current model.
    Gray points represent candidate models, blue points indicate the best model at each elimination step, and orange stars mark the selected models. Each panel uses a broken vertical axis with different scales above and below the break.}
    \label{fig:FP_model_selection}
\end{figure}

Following \cref{alg:B-VSI}, we perform sequential operator elimination using the residual-space BIC in \cref{eq:residual_bic}.
During model selection, only the residual-space MLE is needed for each candidate model.
\Cref{fig:FP_model_selection} shows the BIC values along the elimination path for both covariance structures.
The BIC decreases as redundant polynomial terms are removed and increases after essential terms are excluded, indicating the onset of underfitting.

The selected model is the same for both covariance structures and contains the Laplace operator and the drift operators associated with
$x_1$, $x_2$, $x_1^2$, and $x_2^2$.
Full posterior inference is performed after the selected model is fixed.
The estimated coefficients are reported in \cref{tab:FP_parameters}.

\begin{table}[htbp]
\centering
\caption{
Estimated parameter values for the FP equation.
Here, $\hbstheta$ denotes the point estimate for VSI and the posterior mean for B-VSI methods.
The potential error is computed as
$\mathrm{MSE}(\hpsi)=N_g^{-1}\sum_{\ell=1}^{N_g}
(\hpsi(\bsx_\ell)-\psi_{\mathrm{GT}}(\bsx_\ell))^2$
after applying a common additive-constant convention.
Variances are reported only for posterior coefficient distributions.
For the hybrid method, variances are estimated as the sample variances of the SVGD particles.
Aggregate error metrics are computed from the reported point estimate or posterior mean and are therefore listed without posterior variances.
}
\label{tab:FP_parameters}
\fontsize{9pt}{10pt}\selectfont
\setlength{\tabcolsep}{3pt}
\sisetup{round-mode=places, round-precision=2, exponent-product=\!\times\!}
\begin{tabular}{@{}c S[table-format=-2.2] S[table-format=-2.2]
*{4}{S[table-format=-2.2] S[table-format=1.2e-1]}@{}}
\toprule
\multirow{3}{*}{Coefficient}
& {\multirow{3}{*}{\makecell{Ground\\truth}}}
& {\multirow{3}{*}{VSI}}
& \multicolumn{2}{c}{B-VSI-BD-}
& \multicolumn{2}{c}{B-VSI-BD-}
& \multicolumn{2}{c}{B-VSI-BT-}
& \multicolumn{2}{c}{B-VSI-BT-} \\
& & & \multicolumn{2}{c}{LaggedCov}
& \multicolumn{2}{c}{Hybrid}
& \multicolumn{2}{c}{LaggedCov}
& \multicolumn{2}{c}{Hybrid} \\
\cmidrule(lr){4-5}
\cmidrule(lr){6-7}
\cmidrule(lr){8-9}
\cmidrule(l){10-11}
& & & {Mean} & {Variance} & {Mean} & {Variance}
& {Mean} & {Variance}
& {Mean} & {Variance} \\
\midrule
\makebox[0.25cm][l]{$\beta^{-1}$} & 1.00 & 0.81
& 0.95 & 1.20e-06
& 0.96 & 5.28e-07
& 1.03 & 1.18e-07
& 1.03 & 7.59e-08 \\

\makebox[0.25cm][l]{$\theta_{x_1}$} & -5.00 & -5.22
& -5.03 & 3.40e-05
& -5.03 & 2.59e-05
& -4.99 & 6.91e-07
& -4.99 & 6.13e-07 \\

\makebox[0.25cm][l]{$\theta_{x_2}$} & -25.00 & -23.42
& -24.40 & 1.98e-04
& -24.48 & 1.51e-04
& -24.85 & 1.39e-05
& -24.87 & 1.20e-05 \\

\makebox[0.25cm][l]{$\theta_{x_1^2}$} & 1.00 & 0.93
& 0.97 & 6.17e-06
& 0.97 & 4.21e-06
& 1.01 & 2.76e-07
& 1.02 & 1.89e-07 \\

\makebox[0.25cm][l]{$\theta_{x_2^2}$} & 5.00 & 4.64
& 4.86 & 9.50e-06
& 4.88 & 6.87e-06
& 4.97 & 7.25e-07
& 4.97 & 6.55e-07 \\

\midrule

$\norm{\bstheta_{\mathrm{GT}}-\hbstheta}{2}$ & {N/A} & 1.65
& 0.62 & {N/A}
& 0.53 & {N/A}
& 0.15 & {N/A}
& 0.14 & {N/A} \\

$\mathrm{MSE}(\hpsi)$ & {N/A} & 29.60
& 4.26 & {N/A}
& 3.12 & {N/A}
& 0.26 & {N/A}
& 0.21 & {N/A} \\
\bottomrule
\end{tabular}
\end{table}

\subsubsection{Parameter and predictive uncertainty with Gaussian approximate priors}
\label{sec:example_FP_EM}

We first use Gaussian priors so that the lagged-covariance Gaussian posterior update in \cref{sec:lagged_covariance_inference} can be applied efficiently.
The priors are
\begin{subequations}
    \label{eq:FP_EM_priors}
    \begin{align}
        \beta^{-1}
        &\sim
        \CalN(3.08,2.69),
        \\
        \theta_{\phi_i}
        &\sim
        \CalN(0,2),
        \qquad
        i\in\mathcal{I}(\CalM^*).
    \end{align}
\end{subequations}
Here and below, $\mathcal{I}(\CalM^\ast)$ denotes the indices of operators retained in the selected model $\CalM^\ast$. Since model selection precedes coefficient inference, the priors below are defined only over this active set.

These Gaussian priors are obtained by moment matching the non-Gaussian priors used in \cref{sec:example_FP_hybrid}.
Specifically, the log-normal prior on $\beta^{-1}$ has mean $3.08$ and variance $2.69$, while the Laplace prior on each polynomial coefficient has variance $2$.
The Gaussian approximation does not encode positivity of $\beta^{-1}$ directly, so the constraint $\beta^{-1}>0$ is enforced through the constrained lagged-covariance update.

\begin{figure}[htbp]
    \centering
    \captionsetup[subfigure]{justification=centering}
    \begin{subfigure}[t]{0.48\textwidth}
        \centering
        \includegraphics[height=5.3969cm]{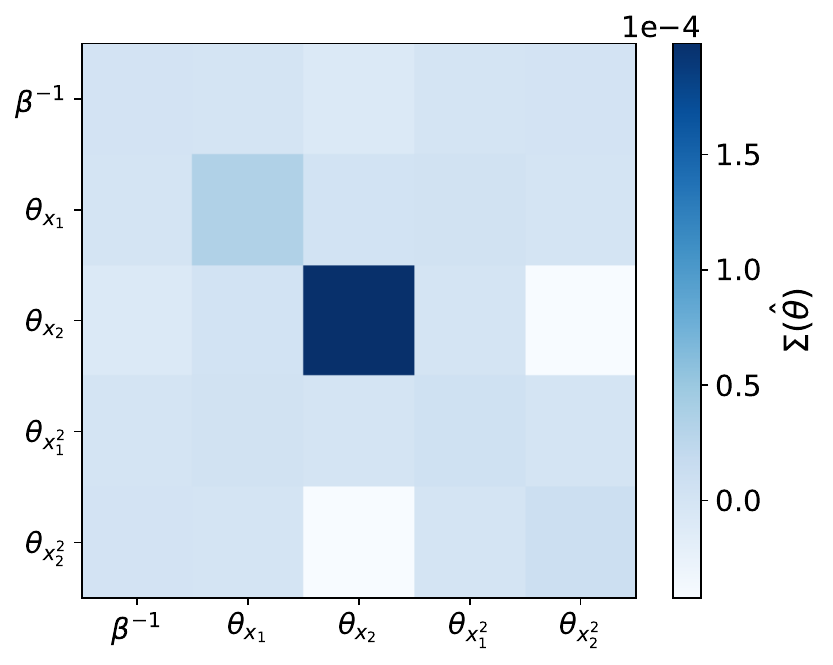}
        \caption{Block-diagonal likelihood covariance}
        \label{fig:FP_EM_covariance_bd}
    \end{subfigure}
    \hfill
    \begin{subfigure}[t]{0.48\textwidth}
        \centering
        \includegraphics[height=5.3969cm]{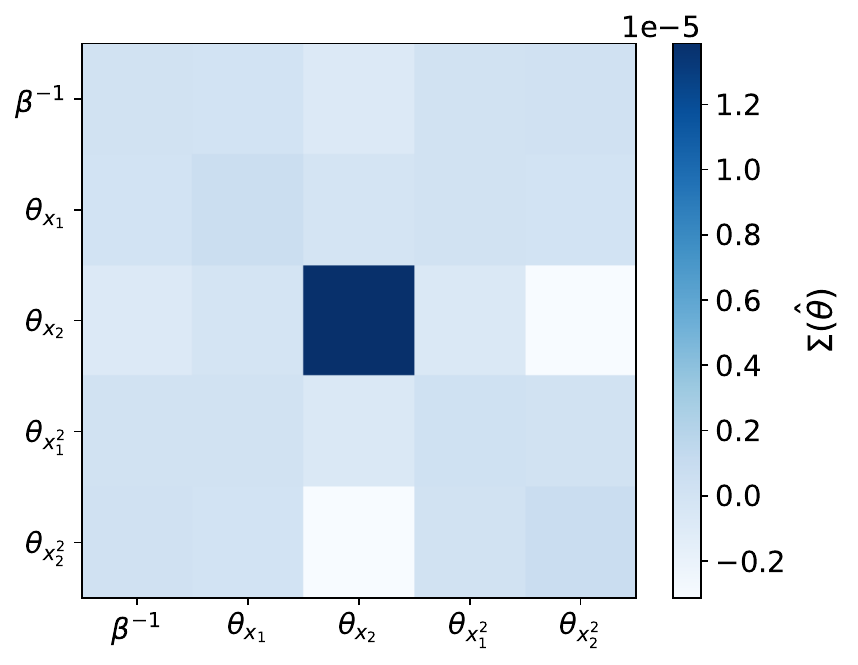}
        \caption{Block-tridiagonal likelihood covariance}
        \label{fig:FP_EM_covariance_bt}
    \end{subfigure}
    \caption{Covariance matrices of the approximate Gaussian posterior $p(\bstheta \mid \CalD,\CalM^*)$ for the selected FP model using (a) block-diagonal and (b) block-tridiagonal likelihood covariance structures.}
    \label{fig:FP_EM_covariance}
\end{figure}

Under the selected model $\CalM^*$, the state-linear residual likelihood and Gaussian prior yield an efficient approximate Gaussian posterior for the active coefficients.
The posterior means and variances are reported in \cref{tab:FP_parameters}, and the posterior covariance matrices obtained using the two types of likelihood covariance structures are shown in \cref{fig:FP_EM_covariance}.
The coefficient $\theta_{x_2}$ has the largest posterior variance among the active polynomial coefficients.
This reflects the combined effects of basis scaling, correlations among polynomial terms, and the fact that the observed density samples only part of the spatial domain over the finite time interval.

\begin{figure}[htbp]
    \centering
    \captionsetup[subfigure]{justification=centering}
    \begin{subfigure}[t]{0.49\textwidth}
        \centering
        \includegraphics[height=5.5268cm]{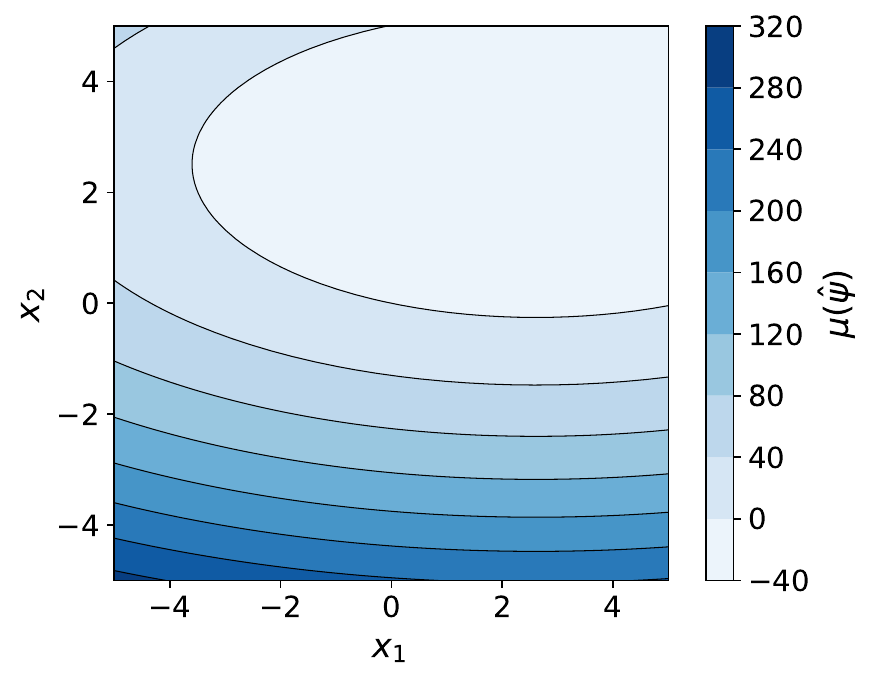}
        \caption{Predictive mean}
        \label{fig:FP_EM_predictive_mean}
    \end{subfigure}
    \hfill
    \begin{subfigure}[t]{0.49\textwidth}
        \centering
        \includegraphics[height=5.6674cm]{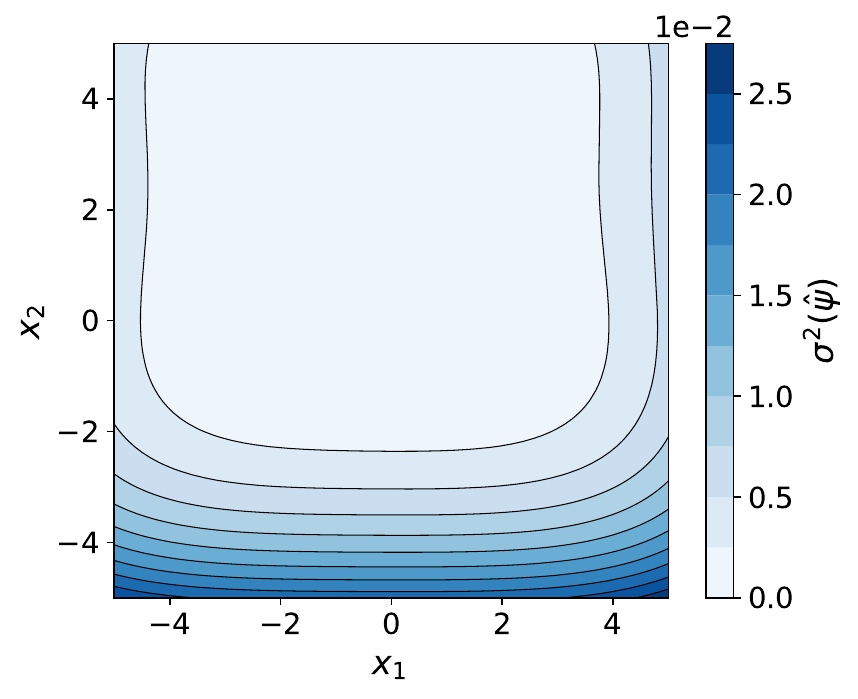}
        \caption{Predictive variance}
        \label{fig:FP_EM_predictive_var}
    \end{subfigure}
    \caption{Mean and variance of the posterior predictive distribution $p(\psi \mid \CalD,\CalM^*)$ for the potential function in the FP equation using the approximate Gaussian posterior with block-diagonal likelihood covariance.}
    \label{fig:FP_EM_predictive}
\end{figure}

\begin{figure}[htbp]
    \centering
    \captionsetup[subfigure]{justification=centering}
    \begin{subfigure}[t]{0.49\textwidth}
        \centering
        \includegraphics[height=5.5268cm]{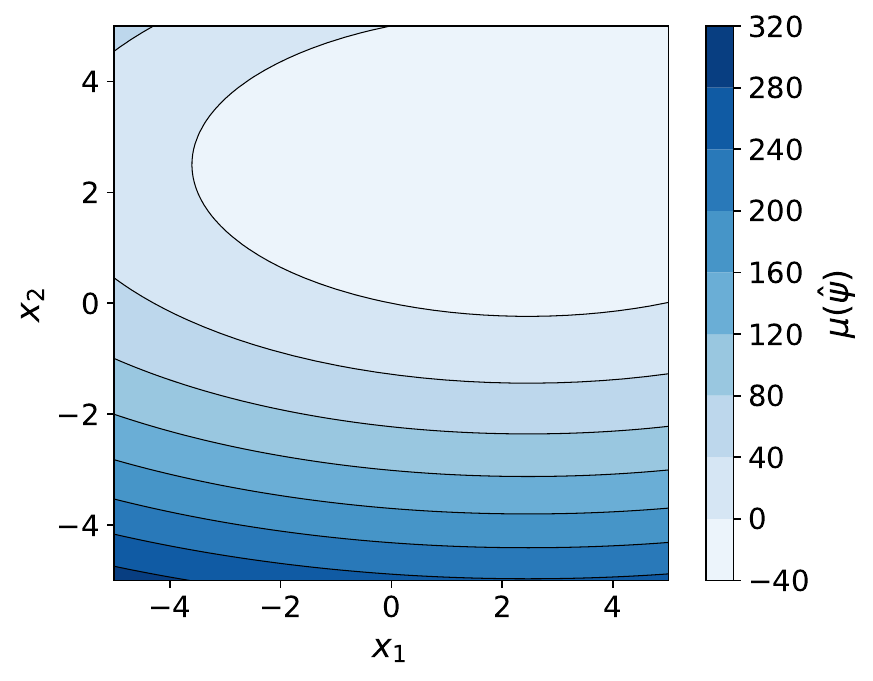}
        \caption{Predictive mean}
        \label{fig:FP_EM_predictive_mean_bt}
    \end{subfigure}
    \hfill
    \begin{subfigure}[t]{0.49\textwidth}
        \centering
        \includegraphics[height=5.6674cm]{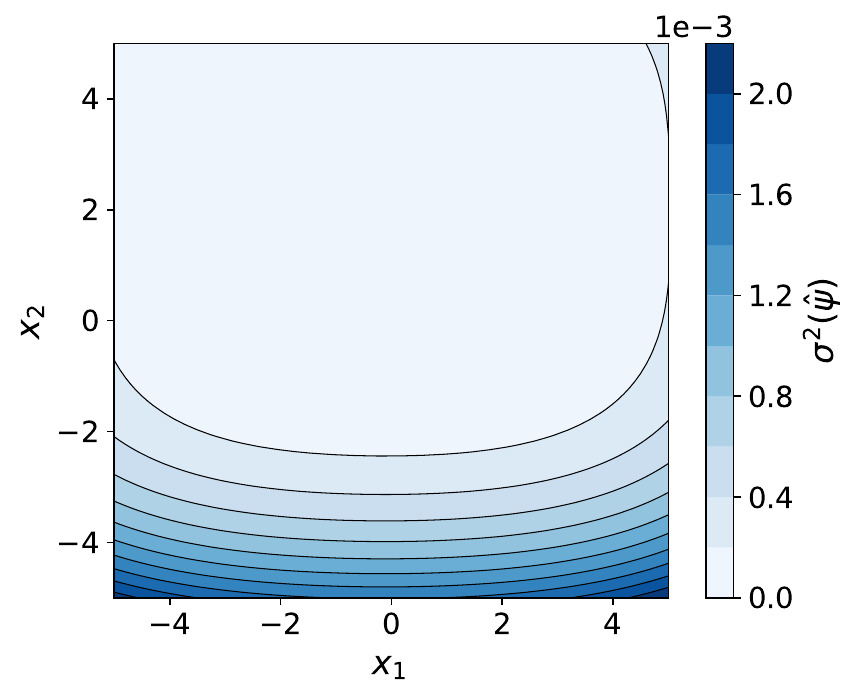}
        \caption{Predictive variance}
        \label{fig:FP_EM_predictive_var_bt}
    \end{subfigure}
    \caption{Mean and variance of the posterior predictive distribution $p(\psi \mid \CalD,\CalM^*)$ for the potential function in the FP equation using the approximate Gaussian posterior with block-tridiagonal likelihood covariance.}
    \label{fig:FP_EM_predictive_bt}
\end{figure}

We next propagate the coefficient posterior to the potential function.
Since the identifiable part of $\psi$ is linear in the coefficients, its values on the plotting grid can be written as
\begin{align}
    \label{eq:FP_predictive_potential_linear_map}
    \bspsi
    =
    \bsPhi\bstheta_{\psi},
\end{align}
where $\bspsi$ collects the potential values on the plotting grid and $\bsPhi$ evaluates the active polynomial basis functions at those grid points.
Therefore, the Gaussian predictive formula in \cref{sec:posterior_predictive} applies.
The posterior predictive mean and variance of $\psi$ are shown in \cref{fig:FP_EM_predictive,fig:FP_EM_predictive_bt} for both covariance structures.
The predictive mean closely matches the ground-truth potential, as also reflected by the MSE reported in \cref{tab:FP_parameters}, with a lower MSE for the block-tridiagonal structure.
The block-tridiagonal structure also gives smaller predictive variances.
The predictive variance is largest in regions where the data provide limited information.
In this example, the density evolves toward the low-potential basin near $\bsc$.
High-potential regions that are sparsely visited by the observed density therefore lead to larger uncertainty in the recovered potential.

\subsubsection{Parameter and predictive uncertainty with non-Gaussian priors}
\label{sec:example_FP_hybrid}

We next consider a non-Gaussian prior specification that directly encodes positivity of the diffusion coefficient and applies continuous shrinkage to the active polynomial coefficients:
\begin{subequations}
    \label{eq:FP_hybrid_priors}
    \begin{align}
        \beta^{-1}
        &\sim
        \operatorname{LogNormal}(1,0.25),
        \\
        \theta_{\phi_i}
        &\sim
        \operatorname{Laplace}(0,1),
        \qquad
        i\in\mathcal{I}(\CalM^*).
    \end{align}
\end{subequations}
Here, $\operatorname{LogNormal}(1,0.25)$ denotes the distribution whose logarithm is Gaussian with mean $1$ and variance $0.25$.
The Laplace prior uses scale parameter $1$.

\begin{figure}[htbp]
    \centering
    \captionsetup[subfigure]{justification=centering}
    \begin{subfigure}[t]{0.48\textwidth}
        \centering
        \includegraphics[height=5.3969cm]{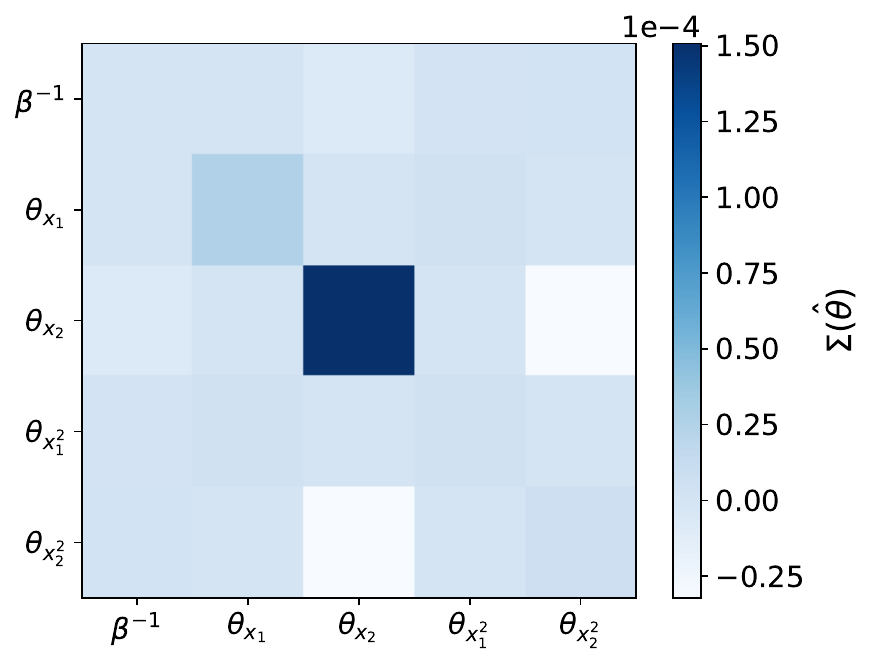}
        \caption{Block-diagonal likelihood covariance}
        \label{fig:FP_SVGD_covariance_bd}
    \end{subfigure}
    \hfill
    \begin{subfigure}[t]{0.48\textwidth}
        \centering
        \includegraphics[height=5.3969cm]{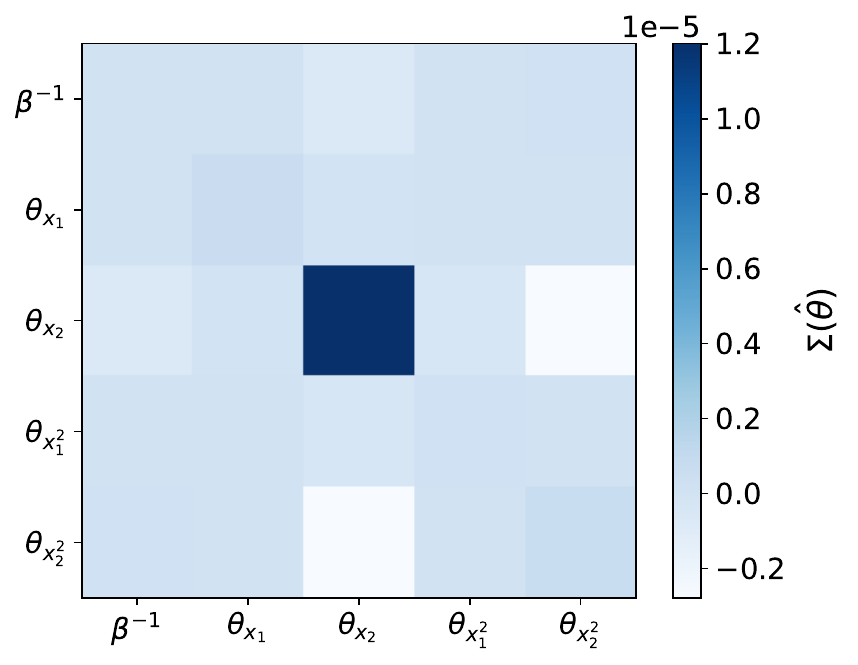}
        \caption{Block-tridiagonal likelihood covariance}
        \label{fig:FP_SVGD_covariance_bt}
    \end{subfigure}
    \caption{Covariance matrices of the particle posterior $p(\bstheta \mid \CalD,\CalM^*)$ for the selected FP model estimated using $n_p=20$ SVGD particles using (a) block-diagonal and (b) block-tridiagonal likelihood covariance structures.}
    \label{fig:FP_SVGD_covariance}
\end{figure}

Because these priors are not conjugate to the Gaussian residual likelihood, we use the hybrid strategy described in \cref{sec:hybrid_constraints}.
We first use the Gaussian moment-matched priors in \cref{eq:FP_EM_priors} to obtain an efficient approximate posterior.
We then initialize $n_p=20$ SVGD particles from this Gaussian approximation and refine them using the residual-space posterior associated with the non-Gaussian priors in \cref{eq:FP_hybrid_priors}.
For each covariance structure, the refinement uses 100 epochs.
The empirical means and variances of the SVGD particles are reported in \cref{tab:FP_parameters} under the corresponding Hybrid columns.
The particle covariance matrices are shown in \cref{fig:FP_SVGD_covariance}.

\begin{figure}[htbp]
    \centering
    \captionsetup[subfigure]{justification=centering}
    \begin{subfigure}[t]{0.49\textwidth}
        \centering
        \includegraphics[height=5.5268cm]{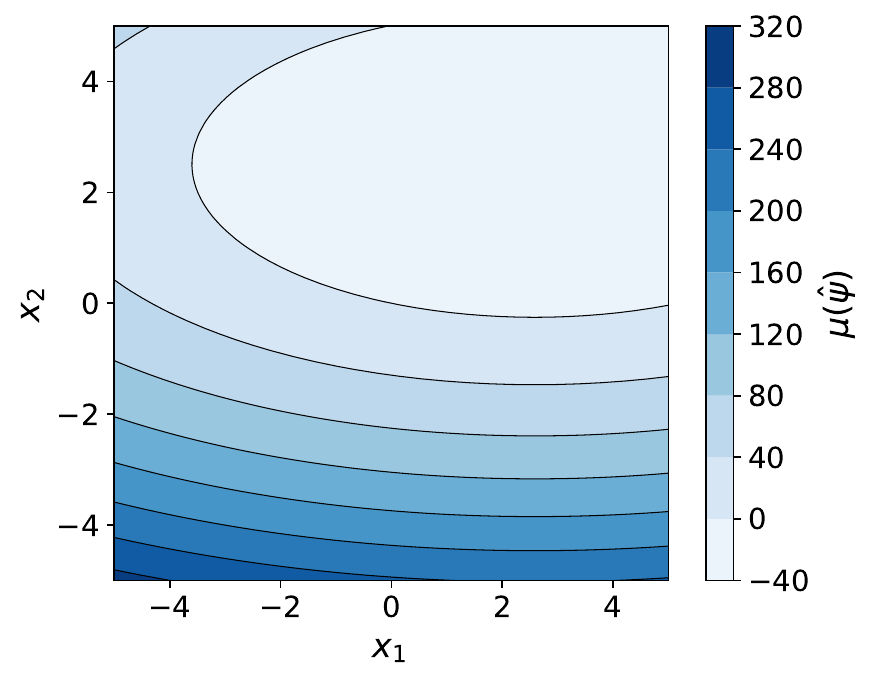}
        \caption{Predictive mean}
        \label{fig:FP_SVGD_predictive_mean}
    \end{subfigure}
    \hfill
    \begin{subfigure}[t]{0.49\textwidth}
        \centering
        \includegraphics[height=5.6674cm]{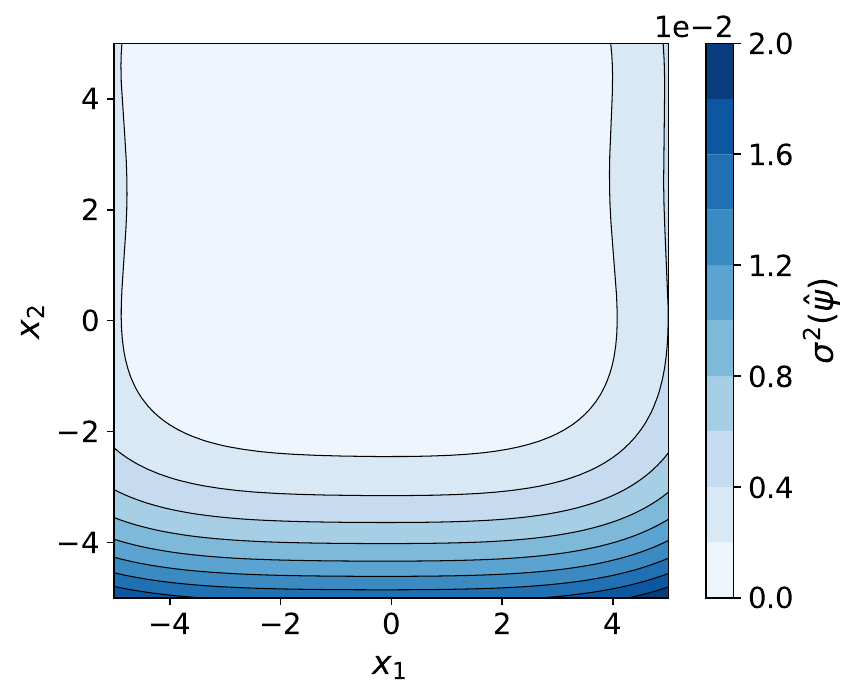}
        \caption{Predictive variance}
        \label{fig:FP_SVGD_predictive_var}
    \end{subfigure}
    \caption{Mean and variance of the posterior predictive distribution $p(\psi \mid \CalD,\CalM^*)$ for the potential function in the FP equation using $n_p=20$ SVGD particles with block-diagonal likelihood covariance.}
    \label{fig:FP_SVGD_predictive}
\end{figure}

\begin{figure}[htbp]
    \centering
    \captionsetup[subfigure]{justification=centering}
    \begin{subfigure}[t]{0.49\textwidth}
        \centering
        \includegraphics[height=5.5268cm]{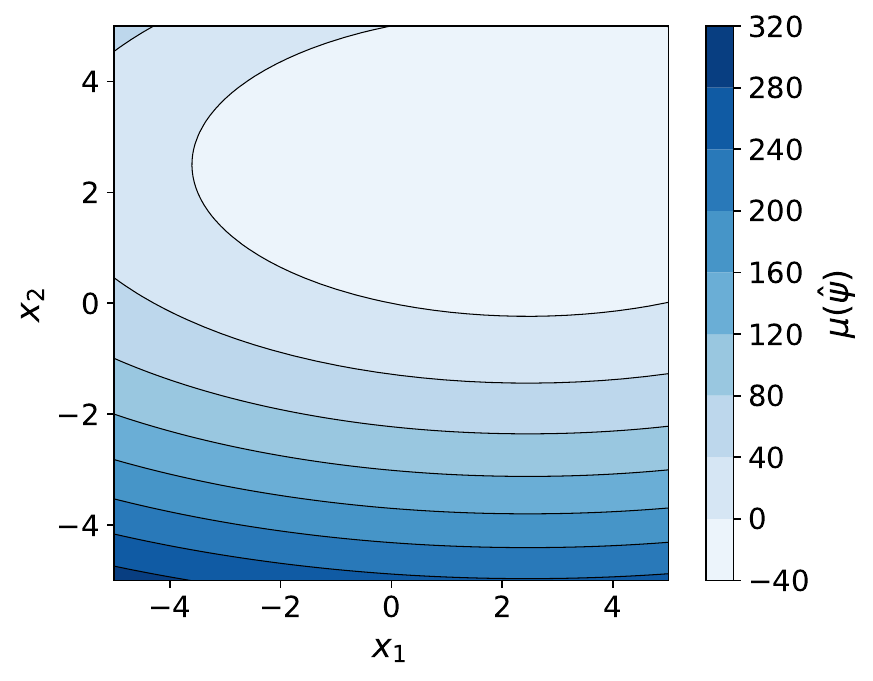}
        \caption{Predictive mean}
        \label{fig:FP_SVGD_predictive_mean_bt}
    \end{subfigure}
    \hfill
    \begin{subfigure}[t]{0.49\textwidth}
        \centering
        \includegraphics[height=5.6674cm]{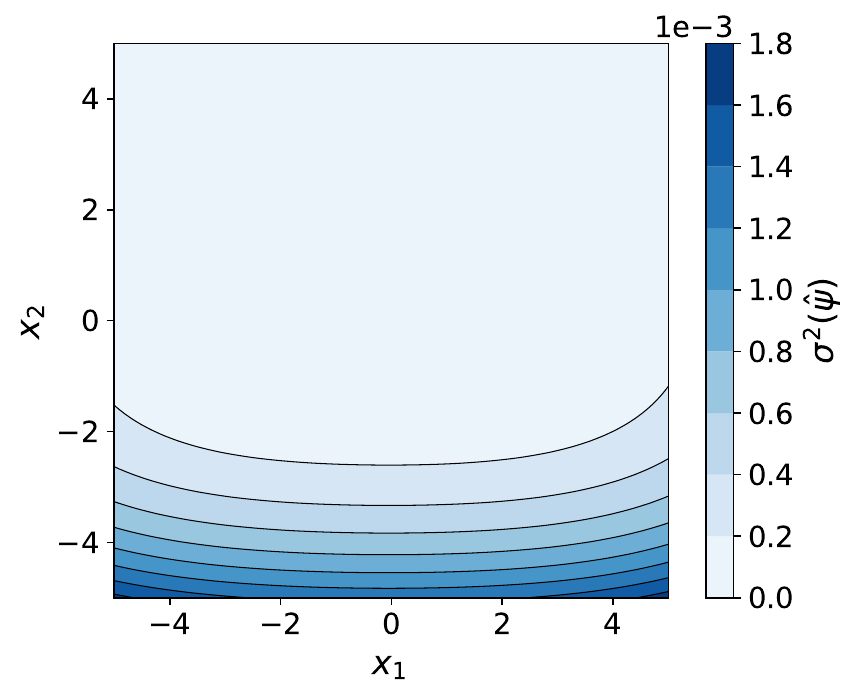}
        \caption{Predictive variance}
        \label{fig:FP_SVGD_predictive_var_bt}
    \end{subfigure}
    \caption{Mean and variance of the posterior predictive distribution $p(\psi \mid \CalD,\CalM^*)$ for the potential function in the FP equation using $n_p=20$ SVGD particles with block-tridiagonal likelihood covariance.}
    \label{fig:FP_SVGD_predictive_bt}
\end{figure}

The posterior predictive distribution of the potential is approximated by Monte Carlo propagation of the SVGD particles:
\begin{align}
    \label{eq:FP_hybrid_predictive_samples}
    \bspsi^{(i)}
    =
    \bsPhi\bstheta_{\psi}^{(i)},
    \qquad
    i=1,\ldots,n_p .
\end{align}
The empirical mean and variance are shown in \cref{fig:FP_SVGD_predictive,fig:FP_SVGD_predictive_bt} for both covariance structures.
The hybrid posterior gives coefficient estimates and predictive uncertainty patterns that are close to those obtained with the Gaussian approximate posterior.
For both covariance structures, the hybrid refinement gives smaller marginal coefficient variances and lower potential MSE than the corresponding Gaussian approximation.
The SVGD refinement remains useful when positivity, sparsity, or other non-Gaussian prior information must be represented explicitly.

\subsection{Nonlinear two-field Cahn--Hilliard equation}
\label{sec:example_CH}

We next consider a nonlinear two-field Cahn--Hilliard equation.
The CH equation models phase separation in multicomponent systems, including alloys and polymer blends~\cite{Cahn1958}.
It describes the evolution of concentration fields driven by chemical-potential gradients and interfacial-energy regularization.
For two concentration fields $c_1$ and $c_2$, we consider
\begin{subequations}
    \label{eq:cahn_hilliard}
    \begin{align}
        \label{eq:cahn_hilliard_field1}
        \pp{c_1}{t}
        &=
        m_1 \nabla^2 \mu_1,
        \qquad
        \mu_1
        =
        \pp{g}{c_1}
        -
        k_1\nabla^2 c_1,
        \\
        \label{eq:cahn_hilliard_field2}
        \pp{c_2}{t}
        &=
        m_2 \nabla^2 \mu_2,
        \qquad
        \mu_2
        =
        \pp{g}{c_2}
        -
        k_2\nabla^2 c_2.
    \end{align}
\end{subequations}
Here, $m_1$ and $m_2$ are mobilities, $\mu_1$ and $\mu_2$ are chemical potentials, $g(c_1,c_2)$ is the homogeneous free-energy density, and $k_1,k_2>0$ are gradient-energy coefficients.
A nonlinear dependence of $g$ on the concentrations makes this a nonlinear PDE system.

\subsubsection{Data generation}
\label{sec:CH_data_generation}

\begin{figure}[htbp]
    \centering
    \begin{subfigure}[t]{0.325\textwidth}
        \centering
        \includegraphics[height=3.45cm]{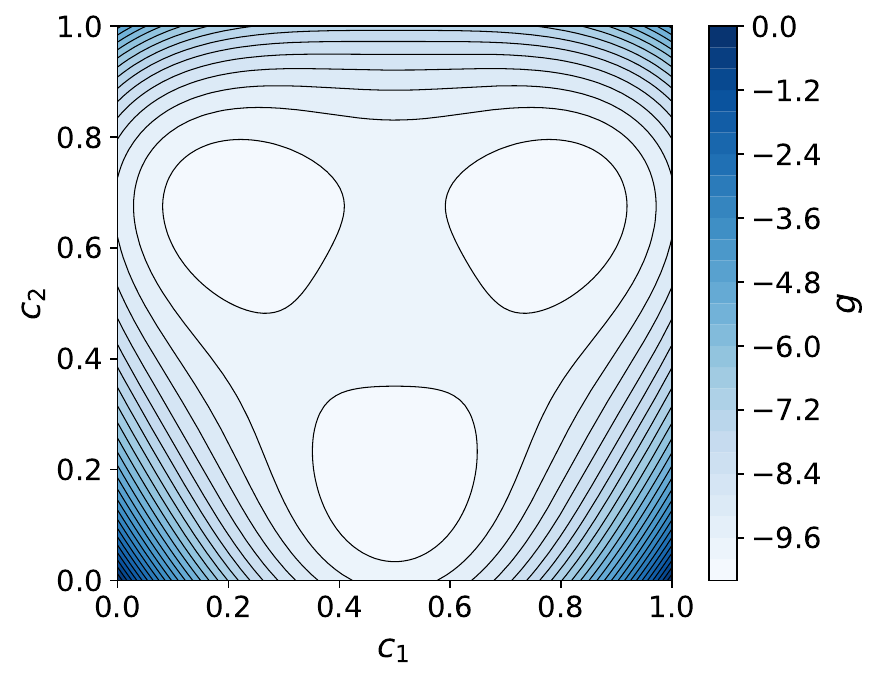}
        \caption{$g(c_1,c_2)$}
        \label{fig:CH_energy_g}
    \end{subfigure}
    \hfill
    \begin{subfigure}[t]{0.325\textwidth}
        \centering
        \includegraphics[height=3.45cm]{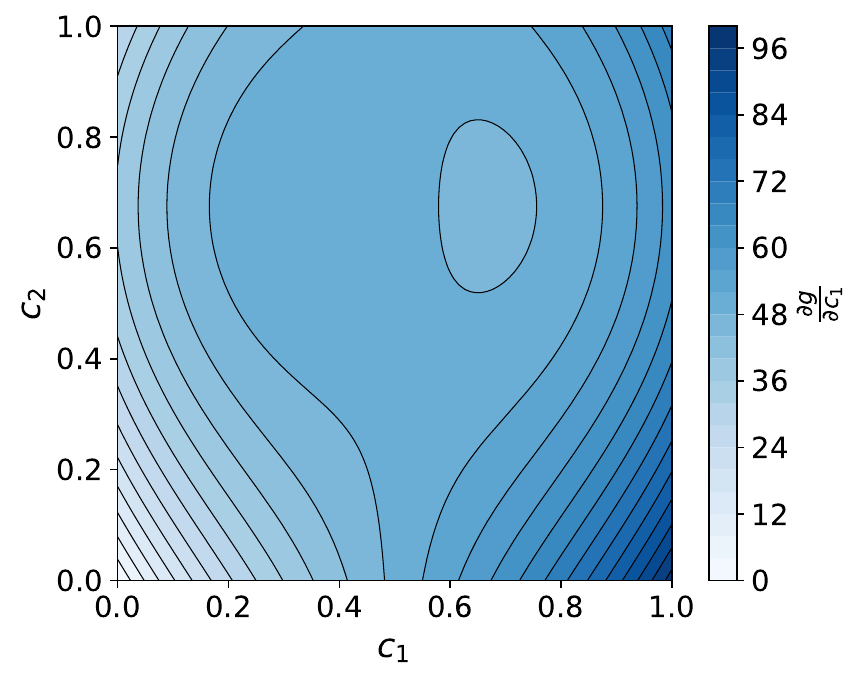}
        \caption{$\pp{g}{c_1}$}
        \label{fig:CH_energy_partialgc1}
    \end{subfigure}
    \hfill
    \begin{subfigure}[t]{0.332\textwidth}
        \centering
        \includegraphics[height=3.45cm]{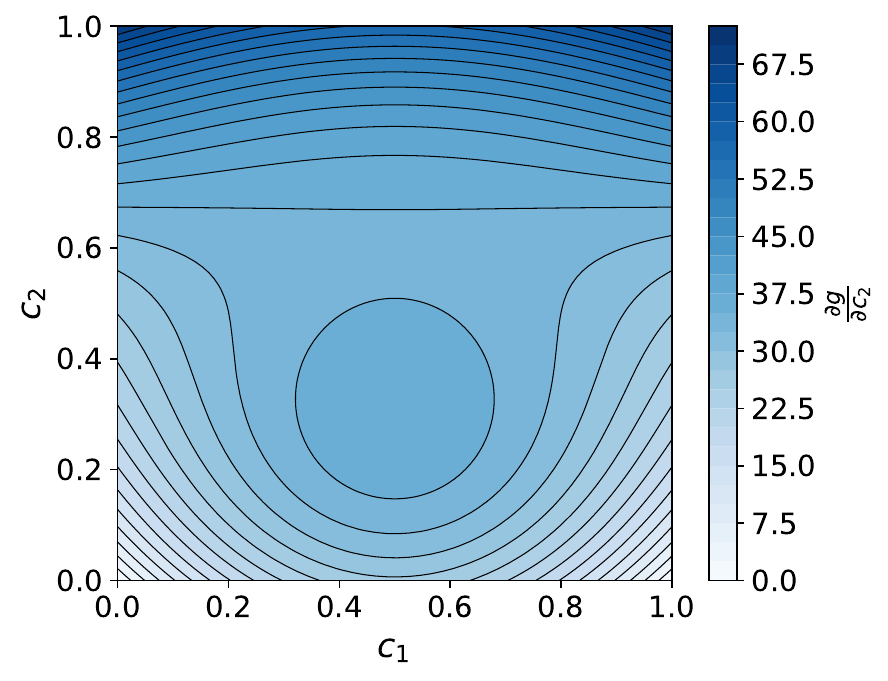}
        \caption{$\pp{g}{c_2}$}
        \label{fig:CH_energy_partialgc2}
    \end{subfigure}
    \caption{Ground-truth free-energy density $g$ and partial derivatives $\partial g/\partial c_1$ and $\partial g/\partial c_2$ used to generate the two-field CH data. The free energy and its derivatives are shown up to additive constants.}
    \label{fig:CH_energy_function}
\end{figure}

Following the setup in~\citet{Wang2021Variational}, we consider the spatial domain $\Omega=[0,20]^2$.
The homogeneous free-energy density is
\begin{align}
    \label{eq:cahn_hilliard_energy}
    \begin{split}
        g(c_1,c_2)
        =&~
        \frac{3d}{2s^{4}}
        \left[
            (2c_1-1)^2
            +
            (2c_2-1)^2
        \right]^2
        \\
        &+
        \frac{d}{s^{3}}
        (2c_2-1)
        \left[
            (2c_2-1)^2
            -
            3(2c_1-1)^2
        \right]
        \\
        &-
        \frac{3d}{2s^{2}}
        \left[
            (2c_1-1)^2
            +
            (2c_2-1)^2
        \right].
    \end{split}
\end{align}
The parameters used to generate the data are
\begin{align}
    \label{eq:CH_generation_parameters}
    m_1=m_2=0.1,
    \qquad
    k_1=k_2=10,
    \qquad
    d=0.4,
    \qquad
    s=0.7.
\end{align}
The free-energy density and its partial derivatives are shown in \cref{fig:CH_energy_function}.

The initial conditions are random perturbations about the homogeneous state, and zero Neumann boundary conditions are applied:
\begin{subequations}
    \label{eq:CH_initial_boundary_conditions}
    \begin{alignat}{3}
        c_1(\bsx,0)
        &=
        0.5+\delta_{c_1},
        \qquad
        &&\delta_{c_1}\sim\CalU(-0.3,0.3),
        \qquad
        &&\bsx\in\Omega,
        \\
        c_2(\bsx,0)
        &=
        0.5+\delta_{c_2},
        \qquad
        &&\delta_{c_2}\sim\CalU(-0.3,0.3),
        \qquad
        &&\bsx\in\Omega,
        \\
        \nabla \mu_1\cdot\bsn
        &=
        0,
        \qquad
        &&
        \nabla c_1\cdot\bsn=0,
        \qquad
        &&\bsx\in\partial\Omega,
        \\
        \nabla \mu_2\cdot\bsn
        &=
        0,
        \qquad
        &&
        \nabla c_2\cdot\bsn=0,
        \qquad
        &&\bsx\in\partial\Omega.
    \end{alignat}
\end{subequations}

\begin{figure}[htbp]
    \centering
    \begin{subfigure}[t]{0.22\textwidth}
        \centering
        \includegraphics[width=\linewidth,trim={45bp 0bp 78bp 25bp},clip]{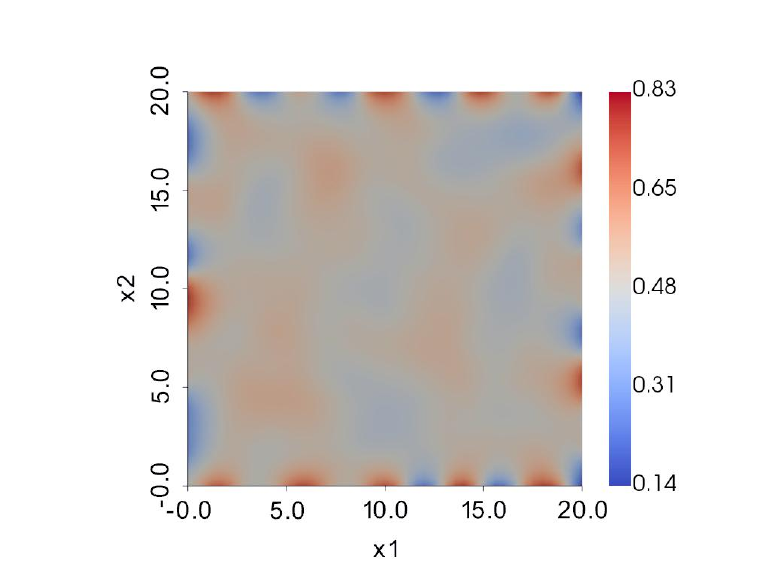}
        \caption{$c_1^{(0)}$ at $t=0$}
    \end{subfigure}
    \hfill
    \begin{subfigure}[t]{0.22\textwidth}
        \centering
        \includegraphics[width=\linewidth,trim={45bp 0bp 78bp 25bp},clip]{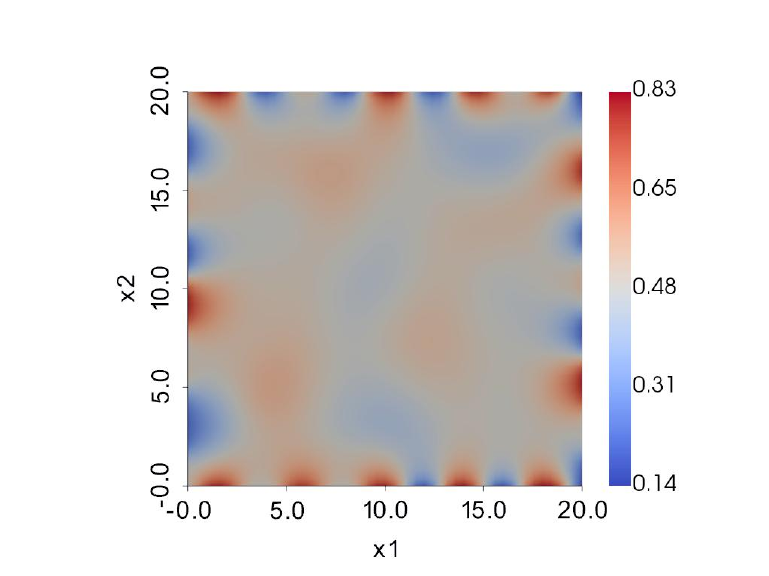}
        \caption{$c_1^{(4)}$ at $t=2.0$}
    \end{subfigure}
    \hfill
    \begin{subfigure}[t]{0.22\textwidth}
        \centering
        \includegraphics[width=\linewidth,trim={45bp 0bp 78bp 25bp},clip]{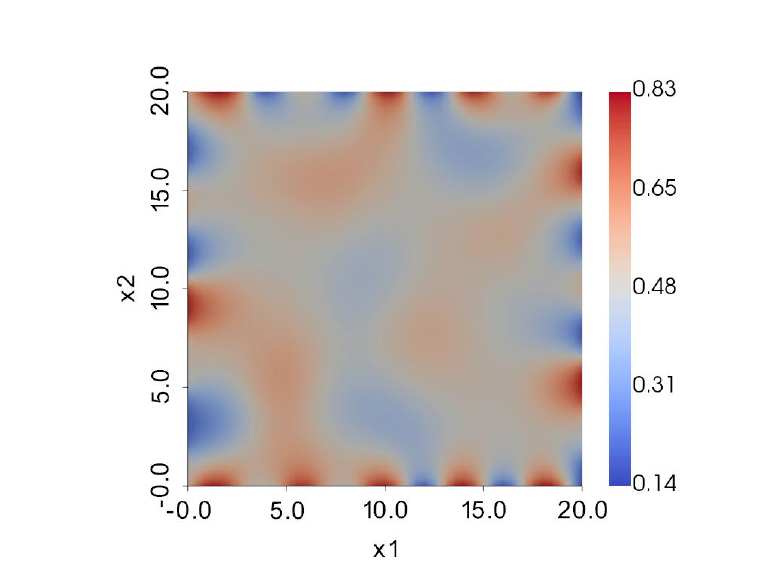}
        \caption{$c_1^{(7)}$ at $t=3.5$}
    \end{subfigure}
    \hfill
    \begin{subfigure}[t]{0.2725\textwidth}
        \centering
        \includegraphics[width=\linewidth,trim={45bp 0bp 20bp 25bp},clip]{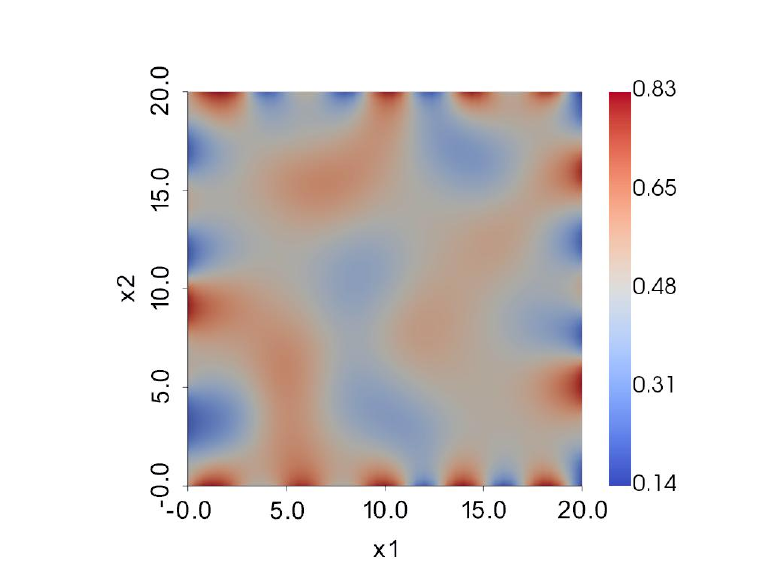}
        \caption{$c_1^{(10)}$ at $t=5.0$}
    \end{subfigure}
    \par\medskip
    \begin{subfigure}[t]{0.22\textwidth}
        \centering
        \includegraphics[width=\linewidth,trim={45bp 0bp 78bp 25bp},clip]{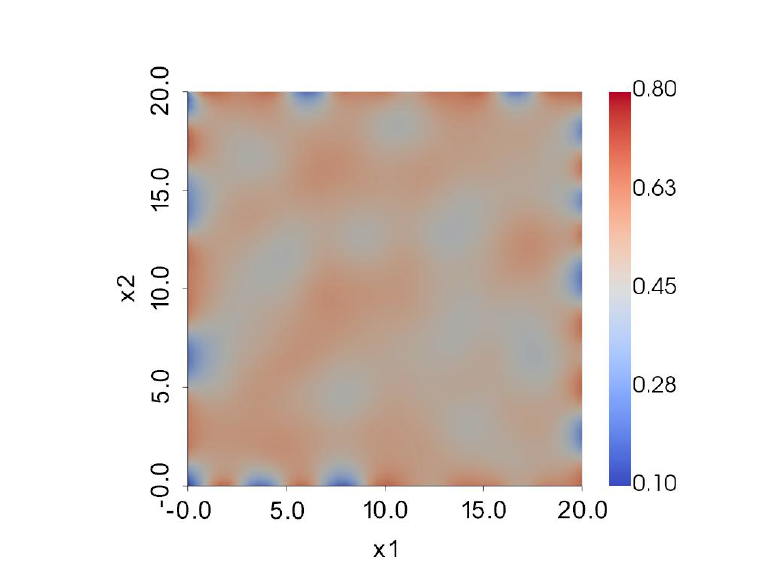}
        \caption{$c_2^{(0)}$ at $t=0$}
    \end{subfigure}
    \hfill
    \begin{subfigure}[t]{0.22\textwidth}
        \centering
        \includegraphics[width=\linewidth,trim={45bp 0bp 78bp 25bp},clip]{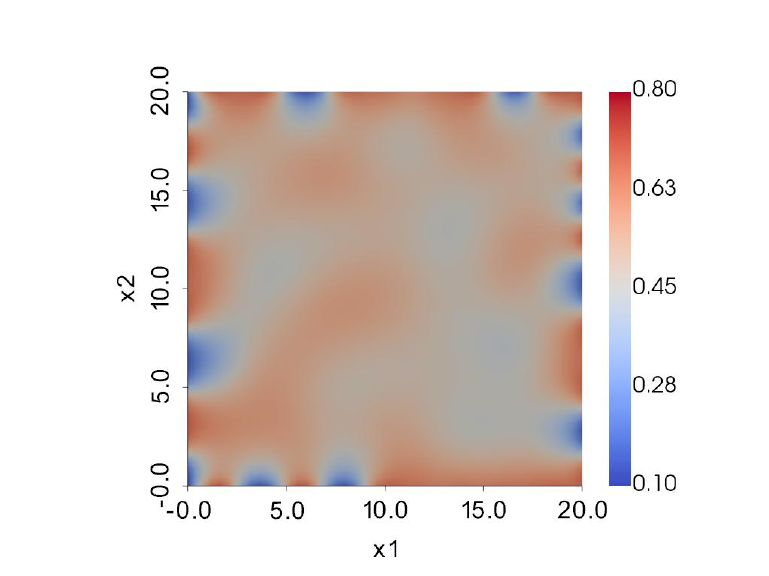}
        \caption{$c_2^{(4)}$ at $t=2.0$}
    \end{subfigure}
    \hfill
    \begin{subfigure}[t]{0.22\textwidth}
        \centering
        \includegraphics[width=\linewidth,trim={45bp 0bp 78bp 25bp},clip]{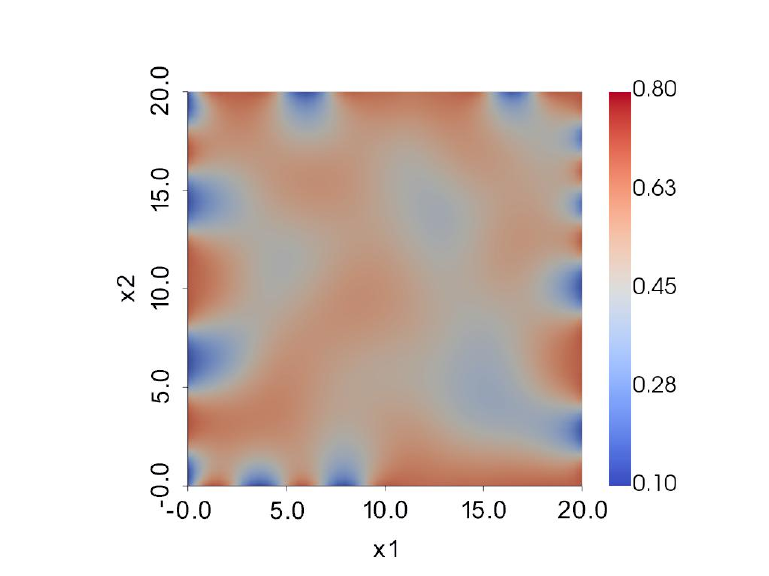}
        \caption{$c_2^{(7)}$ at $t=3.5$}
    \end{subfigure}
    \hfill
    \begin{subfigure}[t]{0.2725\textwidth}
        \centering
        \includegraphics[width=\linewidth,trim={45bp 0bp 20bp 25bp},clip]{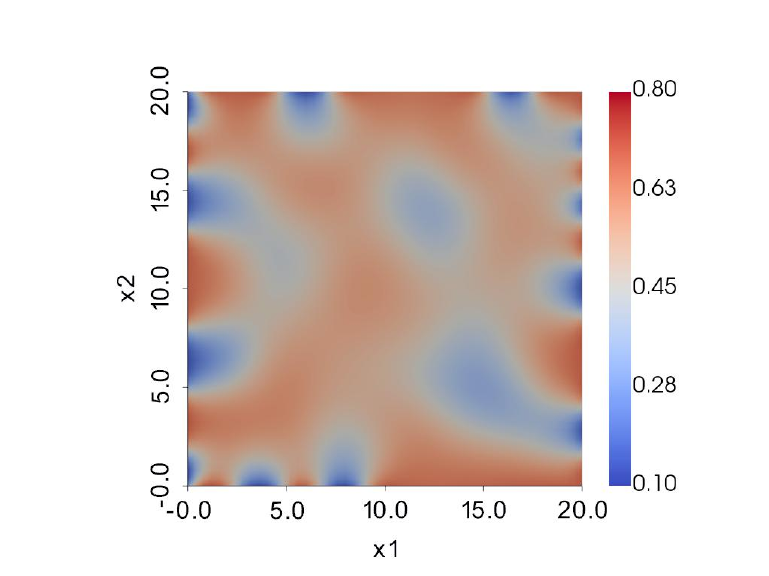}
        \caption{$c_2^{(10)}$ at $t=5.0$}
    \end{subfigure}
    \caption{Time evolution of the concentration fields $c_1$ and $c_2$ in the two-field CH equation at selected time steps.}
    \label{fig:CH_data}
\end{figure}

The CH system is solved using the finite element method described in \cref{sec:FEM}.
The solution is computed on a $25\times25$ mesh using $\mathcal{C}^0$-continuous quadratic Lagrange elements over the time interval $[0, 5]$.
The resulting noise-free concentration fields are shown in \cref{fig:CH_data}.
The observed data are generated by adding independent Gaussian noise to the nodal values of both concentration fields:
\begin{subequations}
    \label{eq:CH_observation_noise}
    \begin{align}
        \hbsc_1^{(k)}
        &=
        \bsc_1^{(k)}
        +
        \bsepsilon_1^{(k)},
        \qquad
        \bsepsilon_1^{(k)}
        \sim
        \CalN(\bszero,\sigma_c^2\bsI),
        \\
        \hbsc_2^{(k)}
        &=
        \bsc_2^{(k)}
        +
        \bsepsilon_2^{(k)},
        \qquad
        \bsepsilon_2^{(k)}
        \sim
        \CalN(\bszero,\sigma_c^2\bsI),
    \end{align}
\end{subequations}
for $k=0,\ldots,n_t$ with $n_t=10$.
In the experiments below, $\sigma_c^2=10^{-6}$.
The dataset is $\CalD
    =
    \left\{
        \hbsc_1^{(k)},
        \hbsc_2^{(k)}
    \right\}_{k=0}^{n_t}$.

\subsubsection{Candidate dictionary and model selection}
\label{sec:CH_model_selection}

The data are generated by a CH system with a single free-energy density $g$.
For system identification, we use a slightly more flexible parameterization in which the two equations are allowed to have separate free-energy representations $g_1$ and $g_2$:
\begin{subequations}
    \label{eq:cahn_hilliard_12}
    \begin{align}
        \pp{c_1}{t}
        &=
        m_1\nabla^2\mu_1,
        \qquad
        \mu_1
        =
        \pp{g_1}{c_1}
        -
        k_1\nabla^2 c_1,
        \\
        \pp{c_2}{t}
        &=
        m_2\nabla^2\mu_2,
        \qquad
        \mu_2
        =
        \pp{g_2}{c_2}
        -
        k_2\nabla^2 c_2.
    \end{align}
\end{subequations}
This relaxation does not enforce that $g_1$ and $g_2$ arise from the same scalar free energy.
It allows us to evaluate how accurately each equation recovers the corresponding free-energy derivative.
If thermodynamic consistency is desired, one could instead enforce a shared free energy $g_1=g_2=g$ during model selection.

We parameterize $g_1$ and $g_2$ using polynomial basis functions in $(c_1,c_2)$.
Let
\begin{subequations}
    \label{eq:CH_polynomial_sets}
    \begin{align}
        \CalP_1
        &=
        \left\{
            c_1^a c_2^b
            ~\middle|~
            2\leq a+b\leq 5,\;
            a\geq 1
        \right\},
        \\
        \CalP_2
        &=
        \left\{
            c_1^a c_2^b
            ~\middle|~
            2\leq a+b\leq 5,\;
            b\geq 1
        \right\}.
    \end{align}
\end{subequations}
Thus, $\CalP_1$ and $\CalP_2$ contain all monomials up to degree five that can contribute to
$\partial g_1/\partial c_1$ and $\partial g_2/\partial c_2$, respectively, after excluding constant and first-order terms.
Both sets have cardinality $14$.
We write
\begin{subequations}
    \label{eq:cahn_hilliard_energy_parameterized}
    \begin{align}
        g_1(c_1,c_2)
        &=
        \sum_{\phi\in\CalP_1}
        \theta_{\phi}^{g_1}
        \phi(c_1,c_2),
        \\
        g_2(c_1,c_2)
        &=
        \sum_{\phi\in\CalP_2}
        \theta_{\phi}^{g_2}
        \phi(c_1,c_2).
    \end{align}
\end{subequations}

Only products of mobilities and free-energy coefficients are identifiable from \cref{eq:cahn_hilliard_12}.
We therefore define the effective coefficients
\begin{subequations}
    \label{eq:CH_effective_coefficients}
    \begin{align}
        \eta_{\phi}^{g_1}
        &=
        m_1\theta_{\phi}^{g_1},
        &
        \eta_{\phi}^{g_2}
        &=
        m_2\theta_{\phi}^{g_2},
        \\
        \lambda_1
        &=
        -m_1k_1,
        &
        \lambda_2
        &=
        -m_2k_2.
    \end{align}
\end{subequations}
The quantities $\lambda_1$ and $\lambda_2$ are signed PDE coefficients multiplying the biharmonic terms.
They should not be confused with the physical positive gradient-energy coefficients $k_1$ and $k_2$.
For the data-generation parameters in \cref{eq:CH_generation_parameters}, the ground-truth signed values are
\begin{align}
    \lambda_1=\lambda_2=-1.
\end{align}

Substituting \cref{eq:cahn_hilliard_energy_parameterized} into \cref{eq:cahn_hilliard_12} gives
\begin{subequations}
    \label{eq:cahn_hilliard_expanded}
    \begin{align}
        \pp{c_1}{t}
        &=
        \sum_{\phi\in\CalP_1}
        \eta_{\phi}^{g_1}
        \nabla\cdot
        \left(
            \ppp{\phi}{c_1}
            \nabla c_1
            +
            \frac{\partial^2\phi}{\partial c_1\partial c_2}
            \nabla c_2
        \right)
        +
        \lambda_1\nabla^4 c_1,
        \\
        \pp{c_2}{t}
        &=
        \sum_{\phi\in\CalP_2}
        \eta_{\phi}^{g_2}
        \nabla\cdot
        \left(
            \frac{\partial^2\phi}{\partial c_1\partial c_2}
            \nabla c_1
            +
            \ppp{\phi}{c_2}
            \nabla c_2
        \right)
        +
        \lambda_2\nabla^4 c_2.
    \end{align}
\end{subequations}
The B-VSI coefficient vector is
\begin{align}
    \label{eq:CH_coefficient_vector}
    \bstheta
    =
    \left[
        \left\{
            \eta_{\phi}^{g_1}
        \right\}_{\phi\in\CalP_1},
        \lambda_1,
        \left\{
            \eta_{\phi}^{g_2}
        \right\}_{\phi\in\CalP_2},
        \lambda_2
    \right]^{\top}.
\end{align}
If the mobilities $m_i$ are known, the physical free-energy coefficients can be recovered by dividing $\eta_{\phi}^{g_i}$ by $m_i$.
If the mobilities are not known, only the effective coefficients in \cref{eq:CH_effective_coefficients} are identifiable.

The candidate dictionary consists of the operators appearing in \cref{eq:cahn_hilliard_expanded}:
\begin{align}
    \label{eq:cahn_hilliard_basis}
    \begin{split}
        \bschi
        =
        \bigg[
        &
        \left\{
        \nabla\cdot
        \left(
            \ppp{\phi}{c_1}
            \nabla c_1
            +
            \frac{\partial^2\phi}{\partial c_1\partial c_2}
            \nabla c_2
        \right)
        \right\}_{\phi\in\CalP_1},
        \nabla^4 c_1,
        \\
        &
        \left\{
        \nabla\cdot
        \left(
            \frac{\partial^2\phi}{\partial c_1\partial c_2}
            \nabla c_1
            +
            \ppp{\phi}{c_2}
            \nabla c_2
        \right)
        \right\}_{\phi\in\CalP_2},
        \nabla^4 c_2
        \bigg]^{\top}.
    \end{split}
\end{align}

For a coefficient function $a(c_1,c_2)$ and concentration field $c$, the weak form of the divergence operator is
\begin{align}
    \label{eq:cahn_hilliard_weak_second_order}
    \int_{\Omega}
    v^h
    \nabla\cdot
    \left(
        a(c_1^h,c_2^h)\nabla c^h
    \right)
    \,\mathrm{d}\bsx
    =
    -
    \int_{\Omega}
    a(c_1^h,c_2^h)
    \nabla v^h\cdot\nabla c^h
    \,\mathrm{d}\bsx
    +
    \int_{\partial\Omega}
    v^h
    a(c_1^h,c_2^h)
    \nabla c^h\cdot\bsn
    \,\mathrm{d}S.
\end{align}
In \cref{eq:cahn_hilliard_basis}, $a$ is one of the second derivatives of a polynomial basis function.
Each summed strong-form candidate term therefore gives a corresponding summed weak-form contribution with its associated effective coefficient.

The biharmonic terms are assembled using a discontinuous Galerkin weak form~\cite{ENGEL2002DG,wells2006discontinuous,Wells2007Analysis}:
\begin{align}
    \label{eq:cahn_hilliard_biharmonic_term}
    \begin{split}
        \int_{\Omega}
        v^h\nabla^4 c^h
        \,\mathrm{d}\bsx
        =
        &
        \sum_{\Omega_e}
        \int_{\Omega_e}
        \nabla^2 v^h
        \nabla^2 c^h
        \,\mathrm{d}\bsx
        \\
        &
        +
        \sum_{E\in E_h^{\mathrm{int}}}
        \int_E
        \left[
            \frac{\alpha}{h_E}
            [\nabla v^h][\nabla c^h]
            -
            \left\langle
                \nabla^2 c^h
            \right\rangle
            [\nabla v^h]
            -
            [\nabla c^h]
            \left\langle
                \nabla^2 v^h
            \right\rangle
        \right]
        \,\mathrm{d}S .
    \end{split}
\end{align}
Here, $\langle\cdot\rangle$ denotes the average across an interior facet, $[\cdot]$ denotes the jump with respect to the outward normals on the neighboring elements, $\alpha\geq0$ is the penalty parameter, $h_E$ is a facet-length scale, and $E_h^{\mathrm{int}}$ is the set of interior facets.
The biharmonic weak form is linear in $c^h$, so no nonlinear residual linearization is required for this operator.

The free-energy operators in \cref{eq:cahn_hilliard_basis} are nonlinear functions of the observed concentration fields.
We therefore construct the residual covariance using the nonlinear local Gaussian approximation in \cref{sec:nonlinear_residual_likelihood}.
Its accuracy for this example is assessed in \cref{app:approximation_numerical}.
Sequential operator elimination is then performed using the residual-space BIC in \cref{eq:residual_bic}.
During model selection, only the residual-space MLE is computed for each candidate model.
Full posterior inference is performed after the selected model $\CalM^*$ is fixed.

\begin{figure}[htbp]
    \centering
    \captionsetup[subfigure]{justification=centering}
    \includegraphics[width=0.65\textwidth]{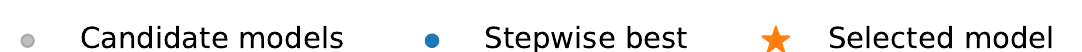}\par\smallskip
    \begin{subfigure}[t]{0.49\textwidth}
        \centering
        \includegraphics[width=\textwidth,trim={0bp 0bp 237bp 0bp},clip]{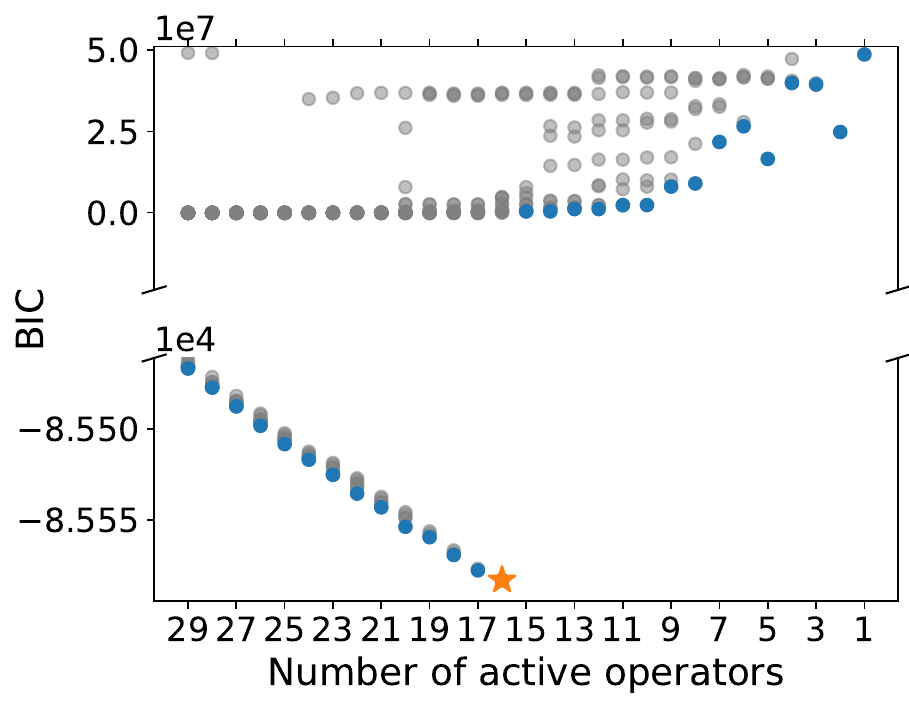}
        \caption{Block-diagonal likelihood covariance}
        \label{fig:CH_model_selection_bd}
    \end{subfigure}
    \hfill
    \begin{subfigure}[t]{0.49\textwidth}
        \centering
        \includegraphics[width=\textwidth,trim={0bp 0bp 237bp 0bp},clip]{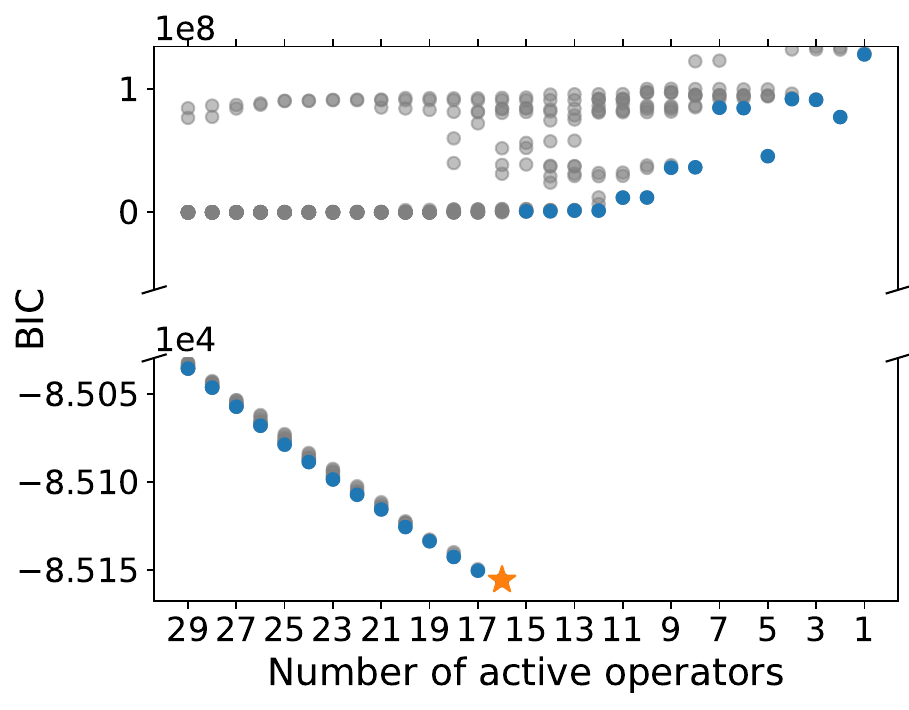}
        \caption{Block-tridiagonal likelihood covariance}
        \label{fig:CH_model_selection_bt}
    \end{subfigure}
    \caption{Residual-space BIC along the sequential operator-elimination paths for the two-field CH equation using (a) block-diagonal and (b) block-tridiagonal likelihood covariance structures.
    Each vertical group corresponds to the candidate models obtained by removing one operator from the current model.
    Gray points represent candidate models, blue points indicate the best model at each elimination step, and orange stars mark the selected 16-operator models. Each panel uses a broken vertical axis with different scales above and below the break.}
    \label{fig:CH_model_selection}
\end{figure}

\Cref{fig:CH_model_selection} shows the BIC values along the elimination paths for both covariance structures.
The BIC decreases as redundant polynomial operators are removed and increases once essential operators are excluded, indicating the onset of underfitting.
Both covariance structures select the same 16-operator model. Posterior inference is performed with this model, and the estimated coefficients are summarized in \cref{tab:CH_parameters}.

\begin{table}[htbp]
\centering
\caption{
Estimated effective coefficients for the CH equation.
The coefficients $\eta_{\phi}^{g_i}=m_i\theta_{\phi}^{g_i}$ are mobility-scaled free-energy coefficients, and $\lambda_i=-m_i k_i$ are signed biharmonic coefficients.
Thus, the rows labeled $\lambda_1$ and $\lambda_2$ correspond to the PDE coefficients multiplying $\nabla^4 c_1$ and $\nabla^4 c_2$, not to the physical positive coefficients $k_1$ and $k_2$.
The notation $\hbstheta$ denotes the point estimate for VSI and the posterior mean for B-VSI methods.
Variances are reported only for posterior coefficient distributions and for $\sigma_u^2$ when it is inferred; the noise variance is listed as $10^{5}\sigma_u^2$, with its variance scaled accordingly.
Aggregate error metrics are computed from the reported point estimate or posterior mean and are therefore listed without posterior variances.
}
\label{tab:CH_parameters}
\fontsize{9pt}{10pt}\selectfont
\setlength{\tabcolsep}{3pt}
\sisetup{round-mode=places, round-precision=2, exponent-product=\!\times\!}
\begin{tabular}{@{}c S[table-format=-2.2] S[table-format=4.2]
*{4}{S[table-format=-2.2] S[table-format=1.2e-1]}@{}}
\toprule
\multirow{3}{*}{Coefficient}
& {\multirow{3}{*}{\makecell{Ground\\truth}}}
& {\multirow{3}{*}{VSI}}
& \multicolumn{4}{c}{Known noise}
& \multicolumn{4}{c}{Unknown noise} \\
\cmidrule(lr){4-7}\cmidrule(l){8-11}
& & & \multicolumn{2}{c}{B-VSI-BD}
& \multicolumn{2}{c}{B-VSI-BT}
& \multicolumn{2}{c}{B-VSI-BD}
& \multicolumn{2}{c}{B-VSI-BT} \\
\cmidrule(lr){4-5}\cmidrule(lr){6-7}\cmidrule(lr){8-9}\cmidrule(l){10-11}
& & & {Mean} & {Variance} & {Mean} & {Variance}
& {Mean} & {Variance}
& {Mean} & {Variance} \\
\midrule
$\lambda_1$ & -1.000 & -0.001 & -0.966 & 1.25e-06 & -0.993 & 2.48e-07 & -0.947 & 4.88e-05 & -0.990 & 9.94e-06 \\
$\eta^{g_1}_{c_1^4}$ & 3.998 & 0.025 & 3.883 & 2.73e-05 & 3.981 & 6.55e-06 & 3.843 & 1.07e-03 & 3.983 & 2.62e-04 \\
$\eta^{g_1}_{c_1^2c_2^2}$ & 7.997 & 0.045 & 7.798 & 3.94e-04 & 7.939 & 1.49e-04 & 7.219 & 1.43e-02 & 7.679 & 5.75e-03 \\
$\eta^{g_1}_{c_1^3}$ & -7.997 & -0.048 & -7.766 & 1.09e-04 & -7.962 & 2.64e-05 & -7.687 & 4.28e-03 & -7.966 & 1.06e-03 \\
$\eta^{g_1}_{c_1^2c_2}$ & -10.796 & -0.025 & -10.514 & 4.70e-04 & -10.721 & 1.74e-04 & -9.847 & 1.70e-02 & -10.428 & 6.74e-03 \\
$\eta^{g_1}_{c_1c_2^2}$ & -7.997 & -0.046 & -7.798 & 3.98e-04 & -7.937 & 1.51e-04 & -7.217 & 1.44e-02 & -7.676 & 5.82e-03 \\
$\eta^{g_1}_{c_1^2}$ & 8.906 & 0.022 & 8.655 & 9.98e-05 & 8.860 & 2.28e-05 & 8.415 & 3.85e-03 & 8.782 & 9.04e-04 \\
$\eta^{g_1}_{c_1c_2}$ & 10.796 & 0.023 & 10.514 & 4.72e-04 & 10.720 & 1.76e-04 & 9.845 & 1.71e-02 & 10.425 & 6.81e-03 \\
$\lambda_2$ & -1.000 & -0.001 & -0.972 & 9.87e-07 & -0.996 & 1.55e-07 & -0.958 & 3.89e-05 & -0.994 & 6.20e-06 \\
$\eta^{g_2}_{c_1^2c_2^2}$ & 7.997 & 0.061 & 7.818 & 2.31e-04 & 7.963 & 9.19e-05 & 7.419 & 8.68e-03 & 7.774 & 3.60e-03 \\
$\eta^{g_2}_{c_2^4}$ & 3.998 & 0.032 & 3.912 & 2.33e-05 & 3.991 & 6.73e-06 & 3.904 & 9.23e-04 & 4.006 & 2.70e-04 \\
$\eta^{g_2}_{c_1^2c_2}$ & -10.796 & -0.041 & -10.552 & 3.38e-04 & -10.752 & 1.26e-04 & -10.066 & 1.27e-02 & -10.529 & 4.94e-03 \\
$\eta^{g_2}_{c_1c_2^2}$ & -7.997 & -0.055 & -7.817 & 2.30e-04 & -7.961 & 9.34e-05 & -7.417 & 8.67e-03 & -7.770 & 3.66e-03 \\
$\eta^{g_2}_{c_2^3}$ & -7.064 & -0.069 & -6.913 & 6.78e-05 & -7.050 & 1.93e-05 & -6.889 & 2.69e-03 & -7.073 & 7.75e-04 \\
$\eta^{g_2}_{c_1c_2}$ & 10.796 & 0.035 & 10.550 & 3.37e-04 & 10.749 & 1.28e-04 & 10.064 & 1.27e-02 & 10.525 & 5.00e-03 \\
$\eta^{g_2}_{c_2^2}$ & 6.107 & 0.051 & 5.975 & 3.72e-05 & 6.090 & 7.59e-06 & 5.858 & 1.45e-03 & 6.055 & 3.02e-04 \\
\midrule
$10^{5}\sigma_u^2$ & 0.10 & {N/A} & {N/A} & {N/A} & {N/A} & {N/A} & 4.02 & 6.22e-04 & 4.02 & 6.21e-04 \\
\midrule
$\norm{\bstheta_{\mathrm{GT}}-\hbstheta}{2}$ & {N/A} & 31.260 & 0.775 & {N/A} & 0.170 & {N/A} & 2.285 & {N/A} & 0.861 & {N/A} \\
$\mathrm{MSE}(\partial g_1/\partial c_1)$ & {N/A} & 2524.233 & 1.937 & {N/A} & 0.090 & {N/A} & 11.837 & {N/A} & 1.245 & {N/A} \\
$\mathrm{MSE}(\partial g_2/\partial c_2)$ & {N/A} & 1322.919 & 0.665 & {N/A} & 0.018 & {N/A} & 3.975 & {N/A} & 0.401 & {N/A} \\
\bottomrule
\end{tabular}
\end{table}

\subsubsection{Parameter and predictive uncertainty with known noise variance}
\label{sec:CH_known_noise}

\begin{figure}[htbp]
    \centering
    \captionsetup[subfigure]{justification=centering}
    \begin{subfigure}[t]{0.49\textwidth}
        \centering
        \includegraphics[height=6.68cm]{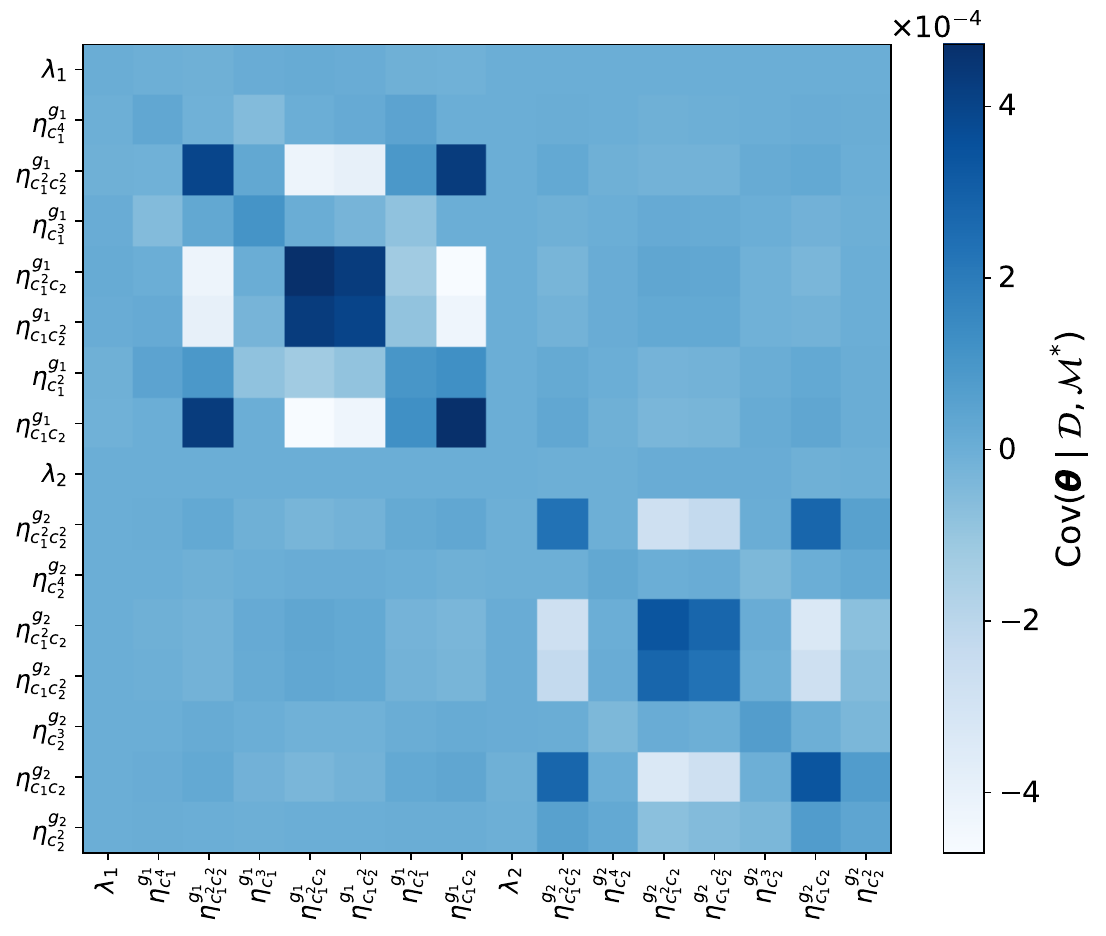}
        \caption{Block-diagonal likelihood covariance}
        \label{fig:CH_knownNoise_posterior_covariance_bd}
    \end{subfigure}
    \hfill
    \begin{subfigure}[t]{0.49\textwidth}
        \centering
        \includegraphics[height=6.68cm]{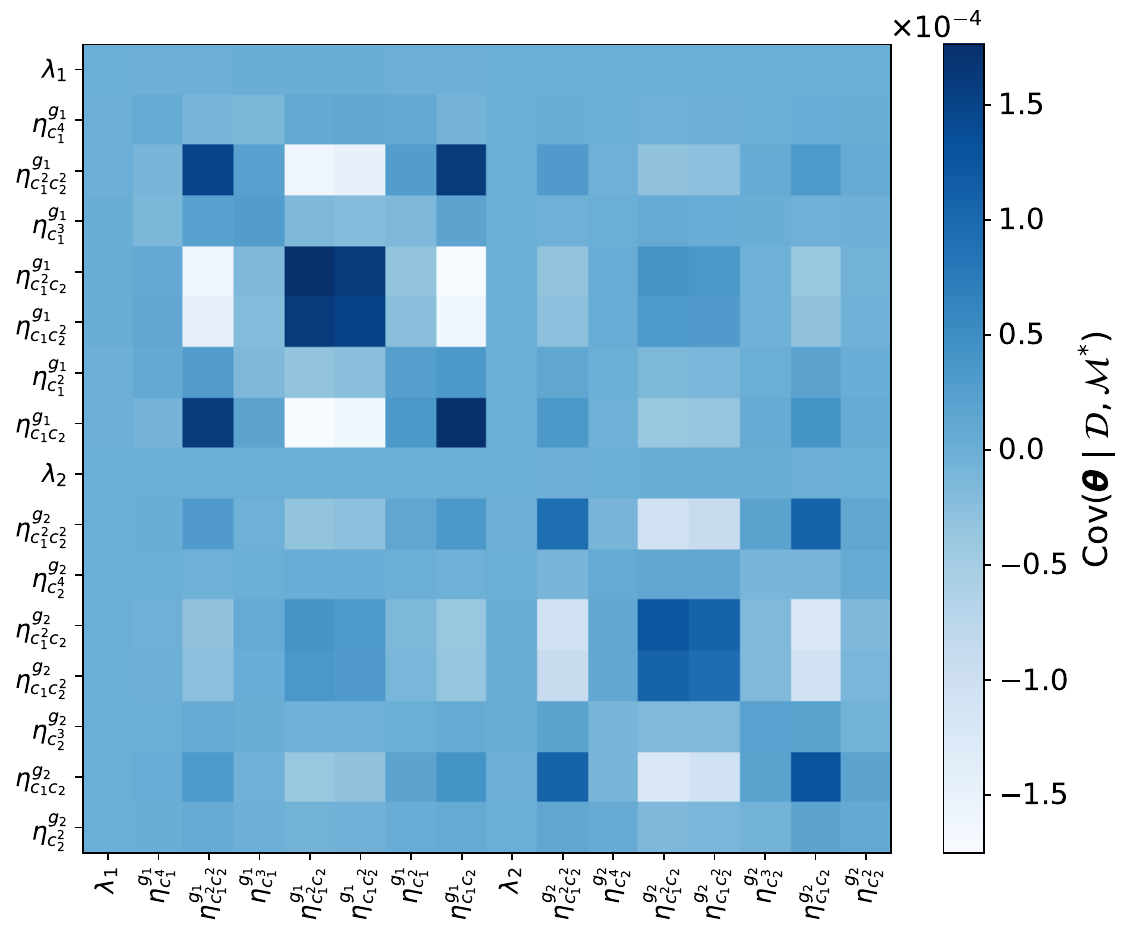}
        \caption{Block-tridiagonal likelihood covariance}
        \label{fig:CH_knownNoise_posterior_covariance_bt}
    \end{subfigure}
    \caption{Covariance matrices of the approximate Gaussian posterior $p(\bstheta \mid \CalD,\CalM^*)$ for the selected CH model with known observation-noise variance, using (a) block-diagonal and (b) block-tridiagonal likelihood covariance structures. The panels use separate color scales.}
    \label{fig:CH_knownNoise_posterior_covariance}
\end{figure}

We first assume that the observation-noise variance is known.
A Gaussian prior is assigned to the effective coefficient vector:
\begin{align}
    \label{eq:CH_parameters_prior}
    \bstheta
    \sim
    \CalN(\bszero,\bsI).
\end{align}
Under the selected model $\CalM^*$, the lagged-covariance Gaussian posterior update in \cref{sec:lagged_covariance_inference} gives an approximate posterior distribution $p(\bstheta \mid \CalD,\CalM^*)$.
The posterior means and variances are reported in \cref{tab:CH_parameters}, and the posterior covariance matrices are shown in \cref{fig:CH_knownNoise_posterior_covariance}.

The posterior covariance shows relatively high uncertainty in mixed concentration terms, including
$c_1^2c_2^2$, $c_1^2c_2$, $c_1c_2^2$, and $c_1c_2$.
These terms couple the two concentration fields and can be more difficult to distinguish from one another when the observed trajectory samples only a limited portion of concentration space.

\begin{figure}[htbp]
    \centering
    \captionsetup[subfigure]{justification=centering}
    \begin{subfigure}[t]{0.49\textwidth}
        \centering
        \includegraphics[height=5.51041cm]{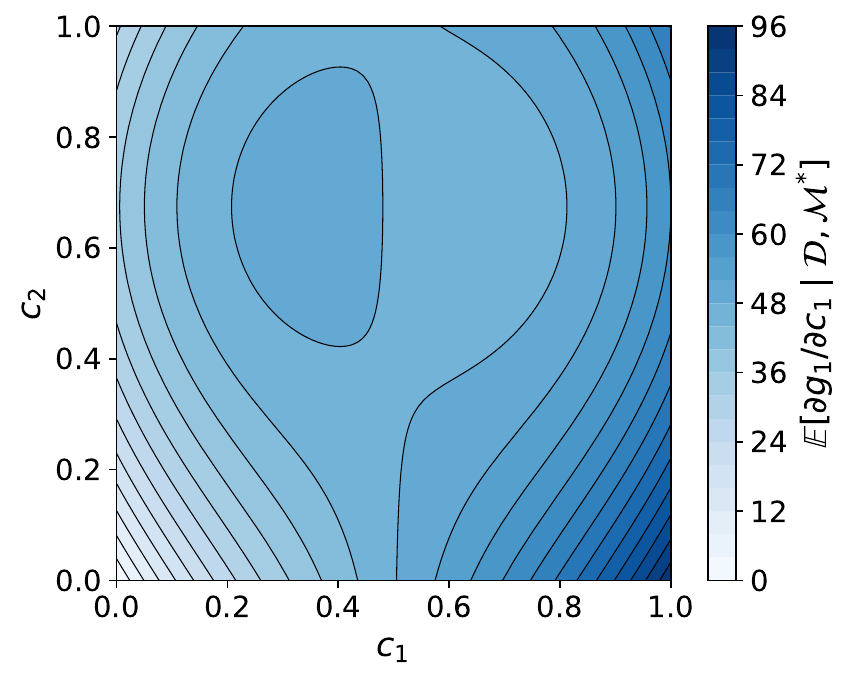}
        \caption{Predictive mean of $\partial g_1/\partial c_1$}
        \label{fig:CH_knownNoise_predictive_partialgc1_mean}
    \end{subfigure}
    \hfill
    \begin{subfigure}[t]{0.49\textwidth}
        \centering
        \includegraphics[height=5.51041cm]{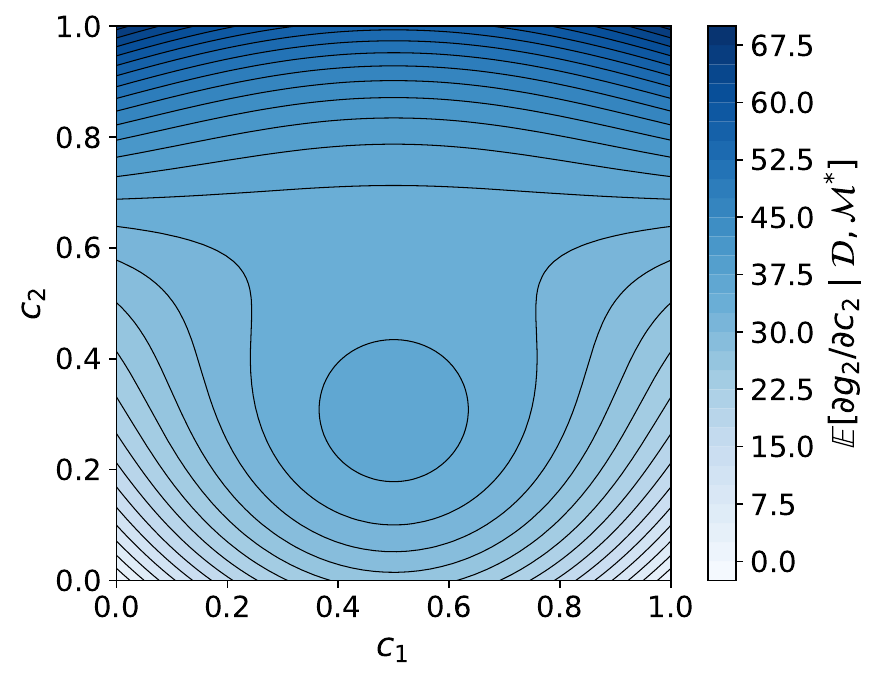}
        \caption{Predictive mean of $\partial g_2/\partial c_2$}
        \label{fig:CH_knownNoise_predictive_partialgc2_mean}
    \end{subfigure}
    \par\medskip
    \begin{subfigure}[t]{0.49\textwidth}
        \centering
        \includegraphics[height=5.65877cm]{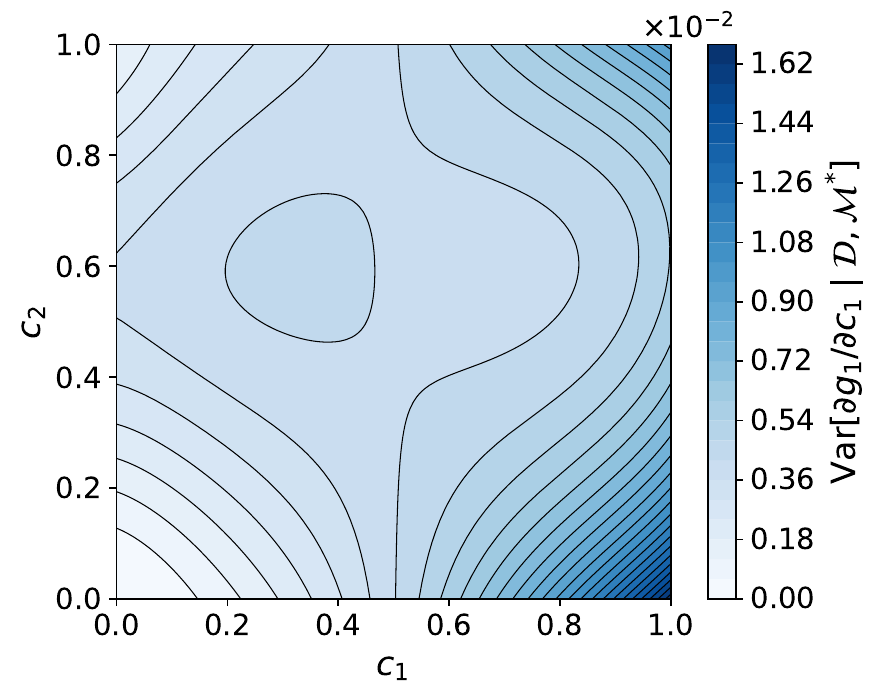}
        \caption{Predictive variance of $\partial g_1/\partial c_1$}
        \label{fig:CH_knownNoise_predictive_partialgc1_var}
    \end{subfigure}
    \hfill
    \begin{subfigure}[t]{0.49\textwidth}
        \centering
        \includegraphics[height=5.65877cm]{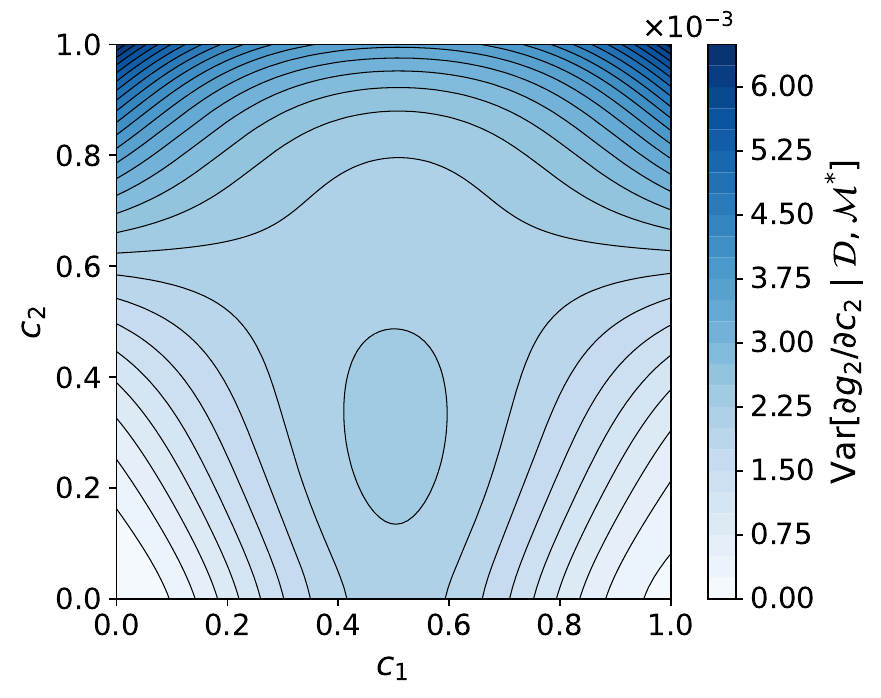}
        \caption{Predictive variance of $\partial g_2/\partial c_2$}
        \label{fig:CH_knownNoise_predictive_partialgc2_var}
    \end{subfigure}
    \caption{Mean and variance of the posterior predictive distribution of the free-energy derivatives $\partial g_1/\partial c_1$ and $\partial g_2/\partial c_2$ for the selected CH model with known observation-noise variance, using the approximate Gaussian posterior with block-diagonal likelihood covariance. Panels (a, b) show the predictive means and panels (c, d) the predictive variances.}
    \label{fig:CH_knownNoise_predictive}
\end{figure}

\begin{figure}[htbp]
    \centering
    \captionsetup[subfigure]{justification=centering}
    \begin{subfigure}[t]{0.49\textwidth}
        \centering
        \includegraphics[height=5.51041cm]{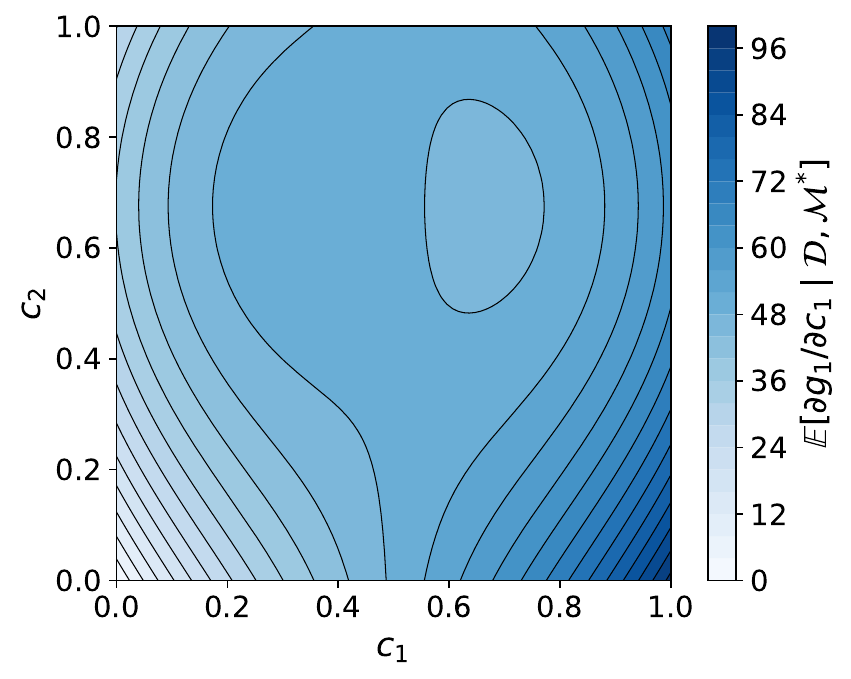}
        \caption{Predictive mean of $\partial g_1/\partial c_1$}
        \label{fig:CH_knownNoise_predictive_partialgc1_mean_bt}
    \end{subfigure}
    \hfill
    \begin{subfigure}[t]{0.49\textwidth}
        \centering
        \includegraphics[height=5.51041cm]{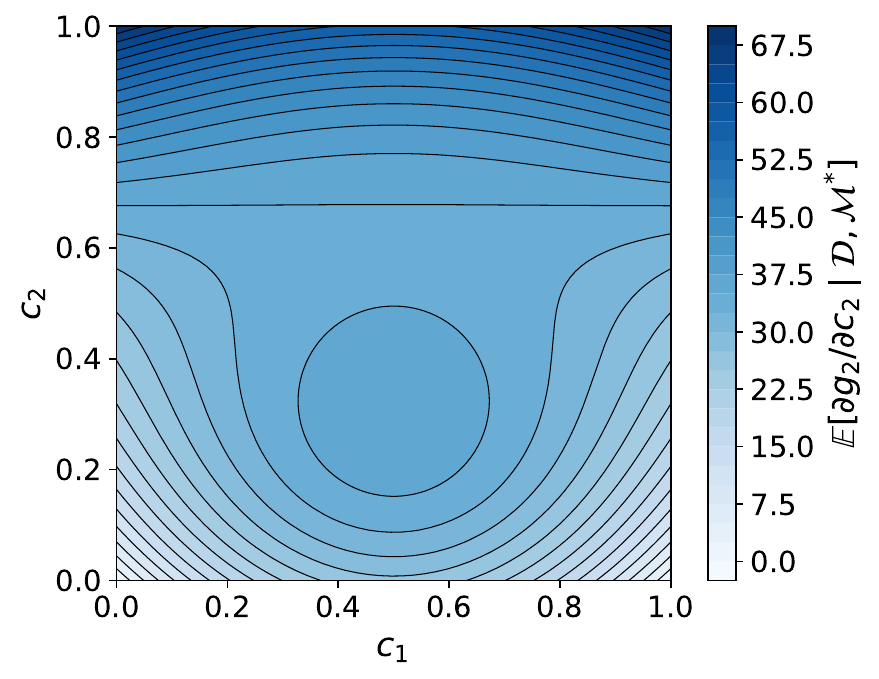}
        \caption{Predictive mean of $\partial g_2/\partial c_2$}
        \label{fig:CH_knownNoise_predictive_partialgc2_mean_bt}
    \end{subfigure}
    \par\medskip
    \begin{subfigure}[t]{0.49\textwidth}
        \centering
        \includegraphics[height=5.65877cm]{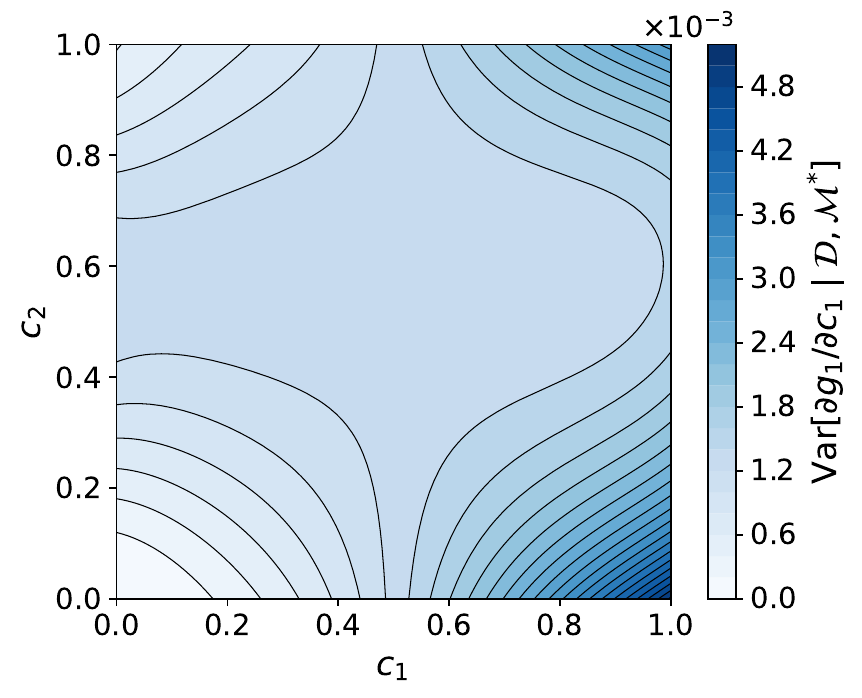}
        \caption{Predictive variance of $\partial g_1/\partial c_1$}
        \label{fig:CH_knownNoise_predictive_partialgc1_var_bt}
    \end{subfigure}
    \hfill
    \begin{subfigure}[t]{0.49\textwidth}
        \centering
        \includegraphics[height=5.65877cm]{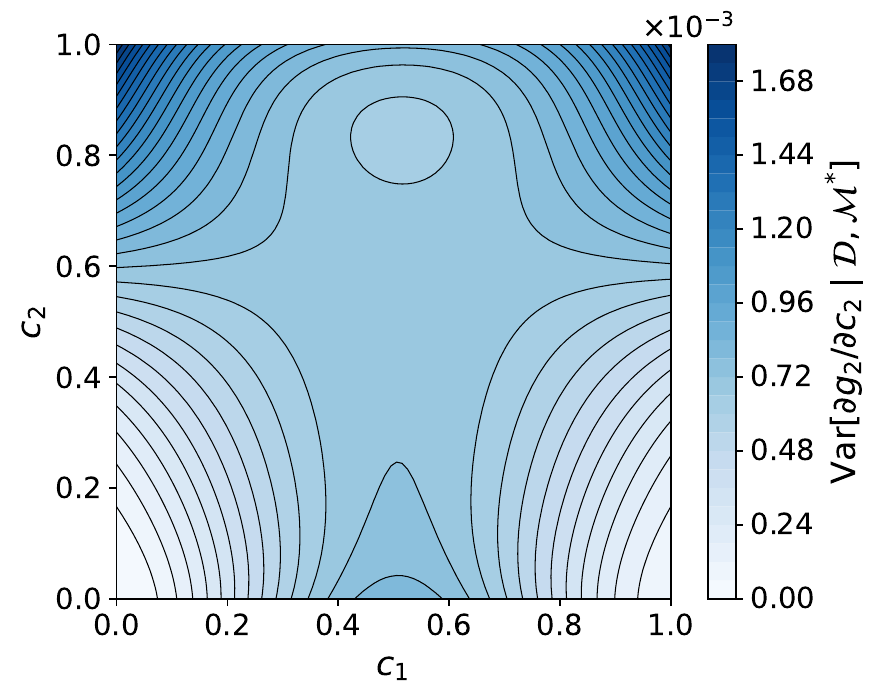}
        \caption{Predictive variance of $\partial g_2/\partial c_2$}
        \label{fig:CH_knownNoise_predictive_partialgc2_var_bt}
    \end{subfigure}
    \caption{Mean and variance of the posterior predictive distribution of the free-energy derivatives $\partial g_1/\partial c_1$ and $\partial g_2/\partial c_2$ for the selected CH model with known observation-noise variance, using the approximate Gaussian posterior with block-tridiagonal likelihood covariance. Panels (a, b) show the predictive means and panels (c, d) the predictive variances.}
    \label{fig:CH_knownNoise_predictive_bt}
\end{figure}

We propagate the coefficient posterior to the free-energy derivatives.
Because the estimated coefficients are mobility-scaled, define the effective derivative fields
\begin{subequations}
    \label{eq:CH_effective_derivatives}
    \begin{align}
        \zeta_1(c_1,c_2)
        &=
        \sum_{\phi\in\CalP_1}
        \eta_{\phi}^{g_1}
        \pp{\phi}{c_1}
        =
        m_1\pp{g_1}{c_1},
        \\
        \zeta_2(c_1,c_2)
        &=
        \sum_{\phi\in\CalP_2}
        \eta_{\phi}^{g_2}
        \pp{\phi}{c_2}
        =
        m_2\pp{g_2}{c_2}.
    \end{align}
\end{subequations}
Since the mobilities are known in this example, posterior predictions for
$\partial g_i/\partial c_i$ are obtained by dividing $\zeta_i$ by $m_i$.
Predictive variances of $\partial g_i/\partial c_i$ are obtained by dividing those of $\zeta_i$ by $m_i^2$. These derivative fields are linear in the coefficient vector, so the Gaussian predictive formula in \cref{sec:posterior_predictive} applies.
The posterior predictive means and variances are shown in \cref{fig:CH_knownNoise_predictive,fig:CH_knownNoise_predictive_bt}.
The predictive means closely match the ground-truth free-energy derivatives shown in \cref{fig:CH_energy_partialgc1,fig:CH_energy_partialgc2}, as also reflected by the MSE values in \cref{tab:CH_parameters}. These derivative MSEs are evaluated on a uniform $100\times100$ grid in $[0,1]^2$, with a common zero additive-constant convention for each derivative. The block-tridiagonal covariance gives lower coefficient error and derivative MSEs than the block-diagonal approximation in this example.
The predictive variances are larger in regions of concentration space that are less strongly sampled by the simulated phase-separation trajectory.
These regions correspond to a wider range of coefficient values that remain plausible under the observed data.

\subsubsection{Joint inference with unknown scalar noise variance}
\label{sec:CH_unknown_noise}

\begin{figure}[htbp]
    \centering
    \captionsetup[subfigure]{justification=centering}
    \begin{subfigure}[t]{0.49\textwidth}
        \centering
        \includegraphics[height=6.68cm]{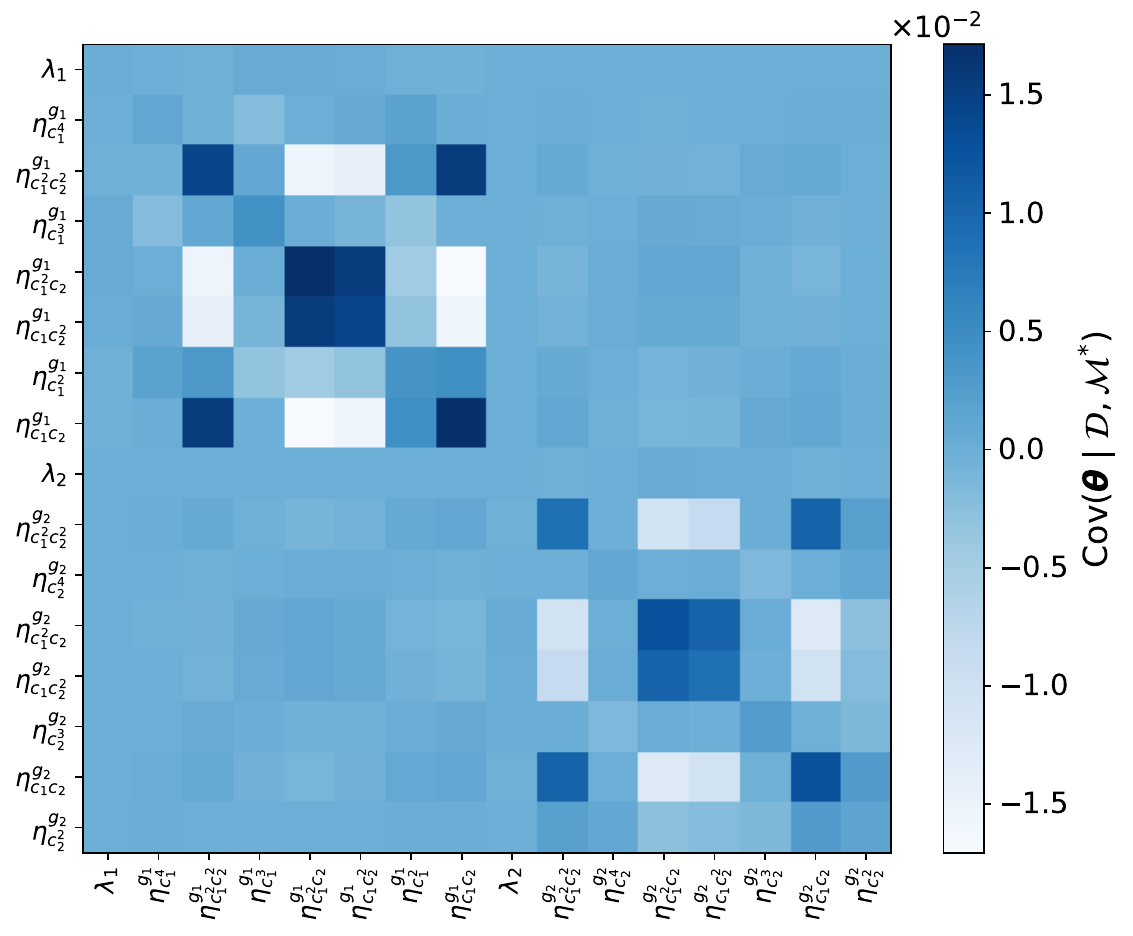}
        \caption{Block-diagonal likelihood covariance}
        \label{fig:CH_unknownNoise_posterior_covariance_bd}
    \end{subfigure}
    \hfill
    \begin{subfigure}[t]{0.49\textwidth}
        \centering
        \includegraphics[height=6.68cm]{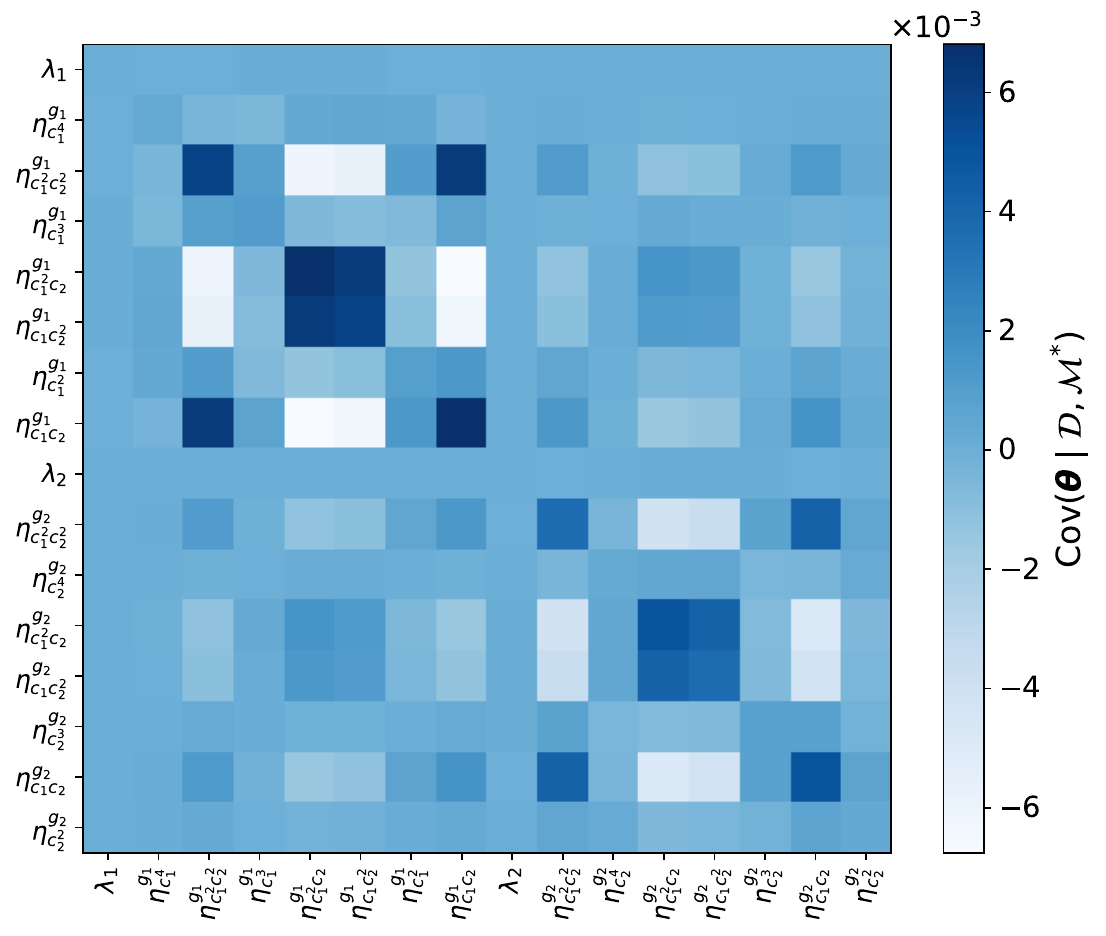}
        \caption{Block-tridiagonal likelihood covariance}
        \label{fig:CH_unknownNoise_posterior_covariance_bt}
    \end{subfigure}
    \caption{Marginal covariance matrices of the approximate posterior $p(\bstheta \mid \CalD,\CalM^*)$ for the selected CH model with unknown observation-noise variance, using (a) block-diagonal and (b) block-tridiagonal likelihood covariance structures. The noise variance is integrated out. The panels use separate color scales.}
    \label{fig:CH_unknownNoise_posterior_covariance}
\end{figure}

We next infer the observation-noise variance jointly with the effective CH coefficients.
To isolate the effect of unknown noise variance, we fix the model form to the selected model $\CalM^*$ rather than repeating model selection.
The observation-noise covariance is assumed to have the scalar form in \cref{eq:unknown_noise_observation_covariance},
\begin{align}
    \bsSigma_{\bsu}^{(k)}
    =
    \sigma_u^2\bsI,
    \qquad
    k=0,\ldots,n_t.
    \nonumber
\end{align}
We assign the conjugate prior
\begin{subequations}
    \label{eq:CH_unknown_noise_prior}
    \begin{align}
        \sigma_u^2
        &\sim
        \operatorname{Inv\text{-}Gamma}(1,1),
        \\
        \bstheta \mid \sigma_u^2,\CalM^*
        &\sim
        \CalN(\bszero,\sigma_u^2\bsI).
    \end{align}
\end{subequations}
The normal-inverse-gamma update in \cref{sec:lagged_covariance_inference} gives an approximate joint posterior over $(\bstheta,\sigma_u^2)$.
The posterior moments of the coefficients and the observation-noise variance are reported in \cref{tab:CH_parameters}, and the posterior coefficient covariance matrices are shown in \cref{fig:CH_unknownNoise_posterior_covariance}.

\begin{figure}[htbp]
    \centering
    \captionsetup[subfigure]{justification=centering}
    \begin{subfigure}[t]{0.49\textwidth}
        \centering
        \includegraphics[height=5.51041cm]{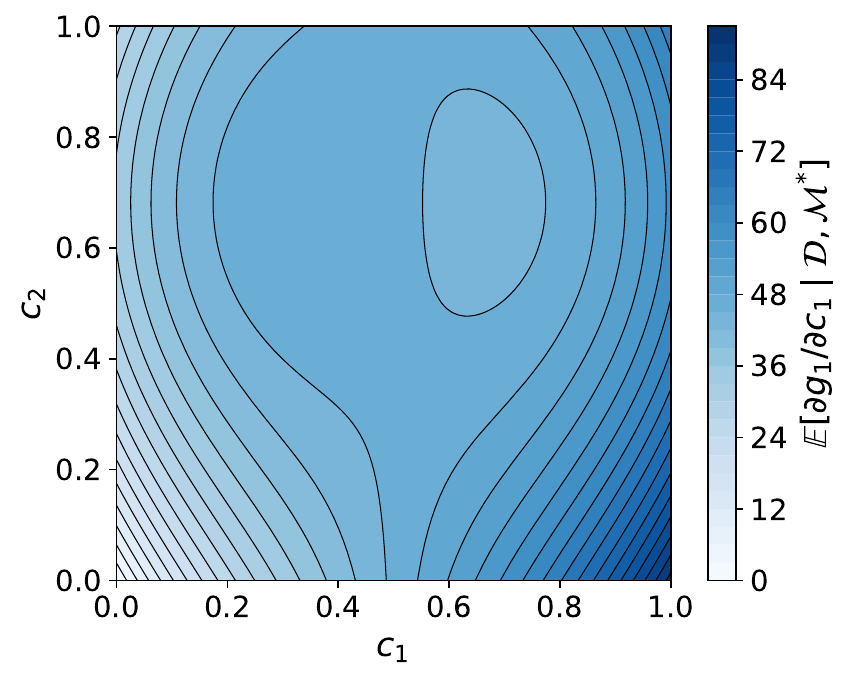}
        \caption{Predictive mean of $\partial g_1/\partial c_1$}
        \label{fig:CH_unknownNoise_predictive_partialgc1_mean}
    \end{subfigure}
    \hfill
    \begin{subfigure}[t]{0.49\textwidth}
        \centering
        \includegraphics[height=5.51041cm]{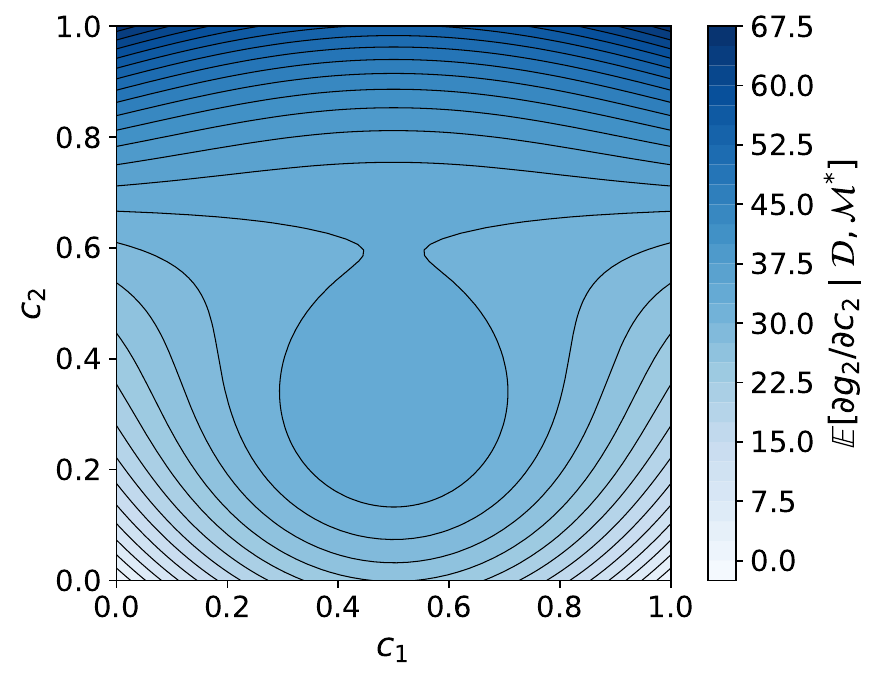}
        \caption{Predictive mean of $\partial g_2/\partial c_2$}
        \label{fig:CH_unknownNoise_predictive_partialgc2_mean}
    \end{subfigure}
    \par\medskip
    \begin{subfigure}[t]{0.49\textwidth}
        \centering
        \includegraphics[height=5.65877cm]{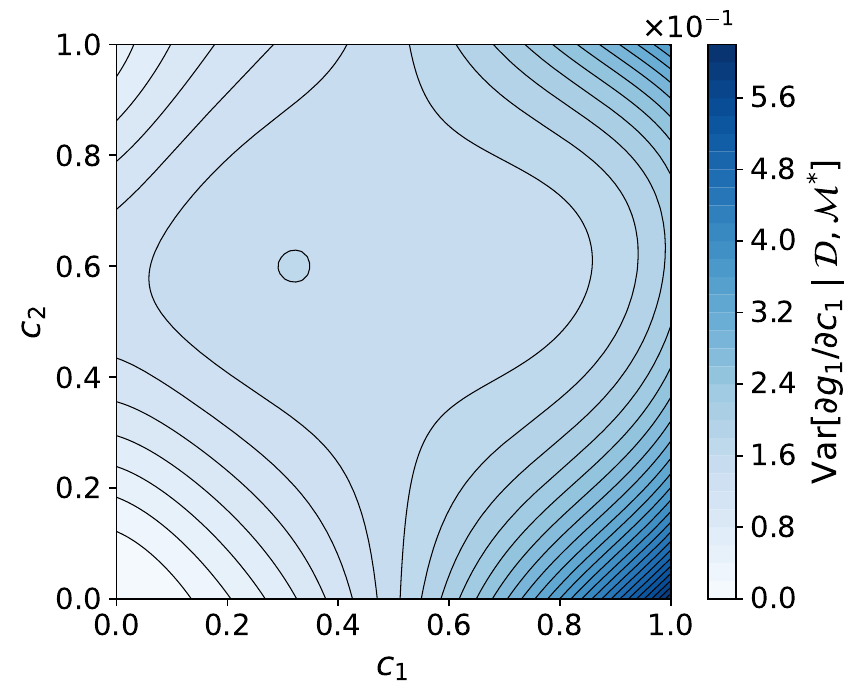}
        \caption{Predictive variance of $\partial g_1/\partial c_1$}
        \label{fig:CH_unknownNoise_predictive_partialgc1_var}
    \end{subfigure}
    \hfill
    \begin{subfigure}[t]{0.49\textwidth}
        \centering
        \includegraphics[height=5.65877cm]{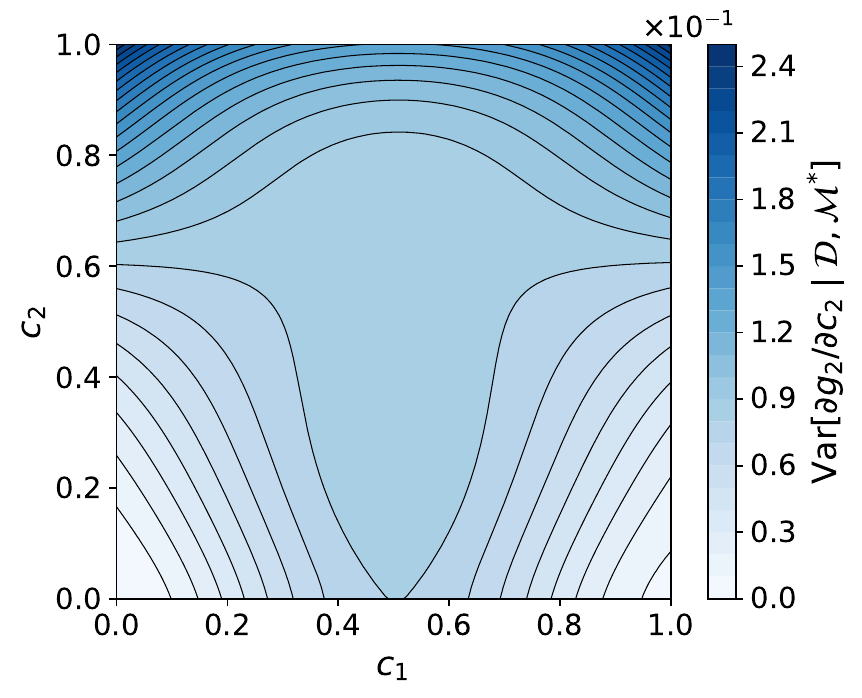}
        \caption{Predictive variance of $\partial g_2/\partial c_2$}
        \label{fig:CH_unknownNoise_predictive_partialgc2_var}
    \end{subfigure}
    \caption{Mean and variance of the posterior predictive distribution of the free-energy derivatives $\partial g_1/\partial c_1$ and $\partial g_2/\partial c_2$ for the selected CH model with unknown observation-noise variance, using the approximate posterior with block-diagonal likelihood covariance. Panels (a, b) show the predictive means and panels (c, d) the predictive variances.}
    \label{fig:CH_unknownNoise_predictive}
\end{figure}

\begin{figure}[htbp]
    \centering
    \captionsetup[subfigure]{justification=centering}
    \begin{subfigure}[t]{0.49\textwidth}
        \centering
        \includegraphics[height=5.51041cm]{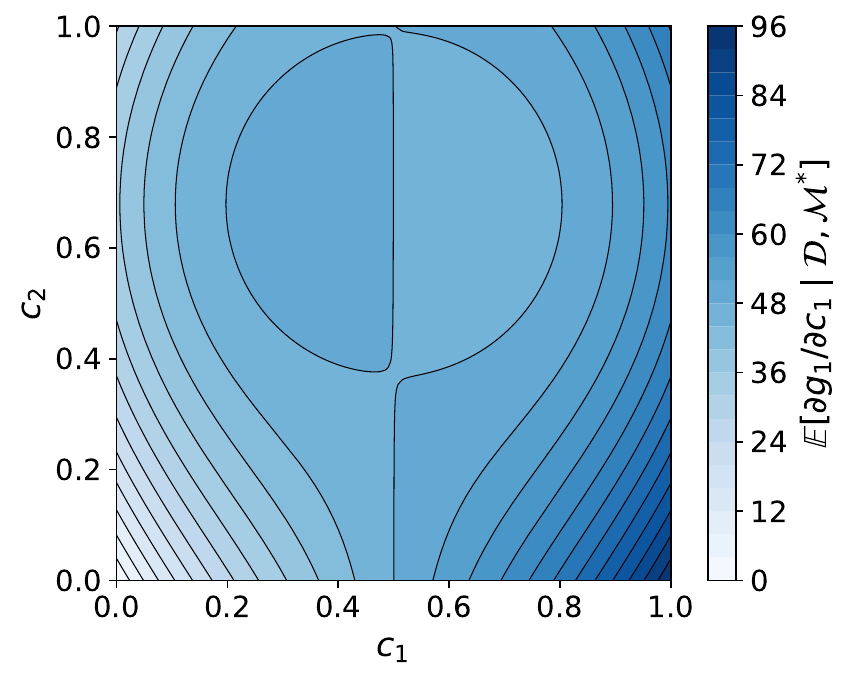}
        \caption{Predictive mean of $\partial g_1/\partial c_1$}
        \label{fig:CH_unknownNoise_predictive_partialgc1_mean_bt}
    \end{subfigure}
    \hfill
    \begin{subfigure}[t]{0.49\textwidth}
        \centering
        \includegraphics[height=5.51041cm]{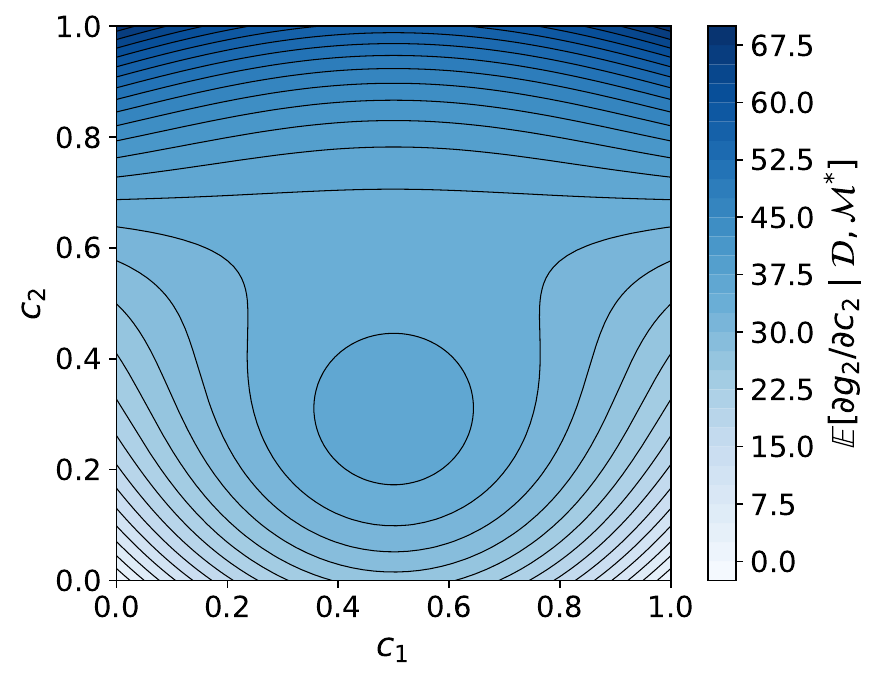}
        \caption{Predictive mean of $\partial g_2/\partial c_2$}
        \label{fig:CH_unknownNoise_predictive_partialgc2_mean_bt}
    \end{subfigure}
    \par\medskip
    \begin{subfigure}[t]{0.49\textwidth}
        \centering
        \includegraphics[height=5.65877cm]{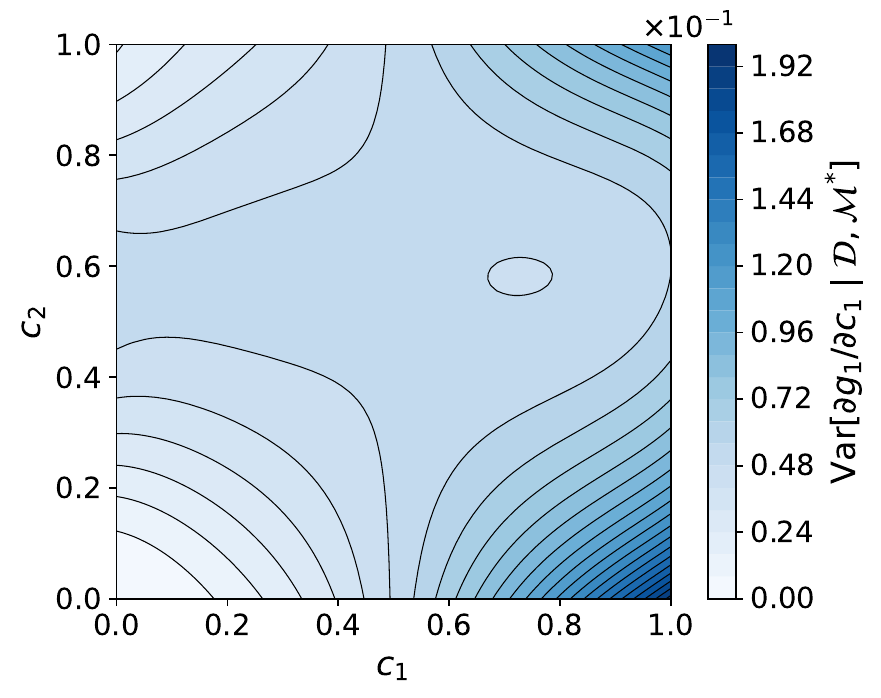}
        \caption{Predictive variance of $\partial g_1/\partial c_1$}
        \label{fig:CH_unknownNoise_predictive_partialgc1_var_bt}
    \end{subfigure}
    \hfill
    \begin{subfigure}[t]{0.49\textwidth}
        \centering
        \includegraphics[height=5.65877cm]{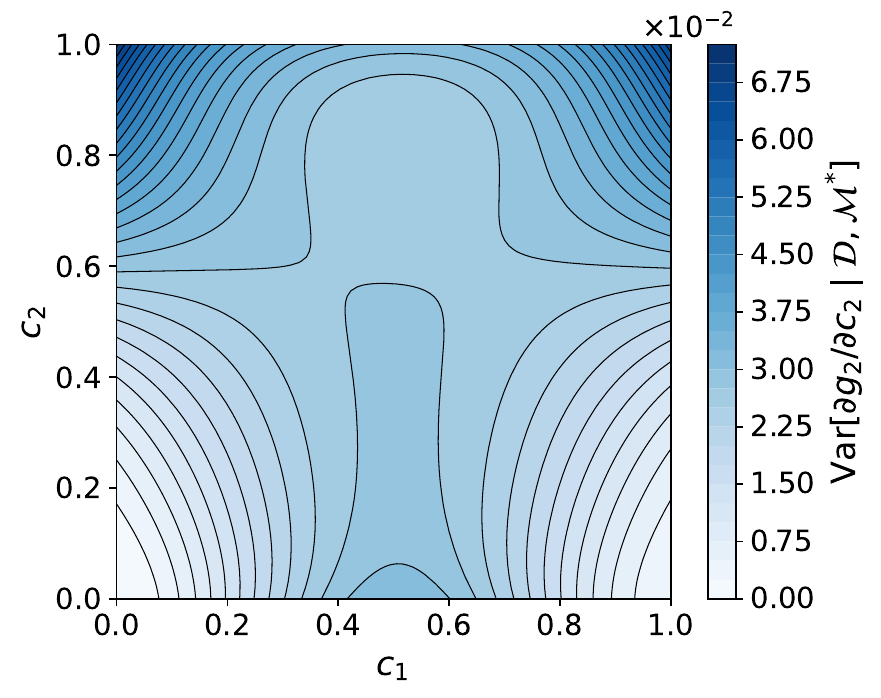}
        \caption{Predictive variance of $\partial g_2/\partial c_2$}
        \label{fig:CH_unknownNoise_predictive_partialgc2_var_bt}
    \end{subfigure}
    \caption{Mean and variance of the posterior predictive distribution of the free-energy derivatives $\partial g_1/\partial c_1$ and $\partial g_2/\partial c_2$ for the selected CH model with unknown observation-noise variance, using the approximate posterior with block-tridiagonal likelihood covariance. Panels (a, b) show the predictive means and panels (c, d) the predictive variances.}
    \label{fig:CH_unknownNoise_predictive_bt}
\end{figure}

The posterior predictive means and variances for the free-energy derivatives are shown in \cref{fig:CH_unknownNoise_predictive,fig:CH_unknownNoise_predictive_bt}.
Compared with the known-noise case, both the coefficient posterior and the posterior predictive distributions have larger variance.
For both covariance structures, the inferred posterior mean of $\sigma_u^2$ is approximately $4.02\times10^{-5}$. The wider coefficient and predictive distributions therefore reflect this higher inferred noise scale as well as joint noise uncertainty under the specified prior.
The predictive means remain close to the ground-truth derivative fields, but the increased variance reflects reduced confidence when the data precision must be inferred rather than prescribed.

\subsection{Ablation Study: Sensitivity to noise level, spatial resolution, and covariance structure}
\label{sec:ablation_study}

We next examine how parameter-estimation accuracy depends on observation noise and spatial resolution.
The purpose of this study is to isolate the effect of the covariance-aware residual likelihood used by B-VSI.
We therefore compare B-VSI against classical VSI~\cite{Wang2019Variational,Wang2021Variational} on the FP problem under a fixed model structure.

For a fair comparison, model selection is not performed in this study.
The operator set is fixed to the ground-truth model, and both methods are evaluated using the same noisy datasets.
Classical VSI is applied without regularization using the unweighted least-squares estimator in \cref{eq:VSI_solution}.
B-VSI is applied using the residual-space MLE with known observation-noise covariance and either the block-diagonal or block-tridiagonal covariance structure.
No prior is used in this comparison.
Thus, the differences among the estimators come from the residual covariance weighting in B-VSI.

\begin{figure}[htbp]
    \centering
    \includegraphics[width=\textwidth]{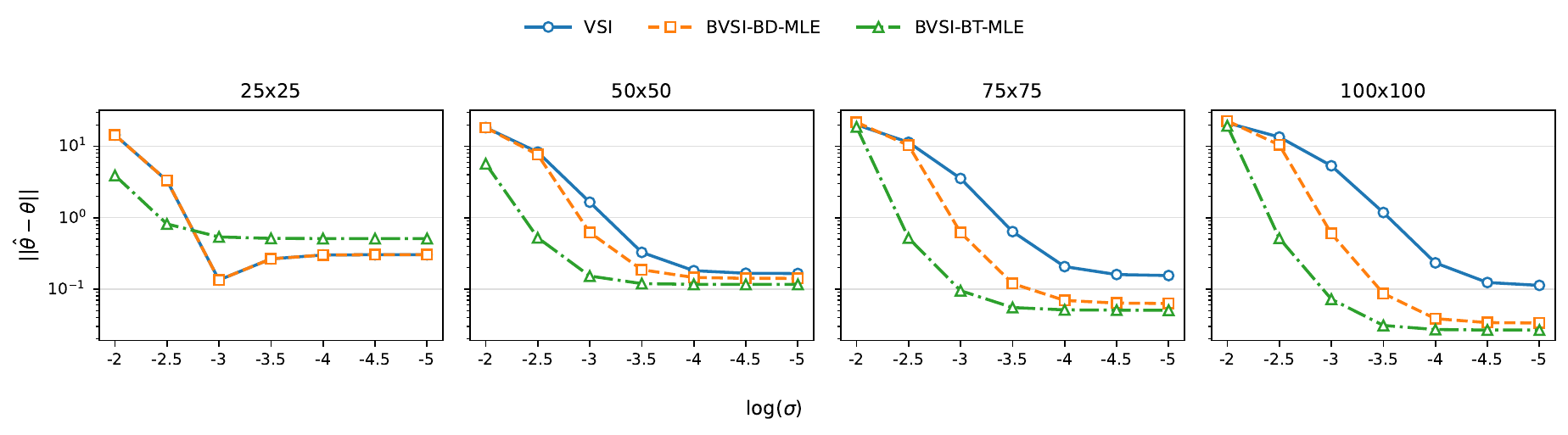}
    \caption{
    Sensitivity of parameter-estimation error to observation-noise level, spatial resolution, and residual covariance structure for the FP equation.
    Each panel corresponds to a fixed spatial resolution, and the horizontal axis gives the observation-noise level $\log_{10}(\sigma)$.
    }
    \label{fig:ablation_study}
\end{figure}

We vary the observation-noise standard deviation over the range
\begin{align}
    10^{-5}
    \leq
    \sigma
    \leq
    10^{-2},
\end{align}
and consider B-VSI grids with spatial resolutions
\begin{align}
    n_x
    \in
    \left\{
        25,
        50,
        75,
        100
    \right\},
\end{align}
where each value of $n_x$ corresponds to an $n_x\times n_x$ spatial grid.
For each case, parameter-estimation error is measured by the absolute coefficient error
\begin{align}
    \label{eq:ablation_parameter_error}
    e_{\bstheta}
    =
    \norm{
        \hbstheta-\bstheta_{\mathrm{GT}}
    }{2},
\end{align}
where $\hbstheta$ is the estimated coefficient vector and $\bstheta_{\mathrm{GT}}$ is the ground-truth coefficient vector.
The results are shown in \cref{fig:ablation_study}.

\subsubsection{Effect of noise level}

On the $50\times50$ and finer grids, both B-VSI covariance structures produce smaller parameter errors than classical VSI for $\log_{10}(\sigma)\leq-2.5$.
This improvement is most evident at intermediate noise levels, $-3.5\leq\log_{10}(\sigma)\leq-2.5$; on the coarsest grid, the block-diagonal estimate coincides with VSI (see \cref{sec:block_diagonal_approx}).
The difference is expected because B-VSI accounts for the covariance structure induced by propagating observational noise through the weak-form residual map.
Classical VSI, by contrast, corresponds to an unweighted least-squares estimator and can be interpreted as assuming an isotropic residual covariance.
A Bayesian interpretation of this connection is discussed in \cref{sec:discussion_VSI_likelihood}.

On the $50\times50$ and finer grids, the errors of all three estimators increase with the noise level.
At sufficiently high noise levels, the benefit of covariance-aware weighting becomes limited because the observed fields enter not only the residual vector but also the operator basis matrices.
Thus, the regression problem itself becomes contaminated by noise in the design matrix.

\subsubsection{Effect of spatial resolution}

B-VSI shows a clear decrease in parameter error under mesh refinement when $\log_{10}(\sigma)<-3$.
Classical VSI exhibits a comparable convergence trend only at lower noise levels, approximately when $\log_{10}(\sigma)\leq-4.5$.
This suggests that covariance-aware residual weighting improves the effective use of spatial resolution when observation noise is not too large.

At higher noise levels, approximately when $\log_{10}(\sigma)>-2.5$, none of the estimators exhibits a clear convergence trend with mesh refinement.
At $\log_{10}(\sigma)=-2.5$, the errors of VSI and of the block-diagonal approximation also grow under refinement, whereas the block-tridiagonal error levels off at about 0.5.
This behavior is consistent with the classical errors-in-variables effect, also known as regression attenuation.
In ordinary least squares, if the regression features are measured with noise and this noise is ignored, the estimated coefficients are biased toward zero.
For example, in the scalar case with $z=x+\epsilon_x$ and $\epsilon_x$ independent of $x$ and $y$, regressing $y$ on the noisy feature $z$ gives an attenuation factor
\begin{align}
    \mathbb{E}[\hat{\theta}]
    \approx
    \frac{\mathrm{Var}(x)}
    {\mathrm{Var}(x)+\mathrm{Var}(\epsilon_x)}
    \theta .
\end{align}
In VSI, the weak-form basis matrix is constructed from noisy observations, so the same errors-in-variables mechanism can bias coefficient estimates, especially for nonlinear operators.
B-VSI accounts for the propagated covariance of the residuals, but it is not a full errors-in-variables correction for noise in the basis functions themselves.
This helps explain the remaining bias observed at higher noise levels or coarser data resolutions \cite{Fuller1987,Carroll2006}.

\subsubsection{Effect of covariance structure}
\label{sec:block_diagonal_approx}

The two B-VSI variants isolate the effect of neglecting the covariance between adjacent residual time steps in the block-diagonal approximation. At intermediate noise levels, retaining this temporal covariance generally reduces the parameter-estimation error, with the improvement becoming more pronounced under mesh refinement; at $\log_{10}(\sigma)=-2.5$, for example, the block-tridiagonal error is about 0.5 on the $50\times50$ and finer grids, compared with 7.6 to 10.4 for the block-diagonal approximation. At sufficiently low noise levels, both covariance structures approach resolution-dependent error plateaus, although retaining the temporal covariance generally yields lower errors on the finer grids.
The $25\times25$ grid is an exception: there, the block-diagonal estimate coincides with VSI, and the block-tridiagonal error levels off at about 0.51 at low noise, above the value of about 0.30 of the other two estimators, although it remains the smallest for $\log_{10}(\sigma)\geq-2.5$.

These results indicate that the block-diagonal approximation can provide comparable accuracy when the neglected temporal correlations have limited influence. The improvement from the block-tridiagonal structure must be balanced against its greater computational cost: the block-tridiagonal likelihood requires sequential block elimination, whereas the block-diagonal likelihood permits independent factorizations across time steps. The corresponding computational costs are reported in \cref{tab:FP_wall_time}.

\section{Discussion}
\label{sec:discussion}

In this section, we compare observation- and residual-space likelihood formulations, examine the computational cost and approximations underlying B-VSI, and provide a Bayesian interpretation of classical VSI.

\subsection{Comparison of likelihood formulations}
\label{sec:discussion_comparison}

We compare the conventional observation-space likelihood with the weak-form residual-space likelihood used in B-VSI.
For clarity, we present the comparison using a state-linear PDE. The same distinction extends to nonlinear systems, although nonlinear residual maps require the local Gaussian approximation developed in \cref{sec:nonlinear_residual_likelihood}.

The observation-space likelihood evaluates model--data agreement by comparing the measured trajectory with that obtained from a forward PDE solve. Under the Gaussian observation model introduced in \cref{eq:observation_model,eq:observation_noise_model}, it has the form in \cref{eq:likelihood_data_discrepancy},
\begin{align}
    p_{\mathrm{obs}}(\CalD \mid \bstheta,\CalM)
    =
    \prod_{k=0}^{n_t}
    \CalN
    \nospaceleft(
        \hbsu^{(k)};
        \barbsu_{\bstheta,\CalM}^{(k)},
        \bsSigma_{\bsu}^{(k)}
    \nospaceright),
    \nonumber
\end{align}
where $\barbsu_{\bstheta,\CalM}^{(k)}$ is obtained by solving the candidate PDE with coefficients $\bstheta$. This formulation provides a global comparison between the measured data and the simulated trajectory, but each likelihood evaluation requires a forward solve.

B-VSI instead evaluates whether the observed states satisfy the candidate weak-form equations directly in residual space. With the block-tridiagonal covariance, the residual-space likelihood developed in \cref{sec:stacked_residual_likelihood,sec:residual_space_objective} is
\begin{align}
    \label{eq:discussion_bt_residual_likelihood}
    p_{\mathrm{res}}^{\mathrm{BT}}
    \nospaceleft(
        \bsr_{\bstheta}\mid\bstheta,\CalM
    \nospaceright)
    =
    \CalN
    \nospaceleft(
        \bsr_{\bstheta};
        \bszero,
        \bsSigma_{r,\bstheta}^{\mathrm{BT}}
    \nospaceright).
\end{align}%
Here, the residuals are constructed locally from consecutive observations over each time interval, while their stacked covariance accounts for the induced spatial and adjacent-time correlations. The resulting likelihood therefore evaluates the observed trajectory through its weak-form residuals without repeated forward PDE solves.

These are distinct statistical formulations and are not generally interchangeable, but offer complementary advantages depending on the observation setting.
The observation-space likelihood is more natural when only sparse sensors or low-dimensional quantities of interest are observed.
The residual-space likelihood is more efficient when full-field spatiotemporal data are available and the governing equation can be evaluated directly on the data.

\subsubsection{Observation-to-residual transformation}
\label{sec:discussion_likelihood_relation}

Under \cref{eq:observation_model,eq:observation_noise_model}, define the complete observation-error trajectory and its covariance by
\begin{align}
    \label{eq:discussion_stacked_observation_discrepancy}
    \bsepsilon
    &=
    \begin{bmatrix}
        {\bsepsilon^{(0)}}^{\top} & \cdots &
        {\bsepsilon^{(n_t)}}^{\top}
    \end{bmatrix}^{\top},
    &
    \bsSigma_{\bsu}
    &=
    \operatorname{blockdiag}\nospaceleft(
        \bsSigma_{\bsu}^{(0)},\ldots,\bsSigma_{\bsu}^{(n_t)}
    \nospaceright).
\end{align}%
Then $\bsepsilon\sim\CalN\nospaceleft(\bszero,\bsSigma_{\bsu}\nospaceright)$, and the observation-space likelihood is equivalently the density of this complete error trajectory:
\begin{align}
    \label{eq:discussion_observation_error_likelihood}
    p_{\mathrm{obs}}(\CalD\mid\bstheta,\CalM)
    =
    \CalN\nospaceleft(\bsepsilon;\bszero,\bsSigma_{\bsu}\nospaceright).
\end{align}%
Because the candidate trajectory satisfies the same discrete weak-form equations, \cref{eq:linear_residual_noise_map} shows that the residual is obtained by transforming the endpoint observation errors. Stacking the time steps gives
\begin{align}
    \label{eq:discussion_trajectory_residual_transformation}
    \bsr_{\bstheta}
    &=\boldsymbol{H}_{\bstheta}\bsepsilon,
    &
    \boldsymbol{H}_{\bstheta}
    &=
    \begin{bmatrix}
        -\bsB & \bsK_{\bstheta} & \bszero & \cdots & \bszero \\
        \bszero & -\bsB & \bsK_{\bstheta} & \ddots & \vdots \\
        \vdots & \ddots & \ddots & \ddots & \bszero \\
        \bszero & \cdots & \bszero & -\bsB & \bsK_{\bstheta}
    \end{bmatrix}.
\end{align}%
Therefore $\bsSigma_{r,\bstheta}^{\mathrm{BT}}=\boldsymbol{H}_{\bstheta}\bsSigma_{\bsu}\boldsymbol{H}_{\bstheta}^{\top}$ is the exact covariance induced by the stacked transformation in the state-linear Gaussian setting.
However, the full-trajectory map is rectangular and dimension-reducing, as
$
\bsepsilon
\in\RR^{(n_t+1)n_{\mathrm{eq}}}
$
and
$
\bsr_{\bstheta}
\in\RR^{n_t n_{\mathrm{eq}}}.
$
The exact distribution induced under the block-tridiagonal covariance structure does not, by itself, imply equality between the residual-only likelihood and the observation-space likelihood.

A direct density relationship is recovered when the initial state is
known exactly and conditioned upon, so that $\bsepsilon^{(0)}=\bszero$ and the
random observation-error trajectory reduces to
\begin{align}
    \label{eq:discussion_reduced_error_trajectory}
    \bsepsilon_{1:n_t}
    =
    \begin{bmatrix}
        {\bsepsilon^{(1)}}^{\top} & \cdots &
        {\bsepsilon^{(n_t)}}^{\top}
    \end{bmatrix}^{\top},
\end{align}%
and the residual transformation becomes
\begin{align}
    \label{eq:discussion_reduced_trajectory_transformation}
    \bsr_{\bstheta}
    &=
    \widetilde{\boldsymbol{H}}_{\bstheta}
    \bsepsilon_{1:n_t},
    &
    \widetilde{\boldsymbol{H}}_{\bstheta}
    &=
    \begin{bmatrix}
        \bsK_{\bstheta} & \bszero & \cdots & \bszero \\
        -\bsB & \bsK_{\bstheta} & \ddots & \vdots \\
        \vdots & \ddots & \ddots & \bszero \\
        \bszero & \cdots & -\bsB & \bsK_{\bstheta}
    \end{bmatrix}.
\end{align}%
Both vectors have dimension $n_t n_{\mathrm{eq}}$, and if
$\bsK_{\bstheta}$ is nonsingular, $\widetilde{\boldsymbol{H}}_{\bstheta}$ is
invertible with
\begin{align}
    \label{eq:discussion_reduced_trajectory_determinant}
    \det\widetilde{\boldsymbol{H}}_{\bstheta}
    =
    \nospaceleft(\det\bsK_{\bstheta}\nospaceright)^{n_t}.
\end{align}%
Under this fixed-initial-state treatment, the initial state is
conditioned upon implicitly and omitted from the conditioning notation for brevity.
The observation- and residual-space likelihoods are then related by the exact change
of variables
\begin{align}
    \label{eq:discussion_reduced_trajectory_density_identity}
    p_{\mathrm{obs}}\nospaceleft(
        \CalD_{1:n_t}\mid \bstheta,\CalM
    \nospaceright)
    =
    \left|\det\bsK_{\bstheta}\right|^{n_t}
    p_{\mathrm{res}}\nospaceleft(
        \bsr_{\bstheta}\mid\bstheta,\CalM
    \nospaceright).
\end{align}%

\subsubsection{Implications for inference}
\label{sec:discussion_inference_implications}

The observation- and residual-space likelihoods are not generally
identical as functions of $\bstheta$, because the Jacobian factor
$\left|\det\bsK_{\bstheta}\right|^{n_t}$ in
\cref{eq:discussion_reduced_trajectory_density_identity} may depend on $\bstheta$.
If $\left|\det\bsK_{\bstheta}\right|$ is independent of $\bstheta$, the likelihoods
are proportional and, with the same prior, yield the same normalized posterior,
MLE, and MAP. A simple sufficient condition is that $\bsK_{\bstheta}$ is independent
of $\bstheta$, as when the inferred coefficients enter only through the forcing or
source term $\bsq_{\bstheta}$.

For point estimation based only on the covariance-weighted quadratic
discrepancy, the fixed-initial-state transformation in
\cref{eq:discussion_reduced_trajectory_transformation} yields the following separate invariance:
\begin{align}
    \label{eq:discussion_quadratic_invariance}
    \bsr_{\bstheta}^{\top}
    \nospaceleft(
        \widetilde{\boldsymbol{H}}_{\bstheta}
        \bsSigma_{\bsu,1:n_t}
        \widetilde{\boldsymbol{H}}_{\bstheta}^{\top}
    \nospaceright)^{-1}
    \bsr_{\bstheta}
    =
    \bsepsilon_{1:n_t}^{\top}
    \bsSigma_{\bsu,1:n_t}^{-1}
    \bsepsilon_{1:n_t},
\end{align}%
where
$\bsSigma_{\bsu,1:n_t}
=
\operatorname{blockdiag}\nospaceleft(
\bsSigma_{\bsu}^{(1)},\ldots,\bsSigma_{\bsu}^{(n_t)}
\nospaceright)$.
Thus, the quadratic-only objective has the same optimizer in both observation and
residual spaces when covariance is propagated consistently and the same penalty is
used, regardless of whether $\left|\det\bsK_{\bstheta}\right|$ depends on
$\bstheta$. This invariance does not extend to the general noisy-initial-state
transformation in \cref{eq:discussion_trajectory_residual_transformation}, which
is rectangular.

\subsubsection{Computational advantage of the residual-space likelihood}
\label{sec:discussion_computational_advantage}

The main computational benefit of the residual-space likelihood is that it avoids repeated forward PDE solves.
For the observation-space likelihood in \cref{eq:likelihood_data_discrepancy}, each proposed coefficient vector requires constructing the model-induced trajectory
$\{\barbsu_{\bstheta,\CalM}^{(k)}\}_{k=0}^{n_t}$.
For time-dependent PDEs, this requires sequential time integration across the full observation window.
For nonlinear PDEs, each time step may additionally require nonlinear iterations.

The residual-space likelihood avoids this step.
For a fixed candidate model, the weak-form residuals are assembled by applying the candidate operators to the observed data.
In the state-linear case, this involves matrix-vector products such as
\begin{align}
    \bsr_{\bstheta}^{(k)}
    =
    \bsK_{\bstheta}\hbsu^{(k)}
    -
    \bsB\hbsu^{(k-1)}
    -
    \bsq_{\bstheta}.
\end{align}
For nonlinear operators, the residual and its local covariance require evaluating the nonlinear weak-form basis functions and their Jacobians at the observed fields.
These operations are performed directly on the data and do not require advancing the PDE solution for each proposed $\bstheta$.

The exact cost depends on the finite element discretization, matrix sparsity, and linear-algebra implementation.
For sparse finite element matrices, residual assembly typically scales with the number of nonzero entries in the relevant operator matrices.
Evaluating the Gaussian residual likelihood also requires covariance-weighted solves and log determinants. The block-diagonal structure provides the simplest factorized implementation through independent diagonal blocks, whereas the block-tridiagonal structure retains the exact adjacent temporal covariance and exploits block-banded factorization or sequential Schur elimination. Neither treatment requires forming the full global covariance. These covariance operations can use direct factorizations, iterative solvers, diagonal or sparse approximations, or precomputed structures when applicable.
Thus, the computational gain should not be interpreted as a universal reduction from $\CalO(\Ndof^3)$ to $\CalO(\Ndof^2)$.
Rather, the central advantage is that B-VSI replaces repeated forward solves and adjoint or sensitivity calculations with weak-form residual assembly and structured covariance-weighted residual evaluation.

This distinction is especially important for gradient-based point estimation and posterior inference.
With an observation-space likelihood, gradients require differentiating through the time integrator and PDE solver.
With the residual-space likelihood, gradients can be computed directly from the residual vector and residual covariance with respect to $\bstheta$.
This avoids backpropagation through a forward PDE solve and makes residual-space MLE, MAP point estimation, and SVGD posterior inference substantially more tractable when full-field data are available.

Finally, the residual-space formulation ties the discretization used for likelihood evaluation to the data representation.
The method still uses finite element shape functions and weak-form assembly, but it does not require choosing a separate forward-simulation mesh for each candidate parameter value.
This avoids the repeated stability and solver-tolerance considerations that arise when evaluating many candidate models through forward time integration.

\subsection{Computational complexity}
\label{sec:discussion_complexity}

We summarize the main computational costs of B-VSI in \cref{tab:complexity}.
The costs depend on the finite element discretization, sparsity pattern, covariance structure, and linear-algebra implementation.
The estimates below should therefore be interpreted as representative costs rather than universal bounds.

The per-time-step residual dimension $n_{\mathrm{eq}}$ is defined in \cref{sec:VSI}; it applies uniformly to scalar and multi-field systems.
Let $n_t$ be the number of time intervals, $d_{\bstheta}$ the number of active coefficients in the current model, $d=|\bschi|$ the size of the full candidate dictionary, $n_p$ the number of SVGD particles, $n_e$ the number of elements, and $n_{\mathrm{loc}}$ the number of local degrees of freedom per element.

We also define two implementation-dependent costs.
Let $C_{\mathrm{fac}}(n_{\mathrm{eq}})$ denote the cost of factoring one $n_{\mathrm{eq}}\times n_{\mathrm{eq}}$ residual covariance or Schur-complement block and computing its log determinant, and let $C_{\mathrm{sol}}(n_{\mathrm{eq}},m)$ denote the cost of solving with one such block for $m$ right-hand sides after factorization.
For dense covariance blocks,
\begin{align}
    C_{\mathrm{fac}}(n_{\mathrm{eq}})=\CalO(n_{\mathrm{eq}}^3),
    \qquad
    C_{\mathrm{sol}}(n_{\mathrm{eq}},m)=\CalO(n_{\mathrm{eq}}^2m).
\end{align}
For diagonal, sparse, low-rank, or otherwise structured covariance blocks, these costs can be substantially smaller.
This reduction applies directly to the block-diagonal blocks $\boldsymbol{D}_k$, but generally not to the block-tridiagonal Schur blocks $\boldsymbol{S}_k$ in \cref{eq:bt_schur_recursion}, because the term $\boldsymbol{E}_{k-1}^{\top}\boldsymbol{S}_{k-1}^{-1}\boldsymbol{E}_{k-1}$ involves an inverse and therefore fills in even when $\boldsymbol{D}_k$ and $\boldsymbol{E}_k$ are sparse.

\paragraph{Weak-form assembly}

The weak-form dataset is assembled from the observed fields and the finite element basis.
For each operator, element-level integration requires looping over elements and forming local contributions.
A typical element-level assembly cost is
\begin{align}
    \CalO(n_e n_{\mathrm{loc}}^2)
\end{align}
per operator, up to factors associated with quadrature and the number of state variables.
For state-linear operators, the corresponding Galerkin matrices can be assembled once and reused across time.
For nonlinear operators, the weak-form basis vectors and local Jacobians are evaluated on the observed fields, typically for each time interval.
Thus, assembling all active nonlinear basis vectors and Jacobians scales approximately as
\begin{align}
    \CalO(n_t d_{\bstheta} n_e n_{\mathrm{loc}}^2),
\end{align}
again up to quadrature and implementation-dependent constants.

The resulting global matrices are usually sparse because finite element basis functions have local support.
They may be banded under particular node orderings, but sparsity is the more general property.
Consequently, applying a preassembled operator matrix to data can often be done in time proportional to the number of nonzero entries rather than $\CalO(n_{\mathrm{eq}}^2)$.

\paragraph{Residual and covariance evaluation}

Once the weak-form basis matrices are available, the stacked residual has the regression form
\begin{align}
    \bsr_{\bstheta}
    =
    \bsy-\bsX\bstheta.
\end{align}
Evaluating the residual over all time intervals costs
\begin{align}
    \CalO(n_t n_{\mathrm{eq}} d_{\bstheta})
\end{align}
when the basis matrix blocks are explicitly available.
If residuals are evaluated through sparse operator applications rather than through assembled dense basis matrices, this cost is better described in terms of the number of nonzero entries in the operator matrices.

The residual covariance for each time interval has the form in \cref{eq:stacked_covariance_diagonal_block},
\begin{align}
    \bsSigma_{r,\bstheta}^{(k)}
    =
    \bsC_{\bstheta}^{(k)}
    \bsSigma_{\bsu}^{(k)}
    {\bsC_{\bstheta}^{(k)}}^{\top}
    +
    \bsB
    \bsSigma_{\bsu}^{(k-1)}
    \bsB^{\top},
    \nonumber
\end{align}
where $\bsC_{\bstheta}^{(k)}$ is the current-state residual sensitivity.
For dense matrices, explicitly forming this covariance block can cost up to $\CalO(n_{\mathrm{eq}}^3)$.
When $\bsSigma_{\bsu}^{(k)}$ is diagonal or scalar and the sensitivity matrices are sparse, the cost can be reduced by exploiting sparsity or by avoiding explicit dense covariance formation.
Under the block-diagonal structure, likelihood evaluation independently factors each $\boldsymbol{D}_k$. Under the block-tridiagonal structure, sequential elimination factors the Schur blocks $\boldsymbol{S}_k$ and applies the adjacent-block transformations in \cref{eq:bt_schur_recursion,eq:bt_residual_recursion,eq:bt_regression_recursion}. Both are linear in $n_t$ at the block level and avoid a dense $\CalO((n_t n_{\mathrm{eq}})^3)$ global factorization. For dense $n_{\mathrm{eq}}\times n_{\mathrm{eq}}$ spatial blocks, both have leading factorization scaling of order $\CalO(n_t n_{\mathrm{eq}}^3)$, although the block-tridiagonal structure has additional block-elimination operations and a larger constant. When the spatial blocks are sparse, the gap can be larger than a constant factor: the block-diagonal structure can use sparse factorizations of each $\boldsymbol{D}_k$, whereas the Schur blocks $\boldsymbol{S}_k$ are generally dense, and each elimination step requires a solve with $\boldsymbol{S}_{k-1}$ for the $n_{\mathrm{eq}}$ columns of $\boldsymbol{E}_{k-1}$, so the block-tridiagonal cost per time step generally remains of order $\CalO(n_{\mathrm{eq}}^3)$. Actual cost depends on sparsity, block structure, and implementation.

\begin{table}[htbp]
    \centering
    \caption{
    Representative computational costs for B-VSI with structured block-diagonal or block-tridiagonal residual covariance.
    Here, $n_{\mathrm{eq}}$ is the number of residual equations per time interval, $d_{\bstheta}$ is the number of active coefficients, $d$ is the dictionary size, and $n_p$ is the number of SVGD particles.
    The functions $C_{\mathrm{fac}}(n_{\mathrm{eq}})$ and $C_{\mathrm{sol}}(n_{\mathrm{eq}},m)$ denote the cost of factoring one residual covariance or Schur block and solving with it for $m$ right-hand sides, respectively.
    }
    \label{tab:complexity}
    \fontsize{10pt}{10pt}\selectfont
    \setlength\extrarowheight{5pt}
    \begin{tabular}{@{}p{0.23\textwidth}p{0.37\textwidth}p{0.34\textwidth}@{}}
    \toprule
    Component
    &
    Dominant operation
    &
    Representative cost
    \\
    \midrule
    Weak-form assembly
    &
    Assemble one operator matrix or Jacobian
    &
    $\CalO(n_e n_{\mathrm{loc}}^2)$
    \\
    Weak-form assembly
    &
    Assemble active nonlinear basis vectors and Jacobians over all time intervals
    &
    $\CalO(n_t d_{\bstheta} n_e n_{\mathrm{loc}}^2)$
    \\
    Residual evaluation
    &
    Compute $\bsr_{\bstheta}=\bsy-\bsX\bstheta$
    &
    $\CalO(n_t n_{\mathrm{eq}} d_{\bstheta})$
    \\
    Residual likelihood
    &
    Factor the selected structured covariance and evaluate the Gaussian density
    &
    Block diagonal: $\CalO(n_t C_{\mathrm{fac}}(n_{\mathrm{eq}}))$;
    Block tridiagonal: $\CalO(n_t C_{\mathrm{fac}}(n_{\mathrm{eq}}))$ plus sequential elimination, with $C_{\mathrm{fac}}$ evaluated for generally dense Schur blocks
    \\
    Lagged-covariance MLE
    &
    Form weighted normal equations and solve for $\bstheta$
    &
    $\CalO\!\left(
    n_t[
    C_{\mathrm{fac}}(n_{\mathrm{eq}})
    +
    C_{\mathrm{sol}}(n_{\mathrm{eq}},d_{\bstheta}+1)\right.$
    $\left.
    +
    n_{\mathrm{eq}} d_{\bstheta}^{2}
    ]
    +
    d_{\bstheta}^{3}
    \right)$
    \\
    Gaussian posterior inference
    &
    Known observation-noise covariance
    &
    Same leading cost as lagged-covariance MLE
    \\
    Normal-inverse-gamma posterior inference
    &
    Unknown scalar observation-noise variance
    &
    Same leading cost as lagged-covariance MLE
    \\
    Full-objective gradient methods
    &
    Evaluate full likelihood and gradient with $\bstheta$-dependent covariance
    &
    Problem-dependent; at least one residual-likelihood evaluation plus covariance-derivative costs per iteration
    \\
    SVGD posterior inference
    &
    Update $n_p$ particles
    &
    $\CalO(n_p T_{\nabla})+\CalO(n_p^2 d_{\bstheta})$
    \\
    Model selection
    &
    Sequential operator elimination
    &
    $\CalO(d^2)\times\CalO(\text{point-estimation fit})$
    \\
    \bottomrule
    \end{tabular}
\end{table}

\paragraph{Lagged-covariance point estimation and Gaussian posterior inference}

In the lagged-covariance update, the selected covariance
$\bsSigma_{r,\bstheta^{(l)}}^{\CalC}$
is assembled or updated using the current parameter value and then held fixed while updating $\bstheta$.
For MLE point estimation, the dominant operations are factoring the covariance blocks and forming the weighted normal equations,
\begin{align}
    \bsX^{\top}\bsW_l\bsX,
    \qquad
    \bsX^{\top}\bsW_l\bsy .
\end{align}
For the block-diagonal structure, each time step requires independent solves with $\boldsymbol{D}_k$ for the columns of $\bsXi^{(k)}$ and ${\bsXi^{\dotu}}^{(k)}$. The block-tridiagonal structure instead performs sequential solves with $\boldsymbol{S}_k$, together with adjacent-block transformations and elimination. For dense blocks and $d_{\bstheta}\ll n_{\mathrm{eq}}$, both structures therefore have a leading-order cost of $\CalO(n_t n_{\mathrm{eq}}^3)$ per lagged-covariance iteration. The block-tridiagonal structure has a larger constant because each time step additionally requires a solve with $n_{\mathrm{eq}}$ right-hand sides and dense block products, with cost $C_{\mathrm{sol}}(n_{\mathrm{eq}},n_{\mathrm{eq}})+\CalO(n_{\mathrm{eq}}^3)$.

The Gaussian posterior update with known observation-noise covariance has the same dominant cost as the lagged-covariance MLE update, with an additional inversion of a
$d_{\bstheta}\times d_{\bstheta}$ matrix.
The normal-inverse-gamma update for an unknown scalar observation-noise variance has the same leading-order cost, with only inexpensive scalar updates for the inverse-gamma parameters.
Thus, the conjugate posterior updates are usually dominated by residual covariance operations rather than by operations in coefficient space.

\paragraph{Full \texorpdfstring{$\bstheta$}{theta}-dependent point estimation and posterior inference}

Gradient-based MLE or MAP point estimation targets the full objective in which $\bsSigma_{r,\bstheta}^{\CalC}$ depends on $\bstheta$.
Each iteration requires evaluating the residual likelihood and its gradient.
The likelihood evaluation has the covariance costs described above.
The gradient additionally requires differentiating the residual vector and, when the full covariance dependence is retained, differentiating the covariance and log-determinant terms with respect to $\bstheta$.
These derivative costs are problem-dependent and may be obtained analytically or through automatic differentiation.

For SVGD posterior inference, the residual-space posterior gradient must be evaluated for each particle.
If $T_{\nabla}$ denotes the cost of one posterior-gradient evaluation for one particle, then one SVGD iteration costs approximately
\begin{align}
    \CalO(n_p T_{\nabla})
    +
    \CalO(n_p^2 d_{\bstheta}),
\end{align}
where the second term is the cost of kernel evaluations and kernel-gradient interactions among particles.
Thus, SVGD is more expensive than a single point-estimation run by roughly a factor of $n_p$, plus the particle-interaction cost.

\paragraph{Model selection}

Sequential operator elimination evaluates a sequence of reduced models.
Starting from a dictionary with $d$ candidate operators, backward elimination visits at most
\begin{align}
    d + (d-1) + \cdots + 1
    =
    \CalO(d^2)
\end{align}
candidate reductions.
If some operators are protected and never removed, this count is reduced accordingly.
During model selection, only an MLE point estimate is required for each candidate model to evaluate the residual-space BIC.
Full posterior inference is performed only after the selected model is fixed.
Therefore, the model-selection cost is approximately
\begin{align}
    \CalO(d^2)\times
    \CalO(\text{point-estimation fit}).
\end{align}

\paragraph{Practical comparison of inference strategies}

The lagged-covariance updates and the full gradient-based methods can have comparable leading-order costs per iteration when dense covariance factorizations dominate.
In practice, however, the lagged-covariance updates are often faster because they avoid automatic differentiation through covariance terms and typically require fewer iterations.
Gradient-based methods are more flexible and can target non-Gaussian priors, constrained parameterizations, or the full $\bstheta$-dependent residual likelihood.
SVGD provides a particle approximation to the posterior, but it incurs an additional factor proportional to the number of particles.

The main computational advantage of B-VSI relative to observation-space likelihood methods is not a universal dense-matrix complexity reduction.
Rather, it is the replacement of repeated forward PDE solves, and their associated sensitivity or adjoint calculations, with weak-form residual assembly and covariance-weighted residual evaluation.

\subsection{Characterization of inference approximations}
\label{sec:successive_approximations}

The inference pipeline involves five approximations, some of which are optional. Their isolated effects are summarized below, and the assumptions and supporting results are discussed in \cref{app:successive_approximations}. The analysis concerns residual-space inference at a fixed discretization and candidate model.
Let $\sigma$ denote the observation-noise amplitude, with fixed covariance shape and trajectory length. For a random quantity $X_\sigma$ and a deterministic rate $a_\sigma$, $X_\sigma=\CalO_p(a_\sigma)$ means that $X_\sigma/a_\sigma$ is bounded in probability as $\sigma\to0$~\cite{vanderVaart1998Asymptotic}, whereas $\CalO(\cdot)$ denotes a deterministic bound, including bounds on moments.

\paragraph{First-order delta-method approximation}
For nonlinear residuals, the neglected residual perturbation is $\CalO_p(\sigma^2)$; under additional smoothness and moment assumptions, the mean correction is $\CalO(\sigma^2)$ and the covariance correction is $\CalO(\sigma^4)$. These errors vanish with noise amplitude at fixed discretization.

\paragraph{Plug-in nonlinear Jacobian}
Evaluating the Jacobian at the observed state perturbs the Jacobian by $\CalO_p(\sigma)$ and the residual covariance by $\CalO_p(\sigma^3)$. Relative to the covariance itself, this error is $\CalO_p(\sigma)$, larger than the relative $\CalO(\sigma^2)$ Taylor correction, so the plug-in step dominates the covariance error at small noise. This is distinct from Taylor truncation: the nonlinear residual is still evaluated on the data, while its covariance uses the observed-state Jacobian.

\paragraph{Block-diagonal temporal covariance}
{The block-diagonal approximation} discards the adjacent $\boldsymbol E_k$ blocks. Its error, measured by the Kullback--Leibler divergence from the block-tridiagonal to the block-diagonal residual density, is bounded by a term quadratic in the normalized adjacent-step residual correlations.

\paragraph{Lagged covariance}
If the covariance is locally Lipschitz in $\bstheta$ and uniformly positive definite, freezing it at the previous iterate perturbs the residual negative log-likelihood by an amount that is, to first order, proportional to the parameter increment; this perturbation vanishes as the updates converge. The converged estimate nevertheless solves the weighted normal equations rather than the full stationarity condition, which also contains the covariance-derivative terms in \cref{eq:full_likelihood_gradient_component}.

\paragraph{Finite-particle SVGD}
SVGD approximates the chosen posterior through a finite interacting particle set and a finite number of iterations. These errors are separate from the residual and covariance approximations.
Existing convergence results require additional target, kernel, initialization, and step-size assumptions, so particle convergence alone does not establish posterior accuracy.

\subsection{Bayesian perspective on VSI}
\label{sec:discussion_VSI_likelihood}

Classical VSI can be interpreted from a Bayesian perspective as a residual-space estimation method with a simplified likelihood model.
This interpretation clarifies when VSI and B-VSI are expected to agree, and when the covariance-aware likelihood in B-VSI becomes important.

\paragraph{Residual likelihood implied by VSI}
The standard VSI objective in \cref{eq:VSI_objective} minimizes an unweighted squared weak-form residual.
Comparing this objective with the MLE objective in B-VSI, such as \cref{eq:lagged_covariance_mle_objective}, shows that VSI corresponds to the likelihood model
\begin{align}
    \bsr_{\bstheta}
    \sim
    \CalN(\bszero,\sigma^2\bsI),
\end{align}
where $\sigma^2$ is a scalar residual variance.
For MLE, the value of $\sigma^2$ does not affect the optimizer because it only rescales the objective.
Thus, the classical VSI estimator can be viewed as an ordinary least-squares estimator in residual space.

This likelihood assumes that the residual errors have constant variance and are mutually uncorrelated across spatial nodes, time steps, and state variables.
Equivalently, it treats all residual equations as equally reliable and statistically independent.
With $\ell_2$ regularization, the same interpretation becomes MAP estimation with a zero-mean Gaussian prior on $\bstheta$.
With $\ell_1$ regularization, it becomes MAP estimation with a zero-mean Laplace prior.
Therefore, regularized VSI can be viewed as a Bayesian point estimator under an isotropic residual likelihood and a sparsity- or shrinkage-promoting prior.

The isotropic residual covariance assumption is generally not implied by the observation model.
As shown in \cref{sec:residual_likelihood}, uncertainty in the measured fields is propagated through the weak-form residual map.
For a state-linear residual, this gives covariance terms of the form in \cref{eq:stacked_covariance_diagonal_block},
\begin{align}
    \bsSigma_{r,\bstheta}^{(k)}
    =
    \bsC_{\bstheta}^{(k)}
    \bsSigma_{\bsu}^{(k)}
    {\bsC_{\bstheta}^{(k)}}^{\top}
    +
    \bsB
    \bsSigma_{\bsu}^{(k-1)}
    \bsB^{\top}.
    \nonumber
\end{align}
Even if the observation noise covariance $\bsSigma_{\bsu}^{(k)}$ is diagonal or isotropic, the residual covariance need not be diagonal or isotropic after multiplication by the finite element residual sensitivity matrices.
Because each weak-form residual depends on local neighborhoods of nodal values, the propagated residual errors can have non-uniform variances and nonzero correlations.
B-VSI accounts for this induced covariance structure, whereas classical VSI replaces it with $\sigma^2\bsI$.

\paragraph{Why VSI remains accurate in low-noise regimes}
The comparative results in \cref{sec:ablation_study} show that VSI can still produce accurate parameter estimates in low-noise conditions, often performing similarly to B-VSI.
This behavior is expected.
When observation noise is small, the measured fields $\hbsu$ are close to the noise-free solution $\barbsu$.
Consequently, the weak-form residuals evaluated at the correct model are also small, and the difference between weighted and unweighted residual minimization has limited practical effect.

More precisely, B-VSI replaces the ordinary least-squares residual objective with a generalized least-squares objective,
\begin{align}
    \bsr_{\bstheta}^{\top}
    \bsSigma_{r,\bstheta}^{-1}
    \bsr_{\bstheta}.
\end{align}
When all candidate parameter values near the optimum produce very small residuals, changing the weighting matrix often produces only a small change in the minimizer.
In this regime, the dominant source of accuracy is the weak-form structure itself, and the covariance correction mainly affects uncertainty quantification rather than the point estimate.

As the noise level increases, however, the covariance structure becomes more consequential.
Residual equations with larger propagated uncertainty should contribute less to the likelihood, and correlated residuals should not be treated as independent information.
Classical VSI cannot make this distinction because it assigns the same variance and independence assumption to every residual equation.
B-VSI therefore provides a statistically better calibrated estimator and posterior approximation when the propagated residual covariance is heteroscedastic or correlated.

In summary, classical VSI is a special case of residual-space Bayesian inference with an isotropic Gaussian residual likelihood.
This approximation is often adequate for point estimation in low-noise settings, but it can become statistically inefficient, and an isotropic-residual posterior built on it can misstate the coefficient uncertainty, when measurement noise is non-negligible.
B-VSI generalizes VSI by using the covariance structure induced by the observation model and the weak-form residual map.

\section{Conclusions}
\label{sec:conclusion}

We presented Bayesian Variational System Identification (B-VSI), a framework for discovering governing PDEs from noisy spatiotemporal data while quantifying uncertainty in the identified model and its derived predictions.
B-VSI extends classical VSI by replacing the unweighted weak-form residual objective with a likelihood defined directly in weak-form residual space.
This likelihood is induced by the observation model, where measurement uncertainty is propagated through the weak-form residual map, producing residual errors that are generally heteroscedastic and correlated.
The framework preserves the main computational advantage of VSI, namely that coefficient estimation and posterior inference can be performed without repeated forward PDE solves.

The proposed formulation provides a unified Bayesian treatment of operator selection, parameter estimation, posterior inference, and posterior prediction.
For a fixed candidate model, the residual-space likelihood supports both point estimates and posterior approximations.
We developed lagged-covariance updates that yield generalized least-squares estimates and conjugate posterior approximations when applicable.
For more general priors, likelihoods, or posterior structures, these updates can be used to initialize gradient-based optimization or particle-based posterior inference.
Model-form uncertainty is handled through sequential operator elimination guided by a residual-space BIC criterion, enabling parsimonious model discovery from candidate operator dictionaries.

The numerical examples on the Fokker--Planck and two-field Cahn--Hilliard equations demonstrate that B-VSI can recover active operators and estimate their coefficients accurately from noisy data.
The results show that accounting for the propagated residual covariance improves robustness relative to classical VSI when measurement noise is non-negligible.
At low noise levels, classical VSI and B-VSI produce similar point estimates, consistent with the interpretation of classical VSI as an isotropic residual-likelihood approximation.
Beyond coefficient estimation, B-VSI provides posterior uncertainty estimates for physically meaningful derived quantities, including potential functions and free-energy derivatives in both the numerical examples presented here.

Overall, B-VSI provides a computationally tractable route to uncertainty-aware PDE discovery from full-field spatiotemporal data.
Future work will assess the framework in realistic and experimental settings, including those with partial or sparse observations and higher-dimensional parameter spaces, and extend it to optimal experimental design for more data-efficient discovery of governing equations.

\appendix
\crefalias{section}{appendix}
\crefalias{subsection}{appendix}
\crefalias{subsubsection}{appendix}
\counterwithin{table}{section}
\counterwithin{figure}{section}
\counterwithin{equation}{section}
\renewcommand{\thetable}{\Alph{section}.\arabic{table}}
\renewcommand{\thefigure}{\Alph{section}.\arabic{figure}}
\renewcommand{\theHequation}{\Alph{section}.\arabic{equation}}

\renewcommand{\thesection}{\Alph{section}} %
\makeatletter
\def\@seccntformat#1{\@ifundefined{#1@cntformat}%
   {\csname the#1\endcsname.\hspace{0.5em}}%
   {\csname #1@cntformat\endcsname}}%
\newcommand\section@cntformat{\appendixname\ \thesection.\hspace{0.5em}}
\makeatother

\clearpage
\section{Sensitivity to observation-noise realizations}
\label{app:repeated-noise}

To assess the sensitivity of the numerical results in \cref{sec:example_FP,sec:example_CH} to individual observation-noise realizations, we repeat the numerical experiments using independently generated noisy datasets.
For each example, the noise-free trajectory is computed once and reused, and each of the $N_{\mathrm{rep}}=5$ noisy datasets adds an independent draw from the observation-noise model in \cref{sec:FP_data_generation} or \cref{sec:CH_data_generation}, generated with random seeds 42--46.
Realization~1 (seed 42) is the dataset reported in \cref{sec:examples} for both examples.
All other numerical and inference settings
are held fixed across repetitions and follow the configuration in \cref{sec:example_FP,sec:example_CH}.
Each noise realization leads to a separate Bayesian inference problem with its own posterior distribution. Thus, the repeated-noise results assess variability across datasets, complementing the posterior uncertainty reported for individual datasets in \cref{sec:example_FP,sec:example_CH}.

\subsection{State-linear Fokker--Planck equation}
\label{app:repeated-noise-fp}

Residual-space model selection recovered the ground-truth operator support in all five realizations under both covariance structures, consistent with the results in \cref{sec:FP_model_selection}.

\begin{figure*}[htbp]
\centering
\includegraphics[width=\textwidth,height=0.8\textheight,keepaspectratio]{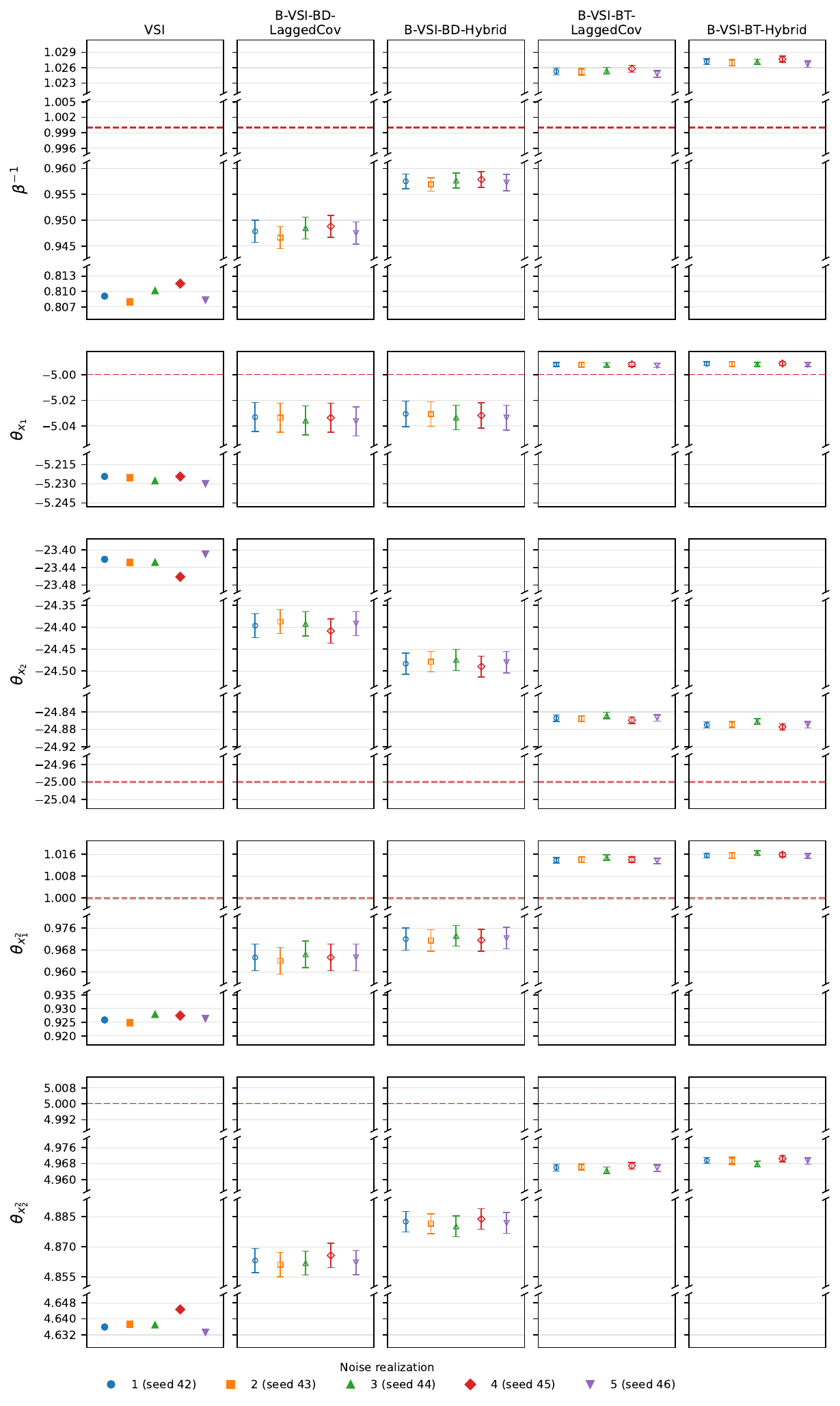}
\caption{Coefficient estimates for the FP equation across five independent observation-noise realizations. Columns show the identification methods and rows show the coefficients. Within each row, all panels share a broken vertical axis: segments around the estimates of each method and around the ground truth are kept, and the empty ranges between them are omitted (marked by break symbols). All segments in a row use the same scale, so interval lengths can be compared across methods, but vertical distances across a break are not to scale; read biases from the tick labels. Dashed red lines mark the ground-truth values. VSI gives point estimates (filled markers). For the B-VSI methods, open markers show the posterior mean and error bars show the marginal 95\% credible interval, the mean $\pm1.96$ posterior standard deviations. For the two hybrid columns, the mean and sample standard deviation are computed from the $n_p=20$ SVGD particles, as in \cref{tab:FP_parameters}.
}
\label{fig:app-fp-repeated-noise}
\end{figure*}

\begin{figure}[htbp]
\centering
\includegraphics[width=\textwidth,height=0.8\textheight,keepaspectratio]{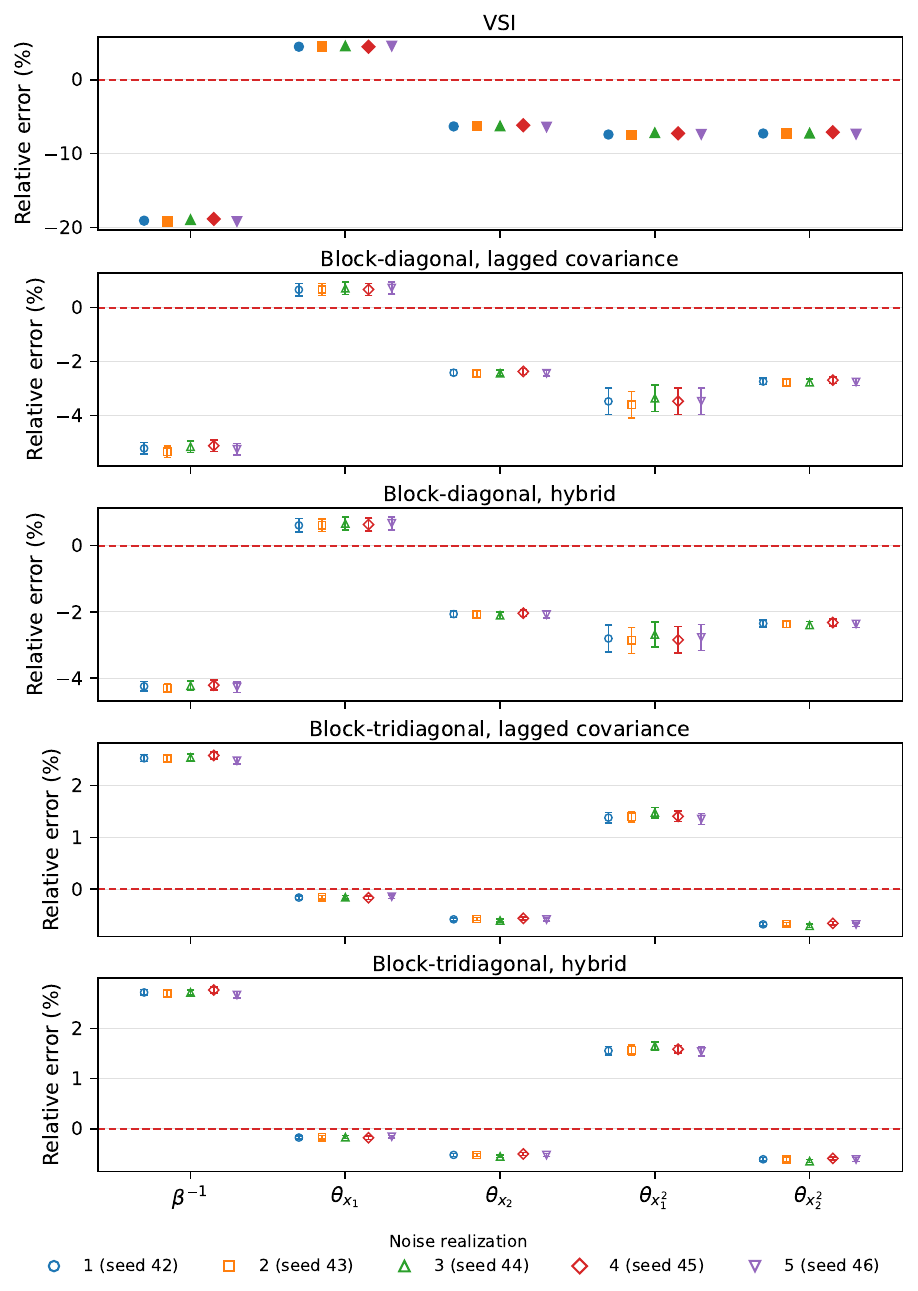}
\caption{Signed relative errors $\hat{\theta}_j/\theta_{j,\mathrm{GT}}-1$, in percent, of the coefficient estimates in \cref{fig:app-fp-repeated-noise}, one panel per method. Markers and credible intervals are as in \cref{fig:app-fp-repeated-noise}, with the intervals divided by $\lvert\theta_{j,\mathrm{GT}}\rvert$. Negative values indicate shrinkage toward zero; dashed red lines mark zero error.}
\label{fig:app-fp-repeated-noise-relative}
\end{figure}

\Cref{fig:app-fp-repeated-noise} shows the coefficient estimates and posterior distributions for each realization separately.
The credible intervals have nearly identical widths across realizations: for both lagged-covariance posteriors, the 95\% half-widths differ by less than 0.2\% for every coefficient.
With the lagged covariance fixed, the Gaussian posterior covariance depends on the data only through the weak-form operators evaluated on the observed fields, which change little between realizations at this noise level, whereas the posterior mean depends linearly on the noisy residual data.
The hybrid half-widths differ by up to 20\% and 34\% for the block-diagonal and block-tridiagonal structures, respectively, reflecting the Monte Carlo variability of variance estimates from $n_p=20$ particles.
The posteriors of different realizations overlap substantially.
For two realizations with marginal posteriors $p_a$ and $p_b$ of a coefficient, we measure their agreement by the Bhattacharyya coefficient $\int\sqrt{p_a(\theta)\,p_b(\theta)}\,\mathrm{d}\theta\in(0,1]$, which equals one for identical posteriors; over all coefficients and pairs of realizations, its median is $0.97$ for both block-diagonal estimates and $0.89$ and $0.86$ for the block-tridiagonal lagged-covariance and hybrid estimates.

Although the posteriors of different realizations overlap substantially, the ground truth lies far in the tail of every posterior.
The posteriors therefore agree with one another but not with the ground truth, which indicates a systematic error shared by all realizations rather than noise-induced uncertainty.
With the block-tridiagonal structure, most of this error is the deterministic error of the finite spatial and temporal resolution: in the sensitivity study of \cref{sec:ablation_study}, the coefficient error on the $50\times50$ mesh is $0.12$ in the noise-free limit, compared with $0.15$ at $\sigma=10^{-3}$, and decreases to $0.03$ on the $100\times100$ mesh.
This error is independent of the observation noise and is not represented in the residual likelihood; the remainder arises from the noise in the weak-form operators, which are assembled from the noisy observations.
With the block-diagonal structure, the coefficient error increases from $0.14$ to $0.62$ between $\sigma=10^{-5}$ and $\sigma=10^{-3}$, so its additional error stems mainly from neglecting the adjacent temporal covariance.

\begin{table}[htbp]
\centering
\caption{Estimation errors for the repeated-noise FP experiment. Each row within a metric corresponds to one independent observation-noise realization, and the last row gives the mean and sample standard deviation over the realizations. The reported quantities are computed from the point estimate for VSI and the posterior mean for the B-VSI methods, using the definitions in \cref{tab:FP_parameters}.}
\label{tab:app-fp-repeated-noise}
\fontsize{10pt}{10pt}\selectfont
\sisetup{round-mode=places, round-precision=2}
\begin{tabular}{@{}cc S[table-format=2.2(1.2), separate-uncertainty, retain-zero-uncertainty] *{4}{S[table-format=1.2(1.2), separate-uncertainty, retain-zero-uncertainty]}@{}}
\toprule
Metric
& Realization
& {VSI}
& {\makecell{B-VSI-BD\\LaggedCov}}
& {\makecell{B-VSI-BD\\Hybrid}}
& {\makecell{B-VSI-BT\\LaggedCov}}
& {\makecell{B-VSI-BT\\Hybrid}} \\
\midrule
\multirow{6}{*}{$\norm{\bstheta_{\mathrm{GT}}-\hbstheta}{2}$}
& 1 & 1.65 & 0.62 & 0.53 & 0.15 & 0.14 \\
& 2 & 1.64 & 0.63 & 0.54 & 0.15 & 0.14 \\
& 3 & 1.64 & 0.63 & 0.54 & 0.16 & 0.15 \\
& 4 & 1.61 & 0.61 & 0.53 & 0.15 & 0.13 \\
& 5 & 1.66 & 0.63 & 0.54 & 0.15 & 0.14 \\
\cmidrule(l){2-7}
& {Mean $\pm$ SD} & 1.64 \pm 0.02 & 0.62 \pm 0.01 & 0.54 \pm 0.01 & 0.15 \pm 0.00 & 0.14 \pm 0.01 \\
\midrule
\multirow{6}{*}{$\mathrm{MSE}(\hpsi)$}
& 1 & 29.60 & 4.26 & 3.12 & 0.26 & 0.21 \\
& 2 & 29.36 & 4.39 & 3.17 & 0.25 & 0.22 \\
& 3 & 29.37 & 4.32 & 3.23 & 0.28 & 0.24 \\
& 4 & 28.16 & 4.10 & 3.05 & 0.24 & 0.20 \\
& 5 & 30.05 & 4.32 & 3.16 & 0.26 & 0.21 \\
\cmidrule(l){2-7}
& {Mean $\pm$ SD} & 29.31 \pm 0.70 & 4.28 \pm 0.11 & 3.15 \pm 0.07 & 0.26 \pm 0.01 & 0.22 \pm 0.02 \\
\bottomrule
\end{tabular}
\end{table}

\Cref{fig:app-fp-repeated-noise-relative} shows the same estimates as signed relative errors, on a common percentage scale within each method, and \cref{tab:app-fp-repeated-noise} reports the coefficient error and potential MSE.
The ordering of the methods in \cref{tab:FP_parameters} is preserved in every realization for both metrics: VSI has the largest errors, followed by the block-diagonal lagged-covariance, block-diagonal hybrid, block-tridiagonal lagged-covariance, and block-tridiagonal hybrid estimates.
The variation across realizations is small compared with the differences between methods, so the representative realization in \cref{tab:FP_parameters} is typical of the five.

\subsection{Nonlinear two-field Cahn--Hilliard equation}
\label{app:repeated-noise-ch}

\begin{figure}[htbp]
\centering
\includegraphics[width=\textwidth,height=0.8\textheight,keepaspectratio]{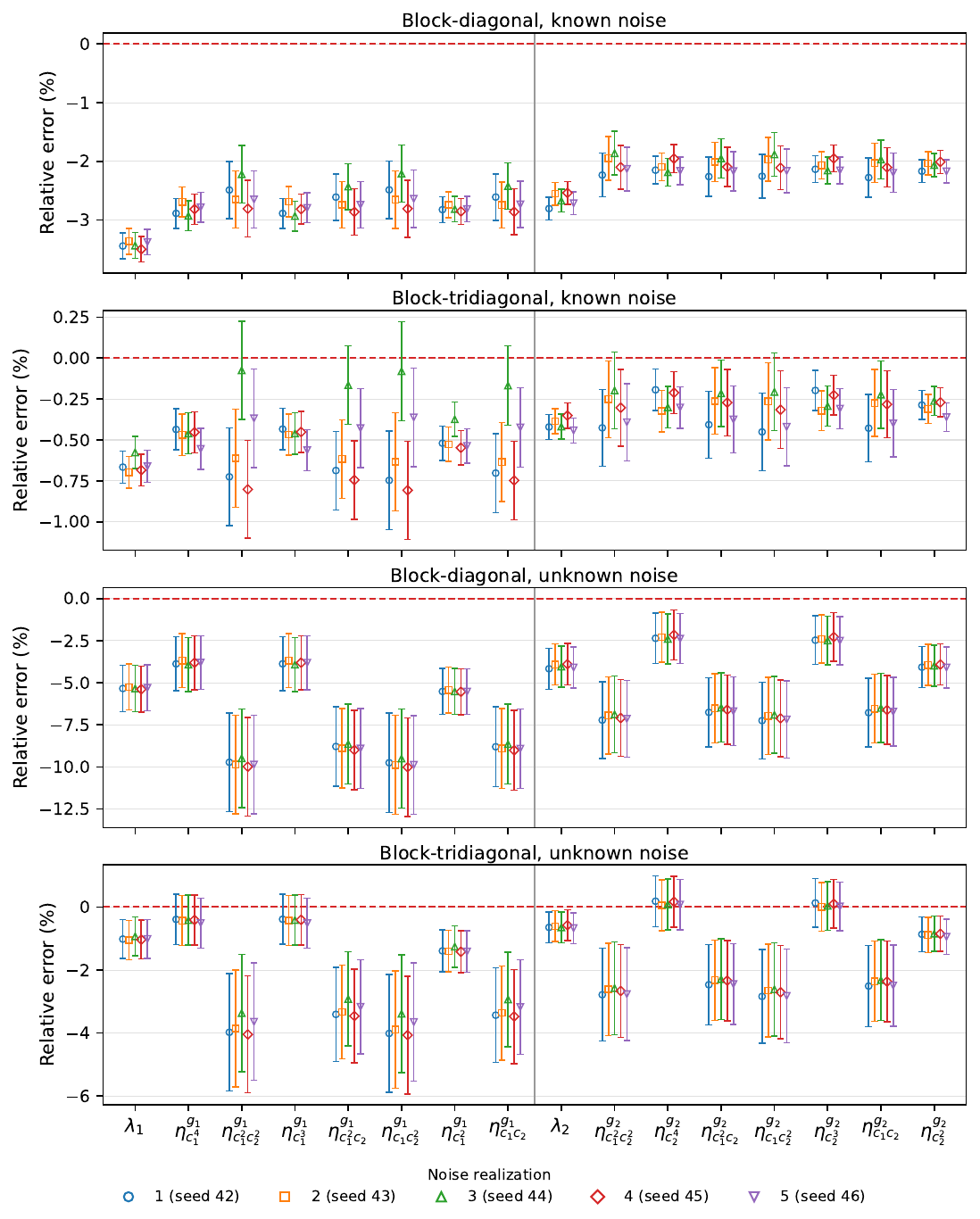}
\caption{Signed relative errors $\hat{\theta}_j/\theta_{j,\mathrm{GT}}-1$, in percent, of the posterior means for the CH equation across five independent observation-noise realizations, one panel per method as in \cref{tab:app-ch-repeated-noise}; the vertical line separates the coefficients of the two fields. Markers, credible intervals, and lines are as in \cref{fig:app-fp-repeated-noise-relative}; unknown-noise intervals use the marginal coefficient covariance including the posterior mean of $\sigma_u^2$. VSI, whose relative errors are close to $-100\%$, is omitted.}
\label{fig:app-ch-repeated-noise}
\end{figure}

\Cref{fig:app-ch-repeated-noise} shows the posterior means and marginal credible intervals for each realization as signed relative errors, obtained from fixed-model inference on the ground-truth support.
The posteriors of different realizations overlap substantially: the median Bhattacharyya coefficient between the marginal posteriors of two realizations is $0.93$ and $0.85$ for the block-diagonal and block-tridiagonal structures with known noise and exceeds $0.99$ with unknown noise.
The near-complete overlap with unknown noise reflects the wider unknown-noise posteriors, consistent with the posterior mean of $\sigma_u^2$, which lies between $4.019\times10^{-5}$ and $4.022\times10^{-5}$ in all unknown-noise runs; the elevated noise-variance estimate discussed in \cref{sec:CH_unknown_noise} is therefore not specific to realization~1.

\Cref{tab:app-ch-repeated-noise} reports the coefficient errors and derivative MSEs.
All B-VSI posterior means except ten block-tridiagonal unknown-noise entries are smaller in magnitude than the ground truth, with relative errors between $-3.5\%$ and $-1.9\%$ for block-diagonal and between $-0.81\%$ and $-0.07\%$ for block-tridiagonal covariance with known noise.
In every realization, the block-tridiagonal structure gives lower coefficient errors and derivative MSEs than the block-diagonal structure under the same noise assumption, and the known-noise posterior mean is more accurate than the unknown-noise posterior mean for the same covariance structure.
The block-tridiagonal known-noise coefficient error varies most across realizations in relative terms, whereas the VSI errors are nearly unchanged.

\begin{table}[htbp]
\centering
\caption{Estimation errors for the repeated-noise CH experiment, arranged as in \cref{tab:app-fp-repeated-noise} and using the definitions in \cref{tab:CH_parameters}.}
\label{tab:app-ch-repeated-noise}
\fontsize{9pt}{10pt}\selectfont
\setlength{\tabcolsep}{4pt}
\sisetup{round-mode=places, round-precision=2}
\begin{tabular}{@{}cc S[table-format=4.2(1.2), separate-uncertainty, retain-zero-uncertainty] *{4}{S[table-format=2.2(1.2), separate-uncertainty, retain-zero-uncertainty]}@{}}
\toprule
Metric
& Realization
& {VSI}
& {\makecell{B-VSI-BD\\Known noise}}
& {\makecell{B-VSI-BT\\Known noise}}
& {\makecell{B-VSI-BD\\Unknown noise}}
& {\makecell{B-VSI-BT\\Unknown noise}} \\
\midrule
\multirow{6}{*}{$\norm{\bstheta_{\mathrm{GT}}-\hbstheta}{2}$}
& 1 & 31.260 & 0.775 & 0.170 & 2.285 & 0.861 \\
& 2 & 31.258 & 0.756 & 0.147 & 2.269 & 0.830 \\
& 3 & 31.273 & 0.717 & 0.078 & 2.225 & 0.756 \\
& 4 & 31.251 & 0.788 & 0.170 & 2.300 & 0.855 \\
& 5 & 31.249 & 0.780 & 0.132 & 2.294 & 0.817 \\
\cmidrule(l){2-7}
& {Mean $\pm$ SD} & 31.26 \pm 0.01 & 0.76 \pm 0.03 & 0.14 \pm 0.04 & 2.27 \pm 0.03 & 0.82 \pm 0.04 \\
\midrule
\multirow{6}{*}{$\mathrm{MSE}\nospaceleft(\pp{g_1}{c_1}\nospaceright)$}
& 1 & 2524.233 & 1.937 & 0.090 & 11.837 & 1.245 \\
& 2 & 2523.972 & 1.961 & 0.086 & 11.845 & 1.230 \\
& 3 & 2525.143 & 1.850 & 0.023 & 11.671 & 0.957 \\
& 4 & 2524.958 & 2.098 & 0.099 & 12.149 & 1.277 \\
& 5 & 2522.218 & 2.007 & 0.065 & 12.024 & 1.157 \\
\cmidrule(l){2-7}
& {Mean $\pm$ SD} & 2524.10 \pm 1.16 & 1.97 \pm 0.09 & 0.07 \pm 0.03 & 11.91 \pm 0.18 & 1.17 \pm 0.13 \\
\midrule
\multirow{6}{*}{$\mathrm{MSE}\nospaceleft(\pp{g_2}{c_2}\nospaceright)$}
& 1 & 1322.919 & 0.665 & 0.018 & 3.975 & 0.401 \\
& 2 & 1324.778 & 0.565 & 0.012 & 3.730 & 0.371 \\
& 3 & 1325.797 & 0.557 & 0.008 & 3.727 & 0.351 \\
& 4 & 1322.188 & 0.570 & 0.010 & 3.747 & 0.363 \\
& 5 & 1321.926 & 0.652 & 0.021 & 3.970 & 0.417 \\
\cmidrule(l){2-7}
& {Mean $\pm$ SD} & 1323.52 \pm 1.69 & 0.60 \pm 0.05 & 0.01 \pm 0.01 & 3.83 \pm 0.13 & 0.38 \pm 0.03 \\
\bottomrule
\end{tabular}
\end{table}
\FloatBarrier

\section{Empirical computational cost}
\label{app:computational_performance}

In this appendix, wall-clock measurements complement the operation-level complexity analysis in \cref{sec:discussion_complexity}.
Classical VSI and B-VSI are compared on the same model structure, so that both methods use the same Galerkin weak-form representation.

All timings are measured along the standard experimental workflow (in \cref{sec:examples}) on a Tesla V100-SXM2-32GB GPU using four CPU threads and double-precision arithmetic.
We divide the computation into weak-form preparation, residual-space model selection, and coefficient inference, and split the time of each lagged-covariance procedure into residual-covariance construction, factorization with solves, and the remaining operations. Weak-form preparation is shared by VSI and B-VSI, whereas residual-covariance operations are specific to B-VSI. After weak-form assembly, classical VSI performs a single ordinary least-squares regression. B-VSI instead computes either the lagged-covariance MLE or the lagged-covariance Gaussian posterior, which additionally quantifies the coefficient uncertainty; these are alternative procedures, and both are timed. The FP example further refines the posterior with hybrid SVGD, and the CH example also computes the normal-inverse-gamma posterior with unknown noise variance.

\Cref{tab:FP_wall_time,tab:ch_wall_time}  report the mean and sample standard deviation of the wall-clock times over the five noise realizations in \cref{app:repeated-noise}. The larger standard deviations of the weak-form preparation entries in \cref{tab:FP_wall_time} arise from one-time initialization overhead in the first run; for example, its operator assembly takes \SI{0.55}{s}, compared with at most \SI{0.03}{s} in the other runs. The standard deviations of the block-tridiagonal CH MLE and known-noise posterior arise from one realization in which these updates converge after six rather than seven iterations.

\begin{table}[hbtp]
\centering
\caption{%
Mean and sample standard deviation of the wall-clock time, in seconds, over five noise realizations for the FP example.
The lagged-covariance MLE and the lagged-covariance Gaussian posterior are alternative coefficient-inference procedures; for each, the rows marked \textit{of which} split its time into residual-covariance construction, factorization with solves, and the remaining operations, which sum to its total up to rounding. These procedures and the least-squares regression were timed in a separate run that records the stages of each procedure, on the same hardware. The hybrid SVGD refinement uses $n_p=20$ particles and 100 epochs.
}
\label{tab:FP_wall_time}
\fontsize{9pt}{10pt}\selectfont
\setlength{\tabcolsep}{3pt}
\begin{tabular}{l *{3}{S[table-format=4.2(1.2), separate-uncertainty, retain-zero-uncertainty]}}
\toprule
Component & {VSI} & {B-VSI-BD} & {B-VSI-BT} \\
\midrule

\multicolumn{4}{l}{\textit{Weak-form preparation}} \\
\quad Form compilation
    & 0.07 \pm 0.05 & 0.07 \pm 0.05 & 0.07 \pm 0.05 \\
\quad Operator assembly
    & 0.13 \pm 0.24 & 0.13 \pm 0.24 & 0.13 \pm 0.24 \\

\addlinespace
\multicolumn{4}{l}{\textit{Model selection}} \\
\quad Sequential operator elimination
    & {--} & 33.41 \pm 0.45 & 437.32 \pm 6.93 \\

\addlinespace
\multicolumn{4}{l}{\textit{Coefficient inference}} \\
\quad Ordinary least-squares regression
    & 0.11 \pm 0.02 & {--} & {--} \\
\quad Lagged-covariance MLE
    & {--} & 0.29 \pm 0.01 & 2.49 \pm 0.00 \\
\qquad \textit{of which} covariance construction
    & {--} & 0.02 \pm 0.00 & 0.01 \pm 0.00 \\
\qquad \textit{of which} covariance factorization and solves
    & {--} & 0.24 \pm 0.01 & 2.46 \pm 0.00 \\
\qquad \textit{of which} other operations
    & {--} & 0.03 \pm 0.00 & 0.02 \pm 0.00 \\
\quad Lagged-covariance Gaussian posterior
    & {--} & 0.28 \pm 0.01 & 2.49 \pm 0.00 \\
\qquad \textit{of which} covariance construction
    & {--} & 0.02 \pm 0.00 & 0.01 \pm 0.00 \\
\qquad \textit{of which} covariance factorization and solves
    & {--} & 0.23 \pm 0.01 & 2.46 \pm 0.00 \\
\qquad \textit{of which} other operations
    & {--} & 0.04 \pm 0.00 & 0.02 \pm 0.00 \\
\quad Hybrid SVGD refinement
    & {--} & 118.33 \pm 3.32 & 4271.11 \pm 0.71 \\

\bottomrule
\end{tabular}
\end{table}

\begin{table}[hbtp]
\centering
\caption{%
Mean and sample standard deviation of the wall-clock time, in seconds, over five noise realizations for the CH example.
Rows are defined as in \cref{tab:FP_wall_time}; the unknown-noise normal-inverse-gamma posterior is a further alternative to the known-noise Gaussian posterior.
}
\label{tab:ch_wall_time}
\fontsize{9pt}{10pt}\selectfont
\setlength{\tabcolsep}{3pt}
\begin{tabular}{l *{3}{S[table-format=2.2(1.2), separate-uncertainty, retain-zero-uncertainty]}}
\toprule
Component & {VSI} & {B-VSI-BD} & {B-VSI-BT} \\
\midrule

\multicolumn{4}{l}{\textit{Weak-form preparation}} \\
\quad Form compilation
    & 0.20 \pm 0.01 & 0.20 \pm 0.01 & 0.20 \pm 0.01 \\
\quad Operator assembly
    & 3.71 \pm 0.13 & 3.71 \pm 0.13 & 3.71 \pm 0.13 \\

\addlinespace
\multicolumn{4}{l}{\textit{Model selection}} \\
\quad Sequential operator elimination
    & {--} & \multicolumn{1}{r}{$5506.24\pm1.04$} & \multicolumn{1}{r}{$10936.45\pm102.81$} \\

\addlinespace
\multicolumn{4}{l}{\textit{Coefficient inference}} \\
\quad Ordinary least-squares regression
    & 0.01 \pm 0.00 & {--} & {--} \\
\quad Lagged-covariance MLE
    & {--} & 11.84 \pm 0.01 & 18.97 \pm 1.08 \\
\qquad \textit{of which} covariance construction
    & {--} & 8.92 \pm 0.01 & 9.98 \pm 0.56 \\
\qquad \textit{of which} covariance factorization and solves
    & {--} & 2.92 \pm 0.00 & 8.90 \pm 0.51 \\
\qquad \textit{of which} other operations
    & {--} & 0.00 \pm 0.00 & 0.09 \pm 0.01 \\
\quad Known-noise lagged-covariance Gaussian posterior
    & {--} & 11.84 \pm 0.00 & 19.11 \pm 1.09 \\
\qquad \textit{of which} covariance construction
    & {--} & 8.92 \pm 0.00 & 10.11 \pm 0.58 \\
\qquad \textit{of which} covariance factorization and solves
    & {--} & 2.91 \pm 0.00 & 8.91 \pm 0.51 \\
\qquad \textit{of which} other operations
    & {--} & 0.01 \pm 0.00 & 0.09 \pm 0.00 \\
\quad Unknown-noise normal-inverse-gamma posterior
    & {--} & 15.93 \pm 0.01 & 24.49 \pm 0.01 \\
\qquad \textit{of which} covariance construction
    & {--} & 8.10 \pm 0.01 & 12.94 \pm 0.01 \\
\qquad \textit{of which} covariance factorization and solves
    & {--} & 7.82 \pm 0.01 & 11.44 \pm 0.01 \\
\qquad \textit{of which} other operations
    & {--} & 0.01 \pm 0.00 & 0.11 \pm 0.00 \\

\bottomrule
\end{tabular}
\end{table}

The timing results show that B-VSI introduces additional computational cost relative to classical VSI. VSI requires an unweighted least-squares solve, whereas B-VSI also constructs the residual covariance and performs covariance-weighted solves iteratively.
The hybrid SVGD refinement is more expensive than the Gaussian posterior update, because each iteration evaluates the full $\bstheta$-dependent likelihood and its gradient for every particle.
This overhead is the computational cost of accounting for propagated observation uncertainty and quantifying posterior parameter uncertainty.

The breakdown further shows that covariance linear algebra, rather than finite-element assembly, is the principal B-VSI-specific cost. Covariance construction, factorization, and solves account for more than 85\% of the mean time of each lagged-covariance procedure in both examples.
Their relative share differs between the examples. In the state-linear FP example, the covariance in \cref{eq:covariance_linear} is formed from coefficient-weighted operator matrices that are the same at every time step, so its construction is cheap and factorization with solves dominates. In the CH example, the Jacobian in \cref{eq:overall_jacobian} depends on the observed state, so each lagged update forms it at every time step from the contributions of all active operators before evaluating \cref{eq:covariance_nonlinear}; construction then dominates.
The block-diagonal structure reduces this cost because its time-step covariance blocks can be factorized independently and in parallel. The block-tridiagonal structure retains adjacent temporal correlations through a sequential Schur-complement recursion. Each elimination step requires additional solves with the Schur blocks for the columns of $\boldsymbol{E}_{k-1}$, and these Schur blocks are generally dense even when $\boldsymbol{D}_k$ and $\boldsymbol{E}_k$ are sparse (see \cref{sec:discussion_complexity}). Consequently, factorization and solves, and hence every lagged-covariance, model-selection, and SVGD stage, are more expensive with the block-tridiagonal structure than with the block-diagonal approximation (\cref{tab:FP_wall_time,tab:ch_wall_time}). Within the block-tridiagonal hybrid SVGD refinement of the FP example, covariance construction and factorization with solves account for \SI{31.20}{s} and \SI{1076.01}{s}, respectively.

\section{Model selection with large candidate dictionaries}
\label{app:large-dictionary}

This appendix examines B-VSI with candidate dictionaries much larger than the one in \cref{sec:example_FP}: enlarged dictionaries that contain the true operators (\cref{app:large-dictionary-nested}) and a non-polynomial dictionary that cannot represent the true potential exactly (\cref{app:large-dictionary-fourier}).
Both studies use the FP problem of \cref{sec:example_FP}, with its noise-free trajectory, identification mesh, and observation model; only the candidate dictionary changes.
In both, the operators are selected by sequential operator elimination with the residual-space BIC (\cref{sec:sequential_operator_elimination}), with the diffusion operator protected from elimination.
Where the selected model is assessed by its identified drift and potential, we use density-weighted relative errors,
\begin{align}
    e_{\nabla\psi}
    &=\left(\frac{\sum_{\ell}\omega_\ell\,\|\nabla\hpsi(\bsx_\ell)-\nabla\psi_{\mathrm{ref}}(\bsx_\ell)\|^2}{\sum_{\ell}\omega_\ell\,\|\nabla\psi_{\mathrm{ref}}(\bsx_\ell)\|^2}\right)^{1/2},
    \nonumber\\
    \label{eq:app-weighted-errors}
    e_{\psi}
    &
    =\left(\frac{\sum_{\ell}\omega_\ell\,\bigl(\hpsi(\bsx_\ell)-\psi_{\mathrm{ref}}(\bsx_\ell)\bigr)^2}{\sum_{\ell}\omega_\ell\,\psi_{\mathrm{ref}}(\bsx_\ell)^2}\right)^{1/2},
\end{align}
where $\bsx_\ell$ are the nodes of the identification mesh, $\omega_\ell$ is the time-averaged density at $\bsx_\ell$, set to zero where the density never exceeds 1\% of its peak, and both potentials are shifted to zero $\omega$-weighted mean.
The reference $\psi_{\mathrm{ref}}$ is the true potential in \cref{app:large-dictionary-nested} and its projection onto the dictionary in \cref{app:large-dictionary-fourier}.
Unlike $\mathrm{MSE}(\hpsi)$ in \cref{tab:FP_parameters}, which averages over the whole domain, these errors exclude the region that the density never visits, where the data do not constrain the potential; with misspecified or substituted operators, the full-domain MSE is dominated by that region and can therefore favor a model that is less accurate where the data constrain the potential.

\subsection{Nested dictionaries containing the true operators}
\label{app:large-dictionary-nested}

We first enlarge the dictionary in \cref{eq:FP_candidate_dictionary} ($d=10$) while keeping the true operators in it.
Two families of nested dictionaries append distractor potentials in the normalized coordinates $z_j=x_j/5$, each multiplied by 25, in a predeclared order (\cref{tab:app-fp-dictionaries}); every dictionary therefore contains the true operators of \cref{sec:FP_model_selection}.

\begin{table}[htbp]
\centering
\caption{Enlarged FP dictionaries, labeled by their total number of candidates $d$. Each column gives the number of distractor potentials from each group appended, in the normalized coordinates $z_j=x_j/5$ and multiplied by 25, to the $d=10$ dictionary in \cref{eq:FP_candidate_dictionary}. The mixed dictionary with $d=30$ contains the first nine Fourier terms, those with $m\leq2$ and $\sin(3\pi z_1)$.}
\label{tab:app-fp-dictionaries}
\fontsize{9pt}{10pt}\selectfont
\begin{tabular}{@{}l c ccc cc@{}}
\toprule
& & \multicolumn{3}{c}{Polynomial family} & \multicolumn{2}{c}{Mixed family} \\
\cmidrule(lr){3-5}\cmidrule(l){6-7}
Distractor group & Size & {$d=21$} & {$d=36$} & {$d=55$} & {$d=30$} & {$d=57$} \\
\midrule
Monomials of degree four and five & 11 & 11 & 11 & 11 & 11 & 11 \\
Monomials of degree six and seven & 15 & -- & 15 & 15 & -- & -- \\
Monomials of degree eight and nine & 19 & -- & -- & 19 & -- & -- \\
Fourier terms $\sin(m\pi z_j)$, $\cos(m\pi z_j)$, $m\leq4$ & 16 & -- & -- & -- & 9 & 16 \\
Products of $\{\sin,\cos\}(\pi z_1)$ and $\{\sin,\cos\}(\pi z_2)$ & 4 & -- & -- & -- & -- & 4 \\
Gaussians on a $4\times4$ grid in $[-0.75,0.75]^2$, width $0.35$ & 16 & -- & -- & -- & -- & 16 \\
\bottomrule
\end{tabular}
\end{table}

We consider the five noise realizations at $\sigma=10^{-3}$ from \cref{app:repeated-noise}, for which both covariance structures recover the true operator support using the polynomial dictionary with $d=10$ (see \cref{sec:example_FP,app:repeated-noise-fp}).
Since posterior inference is identical whenever the true support is recovered, as reported in \cref{sec:example_FP,app:repeated-noise-fp}, we focus here on model-selection results.

\subsubsection{Greedy backward elimination}
\label{app:large-dictionary-search}

We apply the sequential operator elimination of \cref{sec:sequential_operator_elimination} to each enlarged dictionary with both covariance structures.
In this appendix, the elimination is continued until only the protected diffusion operator remains, which requires $d(d-1)/2+1$ candidate-model fits.
Terminating at the first step that does not lower the residual-space BIC would save $0.4\%$ to $23.5\%$ of these fits in the dictionaries of this appendix, because most fits are spent at the early steps with large models.

With the block-tridiagonal structure, greedy elimination recovers the true support in all five realizations for every polynomial dictionary and for the mixed dictionary with $d=30$, but fails in every realization for the mixed dictionary with $d=57$ (\cref{tab:app-fp-nested}).
With the block-diagonal structure, it fails in every realization for $d\geq21$ in both families, although it succeeds for $d=10$.

\begin{table}[htbp]
\centering
\caption{Greedy backward elimination with enlarged dictionaries containing the true operators, over five noise realizations at $\sigma=10^{-3}$. Runs selecting the true model counts the noise realizations (runs) in which the selected model equals the true model. The selected operators give the range of the number of operators in the selected models, and the missed true operators are the true potential terms absent from the selected model in every realization (none when the true model is selected). Candidate fits count the models fitted along the full elimination path, $d(d-1)/2+1$ with the diffusion operator protected (\cref{app:large-dictionary-search}), and the selection time is the mean and sample standard deviation of the wall-clock time. $\Delta\mathrm{BIC}$ is the range over the realizations of the residual-space BIC of the selected model minus that of the true model, evaluated with the same covariance structure.}
\label{tab:app-fp-nested}
\fontsize{9pt}{10pt}\selectfont
\setlength{\tabcolsep}{3pt}
\sisetup{round-mode=places, round-precision=2}
\begin{tabular}{@{}l c l c c c S[table-format=4.0, round-mode=none] c c@{}}
\toprule
Dictionary & $d$ & Covariance
& \makecell{Runs\\selecting\\true model}
& \makecell{Selected\\operators}
& \makecell{Missed true\\operators}
& {\makecell{Candidate\\fits}}
& \makecell{Selection\\time ($10^{3}$\,s)}
& \makecell{$\Delta\mathrm{BIC}$} \\
\midrule
\multirow{6}{*}{Polynomial}
& \multirow{2}{*}{21} & Block-diagonal & 0/5 & 6--7 & $x_1^2$ & 211 & $0.23\pm0.01$ & $(4.62\text{--}6.36)\times10^{4}$ \\
& & Block-tridiagonal & 5/5 & 5 & -- & 211 & $2.33\pm0.05$ & 0.00 \\
\cmidrule(l){2-9}
& \multirow{2}{*}{36} & Block-diagonal & 0/5 & 7 & $x_1^2$ & 631 & $0.78\pm0.04$ & $(6.36\text{--}6.40)\times10^{4}$ \\
& & Block-tridiagonal & 5/5 & 5 & -- & 631 & $8.51\pm0.13$ & 0.00 \\
\cmidrule(l){2-9}
& \multirow{2}{*}{55} & Block-diagonal & 0/5 & 7 & $x_1^2$ & 1486 & $2.18\pm0.14$ & $(6.36\text{--}6.40)\times10^{4}$ \\
& & Block-tridiagonal & 5/5 & 5 & -- & 1486 & $21.27\pm0.14$ & 0.00 \\
\midrule
\multirow{4}{*}{Mixed}
& \multirow{2}{*}{30} & Block-diagonal & 0/5 & 6 & $x_1^2$, $x_2^2$ & 436 & $0.52\pm0.01$ & $(5.17\text{--}5.26)\times10^{4}$ \\
& & Block-tridiagonal & 5/5 & 5 & -- & 436 & $5.84\pm0.02$ & 0.00 \\
\cmidrule(l){2-9}
& \multirow{2}{*}{57} & Block-diagonal & 0/5 & 8--9 & $x_1$, $x_2$, $x_1^2$, $x_2^2$ & 1597 & $2.99\pm0.07$ & $(1.18\text{--}2.12)\times10^{5}$ \\
& & Block-tridiagonal & 0/5 & 7--11 & $x_2^2$ & 1597 & $22.10\pm0.30$ & $(0.64\text{--}2.08)\times10^{5}$ \\
\bottomrule
\end{tabular}
\end{table}

\begin{figure}[htbp]
    \centering
    \captionsetup[subfigure]{justification=centering}
    \includegraphics[width=0.7\textwidth]{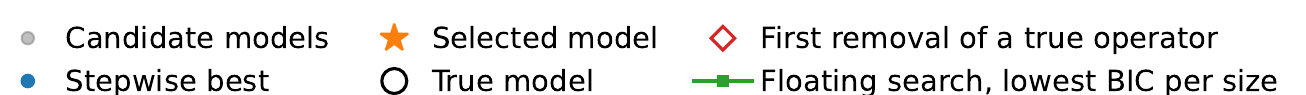}\par\smallskip
    \begin{subfigure}[t]{0.325\textwidth}
        \centering
        \includegraphics[width=\linewidth]{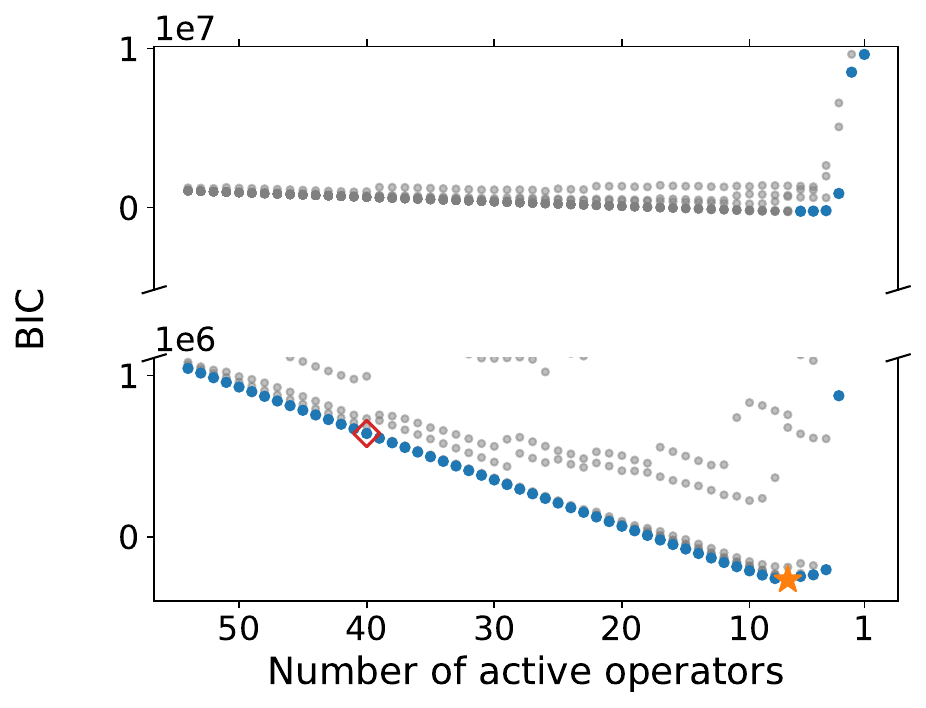}
        \caption{Polynomial dictionary, $d=55$,\\block-diagonal likelihood covariance}
        \label{fig:app-fp-poly-model-selection-bd}
    \end{subfigure}
    \hfill
    \begin{subfigure}[t]{0.325\textwidth}
        \centering
        \includegraphics[width=\linewidth]{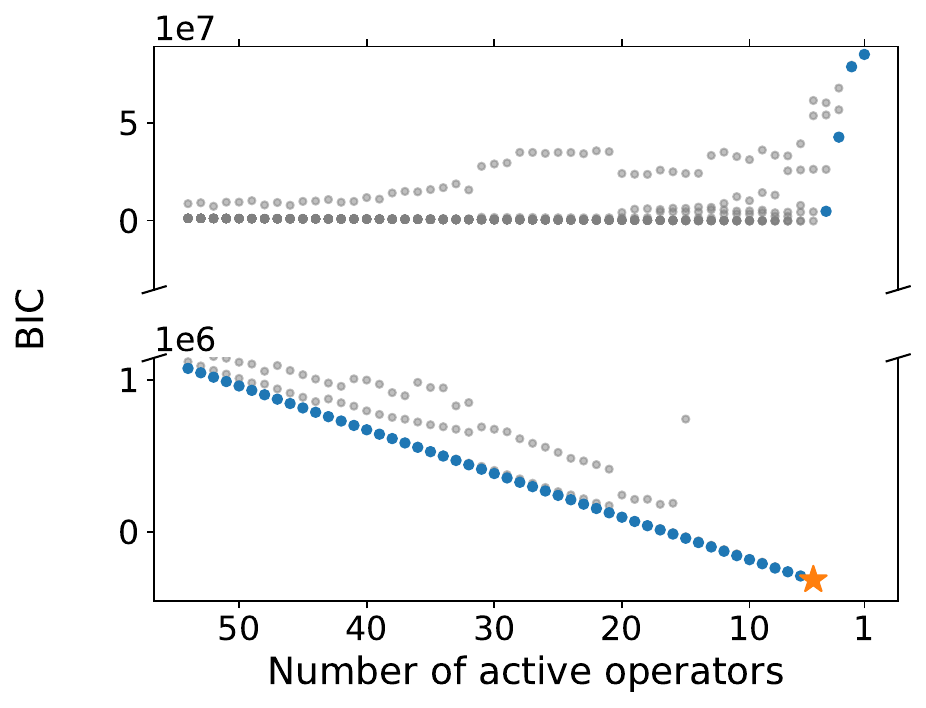}
        \caption{Polynomial dictionary, $d=55$,\\block-tridiagonal likelihood covariance}
        \label{fig:app-fp-poly-model-selection-bt}
    \end{subfigure}
    \hfill
    \begin{subfigure}[t]{0.325\textwidth}
        \centering
        \includegraphics[width=\linewidth]{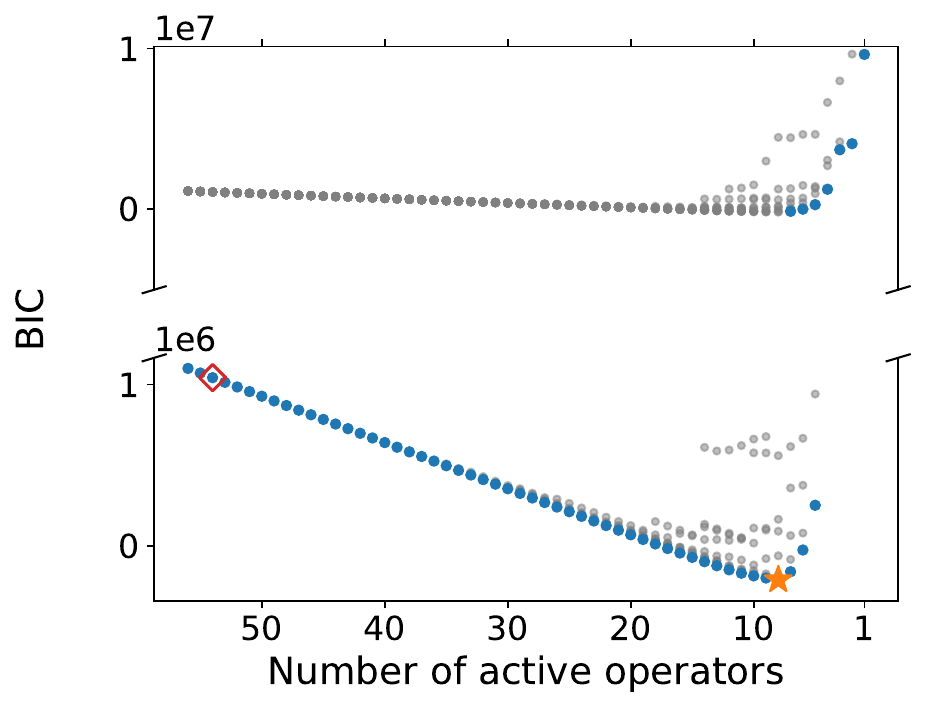}
        \caption{Mixed dictionary, $d=57$,\\block-diagonal likelihood covariance}
        \label{fig:app-fp-nested-model-selection-bd}
    \end{subfigure}
    \par\medskip
    \begin{subfigure}[t]{0.325\textwidth}
        \centering
        \includegraphics[width=\linewidth]{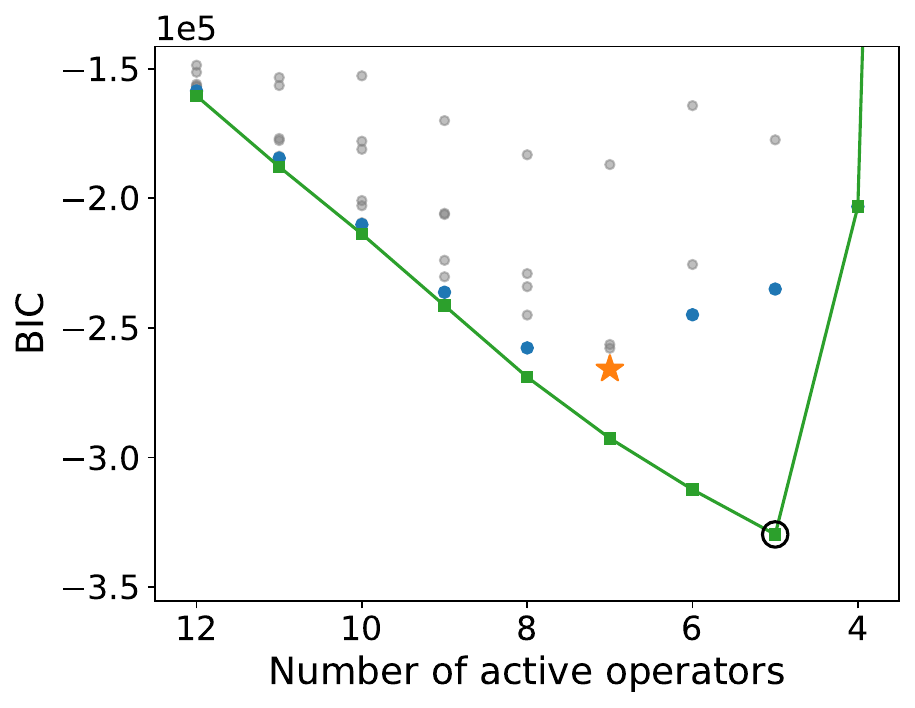}
        \caption{Final steps of (a)}
        \label{fig:app-fp-poly-zoom-bd}
    \end{subfigure}
    \hfill
    \begin{subfigure}[t]{0.325\textwidth}
        \centering
        \includegraphics[width=\linewidth]{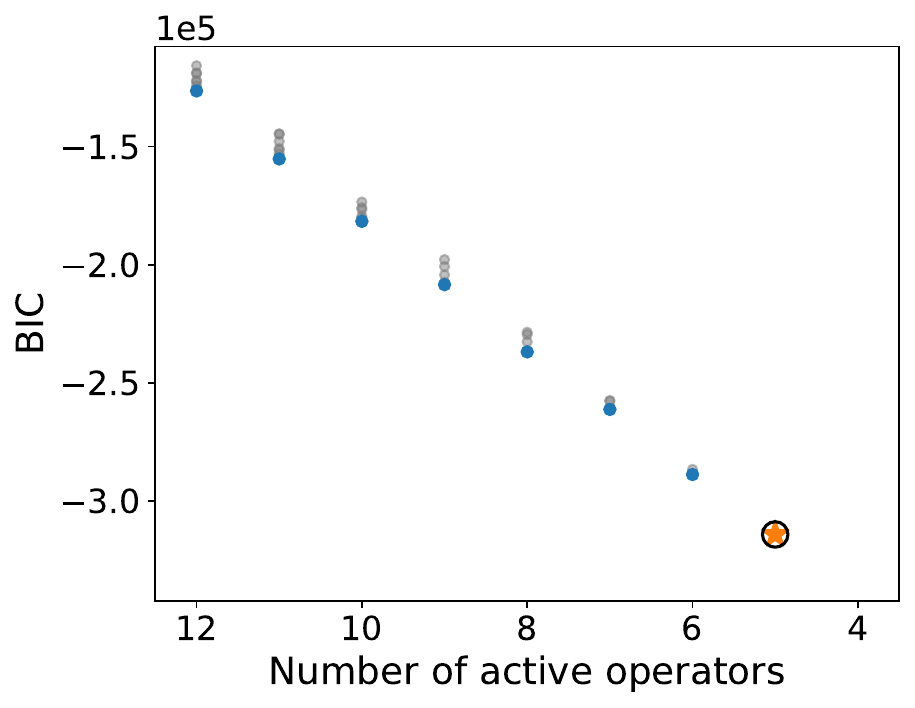}
        \caption{Final steps of (b)}
        \label{fig:app-fp-poly-zoom-bt}
    \end{subfigure}
    \hfill
    \begin{subfigure}[t]{0.325\textwidth}
        \centering
        \includegraphics[width=\linewidth]{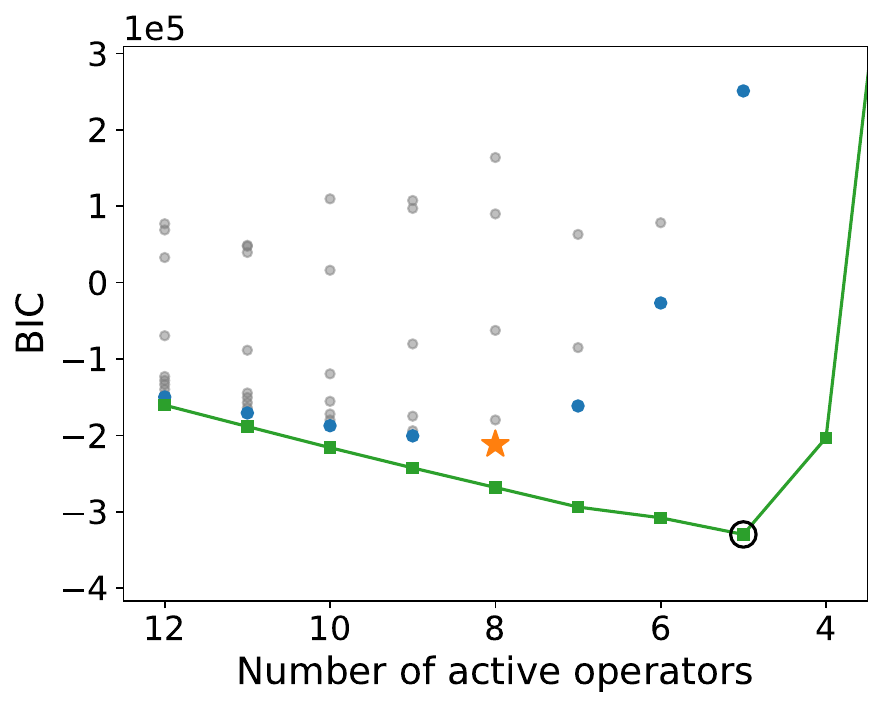}
        \caption{Final steps of (c)}
        \label{fig:app-fp-nested-zoom-bd}
    \end{subfigure}
    \caption{Residual-space BIC along the greedy sequential operator-elimination paths for the FP equation in a representative noise realization (realization~1) at $\sigma=10^{-3}$, with (a, d) the polynomial dictionary with $d=55$ using block-diagonal and (b, e) block-tridiagonal likelihood covariance structures, and (c, f) the mixed dictionary with $d=57$ using block-diagonal likelihood covariance. Panels (a--c) show the full paths, with diamonds marking the step at which a true operator is first removed. Panels (d--f) show the final steps of these paths together with the residual-space BIC of the true model under the same covariance structure and, in (d, f), the lowest residual-space BIC found by the floating search of \cref{app:large-dictionary-floating} for each number of active operators. Other markers and broken vertical axes are as in \cref{fig:FP_model_selection}.}
    \label{fig:app-fp-nested-model-selection}
\end{figure}

\begin{figure}[htbp]
    \centering
    \captionsetup[subfigure]{justification=centering}
    \begin{subfigure}[t]{0.41\textwidth}
        \centering
        \includegraphics[width=\linewidth]{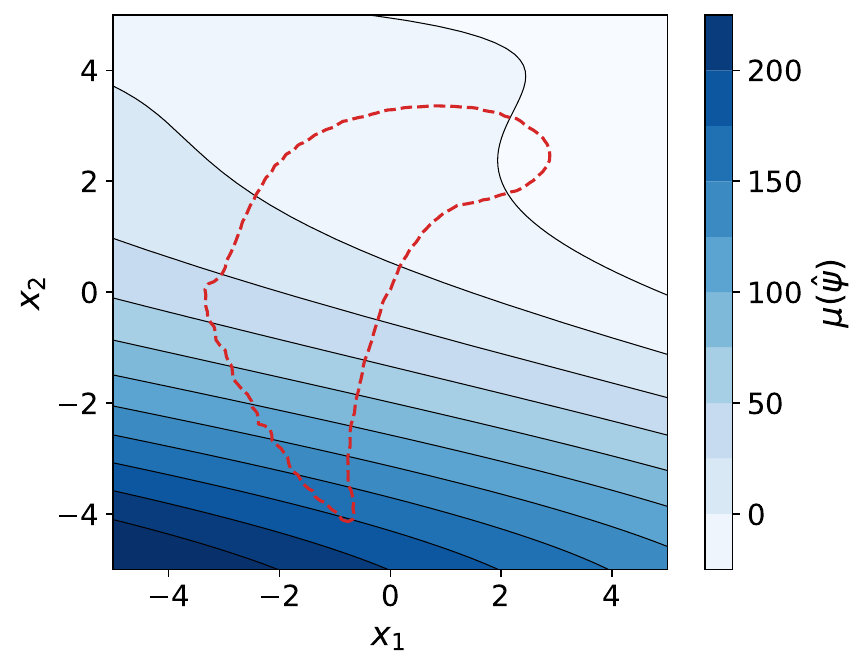}
        \caption{Polynomial dictionary, $d=21$}
        \label{fig:app-fp-failure-poly21}
    \end{subfigure}
    \hspace{0.04\textwidth}
    \begin{subfigure}[t]{0.41\textwidth}
        \centering
        \includegraphics[width=\linewidth]{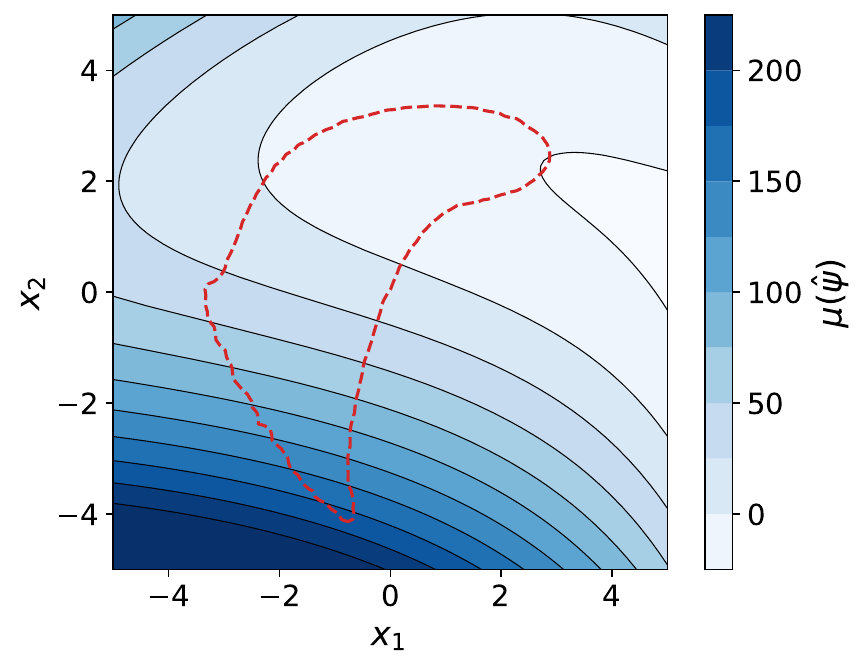}
        \caption{Polynomial dictionaries, $d=36$ and $d=55$}
        \label{fig:app-fp-failure-poly36}
    \end{subfigure}

    \medskip
    \begin{subfigure}[t]{0.41\textwidth}
        \centering
        \includegraphics[width=\linewidth]{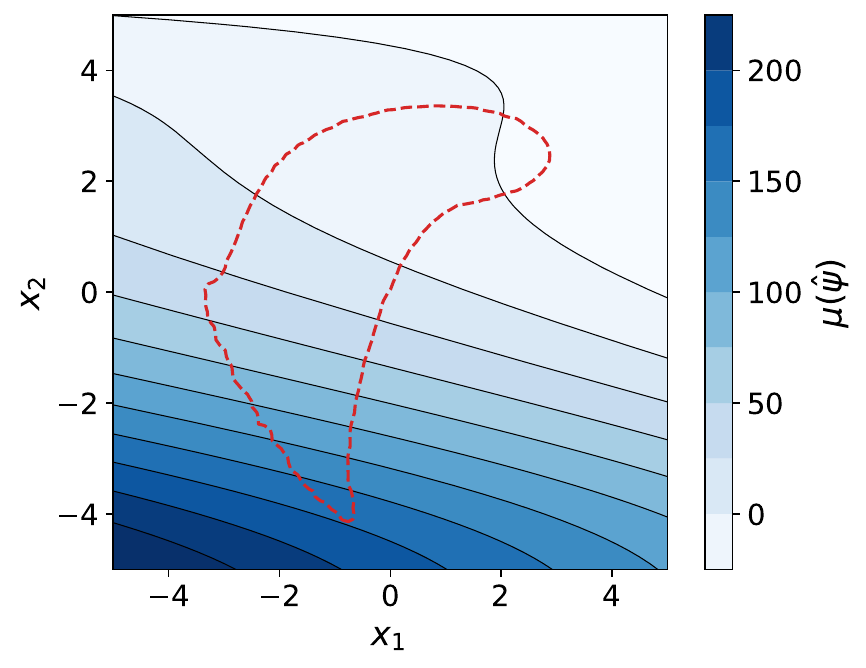}
        \caption{Mixed dictionary, $d=30$}
        \label{fig:app-fp-failure-nested30}
    \end{subfigure}
    \hspace{0.04\textwidth}
    \begin{subfigure}[t]{0.41\textwidth}
        \centering
        \includegraphics[width=\linewidth]{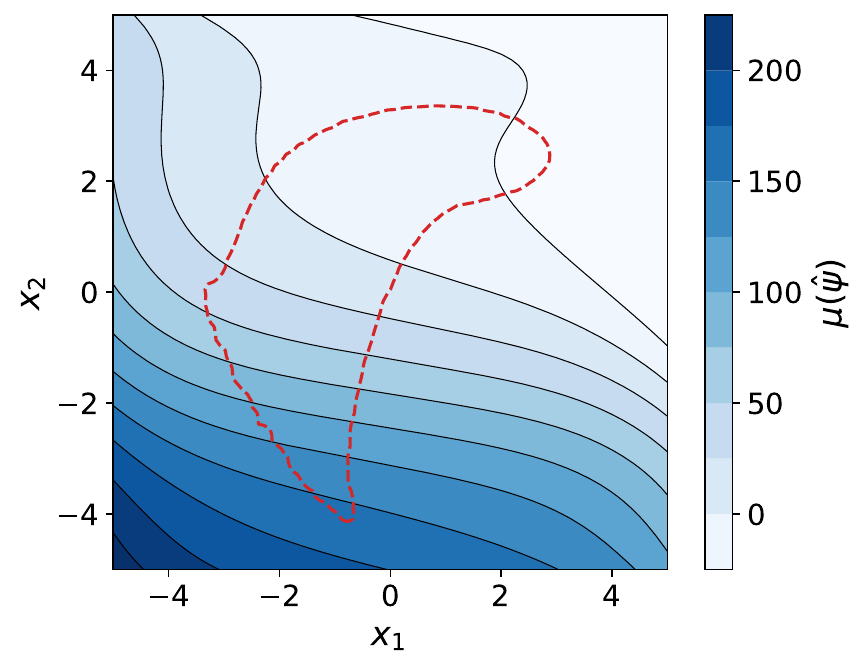}
        \caption{Mixed dictionary, $d=57$}
        \label{fig:app-fp-failure-nested57}
    \end{subfigure}
    \caption{Posterior-predictive means of the potential for the FP equation from the models selected by greedy elimination with block-diagonal likelihood covariance at $\sigma=10^{-3}$, in the representative realization of \cref{fig:app-fp-nested-model-selection}, for (a, b) the polynomial and (c, d) the mixed dictionaries. The polynomial dictionaries with $d=36$ and $d=55$ select the same model, shown once in (b). The true potential is shown in \cref{fig:FP_potential_function}. All maps are shifted by their density-weighted means and use the color levels of the true potential in the region visited by the density, as defined in \cref{app:large-dictionary}; the dashed line marks the boundary of that region, outside which the data do not constrain the potential.}
    \label{fig:app-fp-nested-failure-potential}
\end{figure}

These failures arise from the search rather than from the criterion.
In every failure, the true support has a lower residual-space BIC than the selected model under the same covariance structure, by $4.6\times10^{4}$ to $2.1\times10^{5}$ (\cref{tab:app-fp-nested}), but the elimination path never visits it, as illustrated for a representative realization in \cref{fig:app-fp-nested-model-selection}.
Greedy elimination never re-includes a removed operator: while candidates correlated with a true operator remain in the model, they can partly compensate for its removal, so removing the true operator can be the locally best step, and once these candidates are themselves removed, the true support can no longer be reached.
The excluded true operators and their substitutes are consistent with this mechanism.
With the block-diagonal structure, the term in $x_1^2$ is eliminated in every failure within the first 2 to 29 steps, before the path can reach the five-operator true model, which requires the removal of all $d-5$ distractors.
Correlated candidates then absorb its contribution: the terms in $x_1x_2$, $x_1^2x_2$, and $x_1x_2^2$ for the polynomial dictionaries with $d=36$ and $d=55$, and mainly those in $x_1x_2$ and $z_2^4$ for $d=21$.
For the mixed dictionary with $d=30$, the term in $x_2^2$ is also eliminated, and both quadratic terms are absorbed by the term in $x_1x_2$ and Fourier terms in $z_2$.
With the block-tridiagonal structure and $d=57$, the term in $x_2^2$, and in three realizations also that in $x_1^2$, is replaced by Fourier terms in $z_2$, together with $x_2^3$ and Gaussian functions in those three realizations.
The models selected with the block-diagonal structure have biased coefficients: $\hat{\beta}^{-1}$ ranges from $0.31$ to $0.74$ over the realizations, with drift errors of $9\%$ to $18\%$; with the block-tridiagonal structure and $d=57$, $\hat{\beta}^{-1}=0.99\pm0.03$ and the drift error is $(5.9\pm3.5)\%$.
Nevertheless, they reproduce the potential closely over the region visited by the density, as illustrated for the representative realization in \cref{fig:app-fp-nested-failure-potential}, where the polynomial dictionaries with $d=36$ and $d=55$ share one panel because they select the same model.
In this realization, the density-weighted relative potential errors are $5.5\%$ to $6.1\%$, compared with drift errors of $8.9\%$ to $15.7\%$, because the substituting candidates approximate the excluded terms there.
These results motivate a search in which removed operators can re-enter the model (\cref{app:large-dictionary-floating}).

\subsubsection{Floating backward search}
\label{app:large-dictionary-floating}

To allow removed operators to re-enter, we also use sequential floating backward selection~\cite{Pudil1994Floating}.
It differs from greedy elimination only in a conditional inclusion step: after each removal, the previously removed operator whose re-inclusion gives the lowest BIC is added back if the resulting model has a lower BIC than the best model of that size found so far, and this step is repeated until no re-inclusion improves.
The selected model is the model with the lowest BIC among the best models found for each size.
The criterion and the protected diffusion operator are the same as for greedy elimination, so the floating search changes only how the model space is explored, at the cost of additional model fits for the inclusion steps.

\begin{figure}[htbp]
    \centering
    \captionsetup[subfigure]{justification=centering}
    \includegraphics[width=0.9\textwidth]{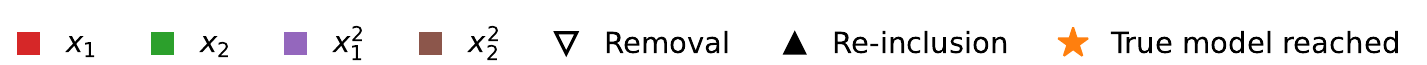}\par\smallskip
    \begin{subfigure}[t]{0.48\textwidth}
        \centering
        \includegraphics[width=0.7\linewidth]{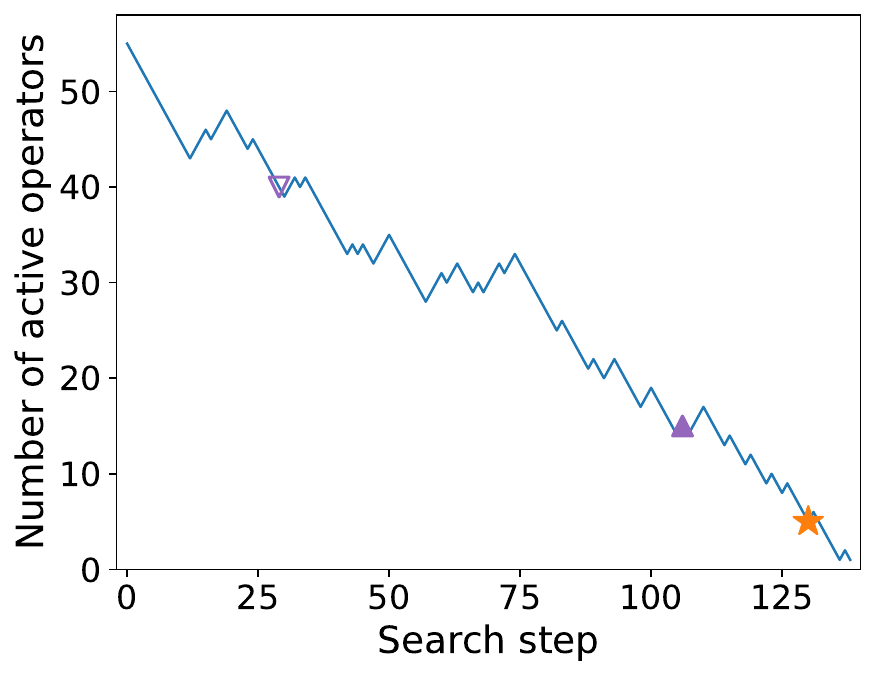}
        \caption{Polynomial dictionary, $d=55$}
        \label{fig:app-fp-floating-poly-bd}
    \end{subfigure}
    \hfill
    \begin{subfigure}[t]{0.48\textwidth}
        \centering
        \includegraphics[width=0.7\linewidth]{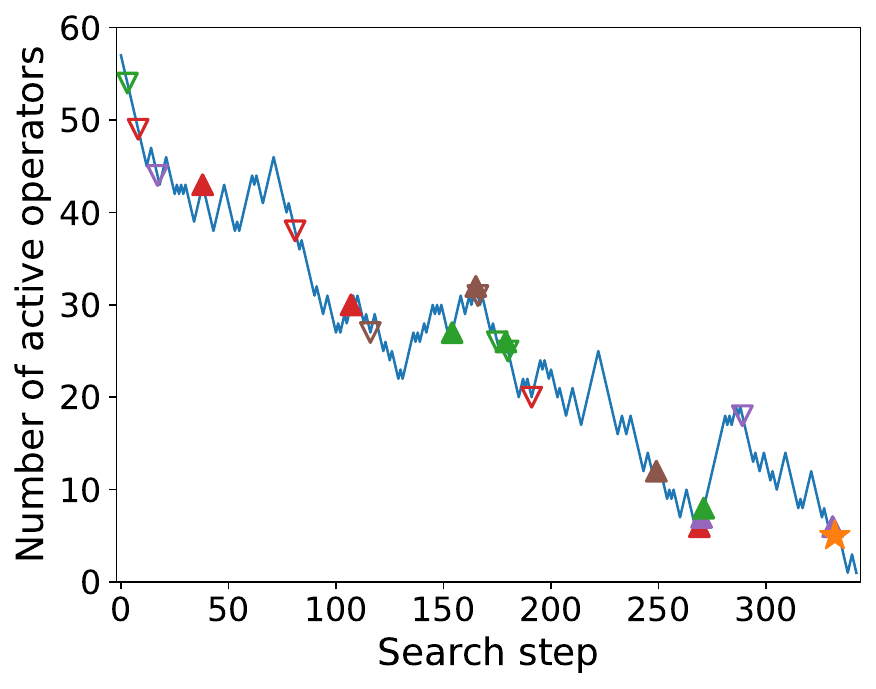}
        \caption{Mixed dictionary, $d=57$}
        \label{fig:app-fp-floating-mixed-bd}
    \end{subfigure}
    \caption{Number of active operators along the floating search with block-diagonal likelihood covariance for the FP equation at $\sigma=10^{-3}$, in the representative realization of \cref{fig:app-fp-nested-model-selection}, with (a) the polynomial dictionary with $d=55$ and (b) the mixed dictionary with $d=57$. Each search step removes an operator or re-includes a previously removed one. Triangles mark the removals and re-inclusions of the true potential operators until the search reaches the true model, marked by the star.}
    \label{fig:app-fp-nested-floating}
\end{figure}

We apply the floating search with both covariance structures (\cref{tab:app-fp-nested-floating}).
With the block-diagonal structure, for which greedy elimination fails for every enlarged dictionary (\cref{tab:app-fp-nested}), it recovers the true support in 24 of the 25 runs.
The exception is one realization with the mixed dictionary with $d=57$, in which the selected model contains eight operators, including the true terms in $x_1$ and $x_1^2$ but not those in $x_2$ and $x_2^2$; as with greedy elimination, the true model has a lower residual-space BIC, by $9.17\times10^{4}$, but is not reached by the search.
With the block-tridiagonal structure, the floating search recovers the true support in every run, including the mixed dictionary with $d=57$, for which greedy elimination fails.
In the recovered cases, a true operator that is removed early, most often the term in $x_1^2$, re-enters the model at a later step, so the search reaches the true model, which has a lower residual-space BIC than the greedy selection.
\Cref{fig:app-fp-poly-zoom-bd,fig:app-fp-nested-zoom-bd} illustrate this for the representative realization: over the final steps, the floating search finds a model with a lower residual-space BIC than the greedy path at every model size, and the difference grows toward the true model, at which the floating search attains its minimum.
The search trajectories in \cref{fig:app-fp-nested-floating} show how the true model is reached.
With the polynomial dictionary with $d=55$, the term in $x_1^2$ is removed at the same model size as in the greedy path, 40 active operators, and re-enters at 15; with the mixed dictionary with $d=57$, each true potential operator is removed and re-included at least once before the true model is reached.
The floating search requires 2.12 to 8.29 times as many candidate-model fits as greedy elimination with the block-diagonal structure and 2.28 to 7.70 times with the block-tridiagonal structure, with the mean ratio increasing with the dictionary size within each family.

\begin{table}[htbp]
\centering
\caption{Floating backward elimination with the dictionaries of \cref{tab:app-fp-nested} and the same five noise realizations. Runs selecting the true model is defined as in \cref{tab:app-fp-nested}. The candidate fits and selection time are defined as in \cref{tab:app-fp-nested} and averaged over the realizations; the ratio compares the candidate fits with those of greedy elimination in \cref{tab:app-fp-nested}.}
\label{tab:app-fp-nested-floating}
\fontsize{9pt}{10pt}\selectfont
\setlength{\tabcolsep}{3.5pt}
\sisetup{round-mode=places, round-precision=2}
\begin{tabular}{@{}l c l c S[table-format=5.2] S[table-format=1.2] c@{}}
\toprule
Dictionary & $d$ & Covariance
& \makecell{Runs\\selecting\\true model}
& {\makecell{Candidate\\fits}}
& {\makecell{Fits relative\\to greedy}}
& \makecell{Selection\\time ($10^{3}$\,s)} \\
\midrule
\multirow{6}{*}{Polynomial}
& \multirow{2}{*}{21} & Block-diagonal & 5/5 & 457.40 & 2.17 & $0.45\pm0.01$ \\
& & Block-tridiagonal & 5/5 & 493.80 & 2.34 & $4.86\pm0.21$ \\
\cmidrule(l){2-7}
& \multirow{2}{*}{36} & Block-diagonal & 5/5 & 2002.20 & 3.17 & $2.11\pm0.55$ \\
& & Block-tridiagonal & 5/5 & 2107.60 & 3.34 & $27.13\pm2.47$ \\
\cmidrule(l){2-7}
& \multirow{2}{*}{55} & Block-diagonal & 5/5 & 5662.20 & 3.81 & $7.04\pm0.39$ \\
& & Block-tridiagonal & 5/5 & 7245.40 & 4.88 & $102.09\pm10.78$ \\
\midrule
\multirow{4}{*}{Mixed}
& \multirow{2}{*}{30} & Block-diagonal & 5/5 & 1629.20 & 3.74 & $1.59\pm0.15$ \\
& & Block-tridiagonal & 5/5 & 1434.40 & 3.29 & $18.26\pm2.16$ \\
\cmidrule(l){2-7}
& \multirow{2}{*}{57} & Block-diagonal & 4/5 & 11619.60 & 7.28 & $15.55\pm1.42$ \\
& & Block-tridiagonal & 5/5 & 9435.00 & 5.91 & $121.40\pm26.31$ \\
\bottomrule
\end{tabular}
\end{table}

Thus, for $d>50$, the residual-space BIC continues to identify the true operators, and the sequential search remains computationally feasible.
Whether greedy elimination reaches the true model depends on the correlation among the candidates; the floating search recovers it in all but one of the runs with either covariance structure, at two to eight times as many candidate-model fits.

\subsection{Misspecified Fourier dictionary}
\label{app:large-dictionary-fourier}

We replace the polynomial candidates in \cref{eq:FP_candidate_dictionary} by 50 Fourier potential candidates in the normalized coordinates $z_j=x_j/5$: the products of $1$, $\cos(k\pi z_j/2)$, and $\sin(k\pi z_j/2)$ for $k=1,2,3$, excluding the constant, together with $\cos(2\pi z_1)$ and $\cos(2\pi z_2)$, each multiplied by 25.
With the diffusion operator, the dictionary has $d=51$ candidates.
The quadratic potential of \cref{sec:FP_data_generation} is not in the span of this dictionary: the least-squares projection of its gradient onto the 14 single-axis candidates has a relative drift error of 1.5\% on a uniform grid of the domain.
This setting therefore combines a large, redundant dictionary with a mildly misspecified one.
Moreover, the nearly collinear single-axis candidates make their individual coefficients unidentifiable over the region visited by the density, so we assess the identified drift and potential rather than the selected support.
We use one noise realization (seed 42) at the noise level $\sigma=10^{-3}$ of \cref{sec:example_FP} and at $\sigma=10^{-4}$.

\begin{table}[htbp]
\centering
\caption{Results for the Fourier dictionary with $d=51$ candidates (noise seed 42). The drift and potential errors are those of \cref{eq:app-weighted-errors}, measured against the projected potential, and the candidate fits and selection time are defined as in \cref{tab:app-fp-nested}.
}
\label{tab:app-fp50}
\fontsize{10pt}{11pt}\selectfont
\setlength{\tabcolsep}{4pt}
\sisetup{round-mode=places, round-precision=2}
\begin{tabular}{@{}cl S[table-format=2.0, round-mode=none] S[table-format=1.2] S[table-format=2.2] S[table-format=1.2] S[table-format=4.0, round-mode=none] S[table-format=5.2]@{}}
\toprule
$\sigma$ & Covariance
& {\makecell{Selected\\operators}}
& {$\hat{\beta}^{-1}$}
& {\makecell{Drift\\error (\%)}}
& {\makecell{Potential\\error (\%)}}
& {\makecell{Candidate\\fits}}
& {\makecell{Selection\\time (s)}} \\
\midrule
\multirow{2}{*}{$10^{-3}$}
& Block-diagonal & 6 & 0.240 & 18.607 & 7.090 & 1276 & 2066.418 \\
& Block-tridiagonal & 10 & 1.004 & 6.201 & 1.315 & 1276 & 16393.241 \\
\midrule
\multirow{2}{*}{$10^{-4}$}
& Block-diagonal & 17 & 0.907 & 4.340 & 1.216 & 1276 & 2051.561 \\
& Block-tridiagonal & 24 & 1.052 & 3.644 & 0.789 & 1276 & 15786.631 \\
\bottomrule
\end{tabular}
\end{table}

\begin{figure}[htbp]
    \centering
    \captionsetup[subfigure]{justification=centering}
    \begin{subfigure}[t]{0.495\textwidth}
        \centering
        \includegraphics[height=5.2cm]{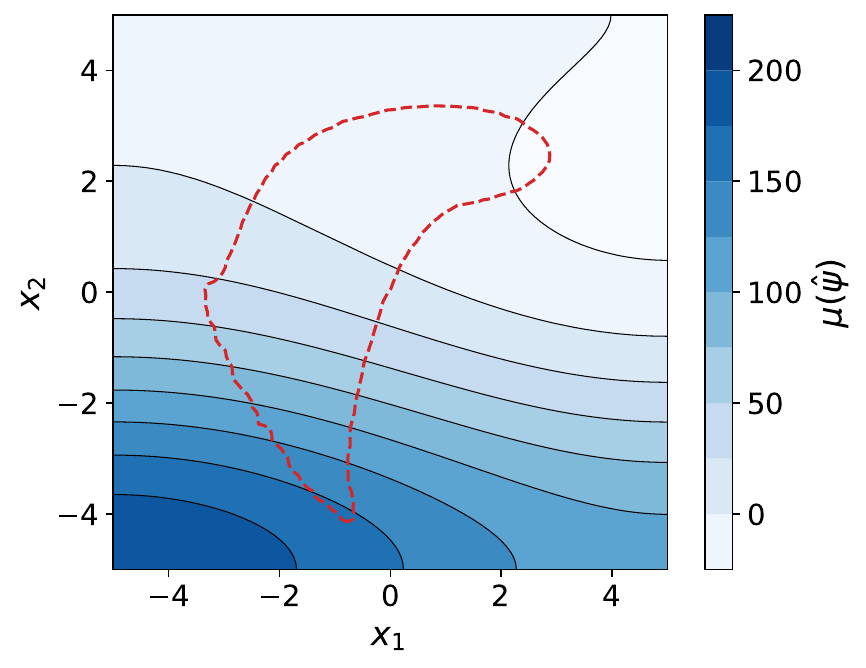}
        \caption{Predictive mean with\\block-diagonal likelihood covariance}
        \label{fig:app-fp50-potential-mean-bd}
    \end{subfigure}
    \hfill
    \begin{subfigure}[t]{0.495\textwidth}
        \centering
        \includegraphics[height=5.2cm]{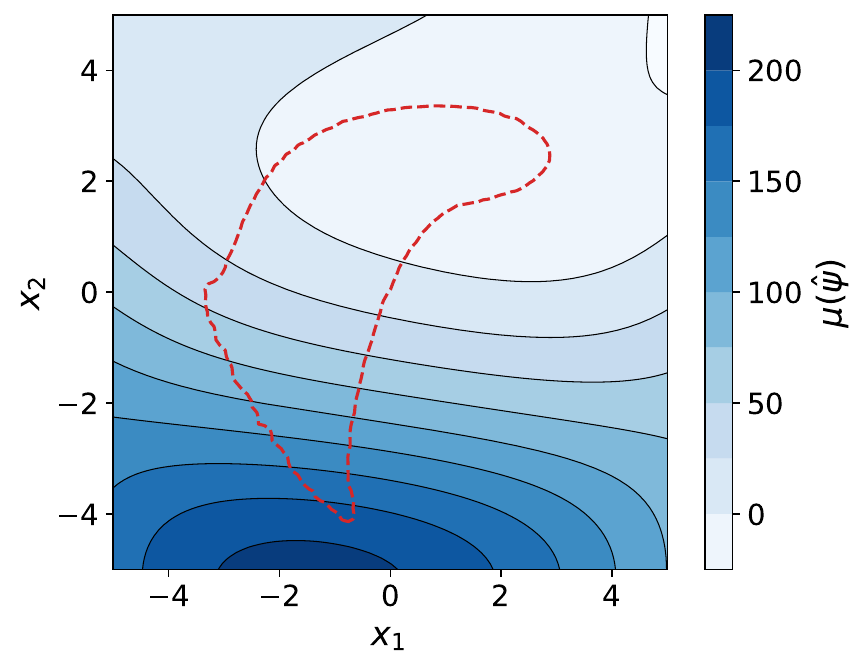}
        \caption{Predictive mean with\\block-tridiagonal likelihood covariance}
        \label{fig:app-fp50-potential-mean-bt}
    \end{subfigure}

    \medskip
    \begin{subfigure}[t]{0.495\textwidth}
        \centering
        \includegraphics[height=5.2cm]{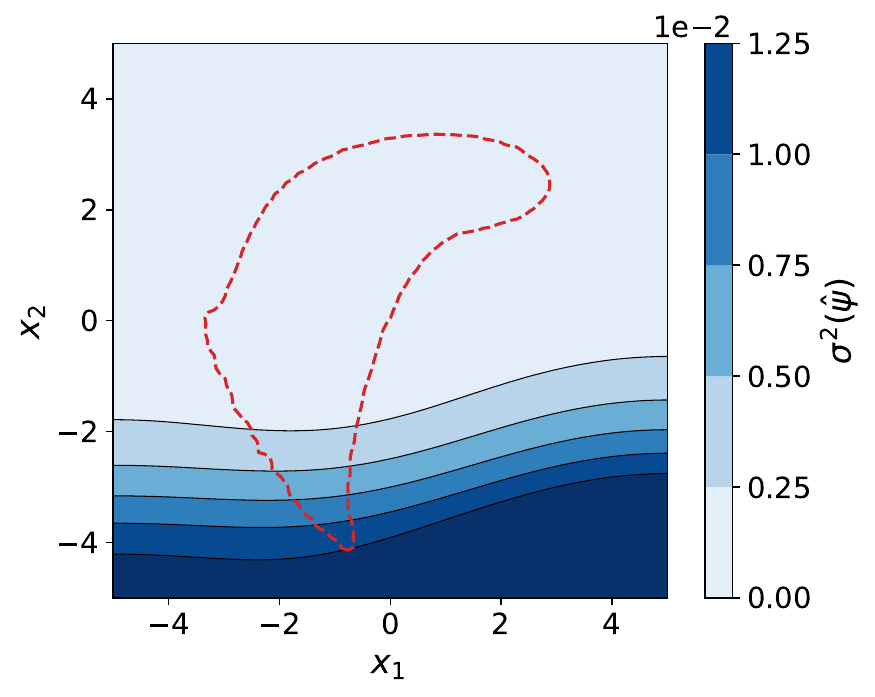}
        \caption{Predictive variance with\\block-diagonal likelihood covariance}
        \label{fig:app-fp50-potential-variance-bd}
    \end{subfigure}
    \hfill
    \begin{subfigure}[t]{0.495\textwidth}
        \centering
        \includegraphics[height=5.2cm]{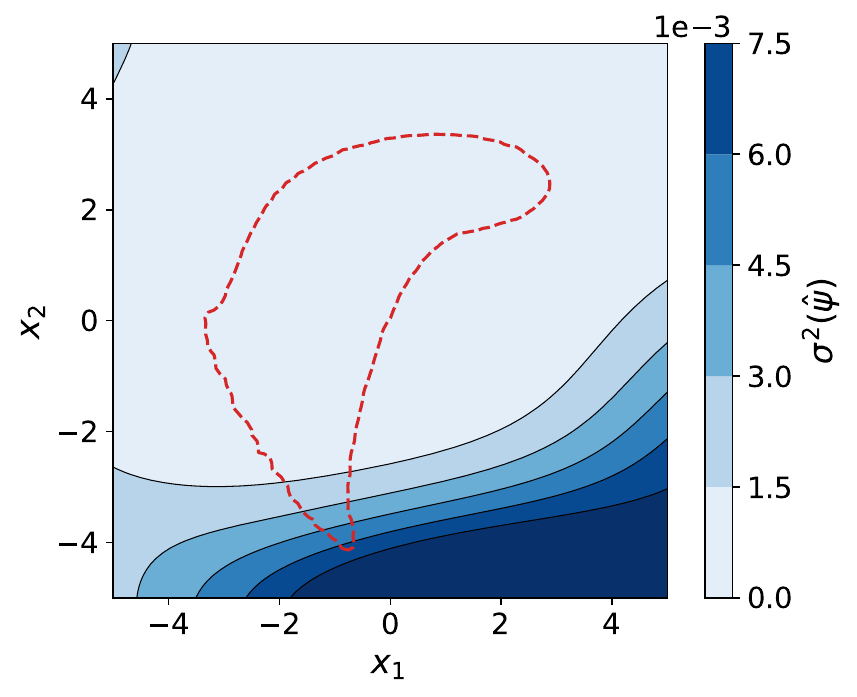}
        \caption{Predictive variance with\\block-tridiagonal likelihood covariance}
        \label{fig:app-fp50-potential-variance-bt}
    \end{subfigure}
    \caption{Mean and variance of the posterior predictive distribution $p(\psi \mid \CalD,\CalM^*)$ for the potential function in the FP equation with the Fourier dictionary of $d=51$ candidates at $\sigma=10^{-3}$, using the approximate Gaussian posterior on the selected models with (a, c) block-diagonal and (b, d) block-tridiagonal likelihood covariance structures. Panels (a, b) show the predictive means and (c, d) the predictive variances; the shift, color levels, and dashed visited-region boundary are as in \cref{fig:app-fp-nested-failure-potential}.}
    \label{fig:app-fp50-potential}
\end{figure}

With block-tridiagonal covariance, the diffusion coefficient is recovered within 6\% at both noise levels, with drift errors below 7\% (\cref{tab:app-fp50}).
At $\sigma=10^{-3}$, the block-diagonal approximation selects only six operators and underestimates $\beta^{-1}$ by a factor of about four, whereas at $\sigma=10^{-4}$ its errors are comparable to those of the block-tridiagonal structure.
Consistent with \cref{sec:block_diagonal_approx}, neglecting the adjacent temporal covariance becomes consequential at the higher noise level, and here it also degrades model selection: the selected model is smaller and represents the drift less accurately (\cref{tab:app-fp50,fig:app-fp50-potential}).
Although the potential is not recovered over the whole domain, its gradient, which determines the drift, generally matches within the visited region (\cref{fig:app-fp50-potential}): the drift errors are $3.64\%$ to $6.20\%$ with block-tridiagonal covariance and $4.34\%$ with block-diagonal covariance at $\sigma=10^{-4}$, the exception being the block-diagonal model at $\sigma=10^{-3}$ ($18.61\%$).

The selected supports, in contrast, differ substantially between the four cases, and no potential candidate is selected in all of them.
At $\sigma=10^{-4}$, the two covariance structures share only 8 of their 16 and 23 selected potential candidates, although their drift errors differ by less than one percentage point.
Nearly collinear candidates represent nearly the same drift over the visited region, so the identified quantity is the potential there rather than the operator support.
The 95\% pointwise credible intervals of the potential contain the projected potential over less than 2\% of the visited region, weighted by the density.
The posterior conditional on the selected model accounts neither for the dictionary misspecification nor for the choice among nearly equivalent supports, so these intervals understate the error in this setting.
Backward elimination required the $\CalO(d^2)$ count of candidate-model fits in \cref{sec:discussion_complexity}, taking about \SI{0.6}{h} with block-diagonal and \SI{4.5}{h} with block-tridiagonal covariance on one V100 GPU.

\section{Characterization of inference approximations: supporting analysis}
\label{app:successive_approximations}

This appendix states the established results underlying \cref{sec:successive_approximations} and their specialization to B-VSI. Each comparison holds the remaining construction fixed. Spatial and temporal discretization, the candidate model, and the observation-noise covariance shape are fixed; discretization error and model-selection uncertainty are separate issues.
\Cref{app:approximation_theory} states these results for the five approximations in turn; their numerical assessment is distributed across the manuscript.
The first-order delta method and the plug-in Jacobian (\cref{app:approximation_delta,app:approximation_plugin}) are assessed by Monte Carlo simulation in the CH example in \cref{app:approximation_numerical}.
The block-diagonal covariance (\cref{app:approximation_bd}) is assessed by comparison with the block-tridiagonal structure in \cref{sec:block_diagonal_approx}, in the examples of \cref{sec:example_FP,sec:example_CH}, and in \cref{app:repeated-noise,app:large-dictionary}.
The finite-particle SVGD approximation (\cref{app:approximation_svgd}) is reflected in the realization-to-realization variability of the hybrid posterior variances in \cref{app:repeated-noise-fp}.
The lagged covariance (\cref{app:approximation_lagged}) is characterized theoretically only; no direct numerical comparison with the full $\bstheta$-dependent likelihood is reported.

\subsection{Theoretical characterization}
\label[appendix]{app:approximation_theory}

\subsubsection{First-order delta method}
\label[appendix]{app:approximation_delta}

For the noise-propagation analysis, fix the data-generating parameter $\bstheta^\ast$ and write $\barbsu_\ast^{(k)}:=\barbsu_{\bstheta^\ast,\CalM}^{(k)}$. The superscript $\ast$ distinguishes this fixed reference from parameters varied during inference. Define the noise-induced residual perturbation by
\begin{align}
    \label{eq:nonlinear_exact_residual_perturbation}
    \delta\bsr^{(k)} = \CalR_{\bstheta^\ast,\CalM}^{(k)} \nospaceleft(\barbsu_\ast^{(k-1)}+\bsepsilon^{(k-1)},\barbsu_\ast^{(k)}+\bsepsilon^{(k)}\nospaceright) -\CalR_{\bstheta^\ast,\CalM}^{(k)} \nospaceleft(\barbsu_\ast^{(k-1)},\barbsu_\ast^{(k)}\nospaceright).
\end{align}
The delta method and second-order moment correction are established Taylor-based approximations~\cite{Oehlert1992DeltaMethod,Sarkka2013Filtering}. Let $\bsJ_\ast^{(k)}$ be the current-state Jacobian at $\barbsu_\ast^{(k)}$ and $\nabla^2\CalR_{\ast,i}^{(k)}$ the corresponding Hessian of residual component $i$. Set $\bsr_{\mathrm{lin}}^{(k)}:=\bsJ_\ast^{(k)}\bsepsilon^{(k)}-\bsB\bsepsilon^{(k-1)}$ and $\bse_{\mathrm{Tay}}^{(k)}:=\delta\bsr^{(k)}-\bsr_{\mathrm{lin}}^{(k)}$. For a twice continuously differentiable residual with Hessian bounded by $M_2$ along the state perturbation segment, the standard Taylor remainder estimate gives
\begin{align}
    \label{eq:nonlinear_taylor_bound}
    \|\bse_{\mathrm{Tay}}^{(k)}\|
    \leq \frac{M_2}{2}\|\bsepsilon^{(k)}\|^2.
\end{align}
Write $\bsSigma_{\bsu}^{(\ell)}=\sigma^2\bar{\bsSigma}_{\bsu}^{(\ell)}$ and $\bsepsilon^{(\ell)}=\sigma\boldsymbol z^{(\ell)}$, where the $\boldsymbol z^{(\ell)}$ are independent centered Gaussian vectors with fixed covariances. Local smoothness gives the stochastic order below as $\sigma\to0$. For the unconditional moment statements, assume additionally that the residual map is globally $C^4$ and its derivatives through order four have at most polynomial growth. This sufficient tail-control condition holds for polynomial residual maps and gives
\begin{align}
    \label{eq:nonlinear_taylor_scaling}
    \bse_{\mathrm{Tay}}^{(k)}=\CalO_p(\sigma^2),
    \qquad
    \mathbb{E}\|\bse_{\mathrm{Tay}}^{(k)}\|^2=\CalO(\sigma^4).
\end{align}
Under these moment assumptions, the established second-order mean correction is
\begin{align}
    \label{eq:nonlinear_mean_correction}
    \mathbb{E}[\delta r_i^{(k)}]
    =
    \frac{\sigma^2}{2}\operatorname{tr}
    \nospaceleft(\nabla^2\CalR_{\ast,i}^{(k)}\bar{\bsSigma}_{\bsu}^{(k)}\nospaceright)
    +\CalO(\sigma^4).
\end{align}
The delta-method expansion and Gaussian moment identities~\cite{Isserlis1918} likewise give
\begin{align}
    \label{eq:nonlinear_covariance_correction}
    \operatorname{Cov}(\delta\bsr^{(k)})
    ={}&\bsSigma_{\mathrm{lin}}^{(k)}+\CalO(\sigma^4).
\end{align}
Here $\bsSigma_{\mathrm{lin}}^{(k)}$ is \cref{eq:covariance_nonlinear} evaluated with $\bsJ_\ast^{(k)}$. These are absolute residual and moment orders, not likelihood-error bounds. The subtraction in \cref{eq:nonlinear_exact_residual_perturbation} excludes deterministic discretization error. State-linear operators incur no Taylor truncation.
These orders are assessed numerically in \cref{app:approximation_numerical}.

\subsubsection{Plug-in nonlinear Jacobian}
\label[appendix]{app:approximation_plugin}

The implementation evaluates the nonlinear Jacobian at the observed state rather than the unknown noise-free state, analogous to linearizing about an estimate rather than the true state in extended Kalman filtering~\cite{Sarkka2013Filtering}. At the same fixed $\bstheta^\ast$, define $\widehat{\bsJ}^{(k)}:=\bsJ_{\bstheta^\ast}^{(k)}$ using \cref{eq:overall_jacobian} and, for comparison only, $\widehat{\bsr}_{\mathrm{lin}}^{(k)}:=\widehat{\bsJ}^{(k)}\bsepsilon^{(k)}-\bsB\bsepsilon^{(k-1)}$. If the residual map is three times continuously differentiable with locally bounded third derivative,
\begin{align}
    \label{eq:plugin_jacobian_expansion}
    \widehat{\bsJ}^{(k)}-\bsJ_\ast^{(k)} ={}\CalO_p(\sigma),\qquad \widehat{\bsr}_{\mathrm{lin}}^{(k)}-\bsr_{\mathrm{lin}}^{(k)} ={} \nospaceleft(\widehat{\bsJ}^{(k)}-\bsJ_\ast^{(k)}\nospaceright)\bsepsilon^{(k)} =\CalO_p(\sigma^2).
\end{align}
Let $\widehat{\bsSigma}^{(k)}$ be \cref{eq:covariance_nonlinear} evaluated with $\widehat{\bsJ}^{(k)}$. Its quadratic dependence on the Jacobian and $\bsSigma_{\bsu}^{(k)}=\CalO(\sigma^2)$ immediately give
\begin{align}
    \label{eq:plugin_covariance_perturbation}
    \widehat{\bsSigma}^{(k)}-\bsSigma_{\mathrm{lin}}^{(k)}=\CalO_p(\sigma^3).
\end{align}
If $\sigma^{-2}\bsSigma_{\mathrm{lin}}^{(k)}$ is nonsingular, this is a relative covariance perturbation of order $\CalO_p(\sigma)$, compared with the relative $\CalO(\sigma^2)$ Taylor covariance correction in \cref{eq:nonlinear_covariance_correction}. The plug-in residual error has the same stochastic order as the Taylor remainder but a different second-order coefficient, so it remains a distinct approximation. The implementation still evaluates the nonlinear residual on the data; $\widehat{\bsr}_{\mathrm{lin}}^{(k)}$ only diagnoses the noise-propagation approximation. Because $\widehat{\bsJ}^{(k)}$ depends on the same observation noise, $\widehat{\bsSigma}^{(k)}$ is a plug-in covariance estimate, not generally the unconditional covariance of $\widehat{\bsr}_{\mathrm{lin}}^{(k)}$. The adjacent-block formula in \cref{eq:stacked_covariance_off_diagonal_block} gives the same absolute order for the fixed-length stacked covariance. No stronger expectation- or likelihood-level claim is made.
These orders are also assessed numerically in \cref{app:approximation_numerical}.

\subsubsection{Block-diagonal temporal covariance}
\label[appendix]{app:approximation_bd}

Within the state-linear Gaussian model, the block-tridiagonal covariance in \cref{eq:block_tridiagonal_residual_covariance} is the exact covariance induced by the stacked residual transformation; for nonlinear residuals, the same statement holds within the first-order local Gaussian approximation. The block-diagonal structure in \cref{eq:block_diagonal_residual_covariance} discards the adjacent blocks $\boldsymbol{E}_k$. Define
\begin{align}
    \label{eq:bd_covariance_discrepancy}
    \boldsymbol{\Delta}_{\mathrm{BD}}
    &:={}
    \bsSigma_{r,\bstheta}^{\mathrm{BT}}
    -\bsSigma_{r,\bstheta}^{\mathrm{BD}},
    &
    \eta_{\mathrm{BD}}
    &:={}
    \frac{\|\boldsymbol{\Delta}_{\mathrm{BD}}\|_{\mathrm{F}}}
         {\|\bsSigma_{r,\bstheta}^{\mathrm{BT}}\|_{\mathrm{F}}}.
\end{align}
Here $\boldsymbol{\Delta}_{\mathrm{BD}}$ has zero diagonal blocks and contains exactly the omitted temporal cross-\hspace{0pt}covariances. Its magnitude has no generic small-noise order relative to the diagonal blocks, because both scale with the observation covariance. The block-diagonal density is the product of the time-step Gaussian marginals, a marginal composite-likelihood construction~\cite{Lindsay1988Composite,Varin2011CompositeLikelihood}. To quantify the covariance omission, fix $\bstheta$ and the sensitivity matrices, assume both covariances are positive definite, and define the dimensionless symmetric perturbation
\begin{align}
    \label{eq:bd_normalized_covariance_perturbation}
    \boldsymbol{A}_{\mathrm{BD}}
    &:={}
    (\bsSigma_{r,\bstheta}^{\mathrm{BD}})^{-1/2}
    \boldsymbol{\Delta}_{\mathrm{BD}}
    (\bsSigma_{r,\bstheta}^{\mathrm{BD}})^{-1/2},
    &
    \rho_{\mathrm{BD}}&:=\|\boldsymbol{A}_{\mathrm{BD}}\|_2.
\end{align}
Its diagonal blocks vanish, so $\operatorname{tr}\boldsymbol{A}_{\mathrm{BD}}=0$. Its only nonzero blocks are $\boldsymbol{D}_k^{-1/2}\boldsymbol{E}_k\boldsymbol{D}_{k+1}^{-1/2}$ and their transposes, whose singular values are the canonical correlations between adjacent residual blocks~\cite{Hotelling1936}; hence $\rho_{\mathrm{BD}}\leq2\max_k\|\boldsymbol{D}_k^{-1/2}\boldsymbol{E}_k\boldsymbol{D}_{k+1}^{-1/2}\|_2$. The standard Gaussian Kullback--Leibler (KL) formula~\cite{Rasmussen2006Gaussian} and matrix-logarithm bounds~\cite{Higham2008Functions} therefore give the specialization
\begin{align}
    \label{eq:bd_gaussian_kl}
    D_{\mathrm{KL}}\nospaceleft(
        \CalN(\bszero,\bsSigma_{r,\bstheta}^{\mathrm{BT}})
        \,\middle\|\,
        \CalN(\bszero,\bsSigma_{r,\bstheta}^{\mathrm{BD}})
    \nospaceright)
    &=-\frac12\log\det(\bsI+\boldsymbol{A}_{\mathrm{BD}}),
    \\
    \label{eq:bd_gaussian_kl_bound}
    0\leq D_{\mathrm{KL}}
    &\leq\frac{\|\boldsymbol{A}_{\mathrm{BD}}\|_{\mathrm F}^2}
                    {4(1-\rho_{\mathrm{BD}})},
    \qquad \rho_{\mathrm{BD}}<1.
\end{align}
Thus the expected log-density discrepancy is second order in the normalized omitted covariance. Common noise scaling cancels from $\boldsymbol A_{\mathrm{BD}}$, so small observation noise alone does not justify the block-diagonal approximation. Nor does a smaller time step necessarily remove the correlations: for state-linear scalar-state residuals with time-invariant observation covariance and positive definite $\bsM\bsSigma_{\bsu}^{(k)}\bsM^\top$, each normalized adjacent block tends to $-\frac12\bsI$ as $\Delta t\to0$ at fixed mesh and $n_t>1$, by \cref{eq:B_matrix_time_derivative,eq:linear_current_sensitivity}. Then \cref{eq:bd_gaussian_kl} tends to $\frac{\Ndof}{2}[n_t\log2-\log(n_t+1)]$, which grows linearly with $n_t$.
For a fixed observed residual, let $\CalL_{\mathrm{res}}(\bstheta;\bsSigma)$ denote \cref{eq:theta_dependent_negative_log_likelihood} evaluated with covariance $\bsSigma$. The corresponding standard inverse-matrix perturbation bound~\cite{Higham2008Functions,Horn2013Matrix} yields, when $\rho_{\mathrm{BD}}<1$,
\begin{align}
    \label{eq:bd_likelihood_perturbation_bound}
    \left|\CalL_{\mathrm{res}}(\bstheta;\bsSigma_{r,\bstheta}^{\mathrm{BT}}) -\CalL_{\mathrm{res}}(\bstheta;\bsSigma_{r,\bstheta}^{\mathrm{BD}})\right| \leq \frac{\|\boldsymbol{A}_{\mathrm{BD}}\|_{\mathrm F}^2}{4(1-\rho_{\mathrm{BD}})} +\frac{\rho_{\mathrm{BD}}}{2(1-\rho_{\mathrm{BD}})} \bsr_{\bstheta}^{\top}(\bsSigma_{r,\bstheta}^{\mathrm{BD}})^{-1}\bsr_{\bstheta}.
\end{align}
Unlike the expected discrepancy, this pointwise bound contains a first-order quadratic-form contribution. These results isolate the block-diagonal treatment as a covariance approximation within the same residual-space construction. They do not imply a universal direction of change in coefficient estimates or posterior widths; those also depend on the residual's parameter dependence and the prior. In composite-likelihood theory, the corresponding uncertainty effect is characterized by the Godambe (sandwich) information rather than the inverse Hessian, and unadjusted composite posteriors can be miscalibrated~\cite{Varin2011CompositeLikelihood,Pauli2011BayesianComposite,Ribatet2012BayesianComposite}; the regularity conditions of those asymptotic results are not established here.

\subsubsection{Lagged covariance}
\label[appendix]{app:approximation_lagged}

The lagged update in \cref{eq:lagged_covariance} replaces $\bsSigma_{r,\bstheta}^{\CalC}$ during one parameter update by $\bsSigma_{r,\bstheta^{(l)}}^{\CalC}$. Local Lipschitz continuity gives
\begin{align}
    \label{eq:lagged_covariance_local_error}
    \left\|
      \bsSigma_{r,\bstheta}^{\CalC}
      -\bsSigma_{r,\bstheta^{(l)}}^{\CalC}
    \right\|_2
    \leq{}&L_{\bsSigma}\|\bstheta-\bstheta^{(l)}\|.
\end{align}
A local likelihood bound follows from standard inverse and matrix-logarithm perturbation identities~\cite{Higham2008Functions} and Weyl's eigenvalue inequality~\cite{Horn2013Matrix}. Let $\bsSigma_l^{\CalC}=\bsSigma_{r,\bstheta^{(l)}}^{\CalC}$ as in \cref{eq:lagged_covariance}, $\boldsymbol{\Delta}_l=\bsSigma_{r,\bstheta}^{\CalC}-\bsSigma_l^{\CalC}$, $\lambda_{\min}(\bsSigma_l^{\CalC})\geq m>0$, and $\|\boldsymbol{\Delta}_l\|_2\leq\delta<m$. Holding the residual $\bsr_{\bstheta}$ fixed isolates only the covariance-lagging effect and gives
\begin{align}
    \label{eq:lagged_likelihood_perturbation_bound}
    \left|
      \CalL_{\mathrm{res}}(\bstheta;\bsSigma_l^{\CalC}+\boldsymbol{\Delta}_l)
      -\CalL_{\mathrm{res}}(\bstheta;\bsSigma_l^{\CalC})
    \right|
    \leq
    \frac{1}{2}
    \left[
      -n_t n_{\mathrm{eq}}\log\nospaceleft(1-\frac{\delta}{m}\nospaceright)
      +\frac{\delta\|\bsr_{\bstheta}\|^2}{m(m-\delta)}
    \right].
\end{align}
This specialization controls the log-determinant and quadratic terms locally under positive definiteness; it is not a global likelihood or posterior bound.
Evaluated at $\bstheta=\bstheta^{(l+1)}$, this bound vanishes as successive parameter increments vanish, provided the Lipschitz and lower eigenvalue bounds remain uniform and the residuals remain bounded. It need not vanish at an arbitrary fixed $\bstheta$, and it does not make the converged lagged estimate a minimizer of the full objective. With $\bsW_{\bstheta}^{\CalC}=(\bsSigma_{r,\bstheta}^{\CalC})^{-1}$, an unconstrained fixed point $\widehat{\bstheta}$ of \cref{eq:lagged_covariance_mle_update} satisfies $\bsX^{\top}\bsW_{\widehat{\bstheta}}^{\CalC}\bsr_{\widehat{\bstheta}}=\bszero$. In contrast, an interior stationary point of the full objective must satisfy
\begin{align}
    \label{eq:full_objective_stationarity}
    \nabla_{\bstheta}\CalL_{\mathrm{res}}(\bstheta)=\bszero,
\end{align}
with the gradient given in \cref{eq:full_likelihood_gradient_component}. The two conditions coincide only when its covariance-derivative terms cancel at the solution, for example when the covariance does not depend on $\bstheta$. Otherwise, the converged lagged-covariance estimate is a distinct fixed-point estimator rather than the full residual-space MLE, as in the established distinction between iterated generalized least squares and maximum likelihood when the covariance depends on the regression parameters~\cite{JobsonFuller1980,Carroll1988Transformation}, and the difference is not controlled by \cref{eq:lagged_likelihood_perturbation_bound}. Likewise, a conjugate posterior computed with the final frozen covariance need not equal the posterior retaining its full parameter dependence.
\subsubsection{Finite-particle SVGD}
\label[appendix]{app:approximation_svgd}

SVGD~\cite{Liu2016SVGD} does not alter the physical residual or its likelihood. It approximates the selected target posterior $p_{\mathrm{res}}(\bstheta\mid\CalD,\CalM)$ by the finite particle collection $\{\bstheta_i^{(l)}\}_{i=1}^{n_p}$ in \cref{eq:SVGD}. To distinguish particle approximation from incomplete iteration, let $\widehat q_l$ be its empirical measure and $q_l$ the population SVGD iterate, obtained by replacing particle averages by expectations, with the same prescribed kernel, bandwidth, step sizes, and initialization law. For a bounded test function $f$, write $q(f)=\int f(\bstheta)\,q(\mathrm d\bstheta)$. The error separates as
\begin{align}
    \label{eq:svgd_error_decomposition}
    \left|\widehat q_l(f)-p_{\mathrm{res}}(f\mid\CalD,\CalM)\right| \leq{}\underbrace{\left|\widehat q_l(f)-q_l(f)\right|}_{\text{finite-particle error}} +\underbrace{\left|q_l(f)-p_{\mathrm{res}}(f\mid\CalD,\CalM)\right|}_{\text{population-iteration error}}.
\end{align}
The population iterate is a time discretization of the mean-field SVGD dynamics~\cite{Liu2017SVGDGradientFlow,Lu2019ScalingLimitSVGD}, and this separation underlies existing population and finite-particle analyses~\cite{Korba2020SVGD,Shi2023FiniteParticle}. Quantitative convergence requires assumptions on the target, kernel, initialization, and step sizes; particle-dependent bandwidth adaptation requires additional analysis. Finite interacting particles are not independent posterior draws, so an independent-sampling Monte Carlo rate cannot be assumed. We therefore assert the decomposition, but no unconditional convergence rate or guarantee that a stationary particle configuration represents the target posterior; finite-particle fixed-point theory guarantees moment matching for a kernel-dependent test-function class~\cite{Liu2018SVGDMomentMatching}.

\subsection{Numerical assessment for the Cahn--Hilliard equation}
\label[appendix]{app:approximation_numerical}

We assess the delta-method and plug-in approximations of \cref{app:approximation_delta,app:approximation_plugin} by Monte Carlo simulation in the CH example of \cref{sec:example_CH}.
The noise-free trajectory of \cref{sec:CH_data_generation} is computed once with the ground-truth coefficients $\bstheta^\ast$.
At each of seven noise levels $\sigma\in\{10^{-5},10^{-4.5},\ldots,10^{-2}\}$, which include the noise level $\sigma=10^{-3}$ of \cref{sec:example_CH}, we draw 2750 realizations of the observation noise in \cref{eq:CH_observation_noise}, in 11 independent batches of 250.
For each draw and residual time step, we evaluate three residual perturbations: the exact perturbation $\delta\bsr^{(k)}$ in \cref{eq:nonlinear_exact_residual_perturbation}; its linearization $\bsr_{\mathrm{lin}}^{(k)}$ at the noise-free state, whose distribution is exactly the zero-mean Gaussian approximation with covariance $\bsSigma_{\mathrm{lin}}^{(k)}$; and the plug-in linearization $\widehat{\bsr}_{\mathrm{lin}}^{(k)}$ of \cref{app:approximation_plugin}.
The three perturbations share the same noise draws, so their paired differences cancel the Monte Carlo error that they have in common.

We report four groups of diagnostics.
First, the relative remainders are the sample means of $\|\delta\bsr^{(k)}-\bsr_{\mathrm{lin}}^{(k)}\|$ and $\|\delta\bsr^{(k)}-\widehat{\bsr}_{\mathrm{lin}}^{(k)}\|$, each divided by the sample mean of $\|\bsr_{\mathrm{lin}}^{(k)}\|$; the Euclidean norms are taken over all degrees of freedom, and the means over all draws and time steps.
Second, the mean and covariance are compared on a fixed patch of 78 degrees of freedom, comprising the nodal values of both fields within unit distance of three points in the domain, at the residual time steps $k\in\{1,2,6,7,10\}$.
The paired mean bias is the maximum over the patch of the difference between the sample means of $\delta\bsr^{(k)}$ (or $\widehat{\bsr}_{\mathrm{lin}}^{(k)}$) and $\bsr_{\mathrm{lin}}^{(k)}$, divided by the standard deviations predicted by $\bsSigma_{\mathrm{lin}}^{(k)}$.
The paired covariance difference is the Frobenius norm of the difference between the corresponding sample covariances, divided by $\|\bsSigma_{\mathrm{lin}}^{(k)}\|_{\mathrm F}$; both are averaged over the five time steps.
Third, the plug-in covariance error $\|\widehat{\bsSigma}^{(k)}-\bsSigma_{\mathrm{lin}}^{(k)}\|_{\mathrm F}/\|\bsSigma_{\mathrm{lin}}^{(k)}\|_{\mathrm F}$ compares the plug-in prediction of \cref{eq:plugin_covariance_perturbation} with the noise-free prediction on the same patch, averaged over the five time steps and 55 draws (five per batch).
Fourth, the marginals are assessed through ten scalar projections of $\delta\bsr^{(k)}$, standardized by their predicted standard deviations: six single nodal values, one for each field near each of the three points, and four random unit directions over all degrees of freedom.

\begin{table}[htbp]
\centering
\caption{Relative nonlinear remainders and plug-in covariance error of the first-order residual approximations in the CH example, as defined in \cref{app:approximation_numerical}. The remainders pool 2750 noise draws per noise level.}
\label{tab:app-ch-residual}
\fontsize{9pt}{10pt}\selectfont
\sisetup{round-mode=places, round-precision=2}
\begin{tabular}{@{}c *{3}{S[table-format=1.2e-1]}@{}}
\toprule
$\sigma$
& {\makecell{Remainder of\\$\bsr_{\mathrm{lin}}^{(k)}$}}
& {\makecell{Remainder of\\$\widehat{\bsr}_{\mathrm{lin}}^{(k)}$}}
& {\makecell{Plug-in\\covariance error}} \\
\midrule
$10^{-5}$   & 1.233e-6 & 1.233e-6 & 1.757e-6 \\
$10^{-4.5}$ & 3.899e-6 & 3.899e-6 & 5.555e-6 \\
$10^{-4}$   & 1.233e-5 & 1.233e-5 & 1.757e-5 \\
$10^{-3.5}$ & 3.899e-5 & 3.899e-5 & 5.555e-5 \\
$10^{-3}$   & 1.233e-4 & 1.233e-4 & 1.757e-4 \\
$10^{-2.5}$ & 3.901e-4 & 3.906e-4 & 5.560e-4 \\
$10^{-2}$   & 1.238e-3 & 1.254e-3 & 1.771e-3 \\
\bottomrule
\end{tabular}
\end{table}

\begin{figure}[htbp]
    \centering
    \captionsetup[subfigure]{justification=centering}
    \begin{subfigure}[t]{0.325\textwidth}
        \centering
        \includegraphics[width=\textwidth]{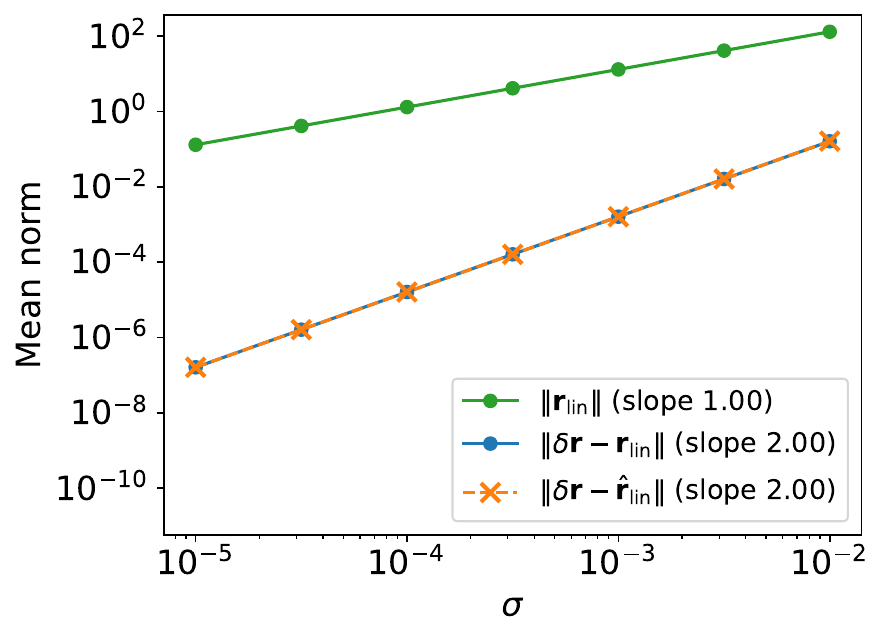}
        \caption{Remainder norms}
        \label{fig:app-ch-residual-remainder}
    \end{subfigure}
    \hfill
    \begin{subfigure}[t]{0.325\textwidth}
        \centering
        \includegraphics[width=\textwidth]{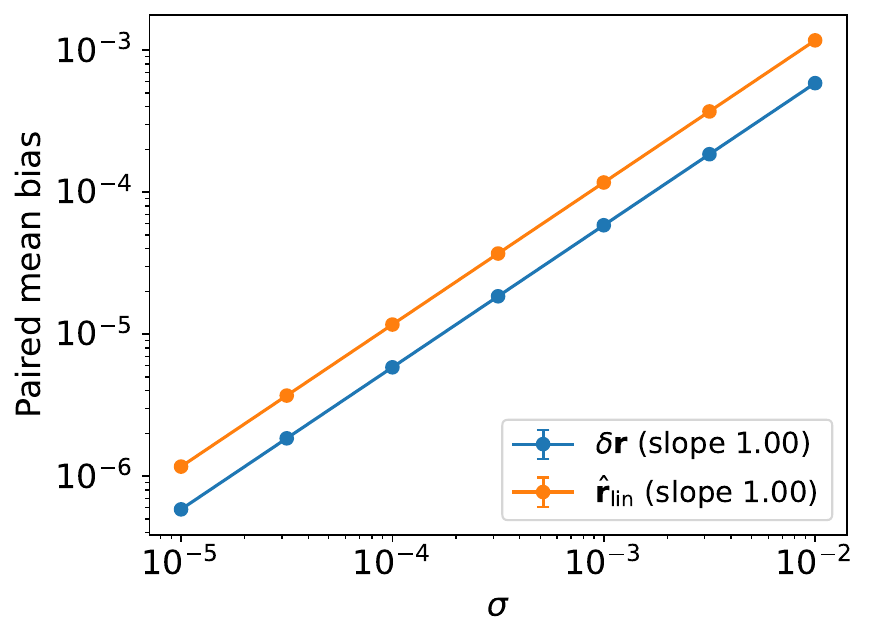}
        \caption{Paired mean bias}
        \label{fig:app-ch-residual-mean}
    \end{subfigure}
    \hfill
    \begin{subfigure}[t]{0.325\textwidth}
        \centering
        \includegraphics[width=\textwidth]{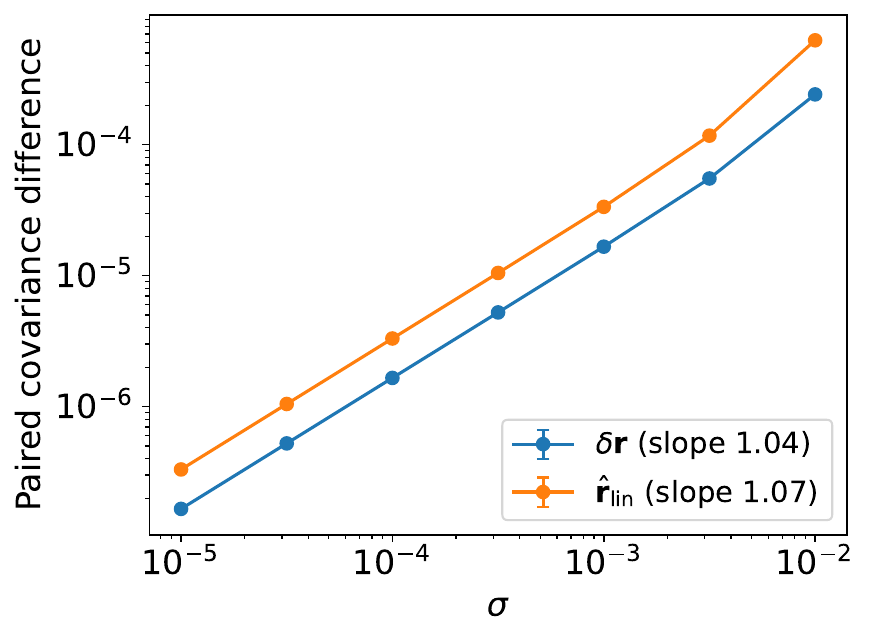}
        \caption{Paired covariance difference}
        \label{fig:app-ch-residual-covariance}
    \end{subfigure}
    \caption{Accuracy of the first-order residual approximations for the two-field CH equation as functions of the noise level $\sigma$, with fitted log--log slopes in the legends. Panel (a) shows the mean norms of the retained linear term and of the remainders of the noise-free and plug-in linearizations. Panels (b) and (c) show the paired mean bias and paired covariance difference of the exact perturbation $\delta\bsr^{(k)}$ and of the plug-in linearization $\widehat{\bsr}_{\mathrm{lin}}^{(k)}$ relative to $\bsr_{\mathrm{lin}}^{(k)}$; error bars show two standard errors across the 11 batches.}
    \label{fig:app-ch-residual}
\end{figure}

\begin{figure}[htbp]
    \centering
    \captionsetup[subfigure]{justification=centering}
    \begin{subfigure}[t]{0.32\textwidth}
        \centering
        \includegraphics[width=0.7\textwidth]{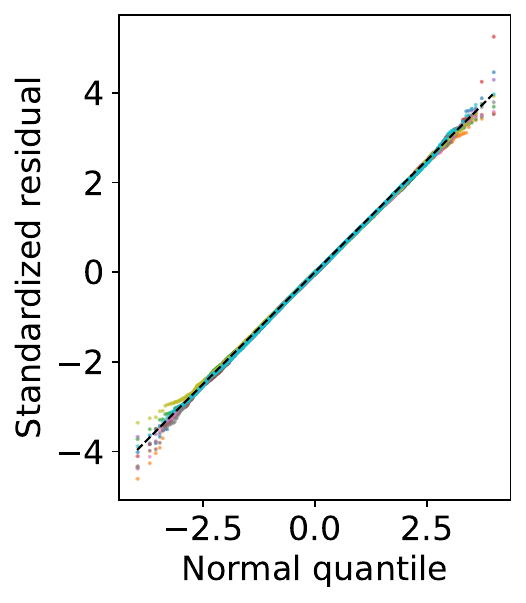}
        \caption{$\sigma=10^{-5}$}
        \label{fig:app-ch-residual-qq-5}
    \end{subfigure}
    \hfill
    \begin{subfigure}[t]{0.32\textwidth}
        \centering
        \includegraphics[width=0.7\textwidth]{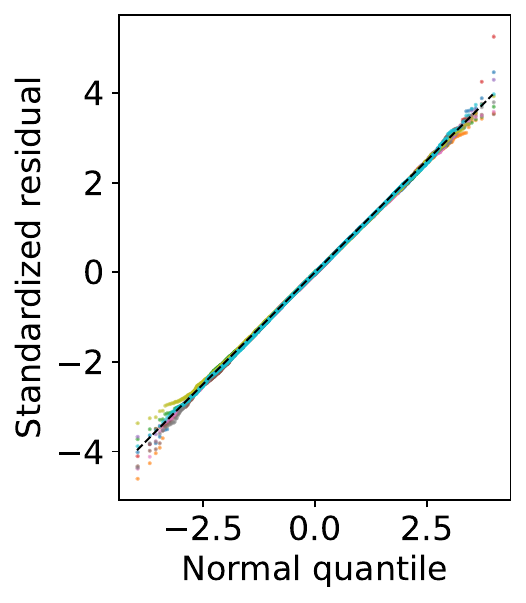}
        \caption{$\sigma=10^{-3}$}
        \label{fig:app-ch-residual-qq-3}
    \end{subfigure}
    \hfill
    \begin{subfigure}[t]{0.32\textwidth}
        \centering
        \includegraphics[width=0.7\textwidth]{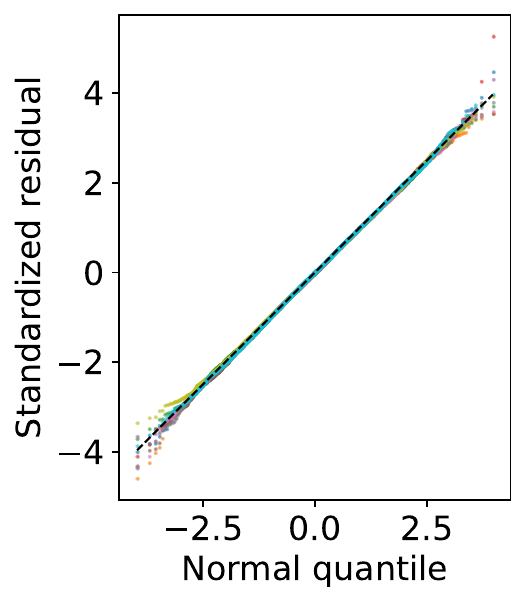}
        \caption{$\sigma=10^{-2}$}
        \label{fig:app-ch-residual-qq-2}
    \end{subfigure}
    \caption{Quantile--quantile comparison of the ten standardized projections of the exact perturbation $\delta\bsr^{(k)}$ with the standard Gaussian distribution for the two-field CH equation, pooled over the 2750 draws and the five residual time steps. Each color shows one projection, and the dashed line marks the standard Gaussian quantiles.}
    \label{fig:app-ch-residual-qq}
\end{figure}

The measured scalings agree with the theoretical orders.
The remainders of both linearizations scale as $\sigma^2$, with fitted log--log slopes of 2.00, whereas the retained linear term scales as $\sigma$, consistent with \cref{eq:nonlinear_taylor_scaling} (\cref{fig:app-ch-residual-remainder}).
Relative to the linear term, the remainders therefore grow linearly in $\sigma$ (\cref{tab:app-ch-residual}).

The paired mean bias of $\delta\bsr^{(k)}$ grows linearly in $\sigma$ in units of the predicted standard deviation, consistent with the $\CalO(\sigma^2)$ mean correction in \cref{eq:nonlinear_mean_correction} relative to the $\CalO(\sigma)$ standard deviation (\cref{fig:app-ch-residual-mean}).
At $\sigma=10^{-3}$, it is $5.9\times10^{-5}$ standard deviations averaged over the five time steps and at most $8.7\times10^{-5}$ at any single time step.
The plug-in linearization has twice this bias, as expected from \cref{eq:plugin_jacobian_expansion}: to leading order, its difference from $\bsr_{\mathrm{lin}}^{(k)}$ is the full second-order term of the residual, whereas the Taylor remainder is half of that term.

The paired covariance difference between $\delta\bsr^{(k)}$ and $\bsr_{\mathrm{lin}}^{(k)}$ is $1.7\times10^{-5}$ at $\sigma=10^{-3}$ and at most $1.9\times10^{-5}$ at any single time step (\cref{fig:app-ch-residual-covariance}).
It grows almost linearly in $\sigma$ rather than as $\sigma^2$, which indicates that it is dominated by finite-sample correlations between the linear and quadratic terms; it is therefore an upper bound on the systematic correction in \cref{eq:nonlinear_covariance_correction}.
The plug-in covariance error is larger, $1.8\times10^{-4}$ at $\sigma=10^{-3}$, and grows linearly in $\sigma$, consistent with the relative $\CalO_p(\sigma)$ order of \cref{eq:plugin_covariance_perturbation} (\cref{tab:app-ch-residual}).
By contrast, the unpaired differences between the sample statistics and the predicted moments are the same for all three perturbations, including the exactly Gaussian $\bsr_{\mathrm{lin}}^{(k)}$, at every noise level: the relative covariance difference is about $0.11$, and the standardized mean bias, about $0.048$, lies below the Monte Carlo level $0.056$ expected for a zero-mean residual. These unpaired differences are therefore Monte Carlo sampling error.

The marginal distributions of $\delta\bsr^{(k)}$ are likewise indistinguishable from those of the Gaussian reference $\bsr_{\mathrm{lin}}^{(k)}$ at every noise level (\cref{fig:app-ch-residual-qq}).
Over all projections and time steps, the largest absolute skewness and excess kurtosis are $0.11$ and $0.26$, and the largest central- and tail-quantile discrepancies are $0.071$ and $0.40$, for both perturbations, without systematic growth in $\sigma$.

Thus, at the noise level of \cref{sec:example_CH}, neglecting the higher-order terms shifts the residual mean by less than $10^{-4}$ standard deviations, changes its covariance by less than $2\times10^{-5}$ in relative norm, and leaves the marginals indistinguishable from the Gaussian reference, while the plug-in Jacobian perturbs the predicted covariance by $1.8\times10^{-4}$.
These conclusions are specific to the tested CH configuration and noise range and do not imply that the nonlinear residual is exactly Gaussian.
Because they are evaluated at the true coefficients, they characterize the residual distribution rather than the coefficient posterior, and a small per-component bias could still accumulate over the many residual components entering the estimate.

\section*{Acknowledgment}
We acknowledge support from the W. M. Keck Foundation Medical Research Grant Program. SS and KG were also supported by a DARPA grant HR0011-25-3-0160 under the COMPASS program. XH was also supported by the Department of Navy award N000142512411 issued by the Office of Naval Research.

During the preparation of this work, the authors used ChatGPT to assist with language editing, organization, clarity, and consistency checks of mathematical notation and presentation. The authors reviewed and revised all AI-assisted text (constituting less than 5\% of the manuscript) and take full responsibility for the content of the manuscript.

\bibliography{references}
\bibliographystyle{elsarticle-num-names}

\end{document}